%% file: main.tex
\documentclass[11pt, a4paper]{article}
\usepackage[utf8]{inputenc}
\usepackage[T1]{fontenc}
\usepackage{fullpage}
\usepackage{hyperref}
\usepackage{subfigure}
\usepackage{amsthm}
\usepackage{amssymb}
\usepackage{amsmath}
\usepackage{eqnarray}
\usepackage{enumerate}
\usepackage{graphicx} 
\usepackage[authoryear,round]{natbib}
\usepackage{booktabs}
\usepackage{textcomp}
\usepackage{multirow}
\usepackage{lmodern}
\usepackage{xcolor}
\usepackage{authblk}
\usepackage{float}

\theoremstyle{plain}
\newtheorem{theorem}{Theorem}[section]

\newtheorem{proposition}[theorem]{Proposition}

\theoremstyle{definition}
\newtheorem{definition}{Definition}

\theoremstyle{remark}
\newtheorem{remark}{Remark}
\numberwithin{equation}{section}
\providecommand{\keywords}[1]
{
  \small	
  \textbf{Keywords:} #1
}
\begin{document}


\title{The implied volatility surface (also) is path-dependent}

\author[1,2]{Hervé Andrès}
\author[1]{Alexandre Boumezoued}
\author[2]{Benjamin Jourdain}
\affil[1]{Milliman R\&D, Paris, France}
\affil[2]{CERMICS, École des Ponts, INRIA, Marne-la-Vallée, France.}
\maketitle

\input{abstract}

\input{introduction}

\input{empirical_study}

\input{ssvi}

\input{path_dependent_ssvi}

\input{conclusion}

\input{acknowledgements}

\input{disclosure_and_funding}

\bibliographystyle{abbrvnat}
\bibliography{bibli}

\appendix
\makeatletter  
\renewcommand{\@seccntformat}[1]{Appendix \csname the#1\endcsname .\quad}
\makeatother
\renewcommand{\thesection}{\Alph{section}} 
\makeatother

\input{appendix}

\end{document}

%% file: abstract.tex
\begin{abstract}
    We propose a new model for the forecasting of both the implied volatility surfaces and the underlying asset price.
    In the spirit of \cite{guyon2022volatility} who are interested in the dependence of volatility indices (e.g. the VIX) on the paths of the associated equity indices (e.g. the S\&P 500), we first study how vanilla options implied volatility can be predicted using the past trajectory of the underlying asset price. Our empirical study reveals that a large part of the movements of the at-the-money-forward implied volatility for up to two years time-to-maturities can be explained using the past returns and their squares. Moreover, we show that this feedback effect gets weaker when the time-to-maturity increases. Building on this new stylized fact, we fit to historical data a parsimonious version of the SSVI parameterization (\citeauthor{gatheral2014arbitrage}, \citeyear{gatheral2014arbitrage}) of the implied volatility surface relying on only four parameters and show that the two parameters ruling the at-the-money-forward implied volatility as a function of the time-to-maturity exhibit a path-dependent behavior with respect to the underlying asset price. Finally, we propose a model for the joint dynamics of the implied volatility surface and the underlying asset price. The latter is modelled using a variant of the path-dependent volatility model of Guyon and Lekeufack and the former is obtained by adding a feedback effect of the underlying asset price onto the two parameters ruling the at-the-money-forward implied volatility in the parsimonious SSVI parameterization and by specifying Ornstein-Uhlenbeck processes for the residuals of these two parameters and Jacobi processes for the two other parameters. Thanks to this model, we are able to simulate highly realistic paths of implied volatility surfaces that are free from static arbitrage.
\end{abstract}

\keywords{Implied volatility modelling; SSVI; Path-dependent volatility; Simulation; Arbitrage}

%% file: introduction.tex
\section{Introduction}
One of the many reasons of the success of the Black-Scholes model (\citeauthor{black1973pricing}, \citeyear{black1973pricing}) is the existence of a one-to-one correspondence between the price $C(K,T)$ of an European call option with strike $K$ and time-to-maturity $T$ and the volatility $\sigma$ of the geometric Brownian motion modelling the dynamics of the underlying asset price $(S_t)_{t\ge0}$ provided that $(S_0-Ke^{-f_T T})^+ < C(K,T) < S_0$ ($f_T$ is the continuously compounded forward rate for the time-to-maturity $T$) which is guaranteed by absence of arbitrage opportunities. When this condition is satisfied, the unique parameter $\sigma$ satisfying $C_{BS}(K,T,\sigma)=C(K,T)$, where $C_{BS}$ denotes the Black-Scholes call option price, is called the implied volatility of the call option. By the put-call parity, the implied volatility of the put option is equal to the one of the call option with same time-to-maturity and strike. Although the implied volatility does not add any new information with respect to the option price, it is commonly used to quote option prices on the markets mainly because it allows to easily compare the value of two options with different underlying assets while the option price heavily depends on the underlying asset price level, making the comparison more difficult. If the Black-Scholes model was an accurate description of financial markets, the implied volatility should be the same for all options on a given asset regardless of the time-to-maturity and the strike. The computation of the implied volatility from market option prices shows that the implied volatility actually depends on the time-to-maturity and the strike which invalidates the Black-Scholes model. The so-called implied volatility surface (IVS) $(K,T)\mapsto \sigma_{BS}(K,T)$ permits to fully describe the option prices on a given asset. \\

It is also well-known that the level and the shape of the IVS varies with time. To be able to jointly model the time evolution of the IVS and the underlying asset price is key for applications covering asset allocation, risk management and hedging. First, such a model allows to backtest or study the P\&L distribution of an investment stragegy involving options and the underlying asset. One can think for example of the strategy consisting in buying a stock and a put of strike $K_1$ and selling a put of strike $K_2$ with $K_2<K_1$ but with same maturity (this is called a put spread). This strategy protects the investor against a drop in the underlying asset price down to the $K_2$ threshold in exchange to a lower premium in comparison to just buying a put of strike $K_1$. By extension, the modelling of the IVS and the underlying asset price makes it possible to optimize an asset allocation strategy involving options. Another application relates to the design and the backtesting of hedging strategies for financial products (e.g. volatility swaps, options on the VIX, etc.) having a volatility risk which is measured by the Black-Scholes vega. To complete this non-exhaustive list, let us finally mention that an IVS-underlying model can also be useful in the insurance industry for:
\begin{enumerate}
    \item computing the equity volatility distribution over a one-year horizon to estimate the capital requirement within Solvency II internal models and
    \item  assessing the time value of options and guarantees within insurance contracts and analyzing the underlying hedging strategies of long-term life insurance contracts embedding path-dependent options. 
\end{enumerate}
\subsection{Literature review}
Inspired by the market models of \cite{heath1992bond} and \cite{brace1997market} for the interest rates term structure, \cite{ledoit1998relative} and \cite{schonbucher1999marketmodel} independently proposed a modelling framework for the joint dynamics of the IVS and the underlying asset price where both are solutions of stochastic differential equations (SDEs) with the drift and volatility coefficients functions of the time, the time-to-maturity and the strike or moneyness variables only. In particular, no-arbitrage conditions on the drift are derived to guarantee the absence of arbitrage opportunities under the risk-neutral probability. A similar approach is adopted by \cite{brace2001market}. More empirical studies include the papers from \cite{skiadopoulos2000dynamics} and \cite{fengler2003dynamics}. The former applies a principal component analysis (PCA) to historical implied volatilities grouped in three time-to-maturity buckets and identifies two factors explaining 78\% of the smiles variation while the latter applies a common PCA and identifies three factors explaining more than 98\% of the variations. To deal with the fact that the study of the dynamics of the IVS is a three-dimensional problem (time, time-to-maturity and strike), \cite{cont2002dynamics} use a Karhunen-Loève decomposition instead of a PCA. They show that the dynamics of IVSs can be well summarized by three orthogonal factors which can be interpreted as the level, the orientation (i.e. a positive shock of this factor increases the volatilities of out-of-the-money calls while decreasing those of out-of-the-money puts) and the convexity of the surface. The associated principal components exhibit persistence (i.e. autocorrelation) and mean reversion close to the one of an $AR(1)$ process. Therefore, Cont and da Fonseca suggest to model each of the principal components as an Ornstein-Uhlenbeck process. \cite{cont2002stochastic} extend this model by specifying the dynamics of the underlying asset price which shares noise terms with the dynamics of the IVS allowing in particular to account for the correlation between the underlying price and the volatility surface level. A second extension, developed by \cite{cont2022simulation}, allows limiting the number of scenarios with static arbitrages by resampling from a given set of IVSs scenarios using smaller weights for scenarios with arbitrages. Another way to address this modelling problem in the literature is to resort to parametric or semi-parametric factors models, see e.g. \cite{hafner2005factor}, \cite{fengler2007semiparametric} or \cite{francois2023joint}. More recently, machine learning techniques such as GANs or neural SDEs have also been used to generate realistic simulations of implied volatility surfaces, see e.g. \cite{wiese2019deep}, \cite{cohen2021arbitrage}, \cite{zhang2023} and \cite{choudhary2024}. Finally, let us mention the paper of \cite{morel2024path} who introduced the so-called Path Shadowing Monte Carlo method which, combined with a statistical model of prices, allows to make state-of-the-art predictions of option smiles using only the distribution of the price process.  \\

In this paper, we develop a new joint model of the IVS and the underlying asset price. Instead of specifying the IVS as the solution of a given SDE or as a linear combination of several factors (whether parametric, semi-parametric or non-parametric), we propose to consider a parameterization of the IVS whose parameters evolution depends on the path of the underlying asset price. The chosen parameterization is the celebrated SSVI parameterization of \cite{gatheral2014arbitrage} that is known to well reproduce observed IVSs and guarantees the absence of static arbitrage under mild conditions. This modelling paradigm consisting in making dynamic the parameters of a model fitting market data at some point in time is similar to the one of \cite{carmon2011tangent} who developed a very general mathematical framework for designing consistent dynamic market models. In \cite{carmona2017levy}, the authors provide a practical implementation of this framework for IVSs allowing to simulate IVSs that are free of both static and dynamic arbitrage. Moreover, they use these simulations of IVSs to find the portfolio with smallest variance for a portfolio consisting of $n$ options of same maturity but different strikes. In the same vein, \cite{bloch2021deep} used an SVI model whose parameters are stochastic processes to model the dynamics of the entire IVS. A convolutional LSTM (Long Short-Term Memory) neural network is used to learn the joint dynamics of these parameters and the underlying forward price. There is one main difference between our approach and the ones of these papers and the literature in general. In our approach, we introduce an explicit modelling of the impact of the underlying asset price onto the level and the shape of the IVS in the spirit of \cite{guyon2022volatility} who focus on volatility indices and realized volatility (hence not on vanilla options implied volatilities). Indeed, in the above literature, the dependence structure between the IVS and the underlying asset price is generally captured through simple assumptions such as a Gaussian copula, common noise terms or using the short-term implied volatility as a term in the underlying asset stochastic volatility dynamics. Moreover, we model the underlying price using the path-dependent volatility framework of \cite{guyon2022volatility} which exhibits high statistical consistency and captures multiple historical stylized facts (leverage effect, volatility clustering, weak and strong Zumbach effects). Before giving more details on our approach, we find useful to dedicate a section to Guyon and Lekeufack's main results.

\subsection{Guyon and Lekeufack's path-dependent volatility model}\label{sec:pdv_model}
\cite{guyon2022volatility} showed that the level of the volatility of major equity indices is essentially explained by the past variations of these equity indices, or in other words, they showed that volatility is mostly path-dependent. To be more specific, they consider two measures of the volatility: the value of an implied volatility index such as the VIX and an estimator of the realized volatility over one day using intraday observations of the equity index. We recall that an implied volatility index is a measure of the expected future variance\footnote{Note that the VIX can also be interpreted as the implied volatility of the 30-days log-contract $-\frac{2}{\tau}\log \frac{S_{t+\tau}}{F_t^{t+\tau}}$ on the S\&P 500 index ($F_t^{t+\tau}$ is the forward price at time $t$ with maturity $t+\tau$). } of a given underlying index (for example the S\&P 500 for the VIX) at a given horizon $T$. Mathematically, the expected future variance writes $\mathbb{E}\left[\frac{1}{T}\int_0^T \sigma_t^2 dt \right]$ where $\sigma$ is the instantaneous volatility of the underlying index and $\mathbb{E}$ here denotes the expectation under the risk-neutral probability. The expected future variance can be estimated from the prices of traded calls and puts on the underlying index using the \cite{carr2001formula} formula. We refer for example to the documentation of the VIX (\citeauthor{vixwp}, \citeyear{vixwp}) or the VSTOXX (\citeauthor{vstoxxwp}, \citeyear{vstoxxwp}) for more details. Note that Guyon and Lekeufack only use short-term implied volatility indices (the horizon $T$ is below 30 days) since they are interested in the modelling of the instantaneous volatility. Let us now introduce the model that they calibrate for both measures of volatility. Let $(S_t)_{t\ge 0}$ be the price process of an equity index and $\text{Volatility}_t$ be one of the two above-mentioned measures of volatility. The Path-Dependent Volatility (PDV) model from the empirical study of \cite{guyon2022volatility} writes as follows:
\begin{equation}\label{eq:guyon_model}
    \text{Volatility}_t = \beta_0+\beta_1 R_{1,t} + \beta_2\Sigma_t.
\end{equation}
The features $R_{1,t}$ and $\Sigma_t$ are defined on a time grid $(t_i)_{i\in \mathbb{N}}$ as follows:
\begin{itemize}
    \item $R_{1,t}$ is a trend feature given by:
\begin{equation}\label{eq:r1_formula}
    R_{1,t} = \sum_{t_i \le t} K_1(t-t_i) r_{t_i}
\end{equation}
where $r_{t_i}= (S_{t_i}-S_{t_{i-1}})/S_{t_{i-1}}$ and $K_1:\mathbb{R}_+\rightarrow \mathbb{R}_+$ is a decreasing kernel weighting the past returns. Since the value of $\beta_1$ estimated by \cite{guyon2022volatility} is negative, this feature allows to capture the leverage effect, i.e. the fact that volatility tends to rise when prices fall.
\item $\Sigma_t$ is an activity or volatility feature given by:
\begin{equation}\label{eq:sigma_formula}
    \Sigma_t = \sqrt{\sum_{t_i \le t} K_2(t-t_i) r_{t_i}^2}
\end{equation}
where $K_2$ also is a decreasing kernel. Since the value of $\beta_2$ estimated by \cite{guyon2022volatility} is positive, this feature allows to capture the volatility clustering phenomenon, i.e. the fact that periods of large volatility tend to be followed by periods of large volatility, and periods of small volatility tend to be followed by periods of small volatility. 
\end{itemize}
In order to capture both the short and long memory of volatility, they propose a time-shifted power law (TSPL) for the two kernels $K_1$ and $K_2$:
\begin{equation}\label{eq:tspl_kernel}
    K_j(\tau) = \frac{Z_{\alpha_j,\delta_j}}{(\tau+\delta_j)^{\alpha_j}},\quad j=1,2,
\end{equation}
with $Z_{\alpha_j,\delta_j}$ the normalization constant such that $\sum_{t-C\le t_i \le t} K_j(t-t_i)\Delta  = 1$ where $\Delta  = 1/252$ (business days frequency) and $C$ is a hyperparameter (called the cut-off lag later in the paper) controlling at which point the sums in $R_1$ and $\Sigma$ are truncated. Note that it is not clear whether the volatility obtained using Equation (\ref{eq:guyon_model}) is positive. For time-continuous versions of the features $R_1$ and $\Sigma$, sufficient conditions guaranteeing the volatility's positivity have been given by \cite{guyon2022volatility} and \cite{nutz2024} in the case of two exponential kernels, i.e. $K_j(\tau) = \lambda_j e^{-\lambda_j\tau}$ for $j\in\{1,2\}$ and by \cite{andres2024existence} for more general choices of $K_2$. However, the framework of the latter does not cover the case when $K_1$ is a TSPL kernel or a convex combination of two exponential kernels, i.e. $K_1(\tau)=\theta \lambda_{1}e^{-\lambda_{1}\tau}+(1-\theta)\lambda_{2}e^{-\lambda_{2}\tau}$, which is a good approximation of TSPL kernels proposed by \citeauthor{guyon2022volatility}. Thus, the volatility's positivity is still an open question for these choices of kernels. Let us nevertheless mention that \citeauthor{guyon2022volatility} observe only non-negative volatilities in their simulations when using two convex combinations of two exponential kernels and "realistic parameter values".   \\

\begin{figure}
    \centering
    \subfigure[VIX vs $R_1$]{ \includegraphics[width=0.45\linewidth]{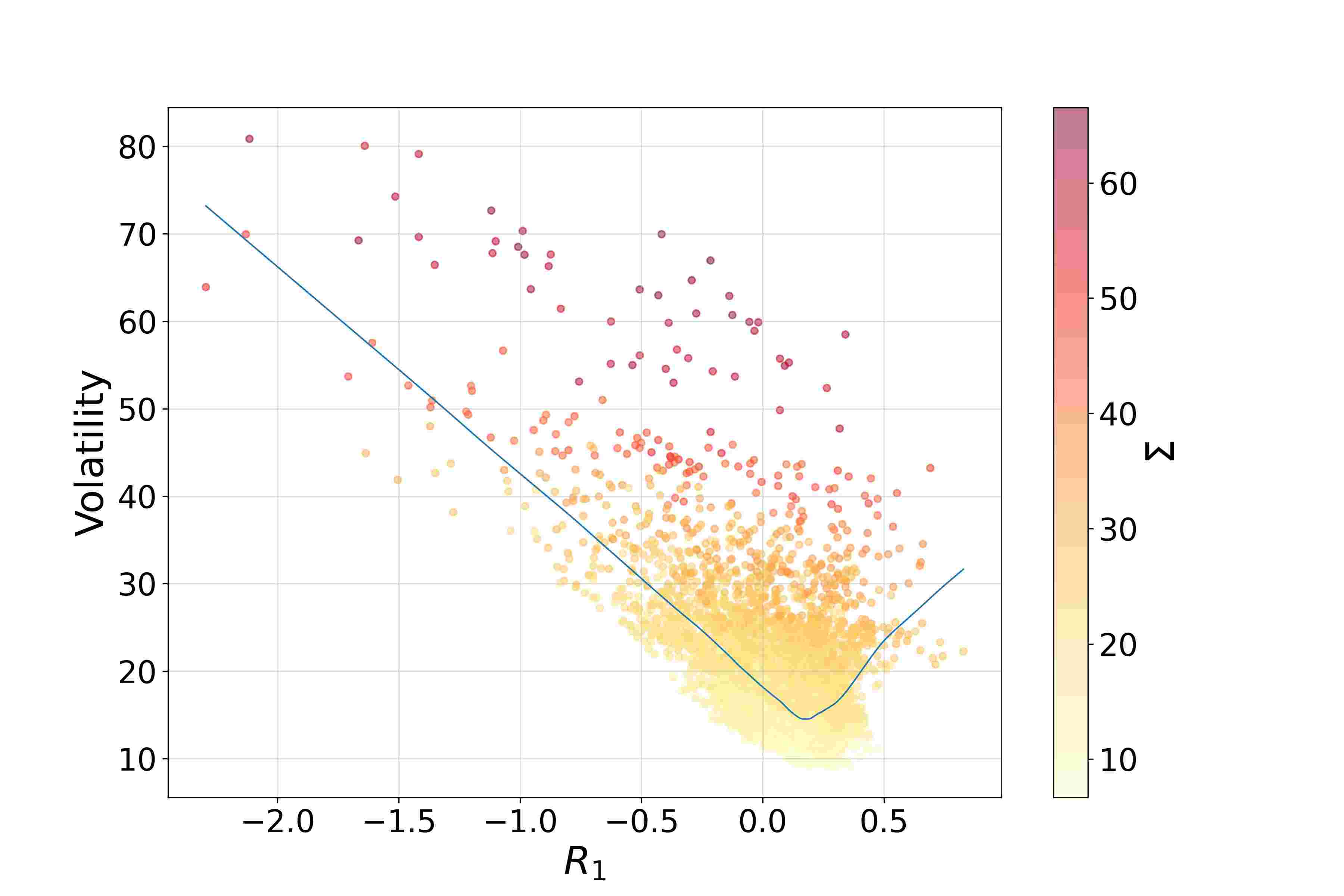}}
    \subfigure[VIX vs $\Sigma$]{ \includegraphics[width=0.45\linewidth]{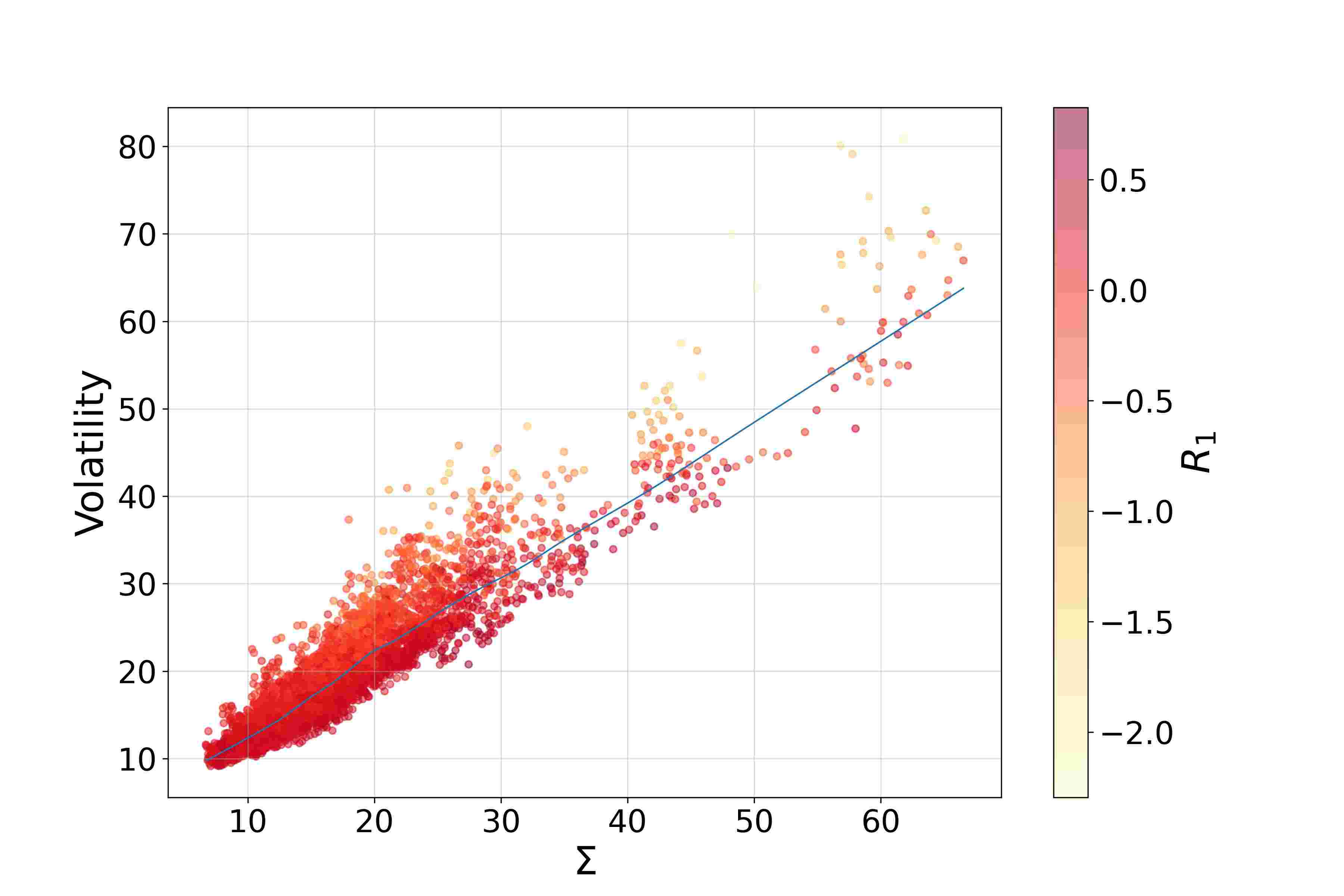}}
    \caption{Values of the VIX against values of the features $R_1$ and $\Sigma$ on the train set. The blue line represents $\mathbb{E}[Y\mid X]$ (obtained by a locally weighted scatterplot smoothing) when displaying $Y$ vs $X$.}
    \label{fig:guyon_graphs}
\end{figure}

In order to measure to which extent the two features of the PDV model allow to explain the variations of the volatility, they use the $R^2$ score:
\begin{equation}\label{eq:r2}
    R^2(y,\hat{y}) = 1-\frac{\sum_{i=1}^{n}(y_i-\hat{y}_i)^2}{\sum_{i=1}^{n}(y_i-\bar{y}_n)^2}
\end{equation}
where $y=(y_i)_{1\le i\le n}$ are the observed data, $\hat{y}= (\hat{y}_i)_{1\le i\le n}$ are the predicted data and $\bar{y}_n=\frac{1}{n}\sum_{i=1}^{n}y_i$. When they calibrate the PDV model on implied volatility indices data, they obtain $R^2$ scores over tested indices that are above 87\% on the train set (January 1, 2000 to December 31, 2018) and above 80\% on the test set (January 1, 2019 to May 15, 2022), which shows that the PDV model explains a large part of the variability observed in the volatility dynamics. In Figure \ref{fig:guyon_graphs}, we reproduce two graphs from their paper that indicate quite clearly the linear relationship between the two features and the VIX. When calibrated on realized volatility data, the performance of the PDV model is reduced: the $R^2$ score is about 70\% on the train set and 60\% on the test set. 

\subsection{Contributions}
The first contribution of the present paper is an empirical study of the dependence of implied volatility on the past movements of the underlying asset price for options on the S\&P 500 and options on the Euro Stoxx 50. This empirical study is inspired by the one of \cite{guyon2022volatility} but there are several differences. First and foremost, we work on vanilla implied volatility instead of implied volatility indices: the former is directly determined by supply and demand while the latter is determined as linear combinations of prices of calls and puts covering the liquid strikes and the two time-to-maturities that are the closest to 30 days. Both are therefore close only if we consider implied volatilities of 1-month time-to-maturity options. Since we consider time-to-maturities up to 24 months, our study can be seen as an extension of the one of \cite{guyon2022volatility}. Second, we analyze the influence of the cut-off lag of the kernel on the performance of the PDV model. Finally, we add a regularization term in the calibration of the model and study its impact. Our study also differs from the numerous studies in the literature around implied volatility forecasting. For example, \cite{bakshi2000call} focus mostly on the frequency with which call (resp. put) prices move in the same direction (resp. opposite direction) as the underlying asset price but they do not try to exhibit a functional relationship between the two and they do not use the past path of the underlying asset price. \cite{cao2020} train a neural network using as input features the S\&P 500 daily return, the time-to-maturity and the Black-Scholes delta to predict the variation in the S\&P 500 implied volatility surface (in a second step, they also use the level of the VIX). Their model provides better predictions than a benchmark model and is consistent with the negative correlation between the asset returns and the implied volatility variations that was already documented in the literature. However, this paper does not explore additional explanatory features beyond the asset price return the day before the prediction. Finally, \cite{wen2024implied} explore how daily implied volatilities up to 20 days in the past allow predicting future implied volatilities. They obtain out-of-sample $R^2$ scores up to 80\% and conclude that implied volatility is (almost) path-dependent. The study we present differs from this one in that we focus solely on the dependence of implied volatility with respect to the past path of the asset price and not to the past path of implied volatilities. Moreover, we use a time window larger than 20 days in our study and we use TSPL kernels to weight the past instead of considering one separate weight for each lagged value of the implied volatility.  \\

The second contribution is to propose a parsimonious version of the Surface Stochastic Volatility Inspired (SSVI) parameterization (\citeauthor{gatheral2014arbitrage}, \citeyear{gatheral2014arbitrage}) of the IVS which relies only on four parameters and provides a reasonable replication of the market IVSs for a wide range of dates. This parsimonious SSVI parameterization is free of static arbitrage provided that a simple inequality constraint involving two parameters is satisfied. We also show that the two parameters governing the ATM implied volatility curve as a function of the time-to-maturity can be well explained by the past path of the underlying asset price. \\

Our final contribution is to introduce a new model for the joint dynamics of the underlying asset price and the implied volatility surface allowing to perform Monte Carlo simulations under the real-world probability. This model is obtained by specifying the time evolution of the four parameters of the parsimonious SSVI parameterization for the IVS and combining it with a variant of the PDV model of \cite{guyon2022volatility} for the underlying price. The dynamics of the two parameters governing the ATM implied volatility curve contains a functional dependence on the past path of the underlying price allowing to embed in the model the feedback effect that we observe on historical data. Moreover, the residuals of these two parameters are modelled using Ornstein-Uhlenbeck processes and the two others parameters of the parsimonious SSVI parameterization are modelled using Jacobi processes. Together with the model specification, we also provide a calibration methodology for all the parameters that are involved in the dynamics. Ultimately, we show through several metrics that the IVSs simulated with our model are highly realistic. \\

This paper is organized as follows: in Section \ref{sec:empirical_study}, we start by the empirical study of the dependence of implied volatility on the past movements of the underlying asset price. Then, we present the SSVI parameterization and its parsimonious version as well as some calibration results in Section \ref{sec:svi}. Finally, Section \ref{sec:pdv_ssvi} is dedicated to the introduction of our new path-dependent SSVI model for simulating implied volatility surfaces and the underlying asset price. 



%% file: empirical_study.tex
\section{Empirical study of the joint dynamics of the implied volatility and its underlying index}\label{sec:empirical_study}

\subsection{Data sets}\label{sec:data}
We consider two data sets of daily implied volatility surfaces sourced from two different data providers. The first one corresponds to options on the S\&P 500 index while the second one corresponds to options on the Euro Stoxx 50. Both data sets start on March 8, 2012 and end on December 30, 2022. They contain the at-the-money-forward (denoted by ATM in the sequel for the sake of simplicity) implied volatilities for time-to-maturities ranging from 1 month to 24 months with a monthly timestep. For the same range of time-to-maturities, the S\&P 500 data set also contains the implied volatilities for values of forward moneyness ranging from 0.6 to 1.4 with a 0.01 step while the Euro Stoxx 50 contains the implied volatilities for Black-Scholes deltas in the following range: \textpm 0.1, \textpm 0.15, \textpm 0.2, \textpm 0.25, \textpm 0.3, \textpm 0.35, \textpm 0.4, \textpm 0.45 (positive deltas correspond to calls while negative deltas correspond to puts). As a remainder, the Black-Scholes delta corresponds to the sensitivity of the option price with respect to the price of the underlying asset. Its formula  (assuming zero dividend) is recalled below:
\begin{equation}\label{eq:delta_formula}
    \Delta^{BS} = \epsilon\mathcal{N}(\epsilon d_1) \text{ with } d_1 = \frac{\ln\frac{S_0}{K}+(f_T+\frac{\sigma^2}{2})T}{\sigma \sqrt{T}}
\end{equation}
where $\mathcal{N}$ is the cumulative normal distribution function, $\epsilon=1$ for call options and $-1$ for put options, $K$ is the strike, $T$ the time-to-maturity, {$f_T$ is the continuously compounded forward rate for the time-to-maturity $T$} and $\sigma$ the Black-Scholes implied volatility. In the sequel of this section, we only focus on the ATM implied volatilities but, in Sections \ref{sec:svi} and \ref{sec:pdv_ssvi}, the away-from-the-money implied volatilities will also be used. Let us emphasize that these two data sets do not consist of raw data of market calls and puts prices but result from a pre-processing which is at the discretion of the data providers. Indeed, options with the above time-to-maturities are not traded every day on the market. For example, on the Chicago Board Options Exchange where calls and puts on the S\&P 500 are traded, the following options are traded at day $t$:
\begin{enumerate}
    \item \textbf{Weekly expiry options:} options expiring every business day between $t$ and $t+28$ business days. Note that before 2022, there was only Monday-, Wednesday- and Friday-expiring options. 
    \item \textbf{End-of-Month options:} options expiring the last business day of the month for up to twelve months after $t$. 
    \item \textbf{Monthly expiry options:} options expiring the third Friday of the month for a given range of future months up to 5 years after $t$. 
\end{enumerate}
A similar decomposition can be found for options written on the Euro Stoxx 50 on the Eurex but with differences in the expiry dates (for example, there are only Weekly expiry options that expire on Fridays). \\

Along with these two IVSs data sets, we also have daily time series of the S\&P 500 and Euro Stoxx 50 indices. The S\&P 500 time series starts on January 2, 1980 while the Euro Stoxx 50 time series starts on December 31, 1986 and both end on December 30, 2022. Note that since we will use at most 12 years of past returns to predict the implied volatility, the whole time series are not used in the following study.  \\

\begin{figure}[t]
    \centering
    \subfigure[S\&P 500]{\includegraphics[width=\linewidth]{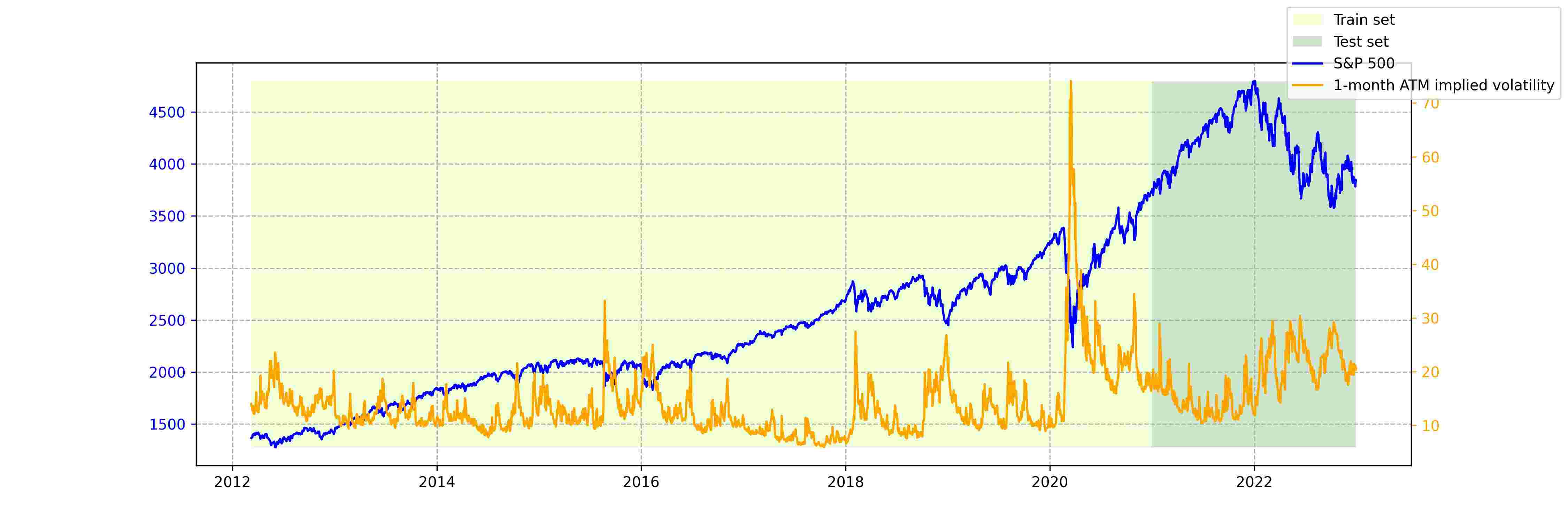}}
    \subfigure[Euro Stoxx 50]{\includegraphics[width=\linewidth]{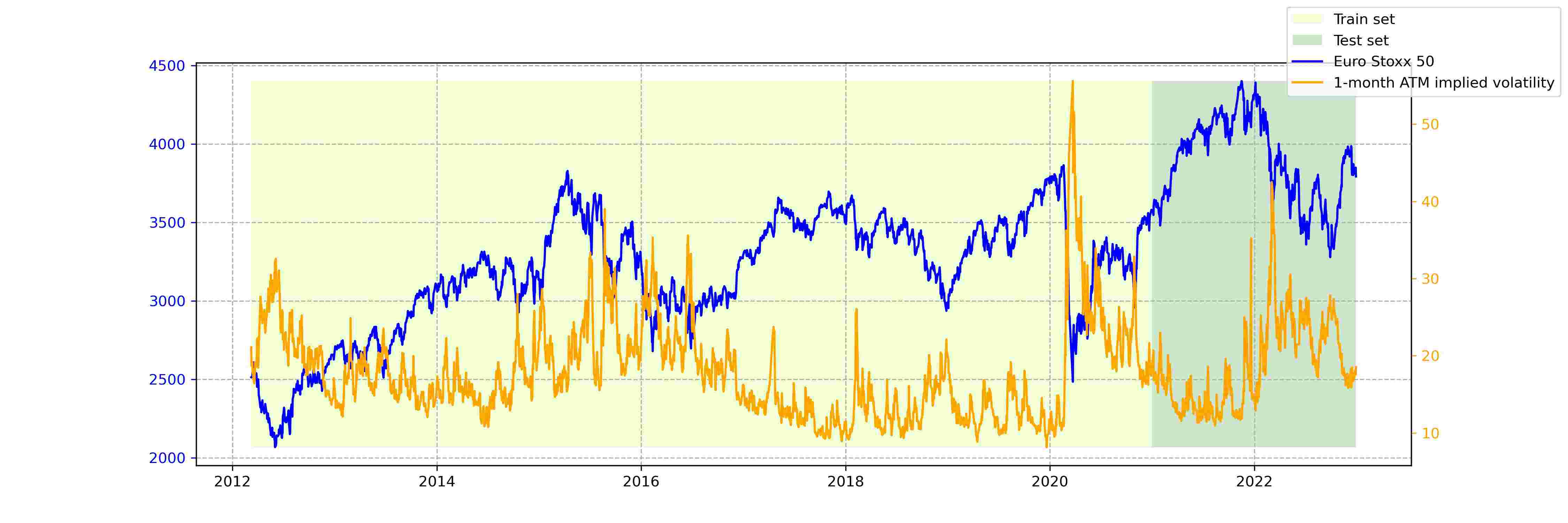}}
    \caption{Joint evolution of the 1-month ATM implied volatility and its underlying index from March 8, 2012 to December 30, 2022. The split between the train and the test sets is represented through the use of different background colors.}
\label{fig:data_sets}
\end{figure}

To measure the out-of-sample performance of the tested model, we split the two data sets into a train set and a test set: the train set spans the period from March 8, 2012 to December 31, 2020 and the test set spans the period from January 1, 2021 to December 30, 2022 so that approximately 80\% of the data is used for the train and 20\% is used for the test. In addition, we will also consider a blocked cross-validation in Section \ref{sec:influence_cutoff}. The 1-month ATM implied volatility along with the underlying asset price are represented in Figure \ref{fig:data_sets} for both data sets. 

\subsection{Calibration methodology} \label{sec:calib_meth}
The PDV model (\ref{eq:guyon_model}) with the TSPL kernel relies on 7 parameters, namely $(\alpha_1,\delta_1,\alpha_2,\delta_2)$ the parameters of the two TSPL kernels $K_1$ and $K_2$ (Equation (\ref{eq:tspl_kernel})) and $(\beta_0,\beta_1,\beta_2)$, respectively the intercept, the sensitivity to the trend feature and the sensitivity to the volatility feature. These 7 parameters are calibrated specifically for each time-to-maturity using the following steps (which are identical to the ones implemented by \cite{guyon2022volatility} to which we refer for more details\footnote{See also the code provided with their paper: \url{https://github.com/Jordylek/VolatilityIsMostlyPathDependent}}):
\begin{enumerate}
    \item We compute four exponentially weighted moving averages (EWMA) with respective spans of 10, 20, 120 and 250 days of the underlying index returns. Then, we run a ridge regression of the ATM implied volatility on the four EWMAs and we fit the TSPL kernel $K_1$ on the optimal linear combination of the exponential kernels which provides us with initial guesses for $\alpha_1$, $\delta_1$. The use of a ridge regression instead of a lasso regression is justified by the fact that we do not need to maximize the number of zeros (i.e. minimize the number of exponential kernels) in view of the subsequent fit of a TSPL kernel. By running a ridge regression of the ATM implied variance on four EWMAs of the underlying index squared returns and fitting the TSPL kernel $K_2$, we obtain similarly initial guesses for $\alpha_2$ and $\delta_2$.  
    \item Initial guesses for $\beta_0$, $\beta_1$ and $\beta_2$ are then obtained using a linear regression of the ATM implied volatility on the features $R_1$ and $\Sigma$ where $\alpha_1$, $\delta_1$, $\alpha_2$ and $\delta_2$ are fixed to the values estimated at step 1. 
    \item Starting from these initial guesses, the 7 parameters are jointly calibrated by solving the following minimization problem using the \texttt{least\_squares} function with the trust-region reflective algorithm from the \texttt{scipy} Python package:
    \begin{equation}\label{eq:target_fun}
        \begin{array}{rrrlll}
        \displaystyle\min_{(\alpha_1,\delta_1,\alpha_2,\delta_2,\beta_0,\beta_1,\beta_2) \in \mathbb{R}^7} & \multicolumn{5}{c}{\displaystyle\sum_{t\in \mathcal{T}_{train}} (IV^{mkt}_t - \beta_0-\beta_1R_{1,t}-\beta_2\Sigma_t)^2 } \\
       \text{s.t.} & \alpha_j, \delta_j &\ge& 0 \text{ for } j \in \{1,2\} \\
       &R_{1,t} &=& \displaystyle\sum_{t-C\le t_i \le t} \frac{Z_{\alpha_1,\delta_1}}{(t-t_i+\delta_1)^{\alpha_1}} r_{t_i} \\
      &\Sigma_{t} &=& \sqrt{\displaystyle\sum_{t-C\le t_i \le t} \frac{Z_{\alpha_2,\delta_2}}{(t-t_i+\delta_2)^{\alpha_2}} r_{t_i}^2}
        \end{array}
    \end{equation}
    where $\mathcal{T}_{train}$ is the set of dates in the train set, $IV^{mkt}_t$ is the market ATM implied volatility observed at time $t$ for some fixed time-to-maturity and $C$ is a cut-off lag. 
\end{enumerate}

\subsection{Numerical results}\label{sec:empirical_study_numerical_results}
\subsubsection{Performance of the PDV model}\label{sec:overall_perf}
\begin{figure}[h]
    \centering
    \subfigure[S\&P 500]{\includegraphics[width=0.45\linewidth]{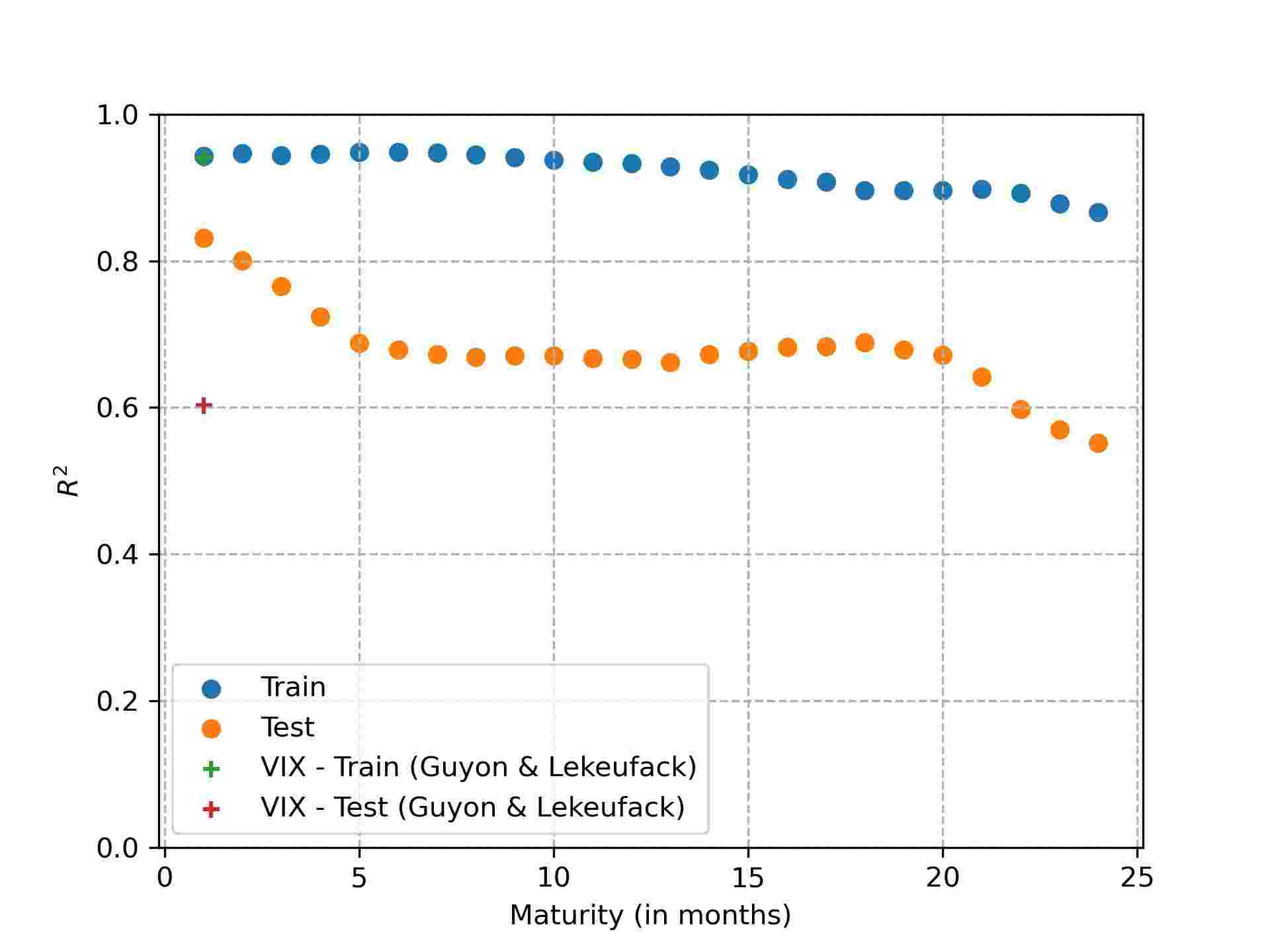}}
    \subfigure[Euro Stoxx 50]{\includegraphics[width=0.45\linewidth]{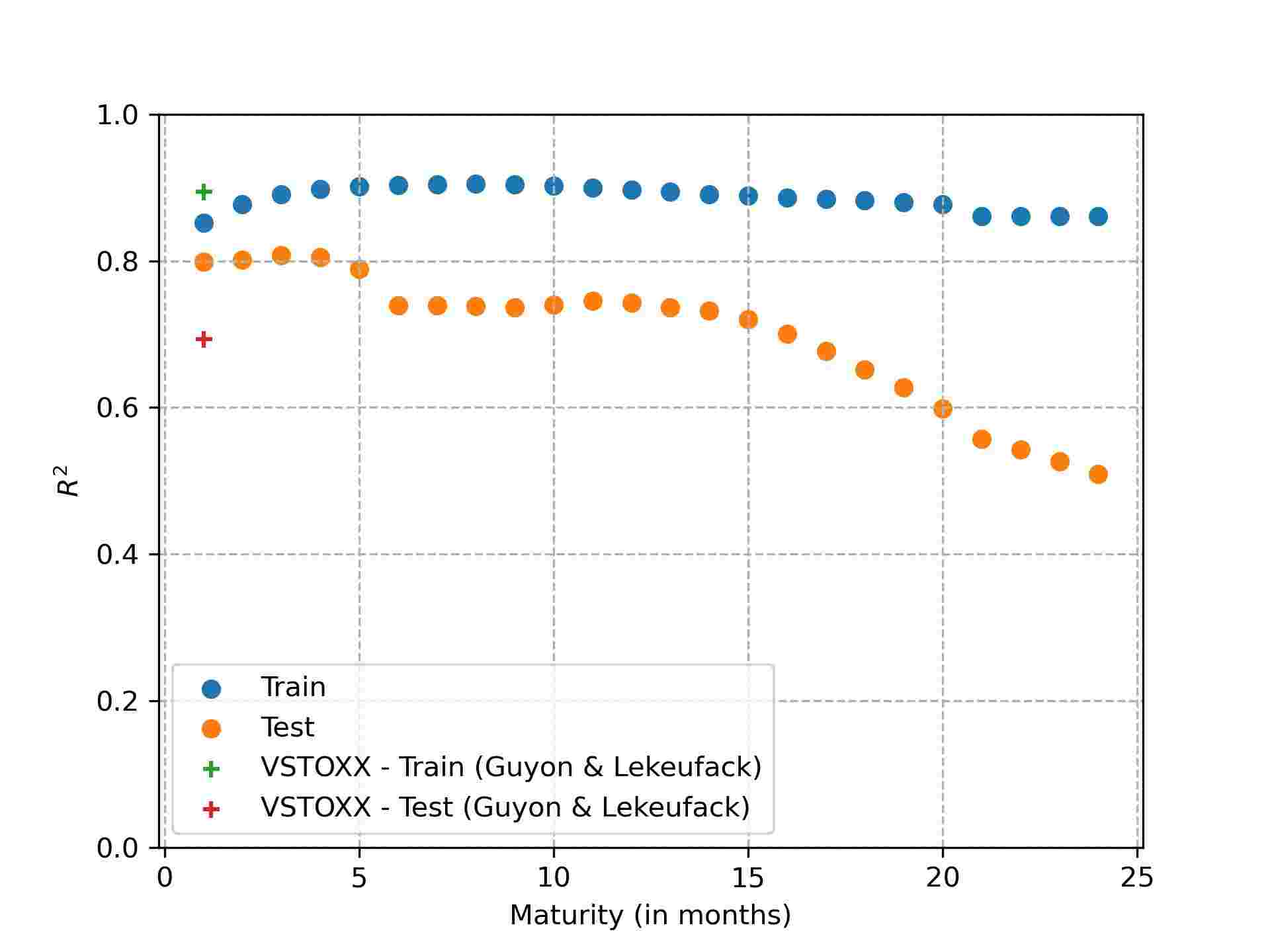}}
    \caption{$R^2$ scores on the train and the test sets as a function of the ATM implied volatility time-to-maturity. The average $R^2$ scores for the VIX and the VSTOXX are also displayed.}
\label{fig:r2_cutoff_1000}
\end{figure}

We start by calibrating the PDV model (\ref{eq:guyon_model}) using the methodology described in Section \ref{sec:calib_meth}. Note that the computation of the features $R_1$ and $\Sigma$ requires to truncate the sums at some point parameterized by $C$. While we use for the moment the previous 1,000 business days (i.e. $C=1000$), consistently with the choice of \cite{guyon2022volatility}, the influence of this hyperparameter will be discussed in Appendix \ref{sec:influence_cutoff}. The performance of the model is measured using the $R^2$ score (the definition is recalled in Equation (\ref{eq:r2})) which allows to assess how much of the variance of the implied volatility is explained by the model. The results are presented in Figure \ref{fig:r2_cutoff_1000}. For the S\&P 500, we obtain $R^2$ scores between 86\% and 95\% on the train set, between 68\% and 83\% for the 15 first time-to-maturities and between 55\% and 68\% for the last time-to-maturities on the test set. For the Euro Stoxx 50, we obtain $R^2$ scores between 85\% and 90\% on the train set, between 70\% and 81\% for the 15 first time-to-maturities and between 50\% and 70\% for the last time-to-maturities on the test set. These results indicate that a large part of the movements of the ATM implied volatility can be explained by the past movements of the underlying asset price. In this regard, they extend those of \cite{guyon2022volatility} to ATM implied volatility data. We also notice that the $R^2$ scores are overall decreasing with the option time-to-maturity: this is quite natural as we expect long-term options to be less sensitive to the variations of the underlying asset price than short-term options. This observation is also consistent with the results of \cite{bakshi2000call} who noticed that 'the longer an option’s remaining life, the more likely its price goes in the opposite direction with the underlying asset' suggesting that there is more exogeneity in the evolution of the prices of long-term options than in those of short-term options. \\

The two following subsections deepen the analysis of Figure \ref{fig:r2_cutoff_1000}. 

\subsubsection{Comment on the gap between the scores on the train and the test sets}\label{sec:comment_gap}
We observe a gap of approximately 24\% for the S\&P 500 and 19\% for the Euro Stoxx 50 between the $R^2$ scores on the train set and the test set. Such gaps are usually symptomatic of overfitted models. However, if we keep only one feature to reduce the complexity of the model, be it the trend feature $R_1$ or the volatility feature $\Sigma$, the $R^2$ scores are lower (especially with the trend feature) and the gap between the train and the test sets widens. Another way to deal with overfitting is to add a regularization term in the objective function. Such a technique is implemented in Appendix \ref{sec:influence_cutoff} but does not reduce the gap between the performance on the train and the test sets. Because of these two arguments, we estimate that the observed gap is not the result of an overfitted model but rather the result of the fact that the test set is both small (only 2 years of data) and of a peculiar nature. Indeed, the test set corresponds to the post-Covid-19 period which is characterized by a lot of uncertainty related to the Russia-Ukraine war, inflation, the rise of interest rates, etc. which may have affected the extent to which the volatility reacts to the underlying index movements. For the S\&P 500, the difference between the evolution on the train set and the test set is very clear: apart from the crash of March 2020, the S\&P 500 has experienced a constant increase with very little variations on the train set while the test set is characterized by a bull market followed by a bear market with high volatility. Note that the difference between the periods is however less clear for the Euro Stoxx 50. To support the claim that the test set is of peculiar nature, we compute the ratio $D$ of the signed distances and the absolute distances between the observed implied volatilities and the predicted implied volatilities on the test set:
\begin{equation}
    D = \frac{\displaystyle\sum_{t\in \mathcal{T}_{test}} IV_t^{mkt} -\beta_0-\beta_1R_{1,t}-\beta_2\Sigma_t }{\displaystyle\sum_{t\in \mathcal{T}_{test}} \left| IV_t^{mkt} -\beta_0-\beta_1R_{1,t}-\beta_2\Sigma_t \right|}
\end{equation}
where $\mathcal{T}_{test}$ is the set of dates in the test set. This indicator allows to measure whether there are more implied volatility observations above ($D>0$) or below ($D<0$) the plane fitted on the train set. Considering the 1-month implied volatilities, we obtain a ratio of 13.7\% for the S\&P 500 and 4.1\% for the Euro Stoxx 50. This suggests that most data points lie above the plane, implying a stronger reaction of the implied volatility to the movements in the underlying index during the post-Covid 19 period. Another possible reason of the gap between the scores on the train and the test sets is the high autocorrelation of the residuals of the PDV model which are given by $IV_t^{mkt-\beta_0-\beta_1R_{1,t}-\beta_2\Sigma_t}$. Indeed, the autocorrelation of the residuals at lag 1 day lies at 89\% (average over the autocorrelations of the residuals for the 24 time-to-maturities) and it is still above 26\% at lag 50 days, indicating a long memory. This high positive autocorrelation of the residuals implies that the predicted implied volatility is sometimes always above the observed implied volatility and sometimes always below. This can lead to low $R^2$ scores if the time period considered for the calculation of these $R^2$ scores is not large enough as it is the case here.



\subsubsection{Comparison with the scores of \cite{guyon2022volatility}}
In Figure \ref{fig:r2_cutoff_1000}, we also represent (with green and red crosses) the $R^2$ scores obtained when calibrating the PDV model on the VIX and the VSTOXX (which are the volatility indices of the S\&P 500 and the Euro Stoxx 50 respectively) using the same historical time period (from March 8, 2012 to December 30, 2022). They allow a consistent comparison between our scores and those of \cite{guyon2022volatility}. Note that the scores are represented at the same abscissa as the scores obtained on the 1-month implied volatilities. This choice is motivated by the fact that both the VIX and the VSTOXX are measures of the 30-days expected variance of their respective underlying index. We observe that the $R^2$ scores on the test set for the volatility indices are: 
\begin{enumerate}
    \item below the scores for the 1-month implied volatility and 
    \item below the scores reported by \cite{guyon2022volatility} for the same indices (the difference being only the train and test periods: their train and test sets respectively span the periods from January 1, 2000 to December 31, 2018 and from January 1, 2019 to May 15, 2022 while our train and test sets respectively span the periods from March 8, 2012 to December 31, 2020 and from January 1, 2021 to December 30, 2022).
\end{enumerate}
The first observation shows that the 1-month ATM implied volatility is even better predicted by the PDV than the volatility indices considered by \cite{guyon2022volatility}. The second observation confirms that the $R^2$ scores of the PDV model are quite sensitive to the choice of the train and test sets as discussed in Section \ref{sec:comment_gap}. \\

The empirical study conducted in this section allowed us to exhibit the dependence of the ATM implied volatility on the past path of the underlying asset price for two major financial indices. We showed that this dependence decreases with the time-to-maturity but remains material even for the largest time-to-maturities. At this stage, it is natural to ask whether these conclusions still hold for away-from-the-money implied volatilities. Instead of reproducing the empirical study for each time-to-maturity and strike (which would increase significantly the dimension of the study), we study the performance of the PDV model in explaining the evolution of the calibrated parameters of the SSVI parameterization of \cite{gatheral2014arbitrage}. The following section is dedicated to the presentation of this parameterization.

%% file: ssvi.tex
\section{A parsimonious version of the SSVI parameterization}\label{sec:svi}
The purpose of this section is to introduce a parsimonious version of the SSVI parameterization and to present some calibration results of this model on the implied volatility historical data that we considered in Section \ref{sec:empirical_study}. We start by some reminders about the SSVI parameterization of \cite{gatheral2014arbitrage}.

\subsection{The SSVI parameterization}\label{sec:ssvi}
Devised at Merill Lynch in 1999 and publicly disseminated by \cite{gatheral2004parsimonious}, the Stochastic Volatility Inspired (SVI) parameterization is a popular parameterization of the implied volatility smile. To be more precise, it is a parameterization of the total implied variance, defined by $w(k,T):=\sigma_{BS}^2(k,T)T$ where $\sigma_{BS}(k,T)$ is the Black-Scholes implied volatility associated to the log-strike\footnote{We recall that the log-strike $k$ of a vanilla option of strike $K$ and forward price $F=S_0e^{f_T T}$ is defined as $k = \log \left(\frac{K}{F} \right)$} $k$ and the time-to-maturity $T$. The popularity of this parameterization is mainly due to its tractability and its ability to fit market implied volatilities quite well. Moreover, it features nice properties such as consistency with Lee's moment formula (\citeauthor{lee2004moment}, \citeyear{lee2004moment}) or the fact that it corresponds exactly to the large-maturity limit of the Heston implied volatility smile\footnote{\cite{martini2023refined} have shown that this limit is actually a SSVI smile. } (\citeauthor{gatheral2011convergence}, \citeyear{gatheral2011convergence}). Nevertheless, this parameterization has two issues. First, the SVI parameterization is not a parameterization of the full total implied variance surface but only of a slice $k\mapsto w(k,T)$ for a fixed time-to-maturity $T$ since the 5 parameters are all time-to-maturity-dependent. Second, at the time of the publication of their paper, it seemed impossible to find conditions on the SVI parameters that guarantee the absence of butterfly arbitrage (the problem has now been solved by \citeauthor{martini2022svi}, \citeyear{martini2022svi}). In their paper, \cite{gatheral2014arbitrage} proposed a parameterization for the full total implied variance surface, where each slice is a particular case of the SVI parameterization, which allows to address these two issues. This parameterization is called the surface SVI (SSVI) and is defined below. 
\begin{definition}
    Let $\varphi$ be a smooth function from $\mathbb{R}_+^*$ to $\mathbb{R}_+^*$ such that the limit $\lim_{T\to 0}\theta_T \varphi(\theta_T)$ exists in $\mathbb{R}$ where $\theta_T:=\sigma_{BS}^2(0,T)T$ is the ATM total implied variance. The SSVI is the surface defined by:
    \begin{equation}
        w(k,T) = \frac{\theta_T}{2}\left(1+\rho\varphi(\theta_T)k+\sqrt{(\varphi(\theta_T)k+\rho)^2+(1-\rho^2)} \right).
    \end{equation}  
\end{definition}

While the SVI parameterization requires 5 parameters for each slice of the IVS, the SSVI relies on the function $\varphi$ and the parameter $\rho\in(-1,1)$, that do not depend on the time-to-maturity, as well as one parameter $\theta_T$ for each time-to-maturity. The latter could be considered as set prior to the calibration since the price of close-to-ATM options are observed on the market so a good estimate of the ATM total implied variance can be computed from raw market quotes. Note that \cite{hendriks2017extended} propose to consider a time-to-maturity-dependent $\rho$ parameter in order to improve the calibration accuracy for very short time-to-maturities (typically below 1 month). Since our database does not contain short-term data, this extension of the SSVI is not investigated in our paper.\\

A natural question at this stage is how to choose the function $\varphi$ in order to both achieve a good fit to market data and to satisfy the conditions derived by \cite{gatheral2014arbitrage} to prevent static arbitrages (see Appendix \ref{sec:ssvi_arbitrage}). The authors propose three examples of parametric form\footnote{Note that some papers (e.g. \citeauthor{corbetta2019robust}, \citeyear{corbetta2019robust} and \citeauthor{mingone2022no}, \citeyear{mingone2022no}) do not consider a parametric form for the function $\varphi$ but they calibrate instead one parameter $\varphi_T$ per time-to-maturity. Since our objective is to obtain a parsimonious version of the SSVI parameterization, this approach is not considered in our paper.} for $\varphi$:
\begin{itemize}
    \item the Heston-like parameterization $\varphi(\theta):=\frac{1}{\lambda \theta}\left(1-\frac{1-e^{-\lambda \theta}}{\lambda\theta} \right)$ with $\lambda >0$,
    \item the power-law parameterization $\varphi(\theta):= \frac{\eta}{\theta^{\gamma}}$ with $\eta>0$ and $0< \gamma < 1$, and 
    \item  the modified power-law parameterization $\varphi(\theta):= \frac{\eta}{\theta^{\gamma}(1+\theta)^{1-\gamma}}$
    with $\eta>0$ and $0<\gamma < 1$.
\end{itemize}

For simplicity of reference to these parameterizations, we abbreviate the SSVI with the Heston-like parameterization to SSVI-HL, the SSVI with the power-law parameterization to SSVI-PL and the SSVI with the modified power-law parameterization to SSVI-MPL. In Appendix \ref{sec:ssvi_calibration_results}, we compare the fitting accuracy of these parameterizations on the two data sets introduced in Section \ref{sec:data}. This numerical study reveals that the SSVI-HL parameterization yields a quite poor fit to the market implied volatilities while the SSVI-PL and the SSVI-MPL parameterizations yield a satisfying fit. Moreover, the fitting accuracy is quite similar between the SSVI-PL and the SSVI-MPL parameterizations.\\

\begin{remark}\label{rk:atm_skew}
    For $\gamma =1/2$, the SSVI-PL and the SSVI-MPL parameterizations induce a power-law decay of the ATM volatility skew for small time-to-maturities. Indeed, recalling that the ATM volatility skew is defined by $\partial_k \sigma_{BS}(0,T)$, it is straightforward to show that $\partial_k \sigma_{BS}(0,T)=\frac{1}{2\sqrt{T}}\rho\sqrt{\theta_T}\varphi(\theta_T)$ for the SSVI parameterization (no matter the choice for $\varphi$). Therefore, we have that $\partial_k \sigma_{BS}(0,T) = \frac{\rho \eta}{2\sqrt{T}}$ for the SSVI-PL and $\partial_k \sigma_{BS}(0,T) = \frac{\rho \eta}{2\sqrt{T}\sqrt{1+\theta_T}} = \frac{\rho \eta}{2\sqrt{T}} + o\left(\frac{1}{\sqrt{T}}\right)$ for the SSVI-MPL assuming that $\lim_{T\to 0}\theta_T = 0$ which is a natural assumption: an ATM option with zero time to expiry has no value. This power-law decay of the ATM volatility skew has long been considered as a stylized fact of implied volatility surfaces (see e.g. \citeauthor{gatheral2018volatility}, \citeyear{gatheral2018volatility}). It has been recently challenged by \cite{guyon2023does} and \cite{delemotte2023yet} who showed that some parametric forms fit better the ATM volatility skew for short time-to-maturities (typically below 1 month) than a power-law. However, they both acknowledge that a power-law provides a satisfying fit for longer time-to-maturities (typically above 1 month). Thus, we think that it is still a relevant property to capture. 
\end{remark}

\subsection{The parsimonious SSVI model}\label{sec:pssvi}



\begin{figure}[h]
    \centering
    \subfigure[S\&P 500 ]{\includegraphics[width=0.45\linewidth]{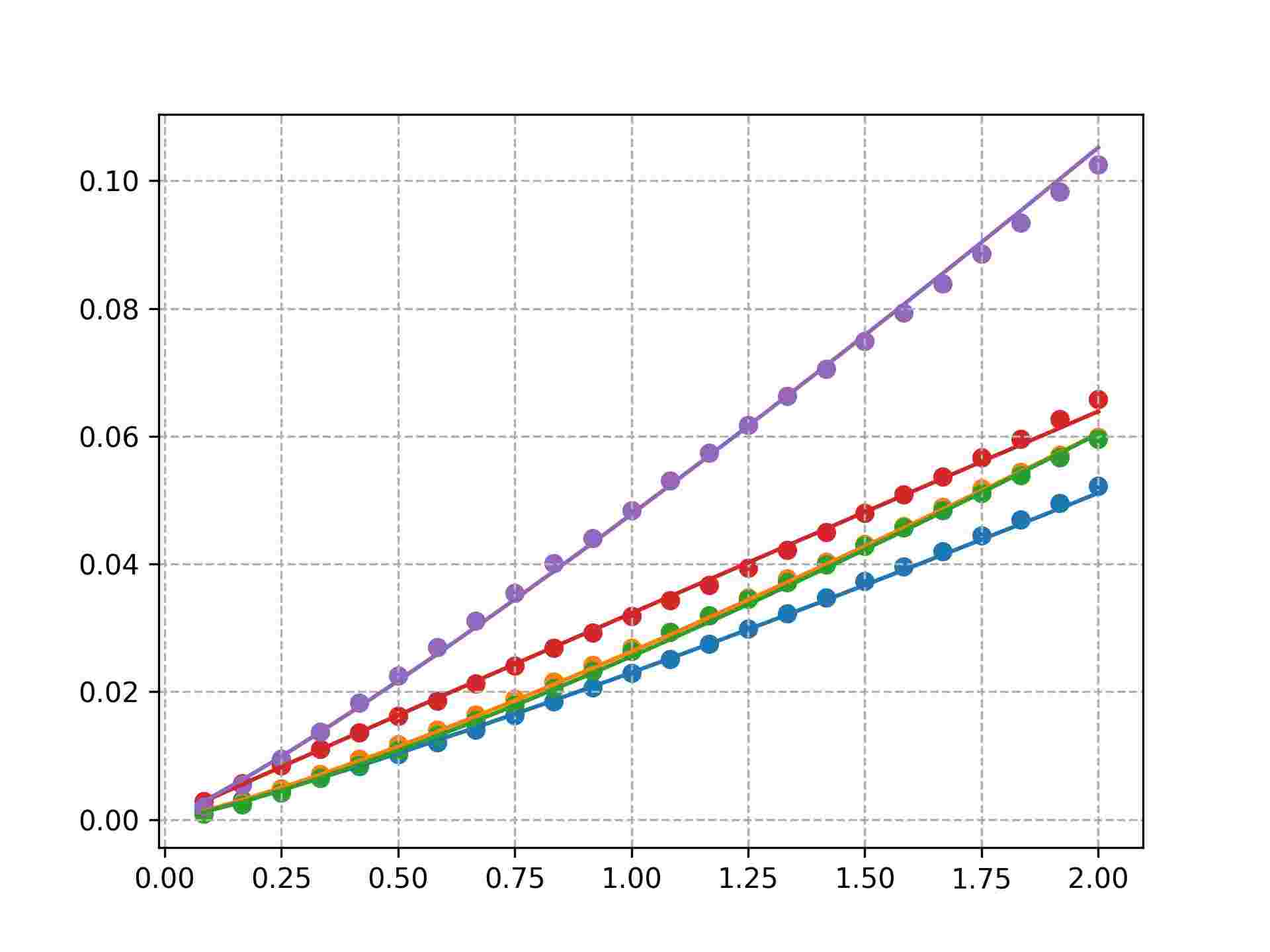}}
    \subfigure[Euro Stoxx 50]{\includegraphics[width=0.45\linewidth]{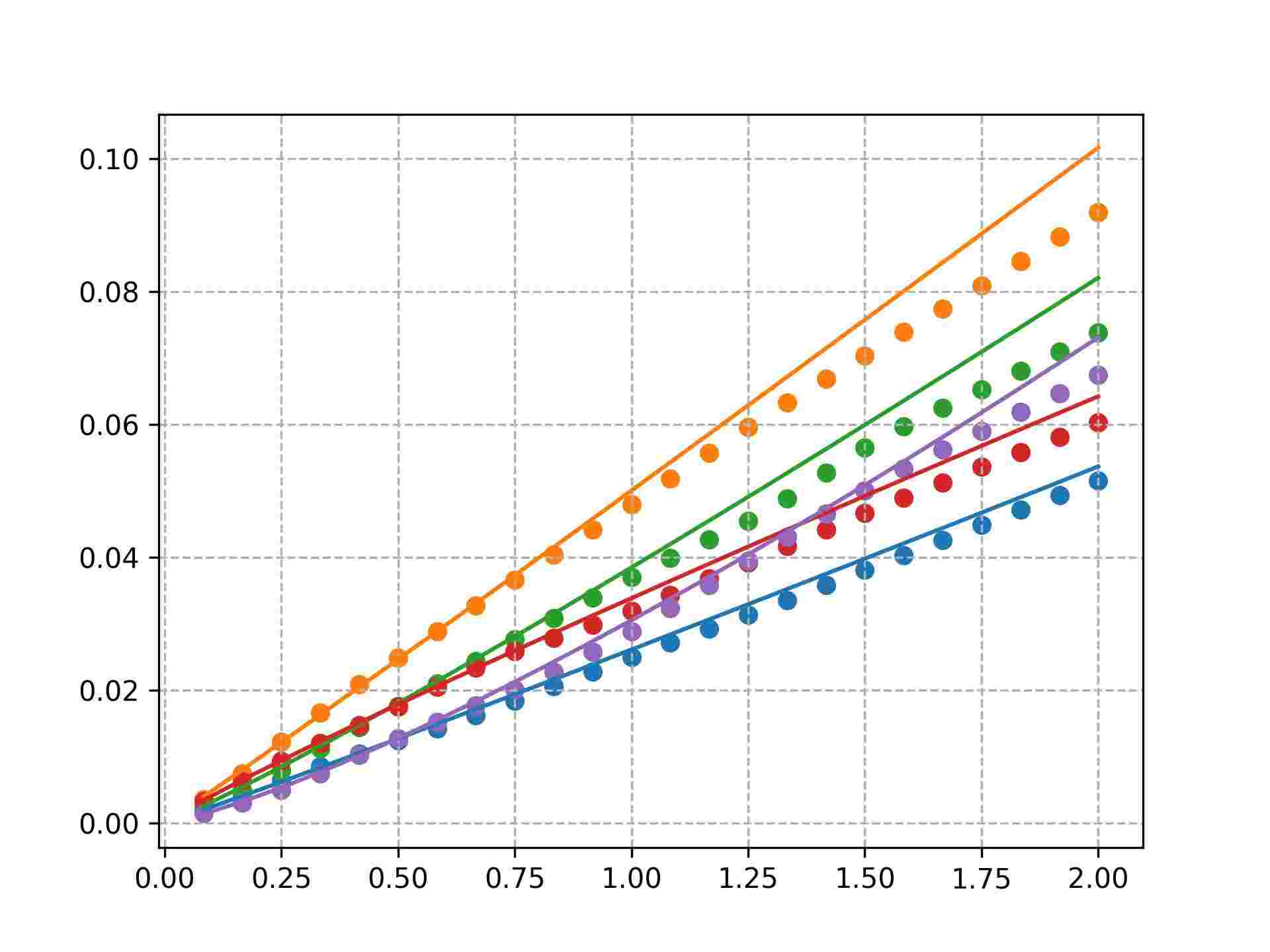}\label{fig:theta_t_worst_case}}
    \caption{Comparison of the S\&P 500 ATM total variances (dots) with the fitted parametric form $T\mapsto aT^p$ (lines) for five dates (each color corresponds to one date). The five dates are those where the average relative error is the closest to the mean over all average relative errors.}
\label{fig:theta_t}
\end{figure}

At this stage, let us recall that our final objective is to design a model to jointly simulate arbitrage-free IVSs and the price of the underlying asset by simulating the evolution of the SSVI parameters as a function of the path of the underlying asset price. However, the number of parameters involved is so large (there are $n+2$ parameters for the SSVI-HL and $n+3$ for the SSVI-PL and the SSVI-MPL to fit an IVS with $n$ maturities) that a model for their joint evolution would be too complicated. Therefore, we need to make the SSVI model more parsimonious. We propose the two following simplifications:
\begin{enumerate}
    \item We consider only the modified power-law parameterization with $\gamma$ fixed to 1/2 because we observe that $\gamma$ is close to this value over the two data sets (more than 84\% of the calibrated $\gamma$'s lie within the $[0.4,0.6]$ interval). Moreover, this choice has the advantage of guaranteeing that the full surface (without any restriction on the ATM total implied variance) is free of static arbitrage on the sole condition that $\eta^2(1+|\rho|) < 4$ (see case (\ref{prop:pssvi_no_arbitrage}) of Proposition \ref{prop:modified_power_law}). Finally, according to Remark \ref{rk:atm_skew}, setting $\gamma=1/2$ implies a power-law decay of the ATM volatility skew which is a relevant property to capture.
    \item We assume that $\theta_T = aT^p$ where $a,p\ge 0$ reducing considerably the number of parameters while ensuring that the no-arbitrage constraint on the ATM total variance is always satisfied ($\partial_T\theta_T \ge 0$). This parametric form is inspired by the calibrated vectors $(\theta_{T_i})_{i=1,\dots,M}$ that exhibit almost a linear behavior with the maturity $T$. 
    In Figure \ref{fig:theta_t}, we show how this parametric form fits the ATM total variances for several dates. Note that the parameters $a$ and $p$ that have been used in this figure are those calibrated on the whole IVS so the quality of the fit is reduced in comparison to a calibration on the ATM volatilities only. Despite this, the fit is overall satisfying. 
\end{enumerate}

The new model obtained after these simplifications is called thereafter the parsimonious SSVI model.
\begin{definition}[Parsimonious SSVI]\label{def:parsimonious_ssvi}
    The parsimonious SSVI is the parameterization of the total implied volatility surface defined by:
\begin{equation}
    w(k,T) = \frac{\theta_T}{2}\left(1+\rho\varphi(\theta_T)k+\sqrt{(\varphi(\theta_T)k+\rho)^2+(1-\rho^2)} \right).
\end{equation}  
where $\theta_T = aT^p$ and $\varphi(\theta) = \frac{\eta}{\sqrt{\theta(1+\theta)}}$ with $a,p\ge 0$ and $\eta>0$. 
\end{definition}

\begin{figure}[h]
    \centering
    \subfigure[S\&P 500]{\includegraphics[width=0.45\linewidth]{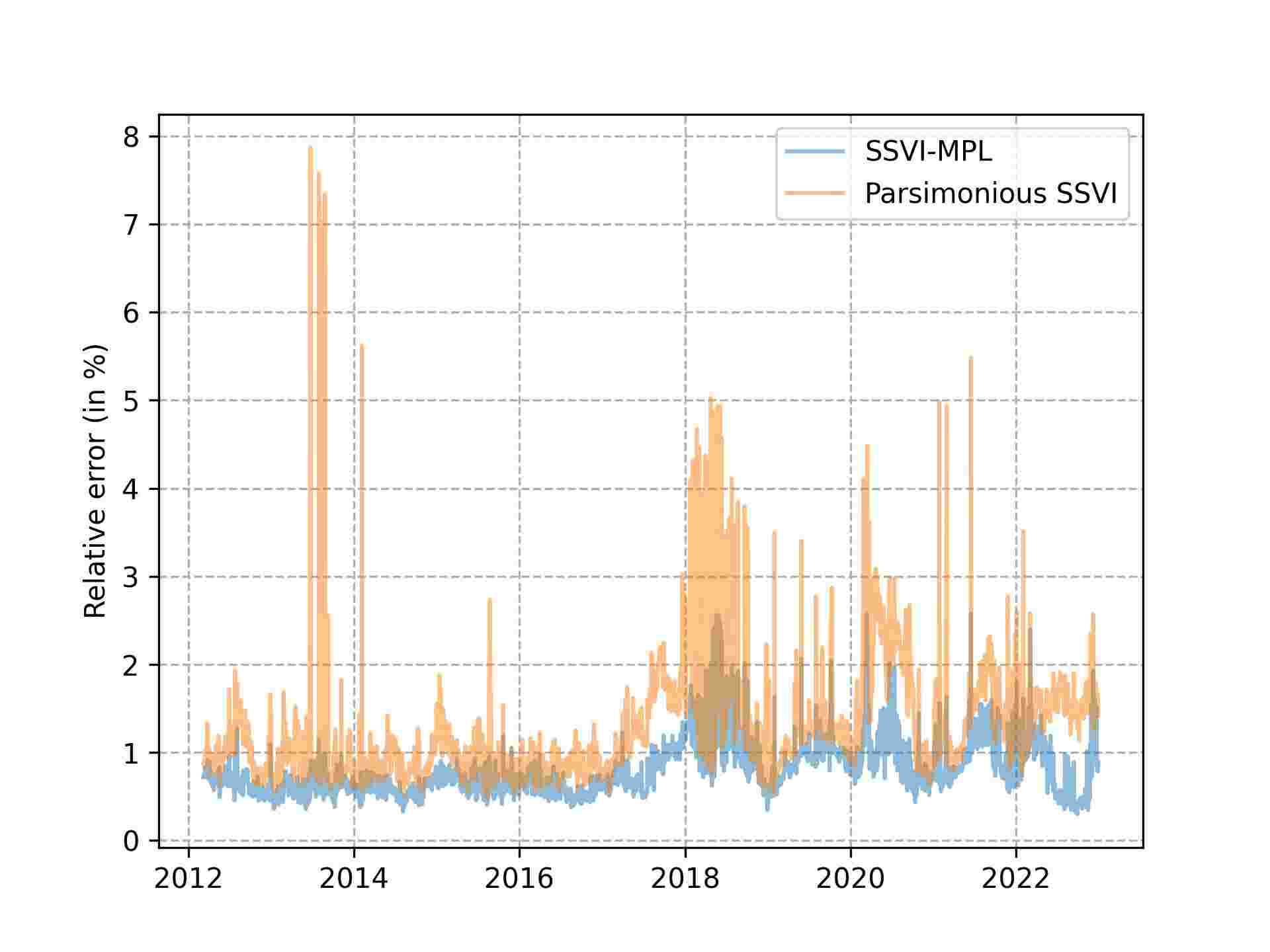}\label{fig:parsimonious_ssvi_relative_error_spx}}
    \subfigure[Euro Stoxx 50]{\includegraphics[width=0.45\linewidth]{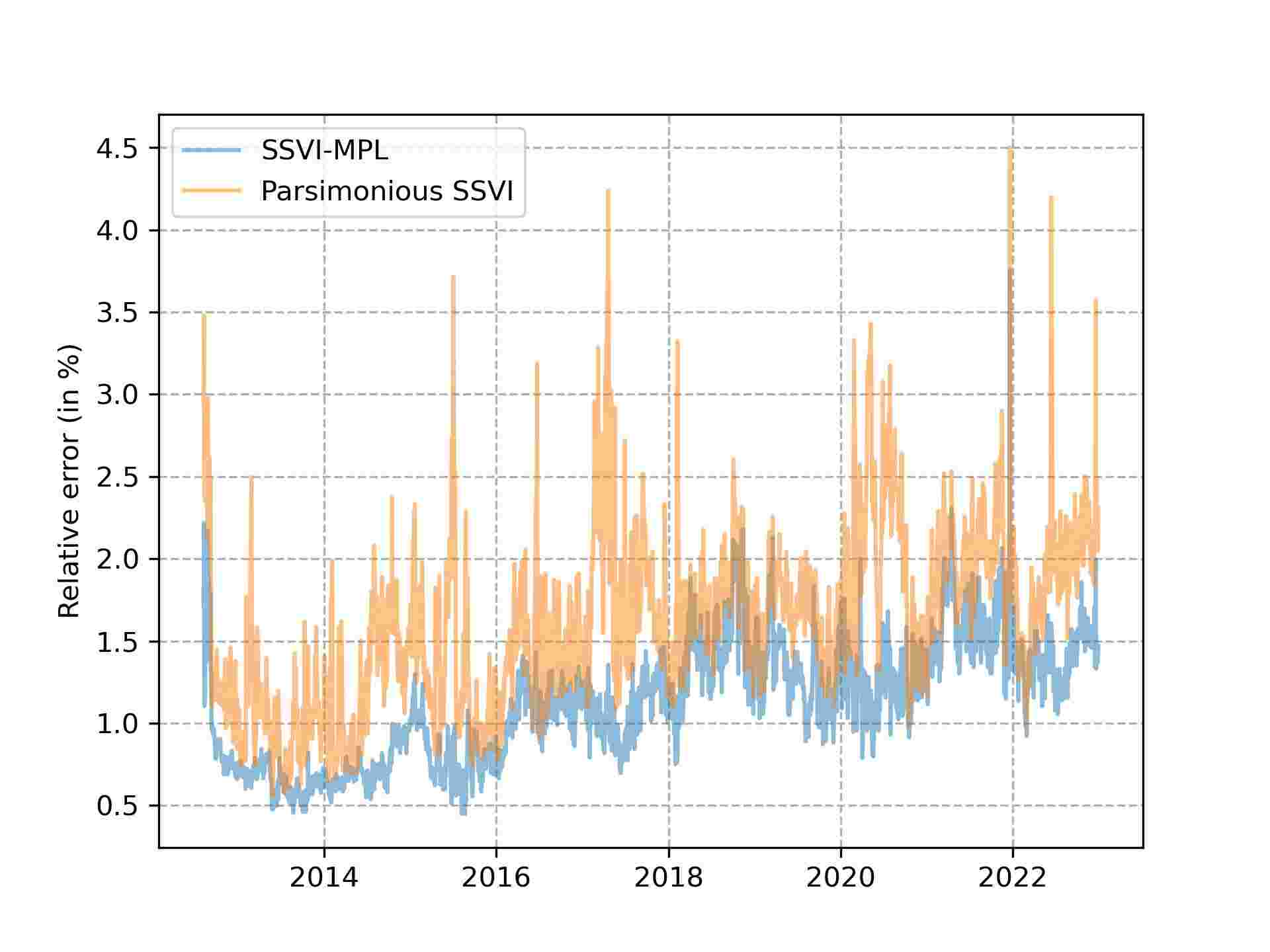}}
    \caption{Average relative errors between the market implied volatilities and the SSVI implied volatilities for the parsimonious SSVI model and the SSVI-MPL model.}
\label{fig:parsimonious_ssvi_relative_error}
\end{figure}

\begin{figure}[h]
    \centering
    \subfigure[S\&P 500]{\includegraphics[width=0.45\linewidth]{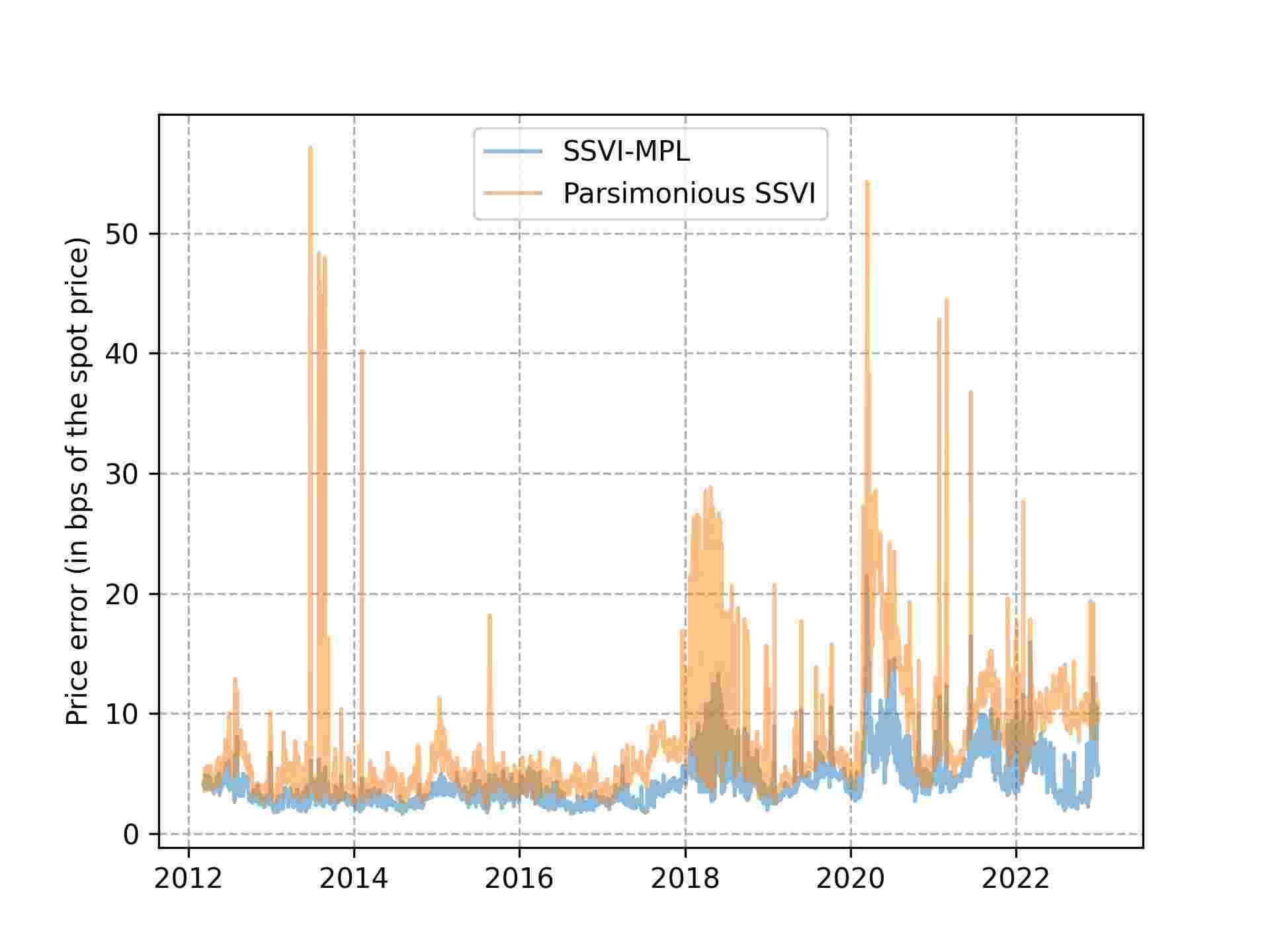}}
    \subfigure[Euro Stoxx 50]{\includegraphics[width=0.45\linewidth]{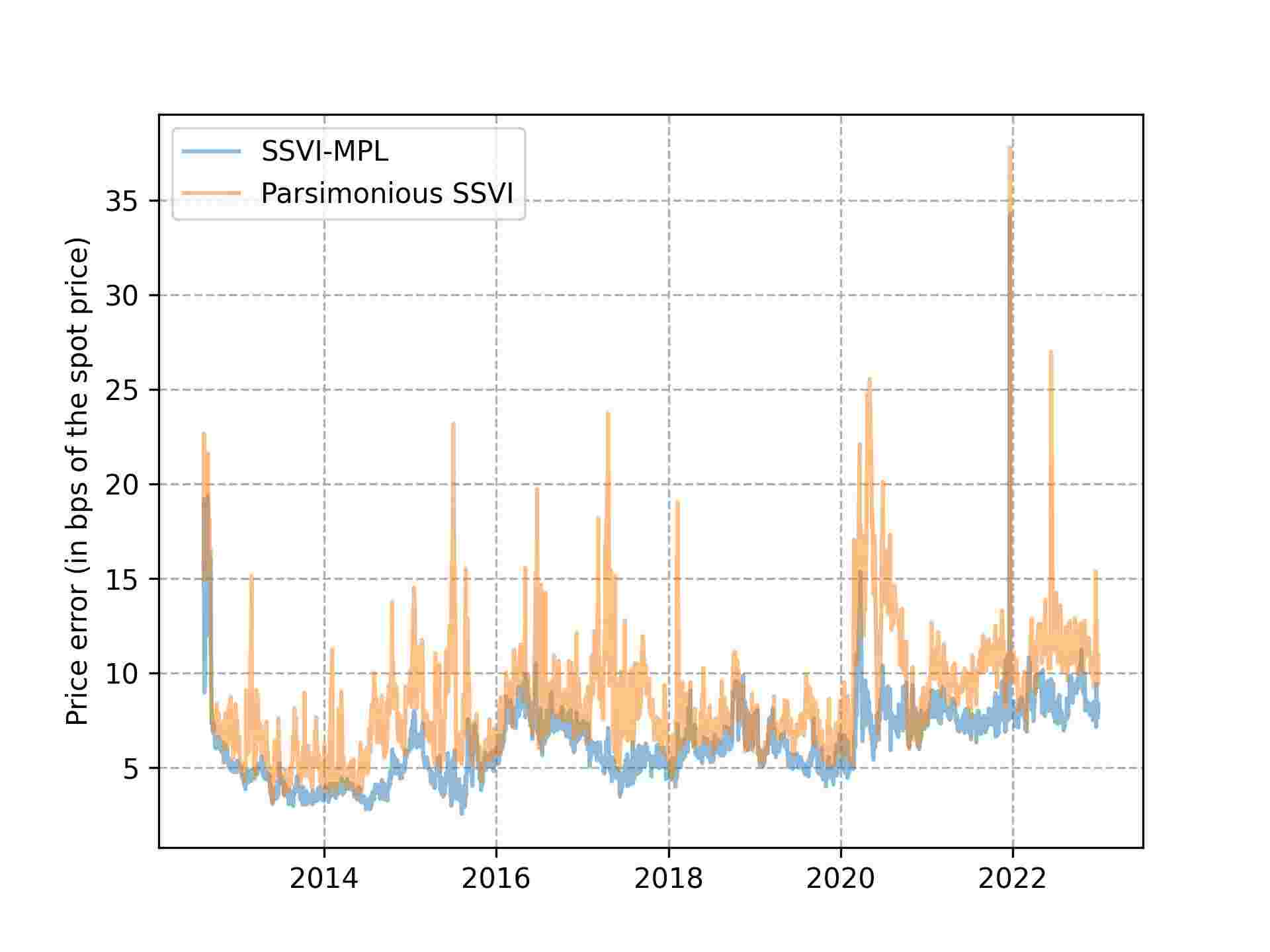}}
    \caption{Average price errors between the market call prices and the SSVI call prices for the parsimonious SSVI model and the SSVI-MPL model.}
\label{fig:parsimonious_ssvi_price_error}
\end{figure}

In Figure \ref{fig:parsimonious_ssvi_relative_error}, the average relative errors between the market implied volatilities and the SSVI implied volatilities for each day of our data sets obtained for the parsimonious SSVI model are compared to those obtained for the SSVI model with the modified power-law parameterization. In Figure \ref{fig:parsimonious_ssvi_price_error}, we also show the average errors between the market call prices and the SSVI call prices\footnote{The call prices are computed assuming zero interest rate and no dividend.} in basis points (bps) of the spot price. 

\begin{figure}[h]
    \centering
    \subfigure[S\&P 500 (February 12, 2021)]{\includegraphics[width=0.45\linewidth]{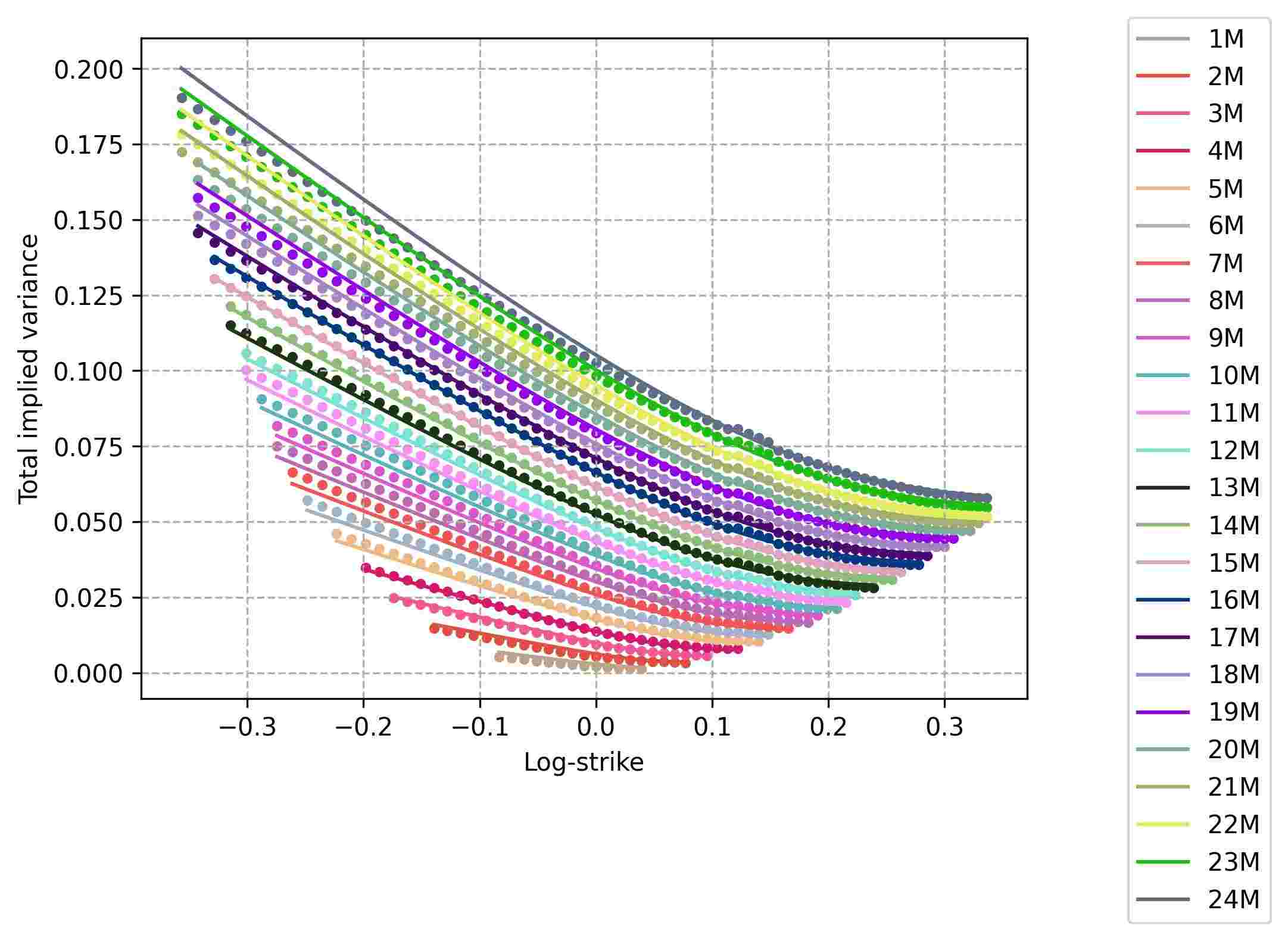}}
    \subfigure[Euro Stoxx 50 (August 30, 2019)]{\includegraphics[width=0.45\linewidth]{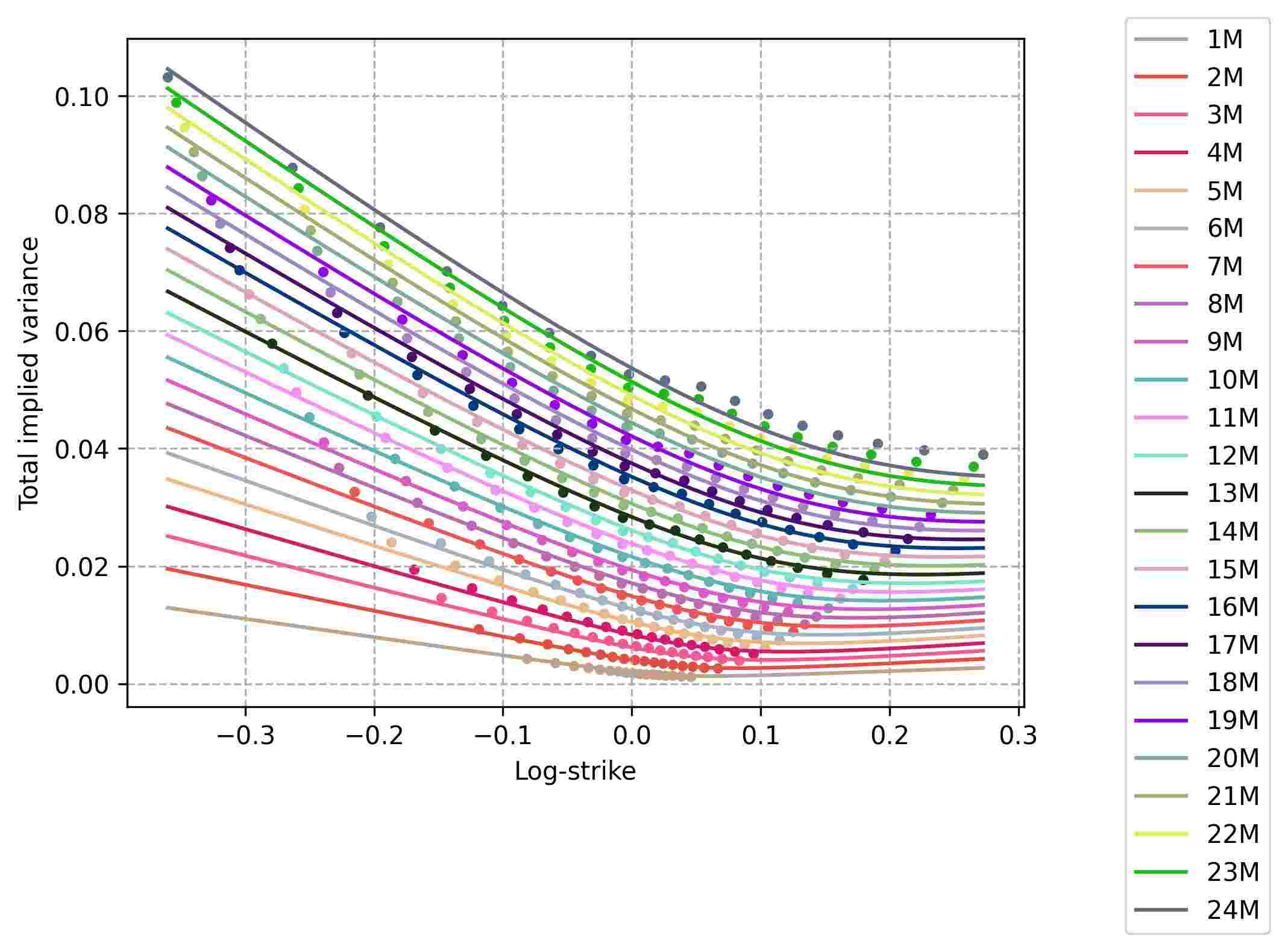}}
    \caption{Illustrations of the fit of the parsimonious SSVI on market total variances for all maturities. The dots (resp. the lines) correspond to the market (resp. the SSVI) total implied variances.}
\label{fig:pssvi_fit}
\end{figure}

As expected, the calibration accuracy is reduced for the parsimonious SSVI. However, it remains overall quite close to the SSVI-MPL calibration in view of the reduction of the number of parameters: 4 parameters for the parsimonious SSVI versus 27 for the SSVI-MPL. The mean of the average relative errors across the whole S\&P 500 (resp. Euro Stoxx 50) data set increases from 0.80\% (resp. 1.18\%) to 1.29\% (resp. 1.60\%). Besides, the mean of the average price errors accross the whole S\&P 500 (resp. Euro Stoxx 50) data set increases from 4 bps (resp. 6 bps) to 7 bps (resp. 8 bps).  In our numerical experiments, we observed that the parametric form for $\theta_T$ is the assumption leading to the largest deterioration of the fit to market implied volatilities, which is consistent with the fact that this assumption is the one limiting the most the number of degrees of freedom of the model. In Figure \ref{fig:pssvi_fit}, we illustrate how the parsimonious SSVI fits the S\&P 500 (resp. the Euro Stoxx 50) total implied variances for a day where the calibrated average relative error is equal to 1.29\% (resp. 1.60\%) corresponding to the mean of the average relative errors across the whole S\&P 500 (resp. Euro Stoxx 50) data set. 

%% file: path_dependent_ssvi.tex
\section{Path-dependent SSVI model}\label{sec:pdv_ssvi}
The present section is dedicated to the introduction of a new model for the joint dynamics of an implied volatility surface and the underlying asset price. The calibration results exposed in Section \ref{sec:ssvi} demonstrate the ability of a particular case of the SSVI parameterization - the parsimonious SSVI - to fit reasonably well historical implied volatility surfaces while guaranteeing the absence of static arbitrage with only 4 parameters. As a consequence, we propose to specify our model as a dynamic version of the parsimonious SSVI: each parameter of the parsimonious SSVI is considered as a stochastic process whose dynamics remains to be determined. One option to jointly model the evolution of the parsimonious SSVI parameters and the underlying asset would be to introduce a correlation between the random noises driving each process. However, in view of the empirical study conducted in Section \ref{sec:empirical_study} which indicates that there is a feedback effect of the past returns and the past squared returns of the underlying index price onto the level of the ATM implied volatility, we prefer another option. Instead of using a correlation, the idea is to explicitly model the response of each of the 4 parameters to the evolution of the underlying asset price. To this end, we measure to which extent the trend feature and the volatility feature of the path-dependent volatility (PDV) model presented in Section \ref{sec:pdv_model} allow to explain the variations of the 4 parameters of the parsimonious SSVI model. This study is presented in Section \ref{sec:pdv_ssvi_params} below. Then, Section \ref{sec:dynamics_specification} introduces the dynamics of the asset price and of the parsimonious SSVI parameters. Finally, Section \ref{sec:model_calibration} and \ref{sec:num_results} detail the calibration and the simulation of the model. \\

\begin{remark}
    It is important to note at this stage that there should be some consistency between the dynamics of the implied volatility surface and the one of the underlying asset price to prevent dynamic arbitrages. More precisely, the process $t\mapsto C_{BS}(K,T,S_t,IV_t(K,T))$, where $C_{BS}$ is the Black-Scholes price of a call option of strike $K$ and maturity $T$ when the spot price is $S_t$ and the implied volatility is $IV_t(K,T)$, should be a martingale. Several authors (e.g. \citeauthor{schonbucher1999marketmodel}, \citeyear{schonbucher1999marketmodel}, \citeauthor{schweizer2008term}, \citeyear{schweizer2008term}, \citeauthor{jacod2010risk}, \citeyear{jacod2010risk}) found conditions guaranteeing the existence of a joint model satisfying the absence of dynamic arbitrage in various settings. However, these conditions are not tractable and we are not aware of any parametric model satisfying those. \cite{amrani2021dynamics} find a necessary condition guaranteeing the absence of dynamic arbitrage for a dynamic version of the SSVI parameterization but this condition is very restrictive. Thus, due to the difficulty of designing a joint model which does not exhibit dynamic arbitrages, we do not look for such a model and we focus on finding one which is as consistent as possible with historical data and which does not exhibit static arbitrages. 
\end{remark}

\subsection{Path-dependency of the parsimonious SSVI parameters}\label{sec:pdv_ssvi_params}
\begin{figure}[h]   
    \centering
    \subfigure[$\rho$]{\includegraphics[width=0.45\linewidth]{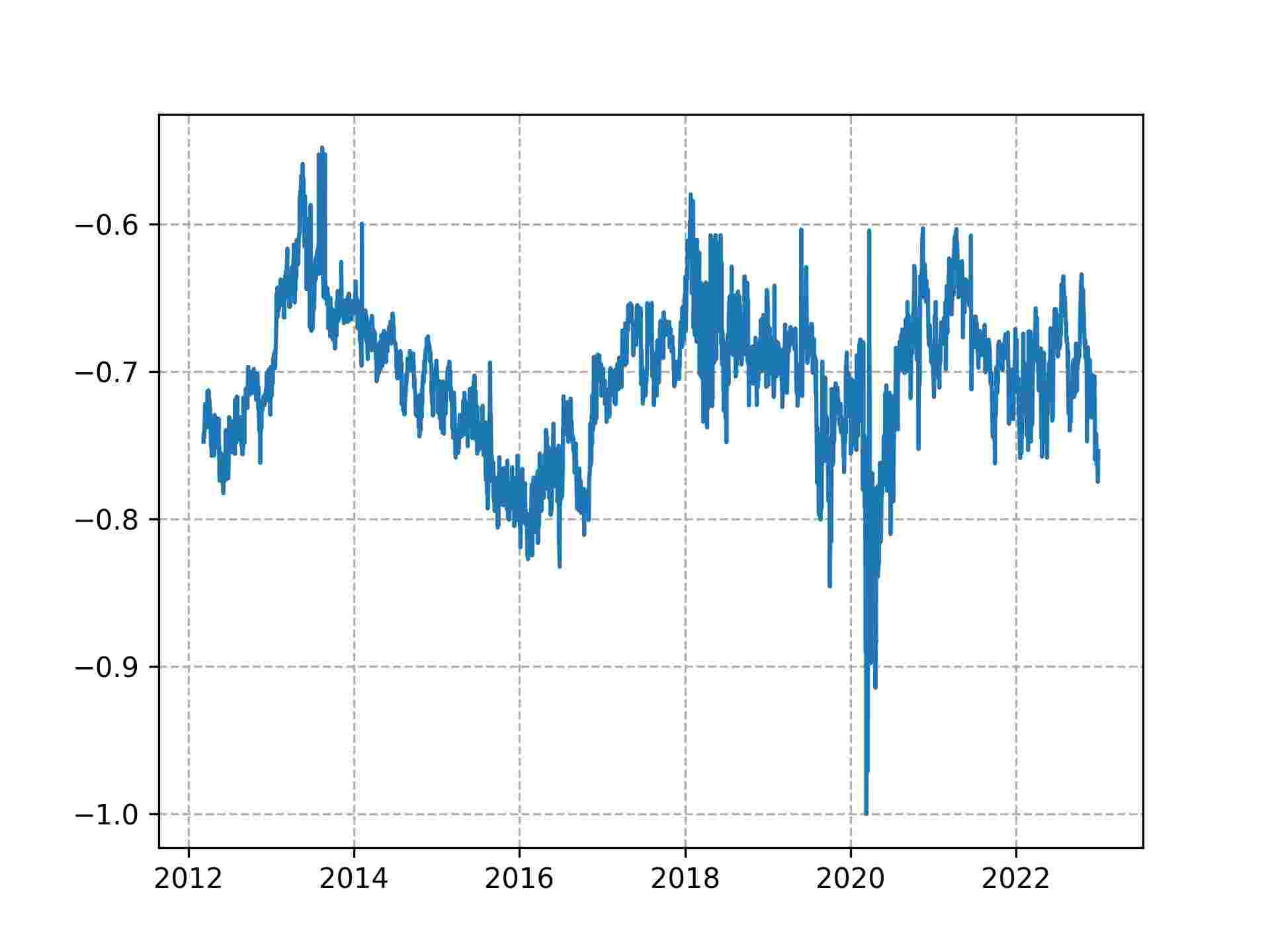}}
    \subfigure[$\eta$]{\includegraphics[width=0.45\linewidth]{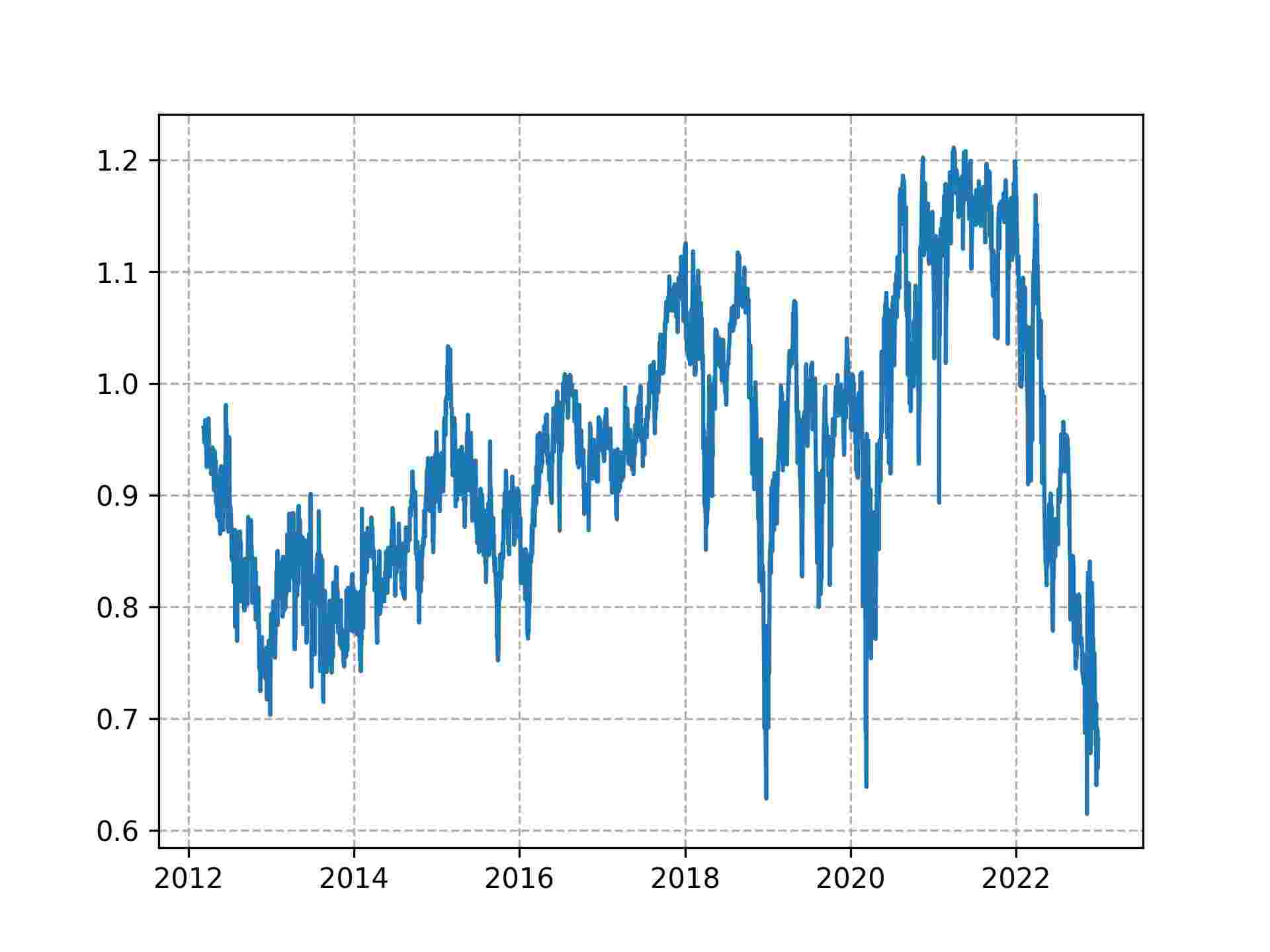}}
    \caption{Historical evolution of the $\rho$ and $\eta$ parameters.}
    \label{fig:rho_eta_evolution}
\end{figure}

The calibration of the parsimonious SSVI in Section \ref{sec:pssvi} provides the daily evolutions of the parameters $a$, $p$, $\rho$ and $\eta$. Based on these daily evolutions, we can calibrate the PDV model (\ref{eq:guyon_model}) where we replace $\text{Volatility}_t$ in Equation (\ref{eq:guyon_model}) by each parameter of the parsimonious SSVI. Note that we consider the logarithm of $p$ instead of $p$ in the PDV model since we observed that this provides a better fit. The calibration methodology is the same as the one exposed in Section \ref{sec:calib_meth}. We refer to Appendix \ref{sec:hyperparams_ssvi for the estimation method of the hyperparameters of the PDV model and their values}. The $R^2$ scores obtained on the train and the test sets for each parameter and for each index are reported in Table \ref{tab:ssvi_r2_scores}. On the one hand, these results show that the time evolution of the parameter $a$ is well explained by the evolution of the underlying asset price both on the train and the test sets, which is line with the results obtained for the ATM implied volatility since $a$ captures the ATM total variance level (whose evolution is similar to the one of the ATM implied volatility and as such we expect the study conducted in Section \ref{sec:empirical_study_numerical_results} to be still valid for the ATM total variance). The same observation holds for $p$. However the fact that the PDV model works well for $p$ was not anticipated as it allows to parameterize how the ATM total variance increases with the time-to-maturity and it is not homogenous to the ATM total variance level. Thus, we emphasize that this is a key finding. On the other hand, the $R^2$ scores for the parameters $\rho$ and $\eta$ are decent on the train set but small or negative on the test set. This indicates that the trend and the volatility features are not good predictors for these two parameters. Let us however point out that the evolution of these two parameters does not appear as completely independent from the one of the underlying asset price as one can see in Figure \ref{fig:rho_eta_evolution}. In particular, we see a large variations of these parameters during the market drop of March 2020. Despite this, we did not find features allowing a good prediction (i.e. material $R^2$ scores on the test set) of $\rho$ and $\eta$ (we tried for example skewness, kurtosis and correlation features). As a consequence, the Browian motions of these parameters will only be correlated to the Brownian motion of the asset price. 

\begin{figure}[h]   
    \centering
    \subfigure[Influence of $\rho$]{\includegraphics[width=0.45\linewidth]{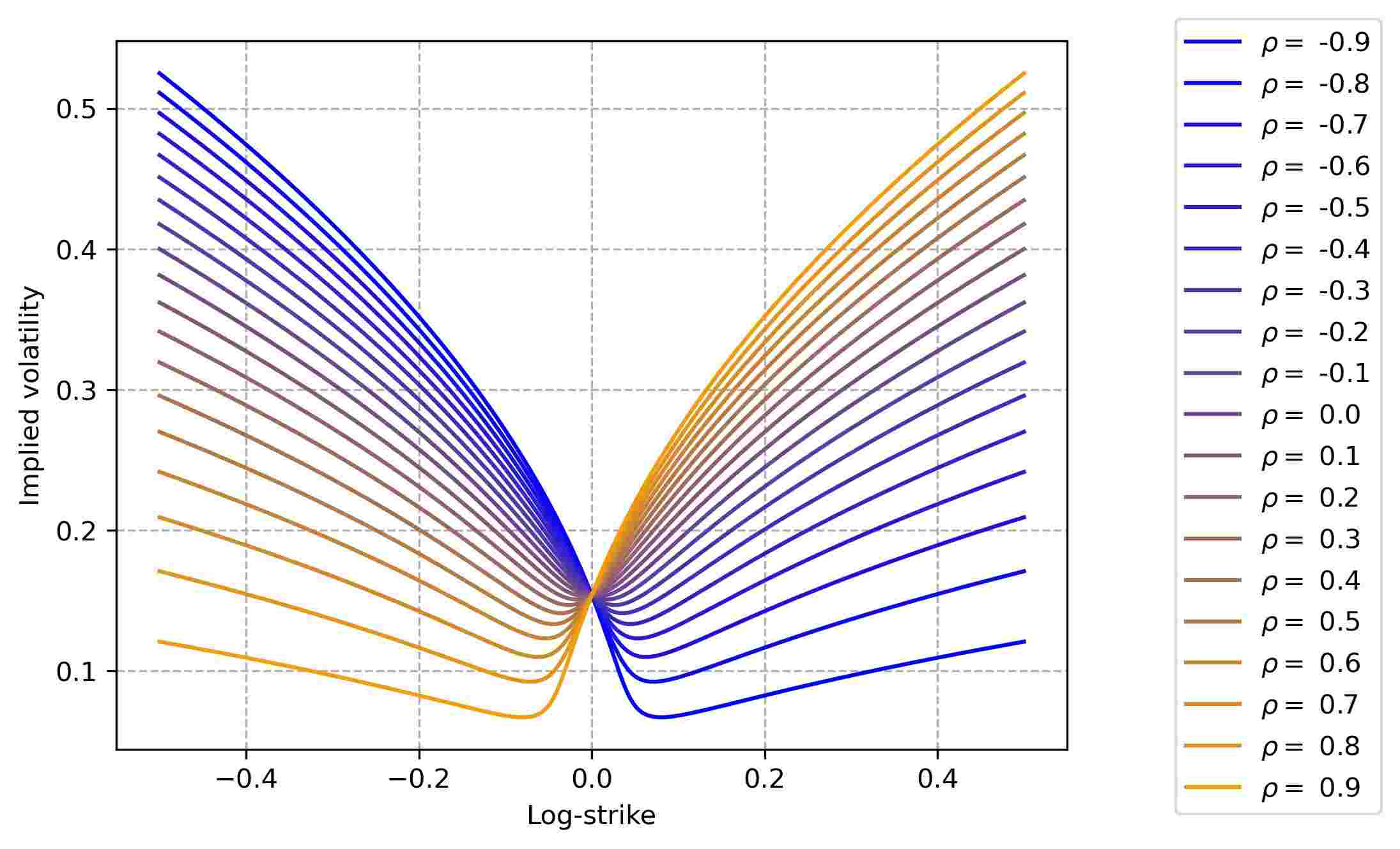}}
    \subfigure[Influence of $\eta$]{\includegraphics[width=0.45\linewidth]{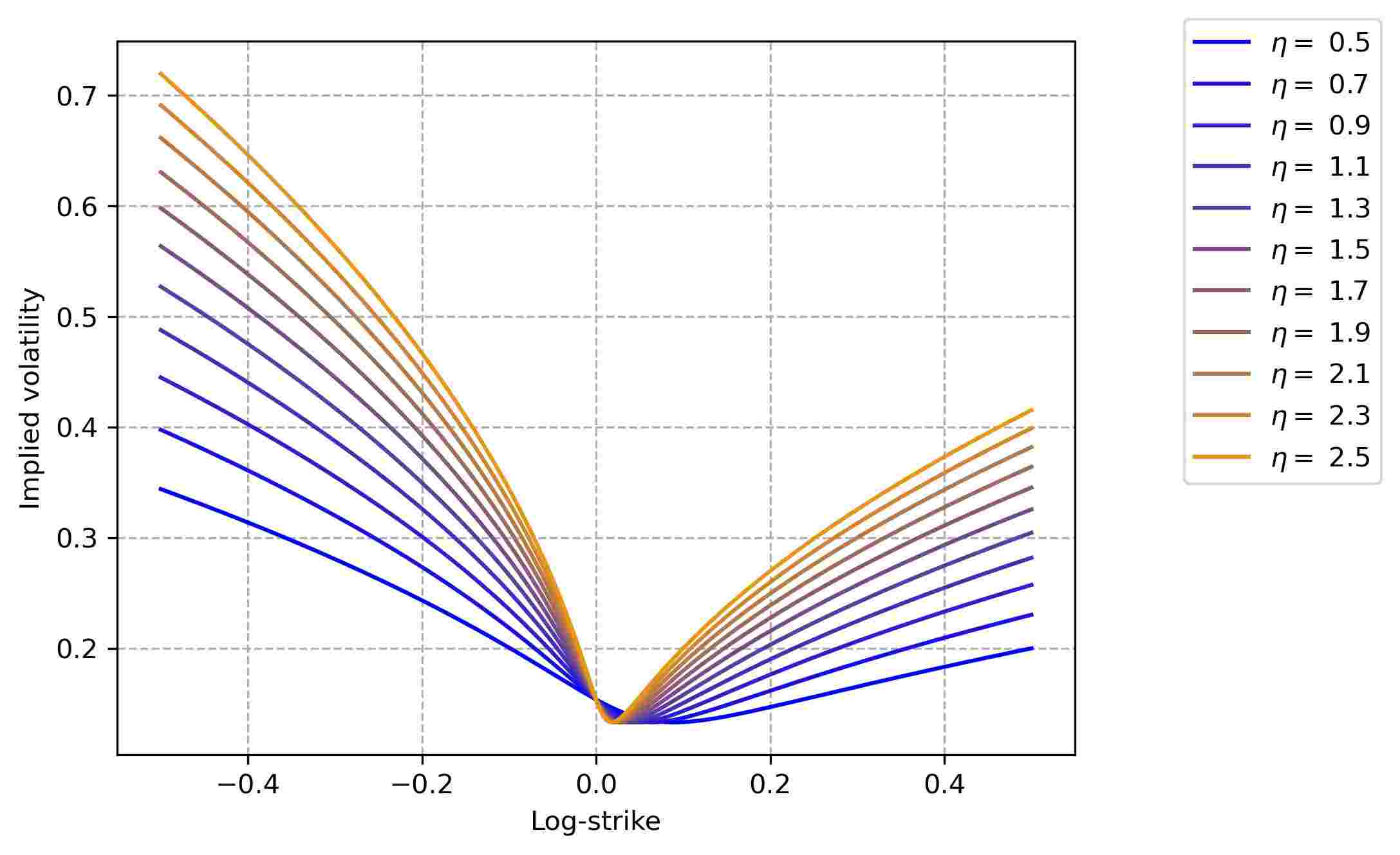}}
    \caption{Influence of the $\rho$ and $\eta$ parameters on the shape of the implied volatility smile computed from the parsimonious SSVI ($a$ and $p$ are respectively fixed to 0.05 and 1.3).}
    \label{fig:influence_rho_eta}
\end{figure}

\begin{remark}
    $\rho$ and $\eta$ parameterize respectively the orientation and the convexity of the implied volatility smile as illustrated in Figure \ref{fig:influence_rho_eta}.
\end{remark}

\begin{remark}
    \cite{guyon2022volatility} and \cite{gazzani2024pricing} considered a third feature given by $R_1^2 \mathbb{1}_{\{R_1\ge 0\}}$ in the PDV model in order to achieve a satisfying joint SPX/VIX fit. Adding this feature to explain the variations of our 4 parameters does not improve the $R^2$ scores.  \\
\end{remark}

\begin{table}[h]
    \centering
    \caption{$R^2$ scores of the PDV model on the parameters of the parsimonious SSVI on the train and the test sets.}
    \label{tab:ssvi_r2_scores}
    \begin{tabular}{@{}ccccccc@{}}
    \toprule
    & \multicolumn{2}{c}{S\&P 500} & \multicolumn{2}{c}{Euro Stoxx 50} \\ \toprule
        &  Train & Test &  Train & Test   \\ \midrule
    $a$   &    93.7\% & 72.9\%        & 85.7\% & 49.4\%    \\
    $p$   &    76.2\%    & 47.0\% &    76.3\%    & 75.3\%    \\
    $\rho$ &    50.7\%    & -400.1\%    &  35.9\%    & 19.7\%    \\
    $\eta$ &    63.6\%    & 18.6\%    &   58.0\%    & -899.5\%    \\ \bottomrule
\end{tabular}
\end{table}

\subsection{Specification of the model for the underlying asset price and the IVS}\label{sec:dynamics_specification}
\subsubsection{Dynamics of the underlying asset price}
Based on the analysis in the previous section, we provide here the dynamics of the four parameters in the parsimonious SSVI model. Since two of these parameters depend on the past path of the underlying asset price, we also need a model for the dynamics of the underlying asset price in order to be able to simulate IVSs over time. Because we want these simulations to be realistic, the model for the underlying asset price should also be as realistic as possible. We opt for the PDV model by Guyon and Lekeufack (more precisely, a variant of their model as we will see) as it allows to replicate almost all historical stylized facts of equity prices (leverage effect, volatility clustering, weak and strong Zumbach effects). The asset price $(S_t)_{t\ge 0}$ is assumed to evolve as follows:
\begin{equation}\label{eq:underlying_dynamics}
    \left\{ 
    \begin{array}{rcl}
        \displaystyle\frac{dS_t}{S_t} &=& \mu_Sdt + \sigma_t dW^S_t \\
        \sigma_t &=&  \beta_0^{\sigma} + \beta_1^{\sigma} R^{\sigma}_{1,t}+\beta_2^{\sigma} \Sigma_t^{\sigma} \\
        R_{1,t}^{\sigma} &=& \displaystyle\int_{-\infty}^t \frac{Z_{\alpha_1^{\sigma},\delta_1^{\sigma}}}{(t-u+\delta_1^{\sigma})^{\alpha_1^{\sigma}}}\times \frac{dS_u}{S_u} \\
        \Sigma_t^{\sigma} &=& \sqrt{\displaystyle\int_{-\infty}^t \frac{Z_{\alpha_2^{\sigma},\delta_2^{\sigma}}}{(t-u+\delta_2^{\sigma})^{\alpha_2^{\sigma}}}\times\left(\frac{dS_u}{S_u}\right)^2}
    \end{array}
    \right.
\end{equation}
where $\mu_S$ is the drift, $\sigma_t$ is the instantaneous or spot volatility, $(W^S_t)_{t\ge 0}$ is a Brownian motion and $Z_{\alpha,\delta} = \left(\int_{-\infty}^t \frac{du}{(t-u+\delta)^{\alpha}} \right)^{-1}$. This specification is mostly similar to the one proposed by \cite{guyon2022volatility} but there are several differences. First, we do not approximate the TSPL kernels by linear combinations of exponential kernels. Guyon and Lekeufack make this approximation to recover a Markovian model that is very fast to simulate. We choose not to follow suit since we already achieve reasonable simulation times as we will show in the numerical experiments. Second, Guyon and Lekeufack propose to introduce multiplicative residuals in order to capture the exogenous part of the volatility, i.e. they specify the dynamics of the spot volatility as $\sigma_t = \kappa_t(\beta_0^{\sigma} + \beta_1^{\sigma} R^{\sigma}_{1,t}+\beta_2^{\sigma} \Sigma_t^{\sigma})$ where $(\kappa_t)_{t\ge 0}$ is an Ornstein-Uhlenbeck process or the exponential of an Ornstein-Uhlenbeck process. Again, we propose not to follow suit. Indeed, the estimation of the residuals requires to have realized volatility estimates in order to calibrate the PDV model. But when we use such estimates in the calibration, the log-returns simulated by the calibrated model are not as volatile as the historical log-returns. To be more precise, we implemented the following steps.
\begin{enumerate}
    \item We calibrate the PDV model on S\&P 500 realized volatility estimates (we use those of \citeauthor{heber2009oxford}, \citeyear{heber2009oxford} from 2000 to 2020) using the approach described in Section \ref{sec:calib_meth}.
    \item We calibrate an exponential Ornstein-Uhlenbeck process on the residuals $\kappa_t$ obtained as the ratio of the realized volatility estimates and the predicted volatilities.
    \item We simulate 1000 paths of the asset price $S$ over 3 years with a daily time step according to (\ref{eq:underlying_dynamics}) with the calibrated parameters.
\end{enumerate}
In this setting, the (annualized) standard deviation of the simulated log-returns is 13.2\% while the one of the historical log-returns is 19.9\% (estimated on the S\&P 500 daily log-returns from 2000 to 2020). Note that a similar discrepancy is observed if one considers additive residuals instead of multiplicative residuals. There are two reasons that can explain this discrepancy.
\begin{enumerate}
    \item We observe that when we filter the Brownian increments from the log-returns $\log \frac{S_{t_{i+1}}}{S_{t_i}}$ and the realized volatility estimates $\hat{\sigma}$ as:
    \begin{equation*}
        \hat{W}_{t_{i+1}}-\hat{W}_{t_i} =\frac{1}{\hat{\sigma}_{t_i}} \left( \log \frac{S_{t_{i+1}}}{S_{t_i}}+ \frac{1}{2}\hat{\sigma}_{t_i}^2(t_{i+1}-t_i) \right),
    \end{equation*}
    the tails of the obtained increments are fatter than those of a normal distribution (fitting a Student distribution on these increments actually yield a degree of freedom around 3). 
    \item The distribution of the log-returns does not appear in the target functions that are used in the above-mentioned calibration procedure so there are \textit{a priori} no reason that it will be reproduced by the calibrated model. 
\end{enumerate}
To remedy this problem, we propose to calibrate the parameters of the PDV model by log-likelihood maximization of the log-returns without using any realized volatility data and without considering any residuals in the PDV model. This calibration procedure will be described in more details in Section \ref{sec:model_calibration}. With this calibration procedure, the standard deviation of the simulated log-returns is 18.7\% so slightly below the historical one. The third and last difference with the model of \cite{guyon2022volatility} is the introduction of a drift in the dynamics of the asset price. This is first motivated by the fact that the S\&P 500 and the Euro Stoxx 50 have both significantly rised on the train set: the average annual log-return on the train set (i.e. between March 8, 2012 and December 31, 2020) is 11.5\% for the S\&P 500 and 3.9\% for the Euro Stoxx 50 so it is difficult to achieve a good fit to the data without adding a drift. Besides, we observed that without incorporating a drift, the model was unable to produce persistent periods of low volatility historically observed (for example from 2017 to 2018 for the S\&P 500, see Figure \ref{fig:data_sets}) which resulted in a simulated implied volatility that was, on average, higher than the historical one. When a positive drift is added\footnote{In practice, the calibrated drift $\mu_S$ is positive for both the S\&P 500 and the Euro Stoxx 50.}, the feature $R_1$ is indeed more likely to take positive values which is then reducing the volatility given that $\beta_1^{\sigma}<0$. As a consequence of this drift, the asset price is not a martingale. However, this is not an issue here as we are working under the real-world probability measure instead of a risk-neutral one. 

\begin{remark}
    As already mentioned in the introduction, for now, there does not exist conditions guaranteeing that the volatility $\sigma$ is positive in the PDV model when the two kernels are of the TSPL kind. However, with the calibrated parameters (see Section \ref{sec:model_calibration}), we did not observe any negative value in our simulations. 
\end{remark}

\subsubsection{Dynamics of the parsimonious SSVI parameters}
Since both parameters $a$ and $p$ in the parsimonious SSVI model exhibit a path-dependent behavior with respect to the underlying asset price, we propose the following dynamics for these parameters:
\begin{equation}\label{eq:a_p_dynamics}
    \left\{ 
    \begin{array}{rcll}
        a_t &=& \left|\beta_0^a+\beta_1^aR_{1,t}^a+(\beta_2^a+\varepsilon_t^a)\Sigma_t^a \right| \\
        p_t &=& \exp\left(\beta_0^p+\beta_1^pR_{1,t}^p+(\beta_2^p+\varepsilon_t^p)\Sigma_t^p\right) \\
        R_{1,t}^{i} &=& \displaystyle\int_{-\infty}^t \frac{Z_{\alpha_1^i,\delta_1^i}}{(t-u+\delta_1^i)^{\alpha_1^i}}\times \frac{dS_u}{S_u} & \text{for } i \in \{a,p\}\\
        \Sigma_t^{i} &=& \sqrt{\displaystyle\int_{-\infty}^t \frac{Z_{\alpha_2^i,\delta_2^i}}{(t-u+\delta_2^i)^{\alpha_2^i}}\times\left(\frac{dS_u}{S_u}\right)^2} & \text{for } i \in \{a,p\}
    \end{array}
    \right.
\end{equation}
where $\varepsilon^a$ and $\varepsilon^p$ are time-dependent residuals allowing to capture the variations in $a$ and $p$ that are not due to the past movements in the underlying asset price. Note that we consider that the residuals $\varepsilon^a$ and $\varepsilon^p$ are both multiplying the volatility feature of the PDV model instead of being purely additive terms of the linear regressions. Such a specification allows taking into account the fact that we observe historically larger deviations
of $\beta_0^a+\beta_1^aR_{1,t}^a+\beta_2^a\Sigma_t^a$ (resp. $\beta_0^p+\beta_1^pR_{1,t}^p+\beta_2^p\Sigma_t^p$) from $a_t$ (resp. $\log p_t$) when the volatility feature $\Sigma_t^a$ (resp. $\Sigma_t^p$) is large. Given the high autocorrelation of these residuals, we model both of them as centered Ornstein-Uhlenbeck processes:
\begin{equation}\label{eq:ou_processes}
    d\varepsilon_t^i = -\kappa_i\varepsilon_t^i dt + \gamma_i dW_t^i, \quad i\in \{a,p\}. 
\end{equation}
where $W^a$ and $W^p$ are two correlated Brownian motions that are also correlated to the Brownian motion $W^S$ driving the asset price. 

\begin{remark}
    Note that we consider different TSPL kernels parameters for $\sigma$, $a$ and $p$. Choosing to have common features $R_1$ and $\Sigma$ for $\sigma$, $a$ and $p$ with parameter-specific $\beta$'s is also an option but it requires to simultaneously calibrate the PDV model on the three time series and it would probably reduce the $R^2$ scores in comparison to the ones obtained with a calibration of the PDV model for each of the three variables. \\
\end{remark}

\begin{remark}
    The above specification guarantees that both $a$ and $p$ are always positive. Since the ATM total variance is parameterized as $\theta_T=aT^p$, it implies that $\partial_T \theta_T\ge 0$ so that there is no calendar arbitrage (see Appendix \ref{sec:ssvi_arbitrage}). In our simulations, we observe that $\beta_0^a+\beta_1^a+(\beta_2^a+\varepsilon^a_t)\Sigma_t^a$ is actually always positive so the absolute value is not necessary. 
\end{remark}

The two remaining parameters of the parsimonious SSVI model, namely $\rho$ and $\eta$, are more difficult to predict using the past returns and past squared returns (see Section \ref{sec:pdv_ssvi_params}). Thus, we do not rely on a PDV model to describe their dynamics. We propose to model both parameters as Jacobi processes guaranteeing that $\rho \in [-1,1]$ and $\eta\in [0,\sqrt{2}]$:
\begin{equation}\label{eq:jacobi_processes}
    \left\{
    \begin{array}{rcl}
        d\rho_t &=& \kappa_{\rho}(\mu_\rho -\rho_t)dt  +\gamma_{\rho}\sqrt{(1-\rho_t)(1+\rho_t)} dW_t^{\rho}, \\
        d\eta_t &=& \kappa_{\eta}(\mu_{\eta} - \eta_t)dt +\gamma_{\eta}\sqrt{(\sqrt{2}-\eta_t)\eta_t} dW_t^{\eta}, \\
    \end{array}
    \right.
\end{equation}
where $\kappa_\rho,\kappa_{\eta}>0$, $\mu_{\rho}\in [-1,1]$, $\mu_{\eta}\in[0,\sqrt{2}]$, $\gamma_{\rho},\gamma_{\eta}>0$, $W^{\rho}$ and $W^{\eta}$ are two correlated Brownian motions that are also correlated to $W^S$, $W^a$ and $W^p$. This modelling choice is motivated by two reasons. First, both parameters are highly autocorrelated. Second, we recall that the parsimonious SSVI model is free of static arbitrage provided that $\eta^2(1+|\rho|) \le 4$. A sufficient condition for this condition to be satisfied is therefore: $\rho \in [-1,1]$ (which is anyway the domain of definition of $\rho$) and $\eta \in [0,\sqrt{2}]$. It turns out that this condition is always satisfied by the calibrated $\rho$ and $\eta$. Hence, by using the Jacobi processes as specified above, we make sure that the simulated implied volatility surfaces are free from static arbitrage.

\begin{remark}
    The model obtained by combining Equations (\ref{eq:underlying_dynamics}), (\ref{eq:a_p_dynamics}), (\ref{eq:ou_processes}) and (\ref{eq:jacobi_processes}) is called the path-dependent SSVI model. 
\end{remark}

\subsection{Model calibration}\label{sec:model_calibration}
This section details a calibration methodology for all the parameters involved in the path-dependent SSVI model whose dynamics has been specified in the previous section. \\

We discretize the asset price along a time grid $(t_k)_{k\ge 0}$ by the specifying the log-returns $r_{t_{k+1}}=\log \frac{S_{t_{k+1}}}{S_{t_k}}$ as:
\begin{equation*}
    r_{t_{k+1}} = \left(\mu_S-\frac{\sigma_{t_k}^2}{2}\right)(t_{k+1}-t_k) + \sigma_{t_k} \sqrt{t_{k+1}-t_k} \varepsilon_{t_{k+1}}^S
\end{equation*}
where $\sigma_{t_k} = \beta_0^{\sigma}+\beta_1^{\sigma}R_{1,t_k}^{\sigma}+\beta_2^{\sigma}\Sigma_{t_k}^{\sigma}$ and $(\varepsilon^S_{t_k})_{1\le k \le N}$ are i.i.d. standard normal random variables. The features $R_1^{\sigma}$ and $\Sigma^{\sigma}$ are discretized and truncated as:
\begin{equation*}
    R_{1,t} = \sum_{t-C_{R_1}^{\sigma}\le t_k \le t} \frac{Z_{\alpha_1^{\sigma},\delta_1^{\sigma}}}{(t-t_k+\delta_1^{\sigma})^{\alpha_1^{\sigma}}} r_{t_k} \text{ and } \Sigma_{t} = \sqrt{\sum_{t-C_{\Sigma}^{\sigma}\le t_k \le t} \frac{Z_{\alpha_2^{\sigma},\delta_2^{\sigma}}}{(t-t_k+\delta_2^{\sigma})^{\alpha_2^{\sigma}}} r_{t_k}^2}.
\end{equation*}
The parameters $(\mu_S,\alpha_1^{\sigma},\delta_1^{\sigma},\alpha_2^{\sigma},\delta_2^{\sigma},\beta_0^{\sigma},\beta_1^{\sigma},\beta_2^{\sigma})$ are estimated by log-likelihood maximization of the log-returns. Indeed, with the above discretizations, the distribution of $r_{t_{k+1}}$ conditionally on all the information available at time $t_k$ is a normal distribution with mean $ \left(\mu_S-\frac{\sigma_{t_k}^2}{2}\right)(t_{k+1}-t_k)$ and variance $\sigma_{t_k}^2 (t_{k+1}-t_k)$. Thus, it is possible to write explicitly the log-likelihood of the vector of observed log-returns $(r^h_{t_1},\dots,r^h_{t_N})$ as a function of the drift and the 7 PDV parameters and to maximize it using a numerical optimization routine. The cut-off lags $C_{R_1}^{\sigma}$ and $C_{\Sigma}^{\sigma}$ are estimated on the realized volatility estimates from \cite{heber2009oxford} using the approach described in Appendix \ref{sec:influence_cutoff}.\\

The calibration methodology of the PDV parameters $\alpha_1^i$, $\delta_1^i$, $\alpha_2^i$, $\delta_2^i$, $\beta_0^i$, $\beta_1^i$ and $\beta_2^i$ for $i\in \{a,p\}$ in Equation (\ref{eq:a_p_dynamics}) is the same as the one exposed in Section \ref{sec:calib_meth}. The cut-off lags $C_{R_1}^i$ and $C_{\Sigma}^i$ are those presented in Appendix \ref{sec:hyperparams_ssvi}. We deduce the historical time series of $\varepsilon^a$ and $\varepsilon^p$ from this calibration. The parameters $\kappa_i$ and $\gamma_i$ of the Ornstein-Uhlenbeck processes in Equation (\ref{eq:ou_processes}) are then calibrated on these time series using the maximum likelihood estimators from \cite{tang2009parameter}. The parameters $\kappa_i$, $\mu_i$ and $\gamma_i$ for $i\in\{\rho,\eta\}$ are calibrated using the following estimators\footnote{Note that these estimators generalize the ones developed by \cite{wei2016} in the case of the CIR process. } which are obtained by maximizing the log-likelihood of a Jacobi process discretized using the Euler-Maruyama scheme: 
\begin{equation*}
    \begin{array}{rcl}
        \hat{\kappa}_i &=& \displaystyle\frac{1}{\Delta} \times \frac{
             \displaystyle\sum_{k=0}^{N-1} \frac{x^i_{t_{k+1}} - x^i_{t_k}}{Q_i\left(x^i_{t_k}\right)} 
             \displaystyle\sum_{k=0}^{N-1} \frac{x^i_{t_k}}{Q_i\left(x^i_{t_k}\right)}  
            - \displaystyle\sum_{k=0}^{N-1} \frac{(x^i_{t_{k+1}} - x^i_{t_k}) x^i_{t_k}}{Q_i\left(x^i_{t_k}\right)}
            \displaystyle\sum_{k=0}^{N-1} \frac{1}{Q_i\left(x^i_{t_k}\right)} 
            }{
            \displaystyle\sum_{k=0}^{N-1} \frac{1}{Q_i\left(x^i_{t_k}\right)}  
             \displaystyle\sum_{k=0}^{N-1} \frac{\left(x^i_{t_k}\right)^2}{Q_i\left(x^i_{t_k}\right)} 
            - \left( \displaystyle\sum_{k=0}^{N-1} \frac{x^i_{t_k}}{Q_i\left(x^i_{t_k}\right)} \right)^2
            },\\
    \hat{\mu}_i &=& \displaystyle\frac{1}{\hat{\kappa}_i\Delta} \times \frac{\displaystyle\sum_{k=0}^{N-1}\frac{x^i_{t_{k+1}}-x^i_{t_k}+\hat{\kappa}_i x^i_{t_k}\Delta}{Q_i\left(x^i_{t_k}\right)}}{\displaystyle\sum_{k=0}^{N-1}\frac{1}{Q_i\left(x^i_{t_k}\right)}}, \\
    \hat{\gamma}_i &=& \displaystyle\frac{1}{N\Delta} \displaystyle\sum_{k=0}^{N-1} \frac{
        \left( x^i_{t_{k+1}} - x^i_{t_k} - \hat{\kappa}_i \left( \hat{\mu}_i - x^i_{t_k} \right) \Delta \right)^2
        }{Q_i\left( x^i_{t_k} \right)}
    \end{array}
\end{equation*}
where $i\in\{\rho,\eta \}$, $(x^i_{t_k})_{k=0,\dots,N}$ are the historical observations of $i$, $Q_{\rho}(x)=(1-x)(1+x)$, $Q_{\eta}(x)=x(\sqrt{2}-x)$ and $\Delta$ is the time step of the grid $(t_k)_{k=0,\dots,N}$ which is assumed to be constant from now on. Finally, once all the above parameters are calibrated, the correlations between the Brownian motions $W^S$, $W^a$, $W^p$, $W^{\rho}$ and $W^{\eta}$ are set to the empirical correlations between the residuals $\hat{\varepsilon}^S$, $\hat{\nu}^a$, $\hat{\nu}^p$, $\hat{\varepsilon}^{\rho}$, $\hat{\varepsilon}^{\eta}$ of the models for $S$, $\varepsilon^a$, $\varepsilon^p$, $\rho$ and $\eta$ respectively. These residuals are computed as:
\begin{equation*}
    \begin{array}{rcll}
        \hat{\varepsilon}^S_{t_{k+1}} &=& \displaystyle\frac{r^h_{t_{k+1}}-\left(\hat{\mu}_S-\frac{\sigma_{t_k}^2}{2}\right)\Delta}{\sigma_{t_k}\sqrt{\Delta}}, \\
        \hat{\nu}^i_{t_{k+1}} &=& \displaystyle\frac{\hat{\varepsilon}^i_{t_{k+1}}-\hat{\varepsilon}^i_{t_{k}}e^{-\hat{\kappa}_i\Delta}}{\hat{\gamma}_i\sqrt{\frac{1-e^{-2\hat{\kappa}_i\Delta}}{2\hat{\kappa}_i}}},& \text{for } i \in \{a,p\}, \\
        \hat{\varepsilon}^x_{t_{k+1}} &=& \displaystyle \frac{x_{t_{k+1}}-x_{t_k}-\hat{\kappa}_x(\hat{\mu}_x-x_{t_k})\Delta}{\hat{\gamma}_x\sqrt{\Delta Q_x(x_{t_k})}}, & \text{for } x \in \{\rho,\eta\},  
    \end{array}
\end{equation*}
where the $\hat{q}$ denotes the estimated value of the parameter $q$. In Table \ref{tab:sp500_params} and \ref{tab:sx5e_params}, we report the calibrated parameters for the S\&P 500 and the Euro Stoxx 50 data sets. The calibration period is the same as the one used for the training of the PDV model in Section \ref{sec:empirical_study}, namely from March 8, 2012 to December 31, 2020.

\begin{table}[htbp]
    \centering
    \caption{Calibrated parameters of the path-dependent SSVI model on the S\&P 500 data set. }
    \label{tab:sp500_params}
    \begin{tabular}{@{}lllllllll@{}}
    \toprule
    \multicolumn{1}{c}{\multirow{2}{*}{$S$}} & $\alpha_1^{\sigma}$ & $\delta_1^{\sigma}$ & $\alpha_2^{\sigma}$ & $\delta_2^{\sigma}$ & $\beta_0^{\sigma}$ & $\beta_1^{\sigma}$ &  $\beta_2^{\sigma}$& $\mu_S$ \\ \cmidrule(l){2-9} 
    \multicolumn{1}{c}{}                   &7.74  & 0.12 & 2.42 & 0.06 & 0.03 & -0.04 & 0.84 & 0.07 \\ \cmidrule(r){1-9}
    \multicolumn{1}{c}{\multirow{2}{*}{$a$}} & $\alpha_1^{a}$ & $\delta_1^{a}$ & $\alpha_2^{a}$ & $\delta_2^{a}$ & $\beta_0^{a}$ & $\beta_1^{a}$ &  $\beta_2^{a}$ \\ \cmidrule(l){2-8} 
    \multicolumn{1}{c}{}                   &1.12  & 0.03 & 0.88 & 0.03 & -8.71$\times 10^{-4}$ & -0.02 & 0.22 \\ \cmidrule(r){1-9}
    \multicolumn{1}{c}{\multirow{2}{*}{$p$}} & $\alpha_1^{p}$ & $\delta_1^{p}$ & $\alpha_2^{p}$ & $\delta_2^{p}$ & $\beta_0^{p}$ & $\beta_1^{p}$ &  $\beta_2^{p}$ \\ \cmidrule(l){2-8} 
    \multicolumn{1}{c}{}                   &1.01  & 0.02 & 0.78 & 0.01 & 0.36 & 0.11 & -1.35 \\ \cmidrule(r){1-9}
    \multicolumn{1}{c}{\multirow{2}{*}{$\varepsilon^a$}} & $\kappa_a$ & $\gamma_a$  \\ \cmidrule(l){2-3} 
    \multicolumn{1}{c}{}                   &16.3  & 0.11  \\ \cmidrule(r){1-9} 
    \multicolumn{1}{c}{\multirow{2}{*}{$\varepsilon^p$}} & $\kappa_p$ & $\gamma_p$  \\ \cmidrule(l){2-3} 
    \multicolumn{1}{c}{}                   &26.4  & 3.24  \\ \cmidrule(r){1-9} 
    \multicolumn{1}{c}{\multirow{2}{*}{$\rho$}} & $\kappa_{\rho}$ & $\mu_{\rho}$ & $\gamma_{\rho}$  \\ \cmidrule(l){2-4} 
    \multicolumn{1}{c}{}                   &60.4  & -0.71 & 0.99  \\ \cmidrule(r){1-9} 
    \multicolumn{1}{c}{\multirow{2}{*}{$\eta$}} & $\kappa_{\eta}$ & $\mu_{\eta}$ & $\gamma_{\eta}$  \\ \cmidrule(l){2-4} 
    \multicolumn{1}{c}{}                   &8.86  & 0.93 & 0.62   \\ 
    \bottomrule
    \end{tabular}
    \end{table}

\begin{table}[htbp]
    \centering
    \caption{Calibrated parameters of the path-dependent SSVI model on the Euro Stoxx 50 data set. }
    \label{tab:sx5e_params}
    \begin{tabular}{@{}lllllllll@{}}
    \toprule
    \multicolumn{1}{c}{\multirow{2}{*}{$S$}} & $\alpha_1^{\sigma}$ & $\delta_1^{\sigma}$ & $\alpha_2^{\sigma}$ & $\delta_2^{\sigma}$ & $\beta_0^{\sigma}$ & $\beta_1^{\sigma}$ &  $\beta_2^{\sigma}$& $\mu_S$ \\ \cmidrule(l){2-9} 
    \multicolumn{1}{c}{}                   &2.01  & 0.03 & 2.01 & 0.05 & 0.04 & -0.04 & 0.80 & 0.06 \\ \cmidrule(r){1-9}
    \multicolumn{1}{c}{\multirow{2}{*}{$a$}} & $\alpha_1^{a}$ & $\delta_1^{a}$ & $\alpha_2^{a}$ & $\delta_2^{a}$ & $\beta_0^{a}$ & $\beta_1^{a}$ &  $\beta_2^{a}$ \\ \cmidrule(l){2-8} 
    \multicolumn{1}{c}{}                   &0.35  & 1.85$\times 10^{-3}$ & 1.04 & 0.07 & -0.01 & -2.67$\times 10^-3$ & 0.25 \\ \cmidrule(r){1-9}
    \multicolumn{1}{c}{\multirow{2}{*}{$p$}} & $\alpha_1^{p}$ & $\delta_1^{p}$ & $\alpha_2^{p}$ & $\delta_2^{p}$ & $\beta_0^{p}$ & $\beta_1^{p}$ &  $\beta_2^{p}$ \\ \cmidrule(l){2-8} 
    \multicolumn{1}{c}{}                   &0.57  & 0.01 & 0.99 & 0.01 & 0.25 & 0.27 & -1.05 \\ \cmidrule(r){1-9}
    \multicolumn{1}{c}{\multirow{2}{*}{$\varepsilon^a$}} & $\kappa_a$ & $\gamma_a$  \\ \cmidrule(l){2-3} 
    \multicolumn{1}{c}{}                   &12.1 & 0.10  \\ \cmidrule(r){1-9} 
    \multicolumn{1}{c}{\multirow{2}{*}{$\varepsilon^p$}} & $\kappa_p$ & $\gamma_p$  \\ \cmidrule(l){2-3} 
    \multicolumn{1}{c}{}                   &17.7  & 2.62  \\ \cmidrule(r){1-9} 
    \multicolumn{1}{c}{\multirow{2}{*}{$\rho$}} & $\kappa_{\rho}$ & $\mu_{\rho}$ & $\gamma_{\rho}$  \\ \cmidrule(l){2-4} 
    \multicolumn{1}{c}{}                   &12.7  & -0.56 & 0.31  \\ \cmidrule(r){1-9} 
    \multicolumn{1}{c}{\multirow{2}{*}{$\eta$}} & $\kappa_{\eta}$ & $\mu_{\eta}$ & $\gamma_{\eta}$  \\ \cmidrule(l){2-4} 
    \multicolumn{1}{c}{}                   &3.89  & 0.83 & 0.53  \\ 
    \bottomrule
    \end{tabular}
    \end{table}

\subsection{Numerical results}\label{sec:num_results}

The aim of this section is to provide some evidence of the consistency of the proposed model with historical underlying prices and IVSs data.  \\

\subsubsection{Simulation of the implied volatility surface conditionally on the historical path of the underlying asset price}\label{sec:cond_simus}
\begin{figure}[htbp]
    \centering
    \subfigure[Historical path of the S\&P 500 ATM term structure]{\includegraphics[width=0.45\linewidth]{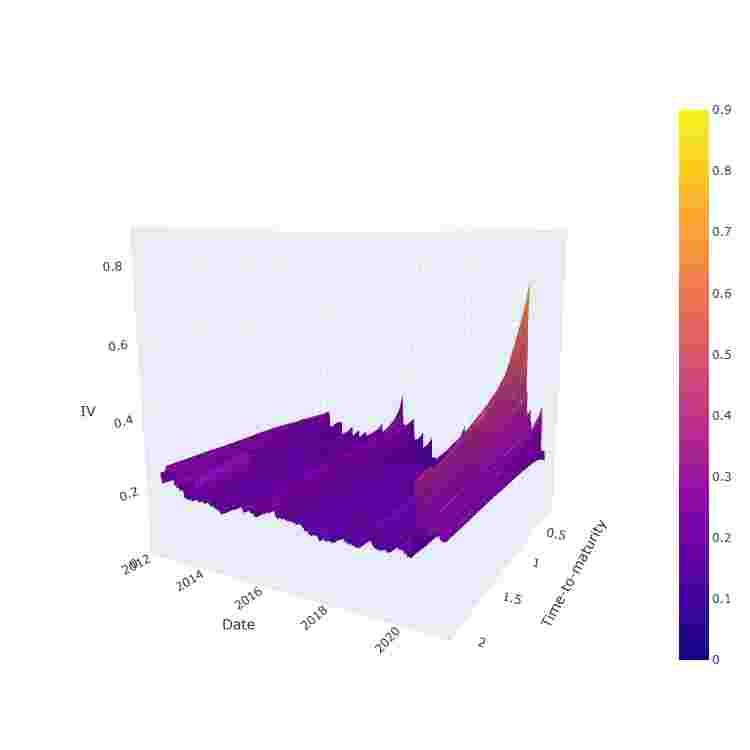}}
    \subfigure[Average path of the S\&P 500 ATM term structure]{\includegraphics[width=0.45\linewidth]{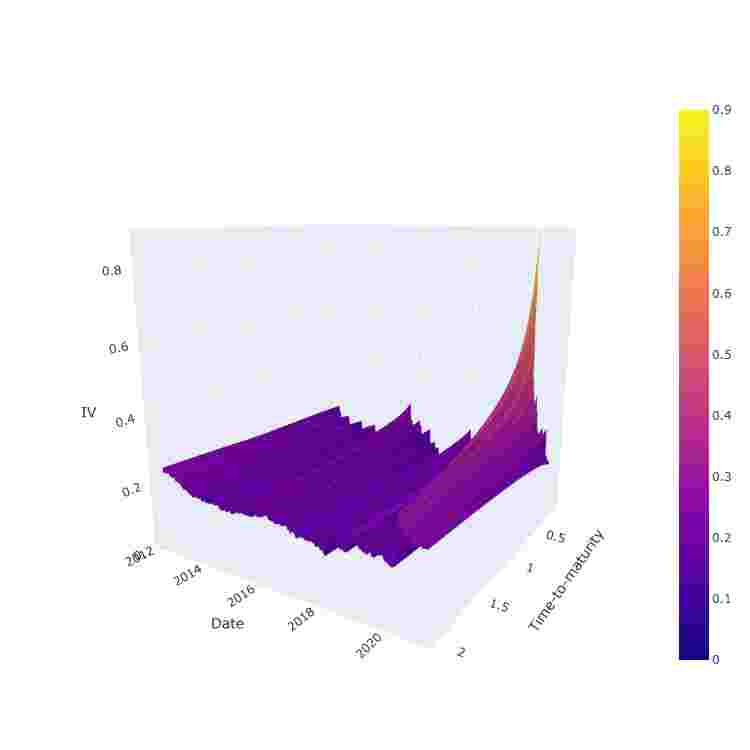}}
    \subfigure[Historical path of the Euro Stoxx 50 ATM term structure]{\includegraphics[width=0.45\linewidth]{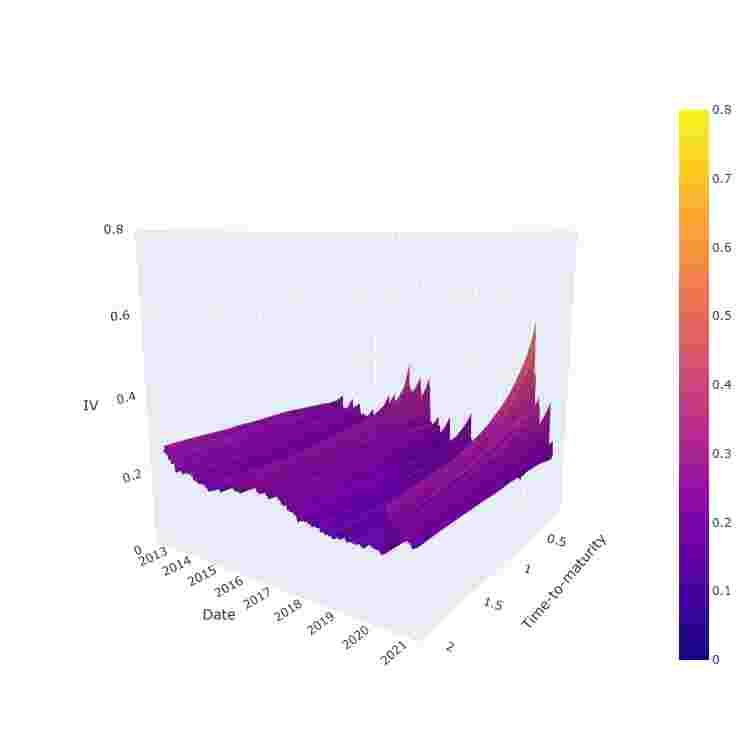}}
    \subfigure[Average path of the Euro Stoxx 50 ATM term structure]{ \includegraphics[width=0.45\linewidth]{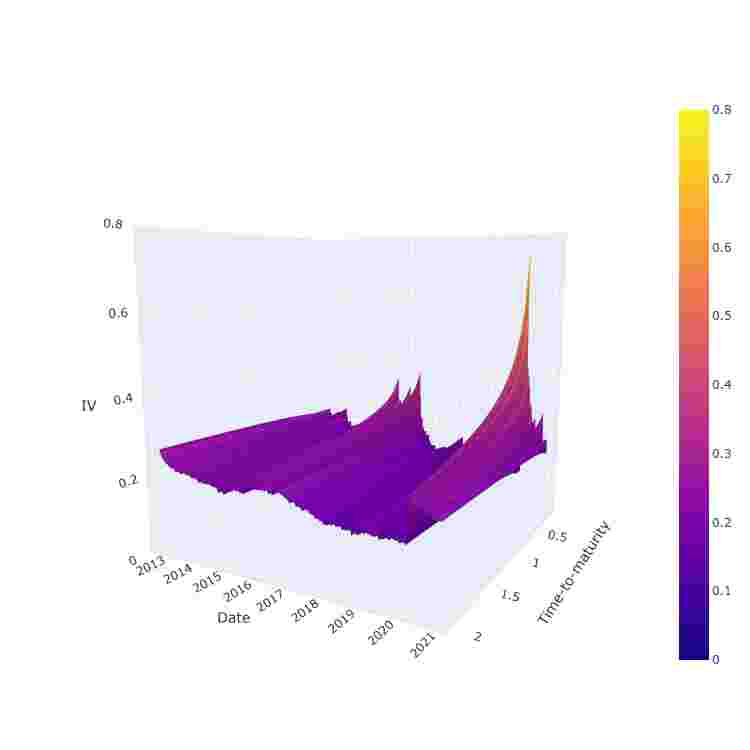}}
    \caption{Comparison of the historical IVS ATM term structure to the average one simulated by the path-dependent SSVI model conditionally on the underlying historical path. }
\label{fig:cond_sample_paths}
\end{figure}
Using the calibrated parameters, we first simulate 1000 daily (i.e. the simulation time step is 1/252) trajectories of $(\varepsilon^a,\varepsilon^p,\eta,\rho)$ over 2185 days (the length of the train data sets) conditionally on the path of the S\&P 500 index between March 22, 2006 and December 31, 2020 and the path of the Euro Stoxx 50 index between September 26, 2006 and December 31, 2020 (the period from March 8, 2012 to December 31, 2020 corresponds to the one being used for the calibration of the path-dependent SSVI model and the period before is the one required for computing the features $R_1$ and $\Sigma$). The Ornstein-Uhlenbeck processes are simulated exactly while the Jacobi process is simulated using the full truncation Euler scheme of \cite{lord2010}. In Figure \ref{fig:cond_sample_paths}, we compare the historical time evolution of the ATM implied volatility curve as a function of the time-to-maturity (in the sequel, we refer to this curve as the IVS ATM term structure) to the one of the average path of the path-dependent SSVI model. We observe that the historical and the simulated paths are visually very close in terms of the level, the amplitude of the variations, the regularity and the overall shape. Moreover, the spikes of the implied volatility due to a drop in the underlying asset price (in particular in March 2020) are well reproduced. \\

\begin{figure}[htbp]
    \centering
    \subfigure[1-month ATM implied volatility (S\&P 500)]{ \includegraphics[width=0.3\linewidth]{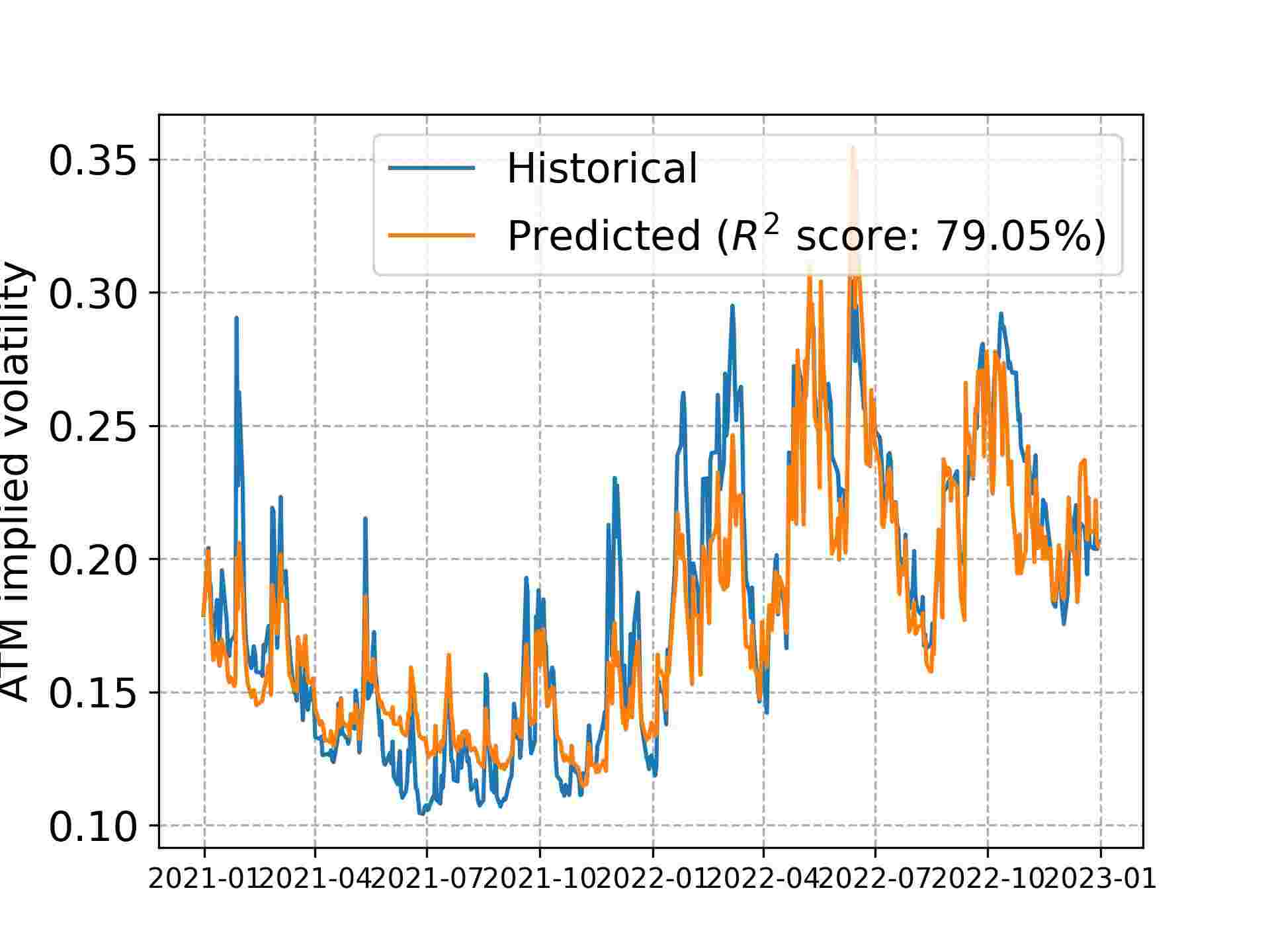}}
    \subfigure[12-months ATM implied volatility (S\&P 500)]{ \includegraphics[width=0.3\linewidth]{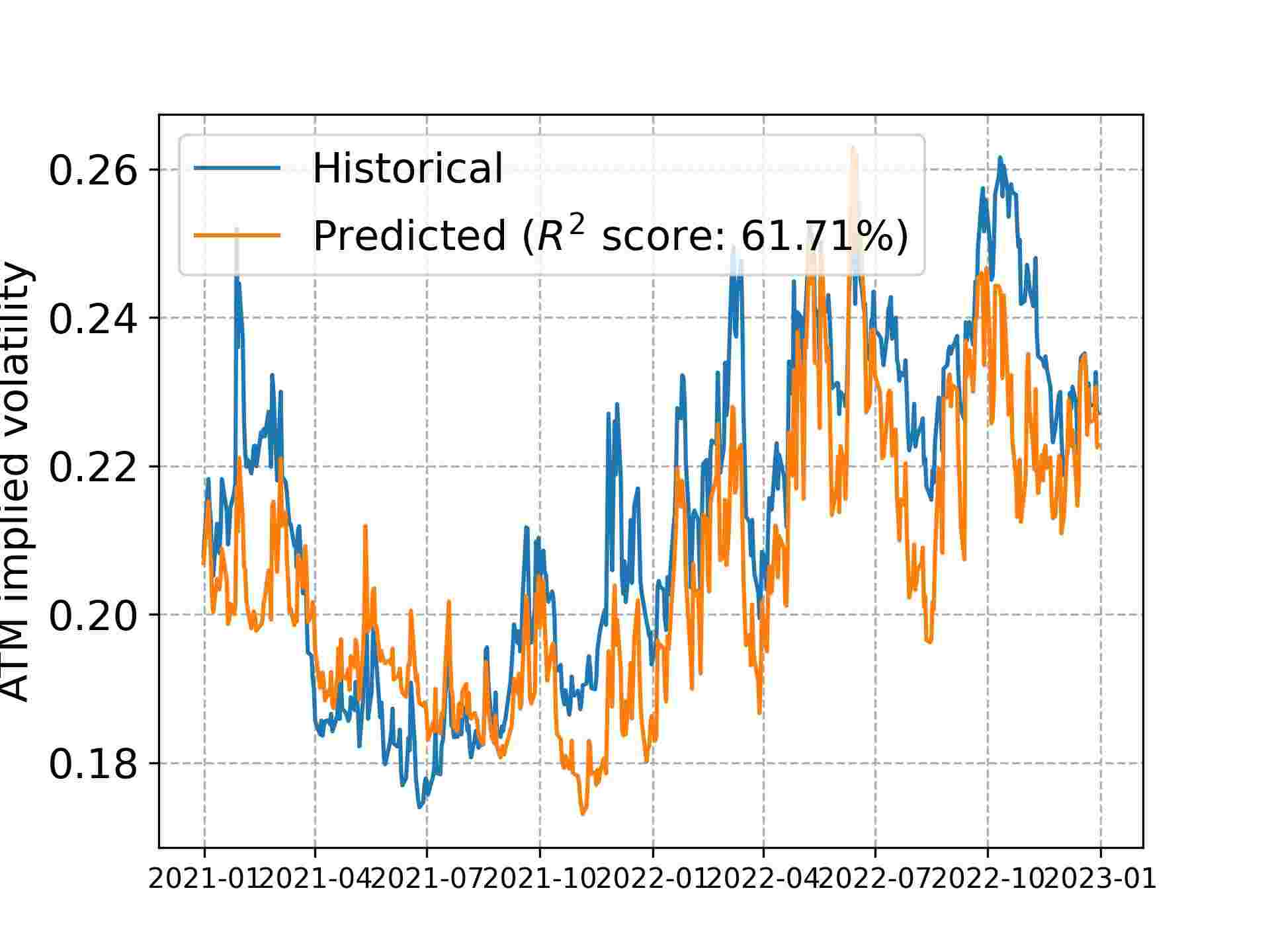}}
    \subfigure[24-months ATM implied volatility (S\&P 500)]{ \includegraphics[width=0.3\linewidth]{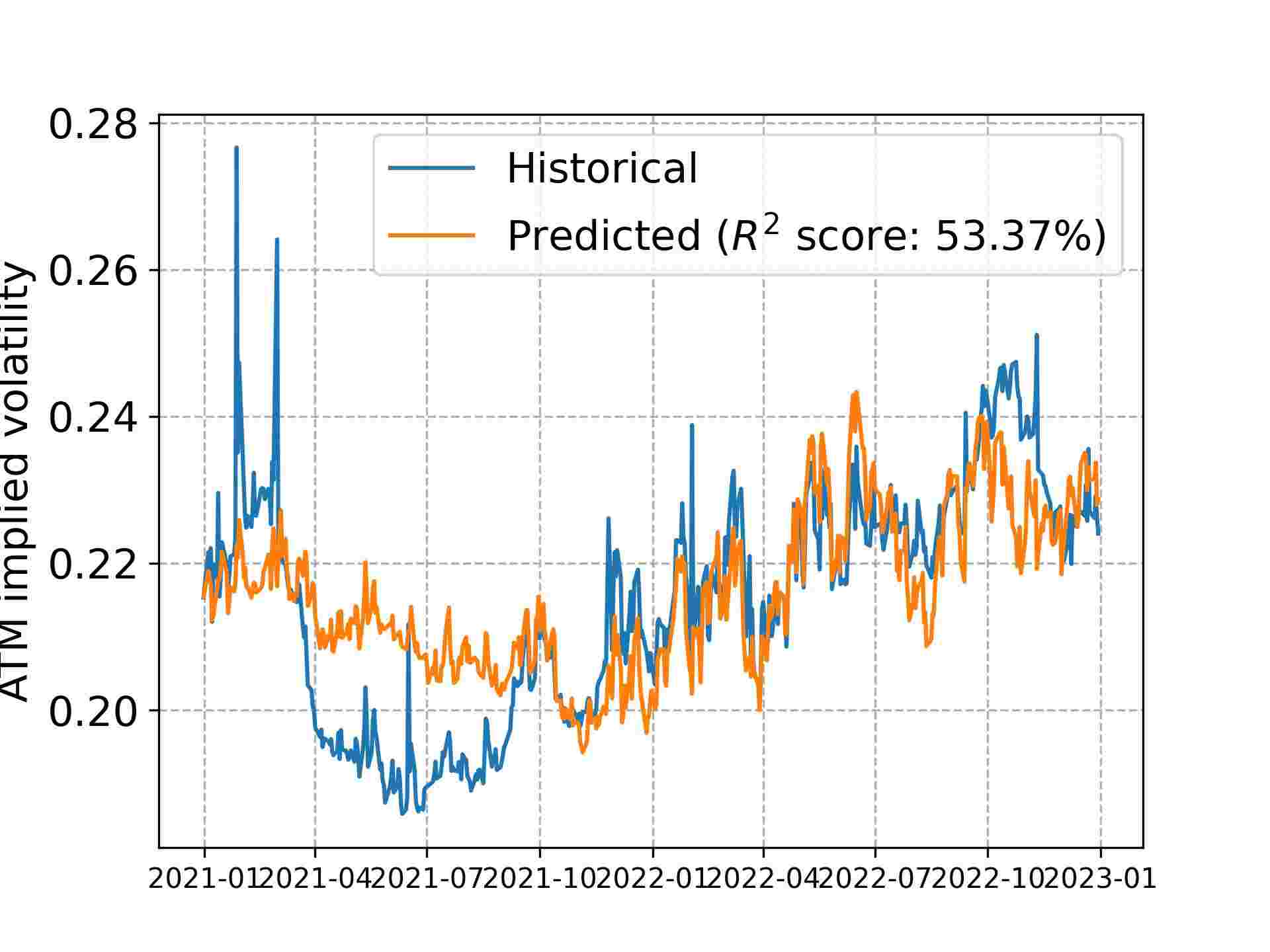}}
    \subfigure[1-month ATM implied volatility (Euro Stoxx 50)]{ \includegraphics[width=0.3\linewidth]{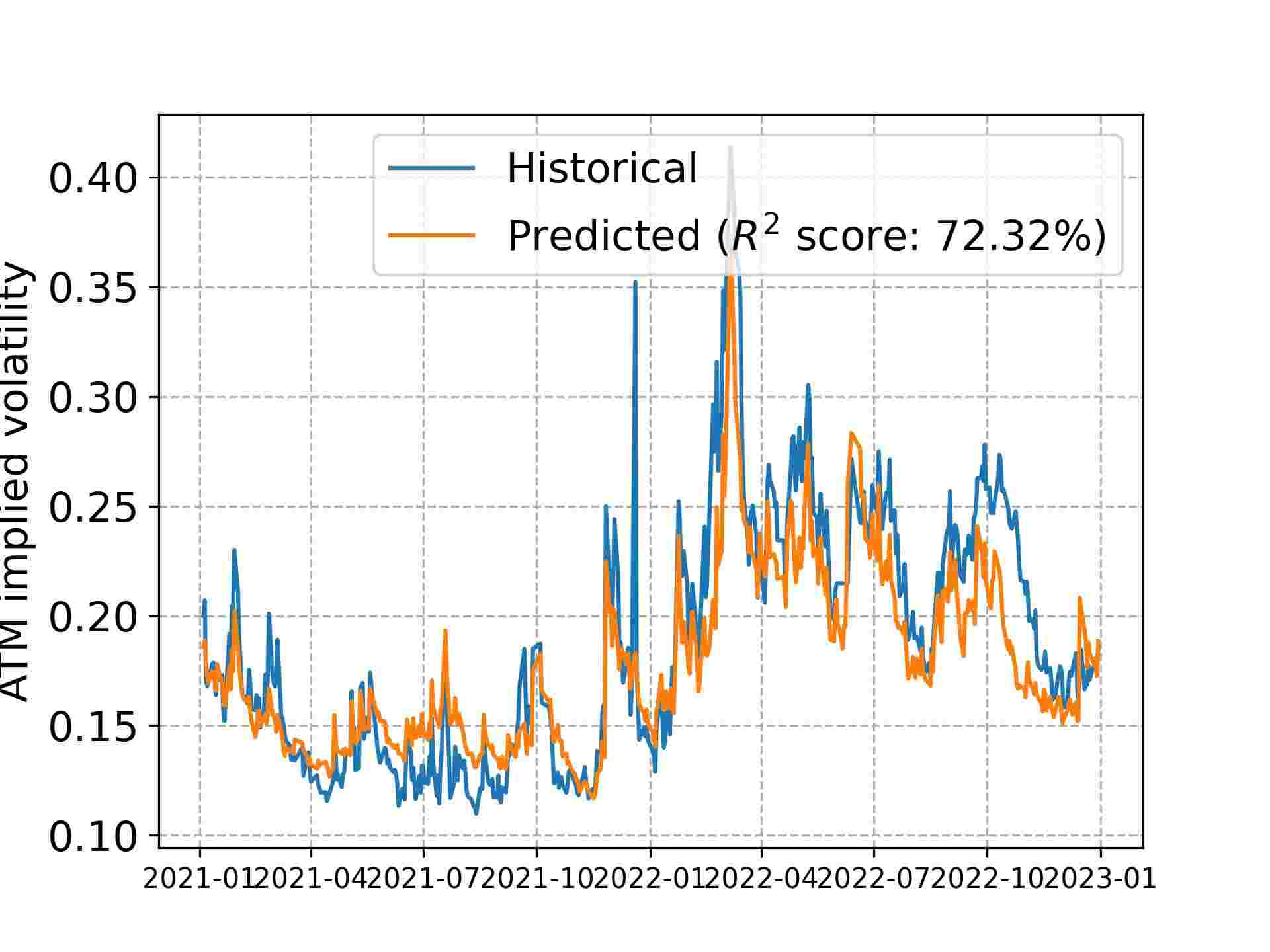}}
    \subfigure[12-months ATM implied volatility (Euro Stoxx 50)]{\includegraphics[width=0.3\linewidth]{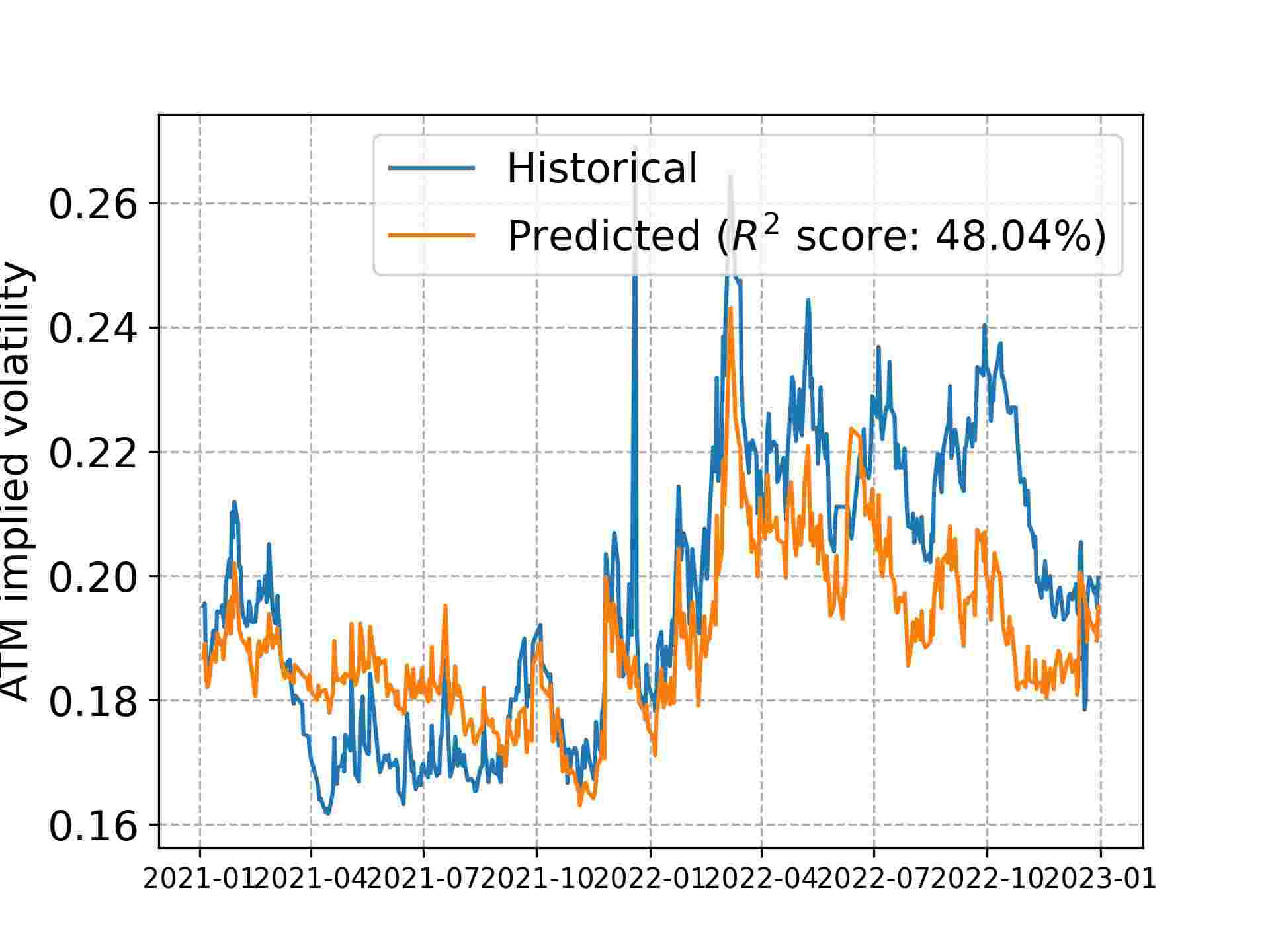}}
    \subfigure[24-months ATM implied volatility (Euro Stoxx 50)]{ \includegraphics[width=0.3\linewidth]{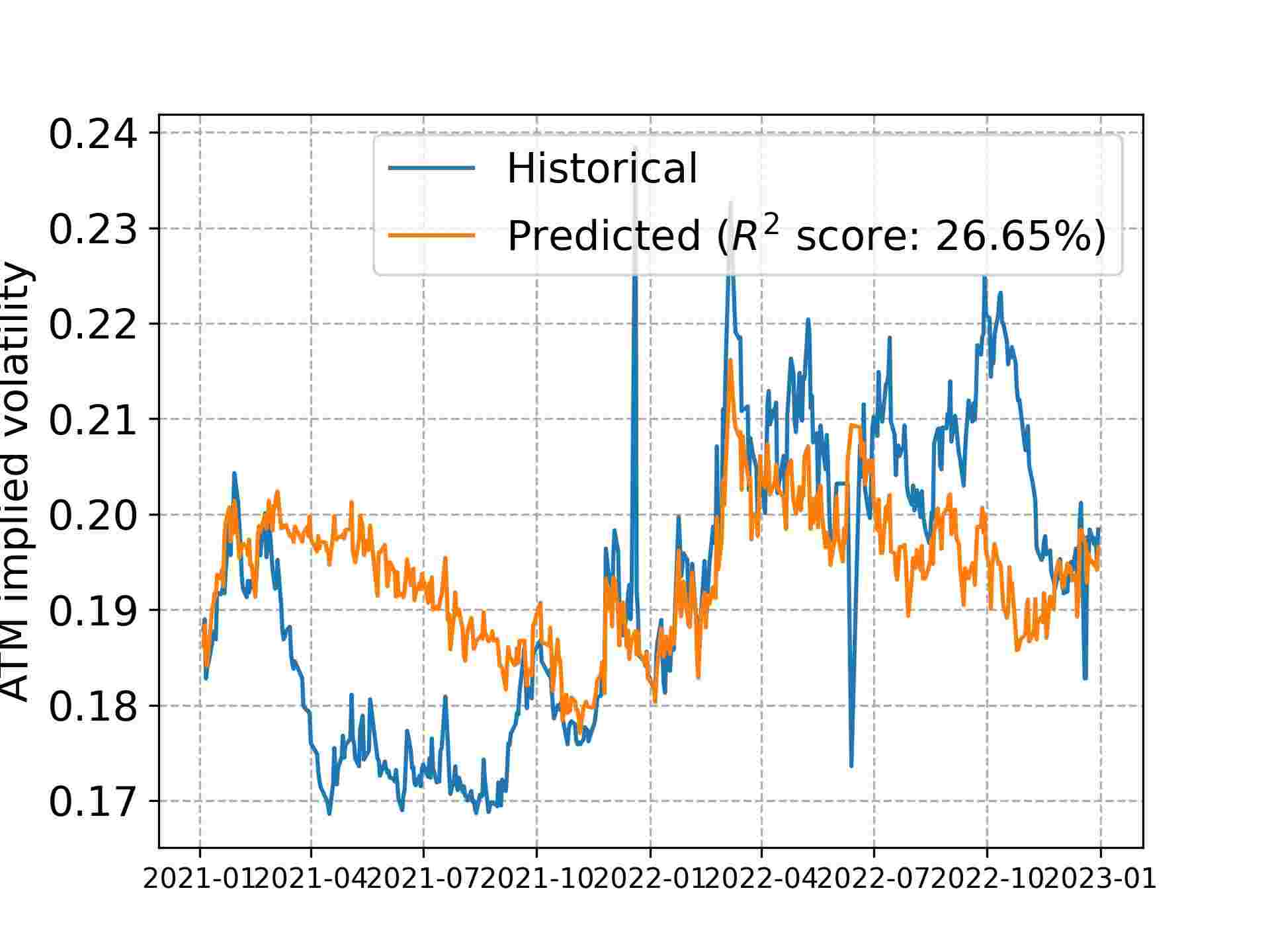}}
    \caption{Out-of-sample prediction of the ATM implied volatility for several time-to-maturities. The predicted ATM implied volatility is obtained by averaging the 1000 paths.}
\label{fig:iv_prediction}
\end{figure}

Second, we simulate 1000 daily trajectories of $(\varepsilon^a,\varepsilon^p,\eta,\rho)$ over 2 years conditionally on the path of the S\&P 500 index between January 14, 2015 and December 31, 2022 and the path of the Euro Stoxx 50 index between February 25, 2015 and December 31, 2022. This second simulation batch allows an out-of-sample comparison while the first one allowed an in-sample comparison. In Figure \ref{fig:iv_prediction}, we demonstrate the out-of-sample predictive power of our model for 1-month, 12-month and 24-month ATM implied volatilities. The obtained $R^2$ scores are a bit smaller than those displayed in empirical study in Section \ref{sec:overall_perf} but let us emphasize here that it is not the ATM implied volatility for each time-to-maturity that is modelled using the PDV model but the parameters $a$ and $p$ of the parsimonious SSVI.

\subsubsection{Simulation of both the asset price and the implied volatility surface}

\begin{figure}
    \centering
    \subfigure[Sample path of the S\&P 500 ATM term structure]{\includegraphics[width=0.45\linewidth]{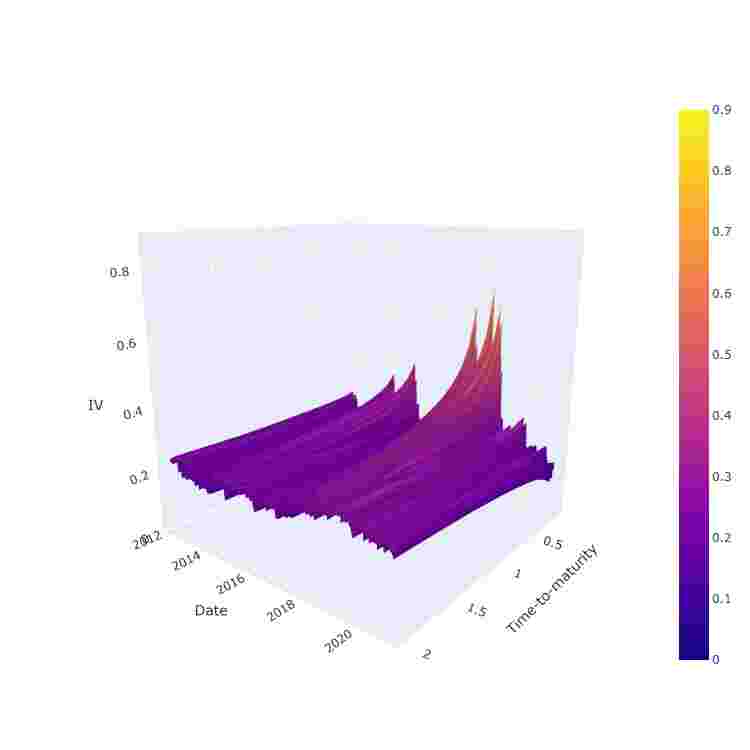}}
    \subfigure[Sample path of the Euro Stoxx 50 ATM term structure]{\includegraphics[width=0.45\linewidth]{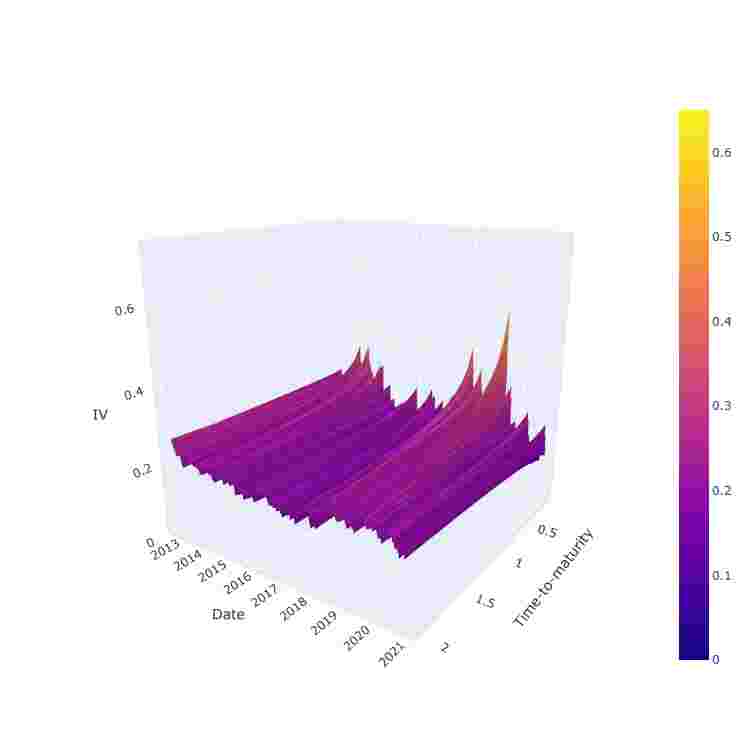}}
    \caption{ATM term structure for two IVS sample paths in the complete path-dependent SSVI model.}
\label{fig:ivs_sample_paths}
\end{figure}

\begin{figure}
    \centering
    \subfigure[1-month ATM implied volatility (S\&P 500)]{ \includegraphics[width=0.25\linewidth]{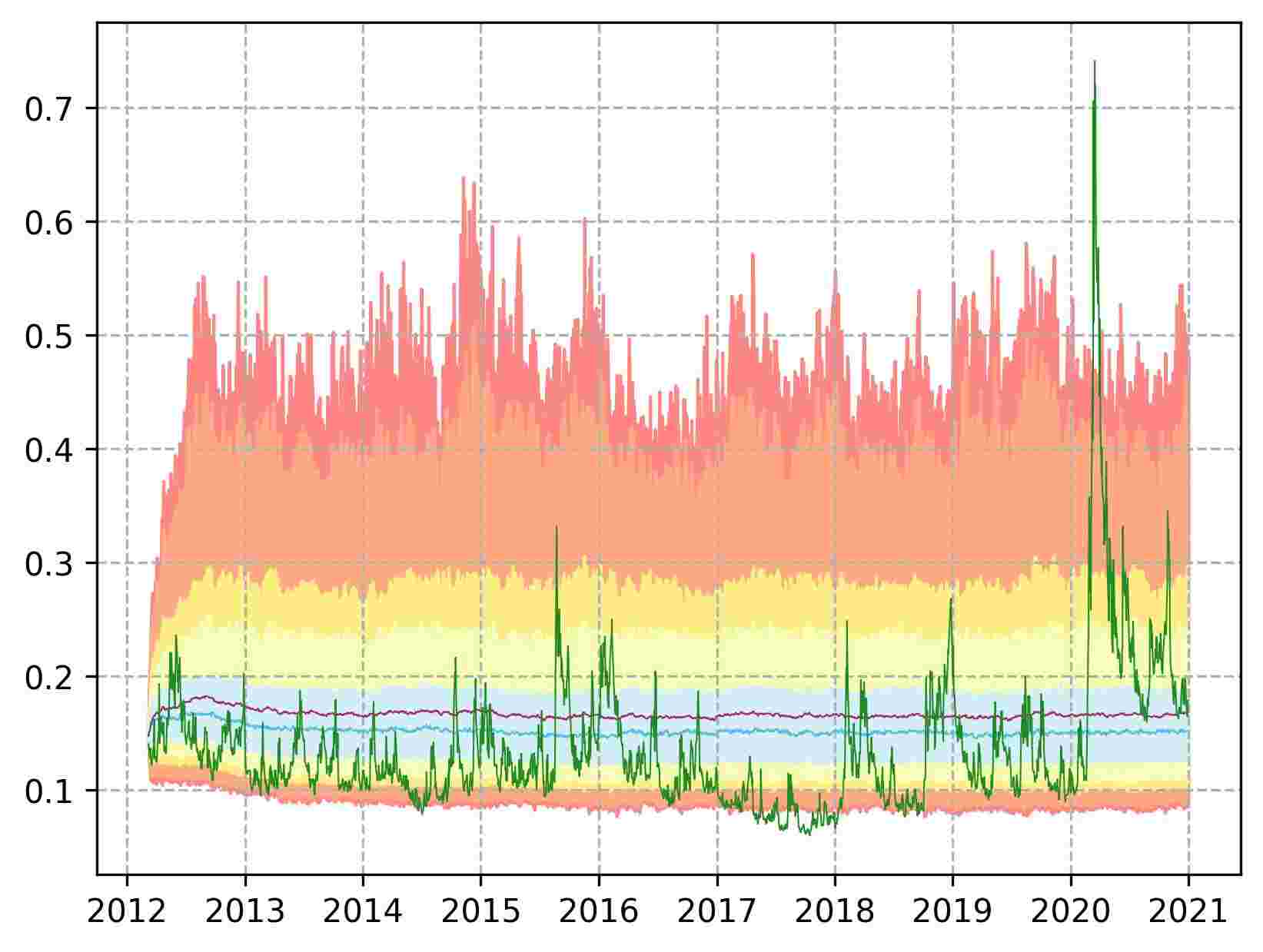}}
    \subfigure[12-months ATM implied volatility (S\&P 500)]{ \includegraphics[width=0.25\linewidth]{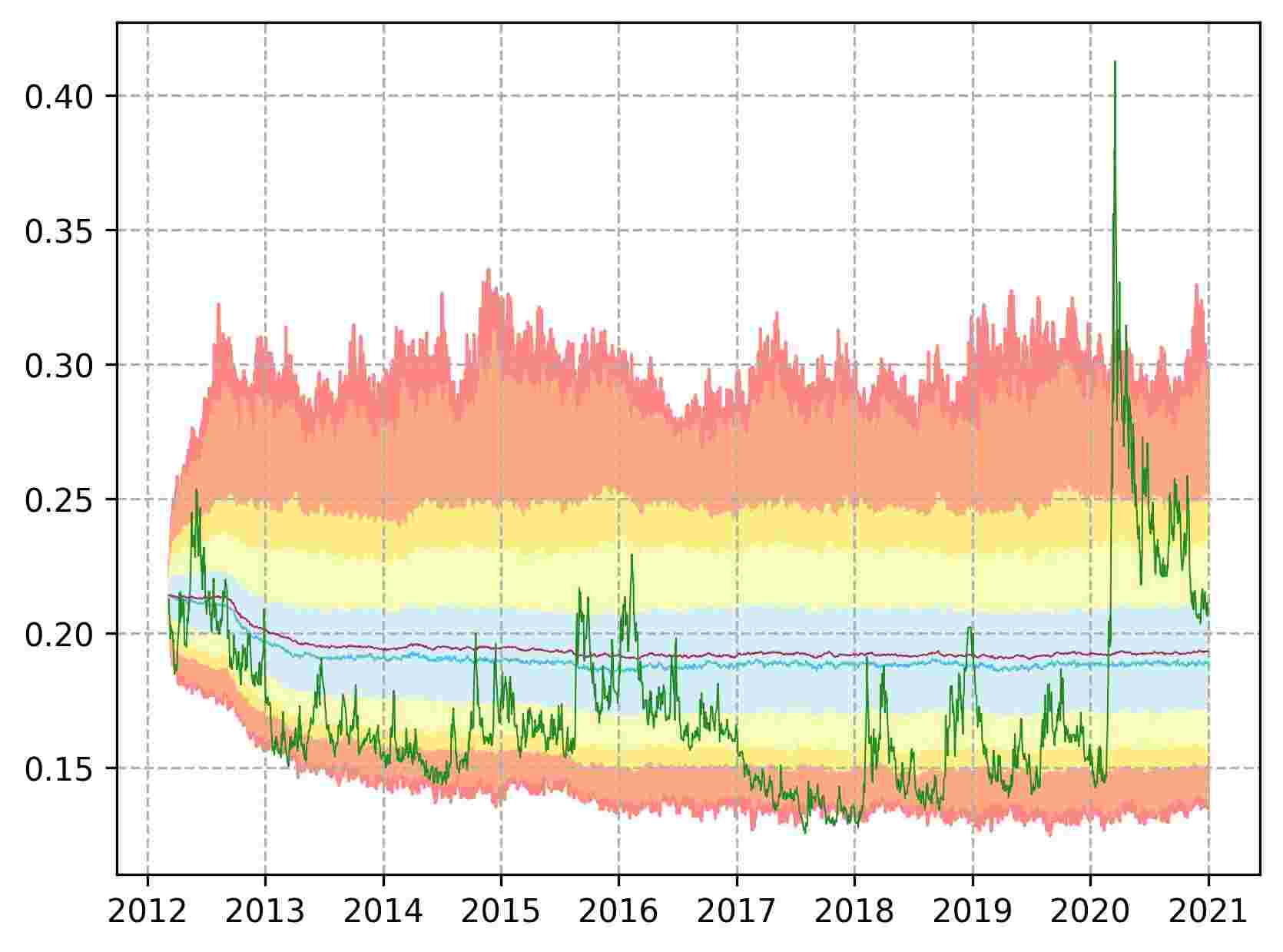}}
    \subfigure[24-months ATM implied volatility (S\&P 500)]{ \includegraphics[width=0.36\linewidth]{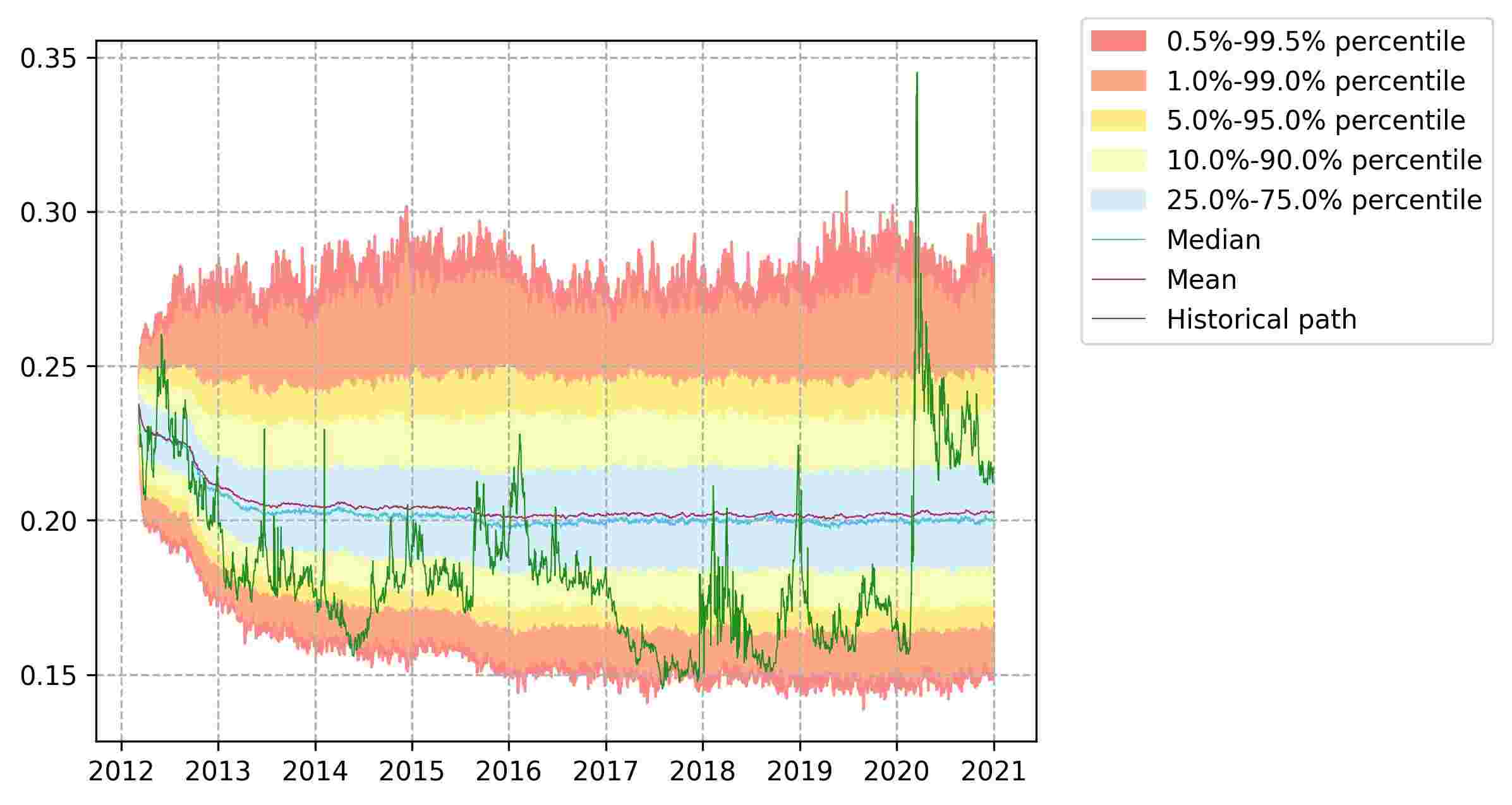}}
    \subfigure[1-month ATM implied volatility (Euro Stoxx 50)]{ \includegraphics[width=0.25\linewidth]{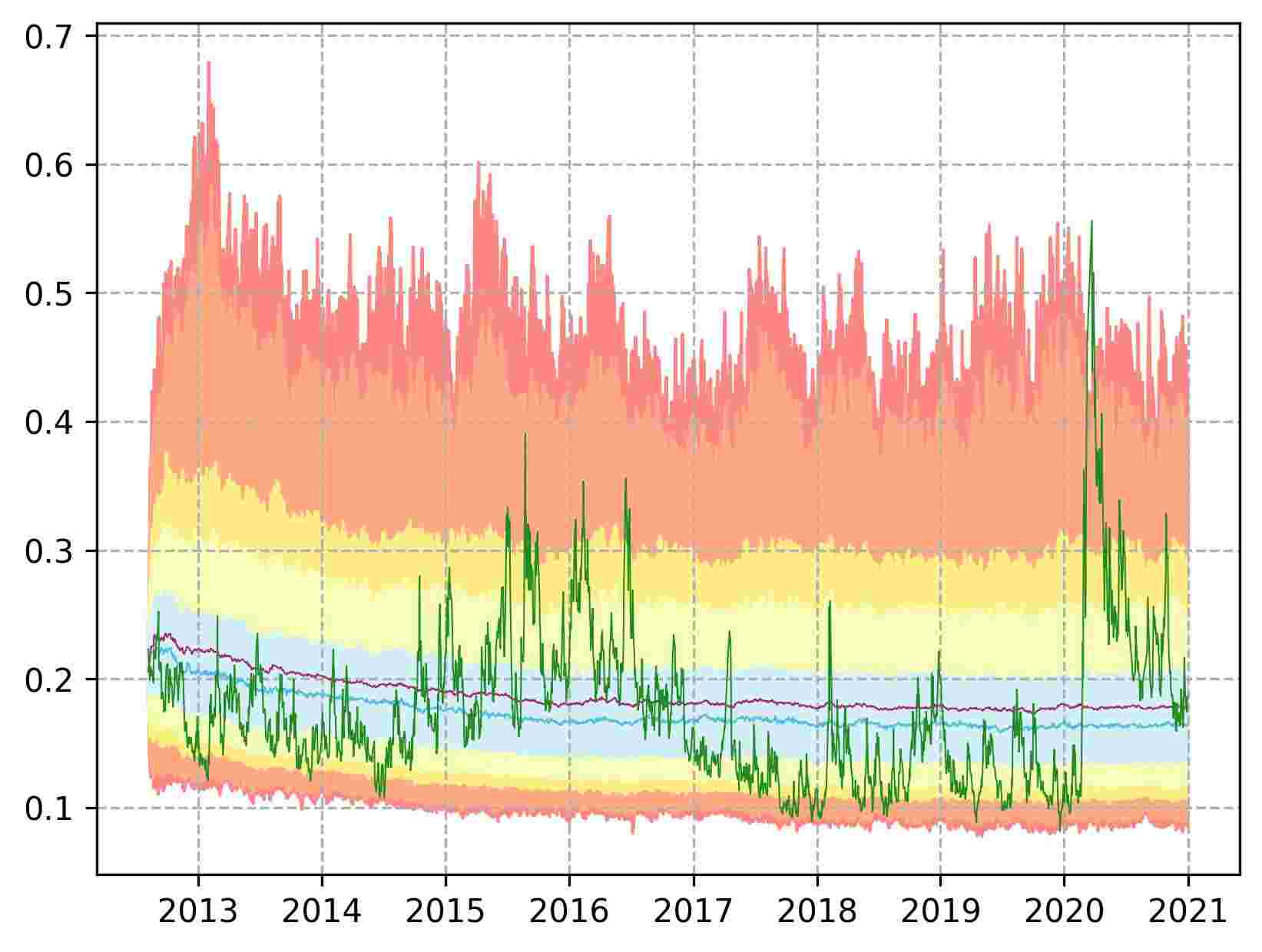}}
    \subfigure[12-months ATM implied volatility (Euro Stoxx 50)]{\includegraphics[width=0.25\linewidth]{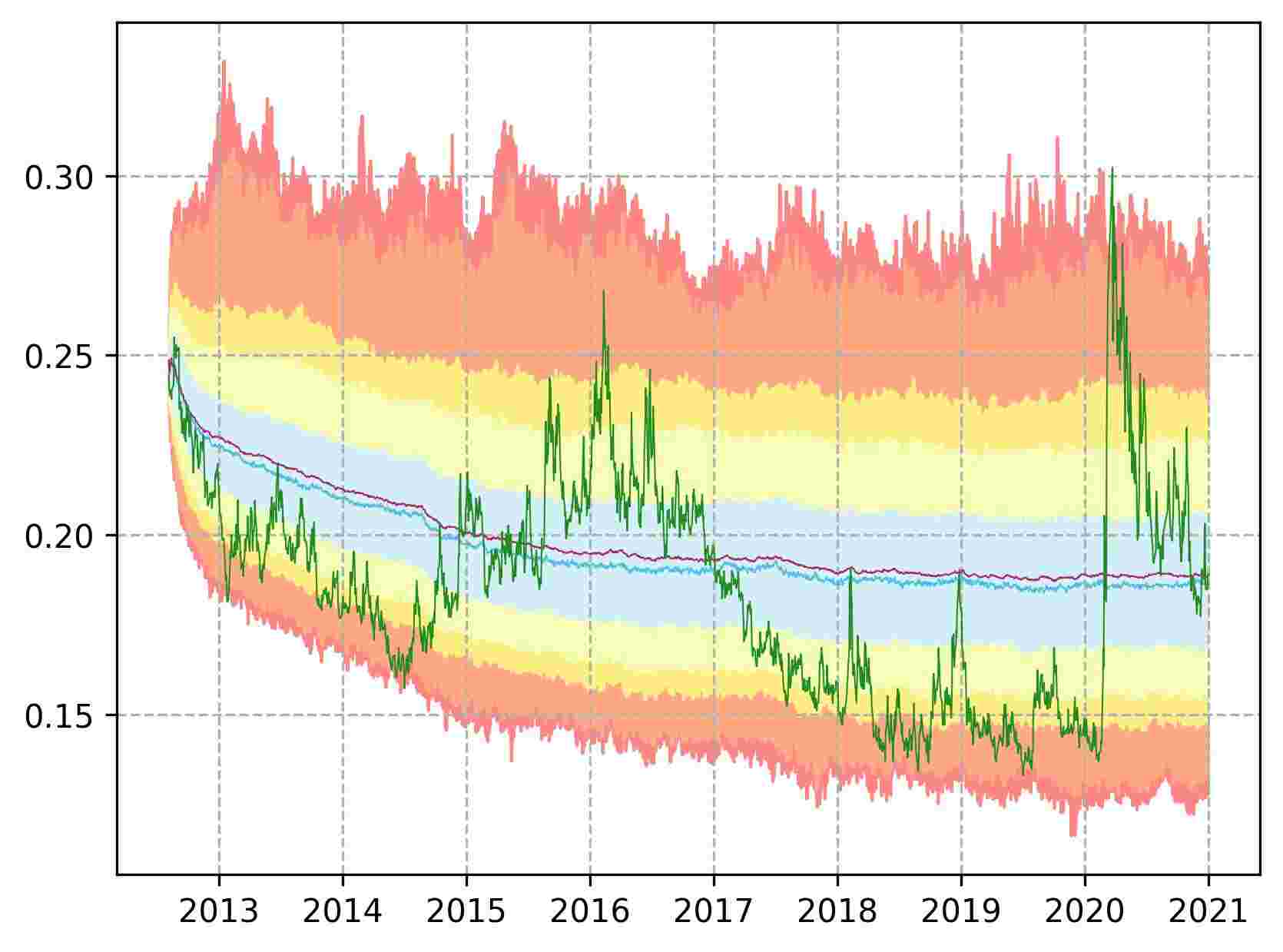}}
    \subfigure[24-months ATM implied volatility (Euro Stoxx 50)]{ \includegraphics[width=0.36\linewidth]{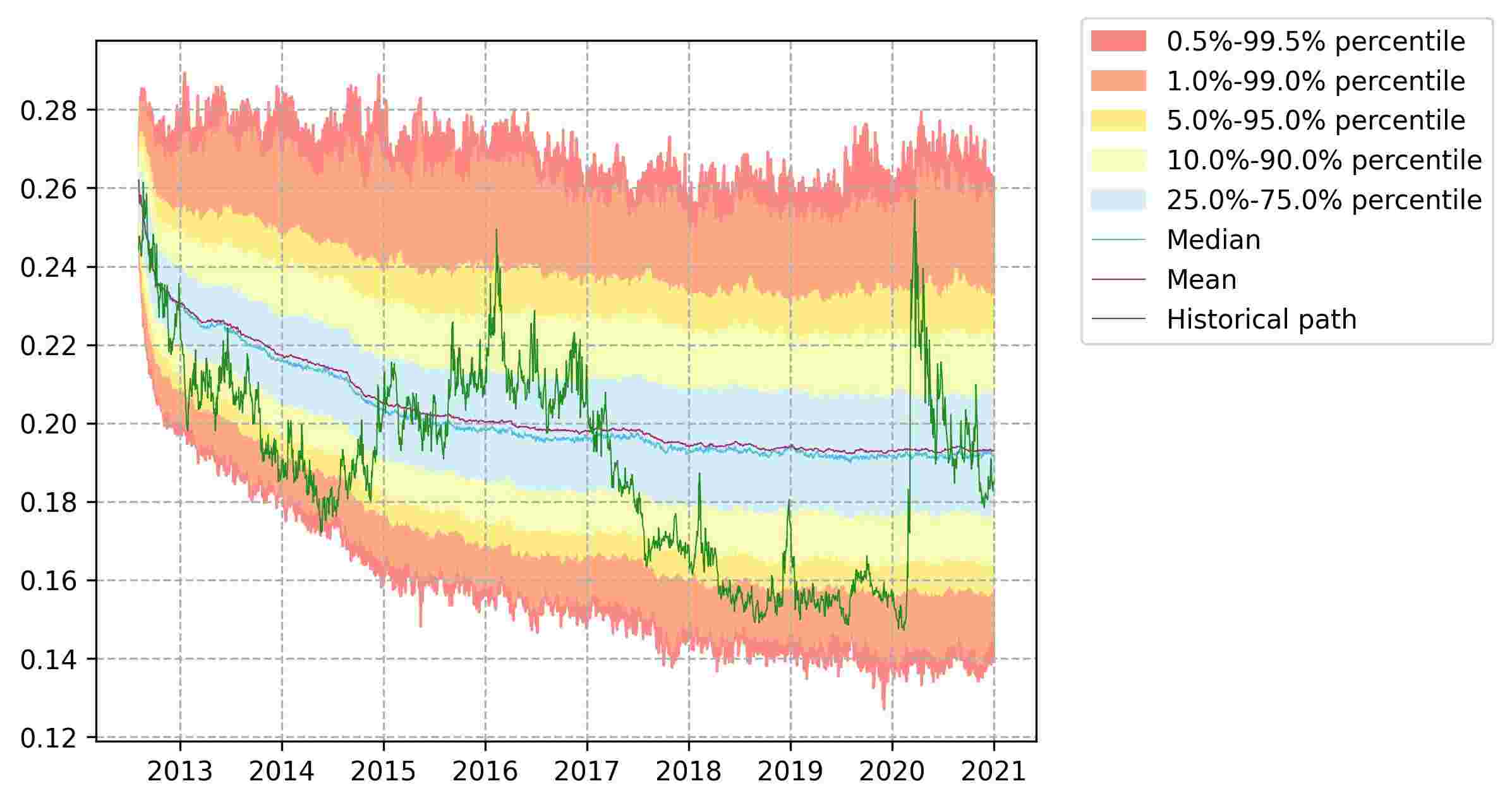}}
    \caption{In-sample quantiles envelopes of the ATM implied volatility for several time-to-maturities. }
\label{fig:quantiles_envelopes}
\end{figure}

\begin{figure}
    \centering
    \subfigure[1-month ATM implied volatility (S\&P 500)]{ \includegraphics[width=0.25\linewidth]{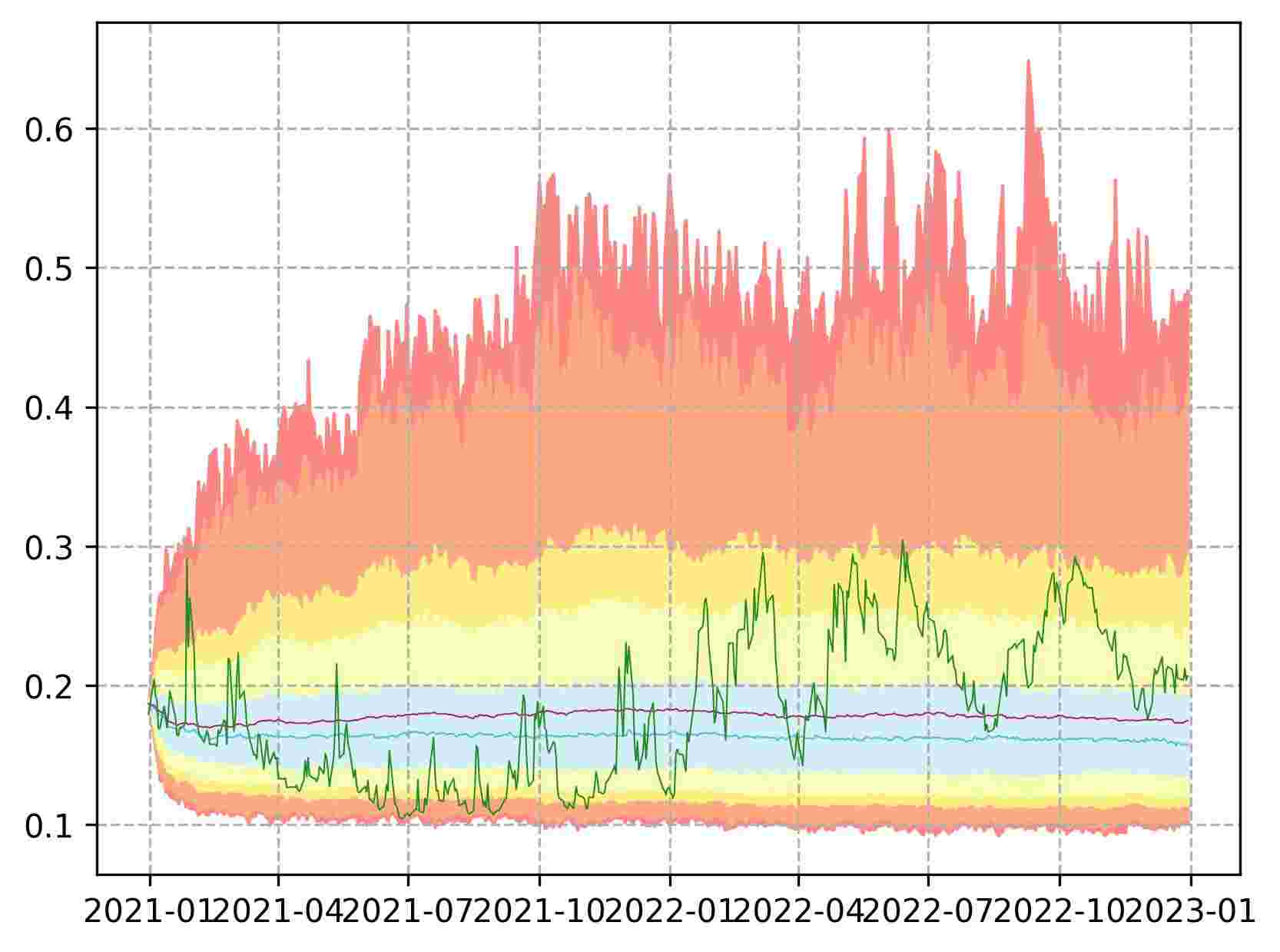}}
    \subfigure[12-months ATM implied volatility (S\&P 500)]{ \includegraphics[width=0.25\linewidth]{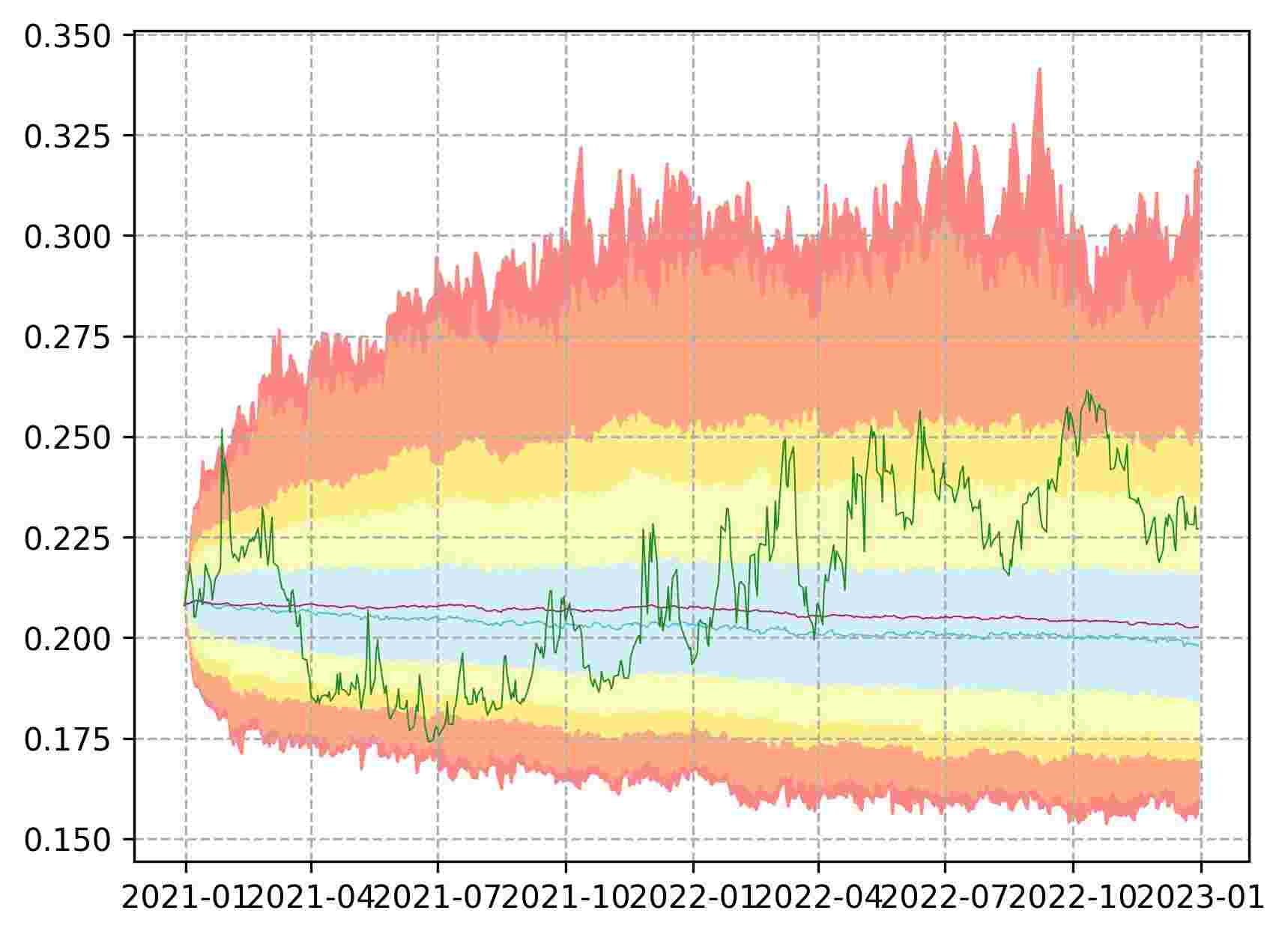}}
    \subfigure[24-months ATM implied volatility (S\&P 500)]{ \includegraphics[width=0.36\linewidth]{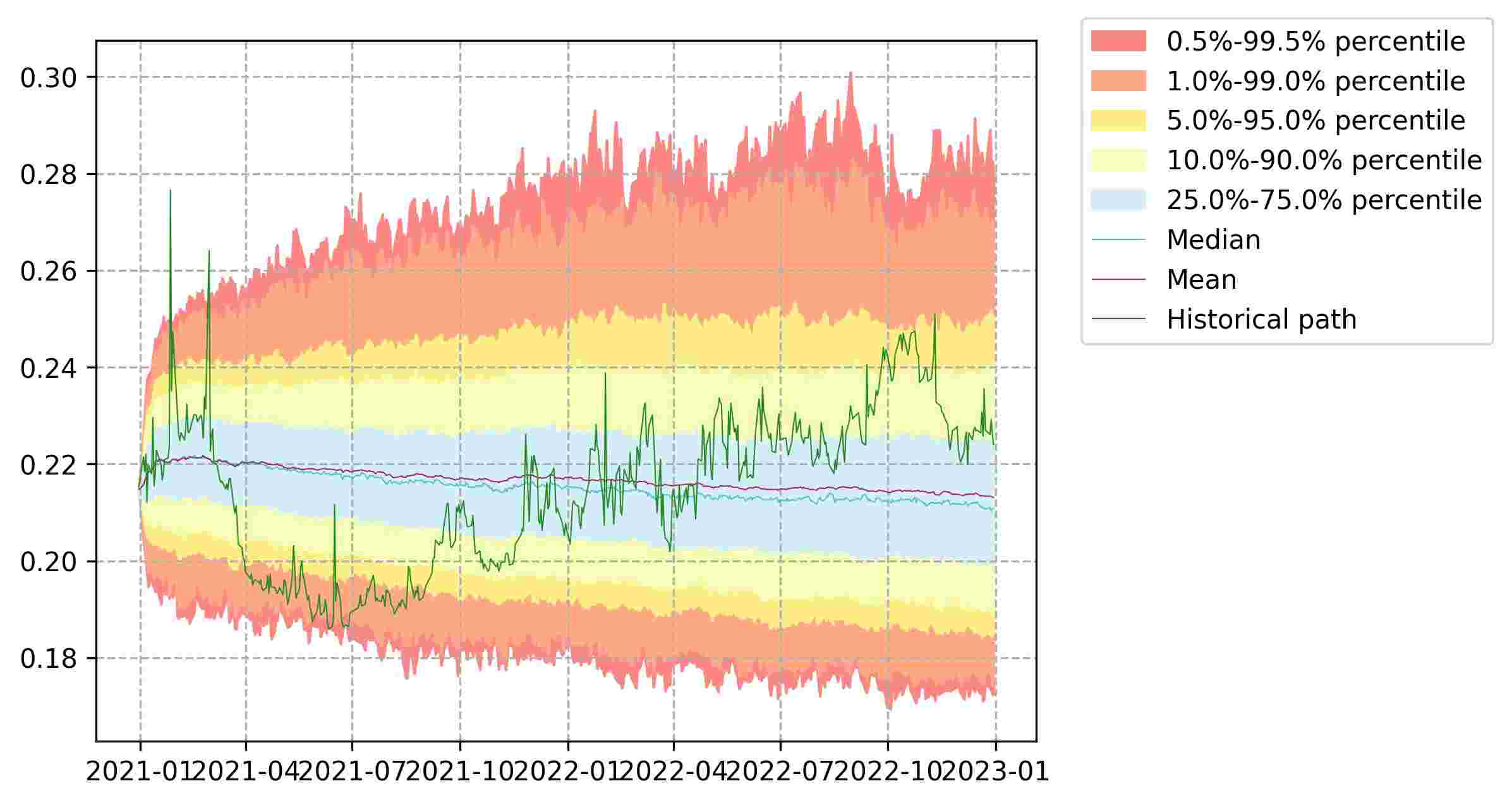}}
    \subfigure[1-month ATM implied volatility (Euro Stoxx 50)]{ \includegraphics[width=0.25\linewidth]{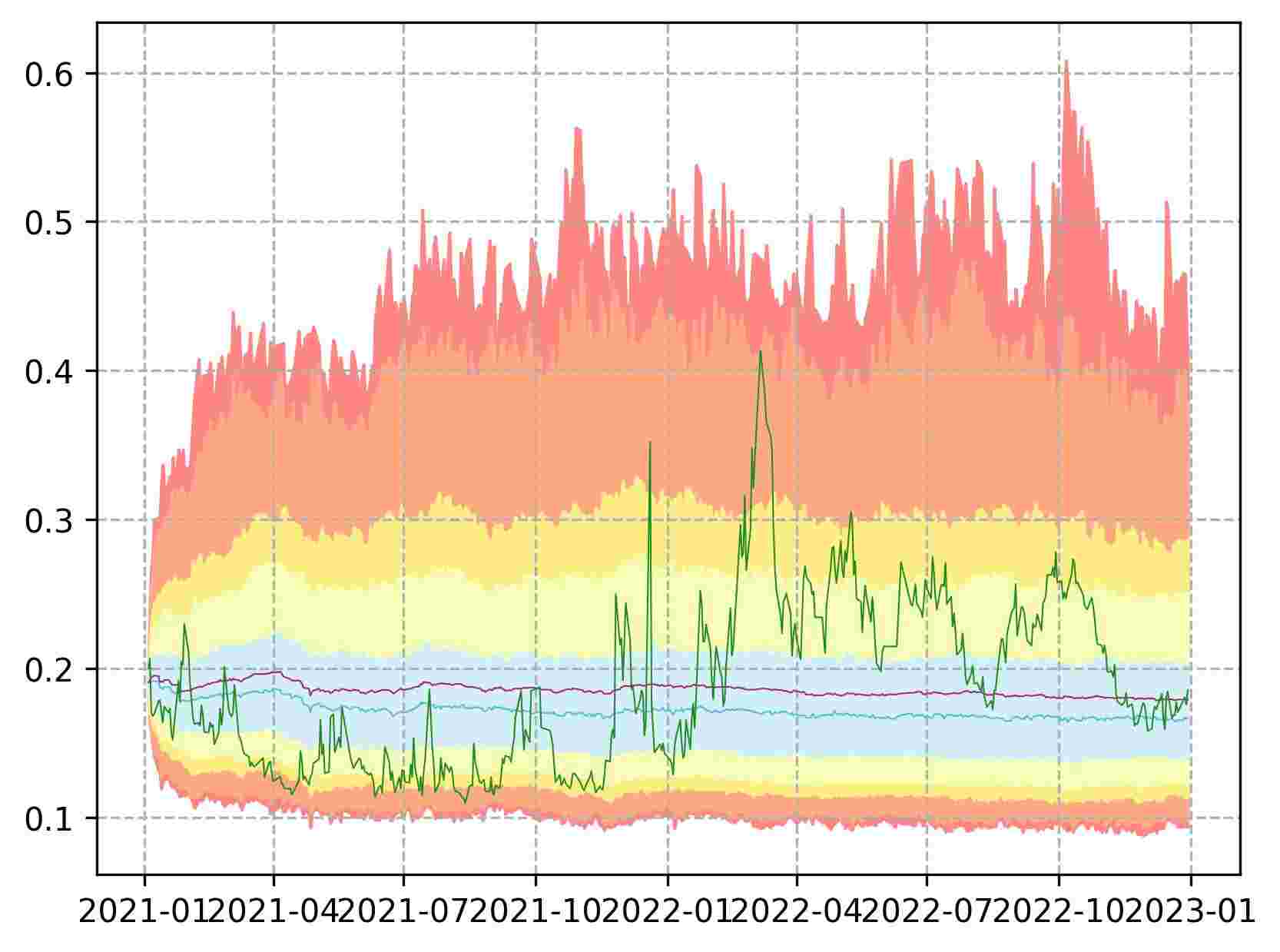}}
    \subfigure[12-months ATM implied volatility (Euro Stoxx 50)]{\includegraphics[width=0.25\linewidth]{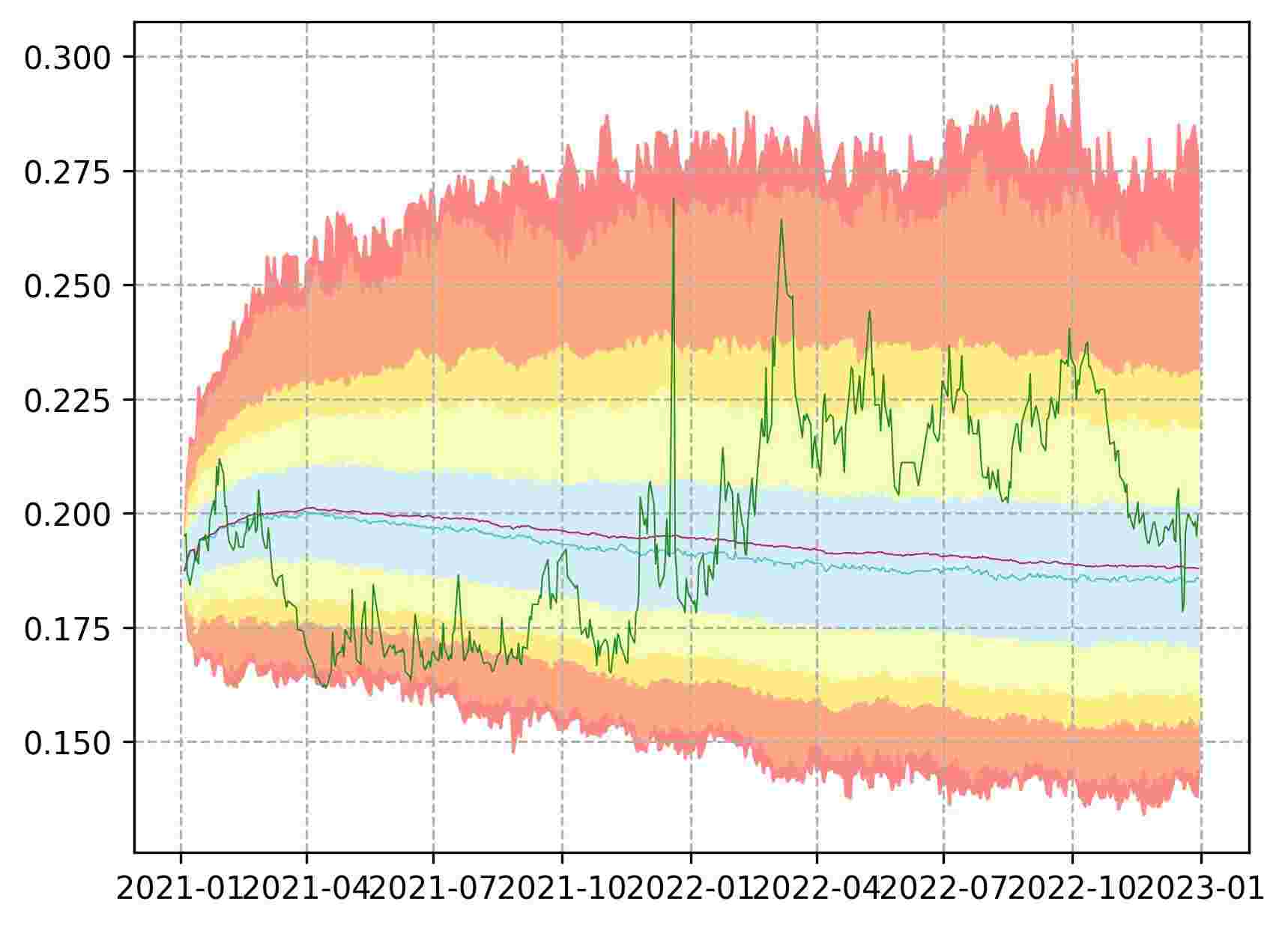}}
    \subfigure[24-months ATM implied volatility (Euro Stoxx 50)]{ \includegraphics[width=0.36\linewidth]{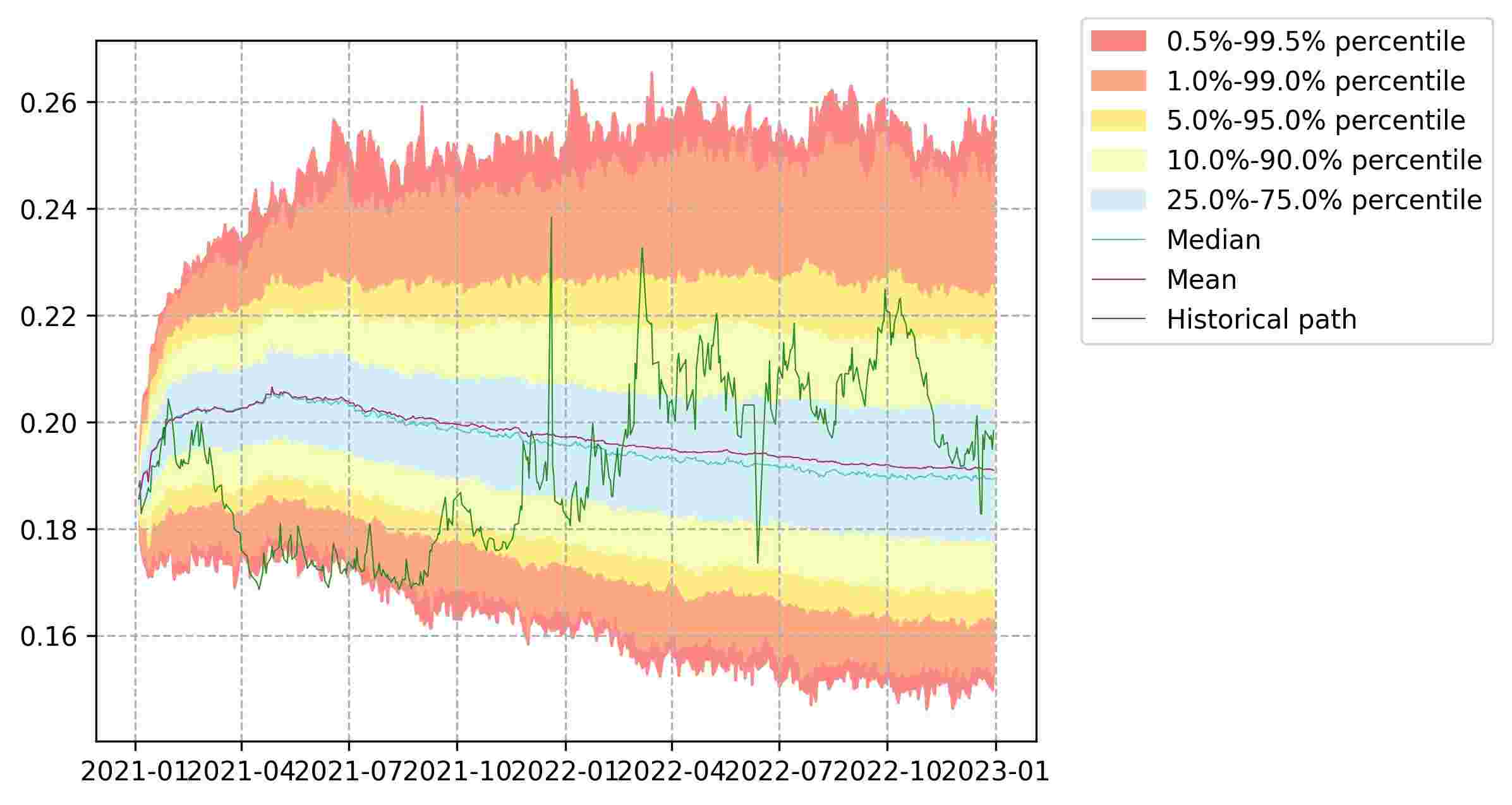}}
    \caption{Out-of-sample quantiles envelopes of the ATM implied volatility for several time-to-maturities.}
\label{fig:quantiles_envelopes_oos}
\end{figure}

\begin{figure}
    \centering
    \subfigure[Historical average IVS of the S\&P 500 ]{\includegraphics[width=0.45\linewidth]{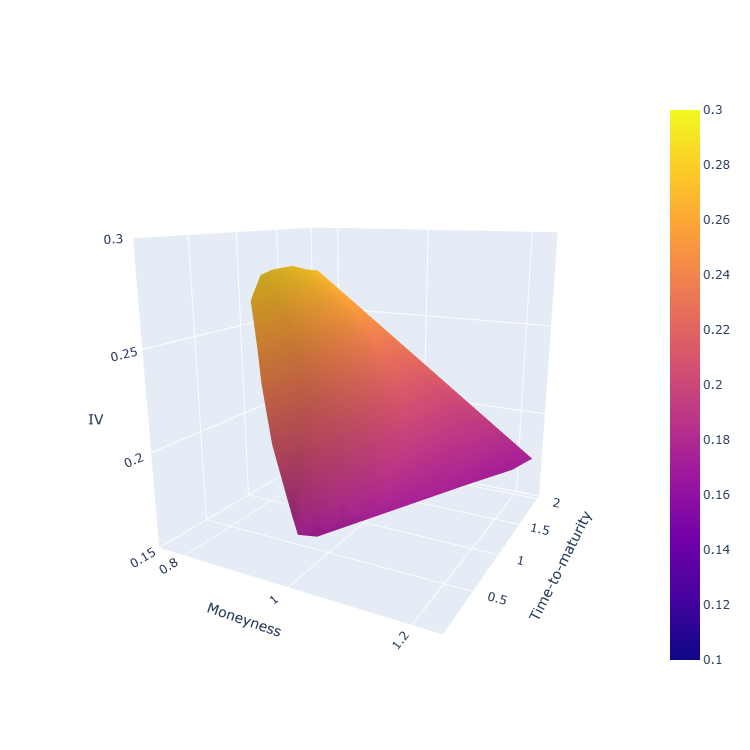}}
    \subfigure[Simulated average IVS of the S\&P 500]{\includegraphics[width=0.45\linewidth]{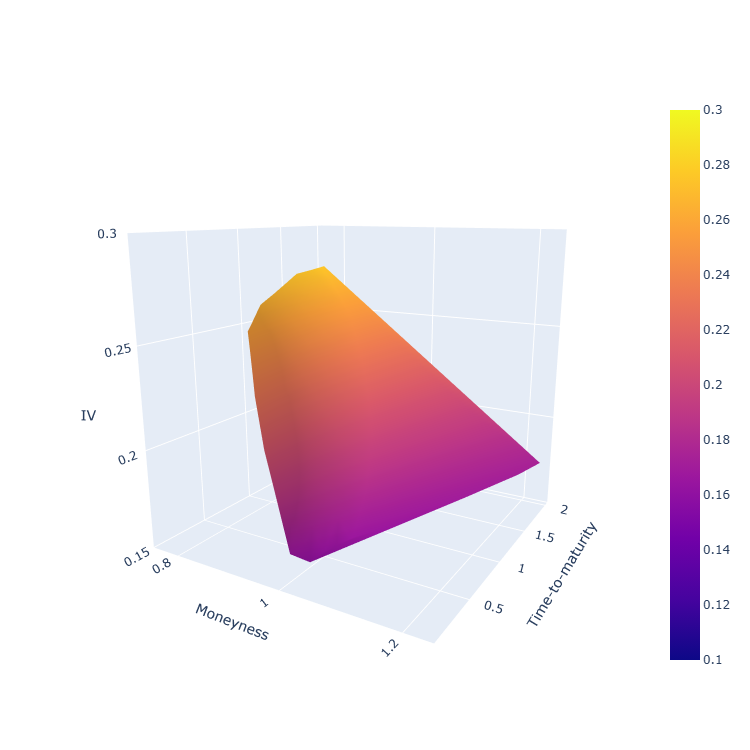}}
    \subfigure[Historical average IVS of the Euro Stoxx 50 ]{\includegraphics[width=0.45\linewidth]{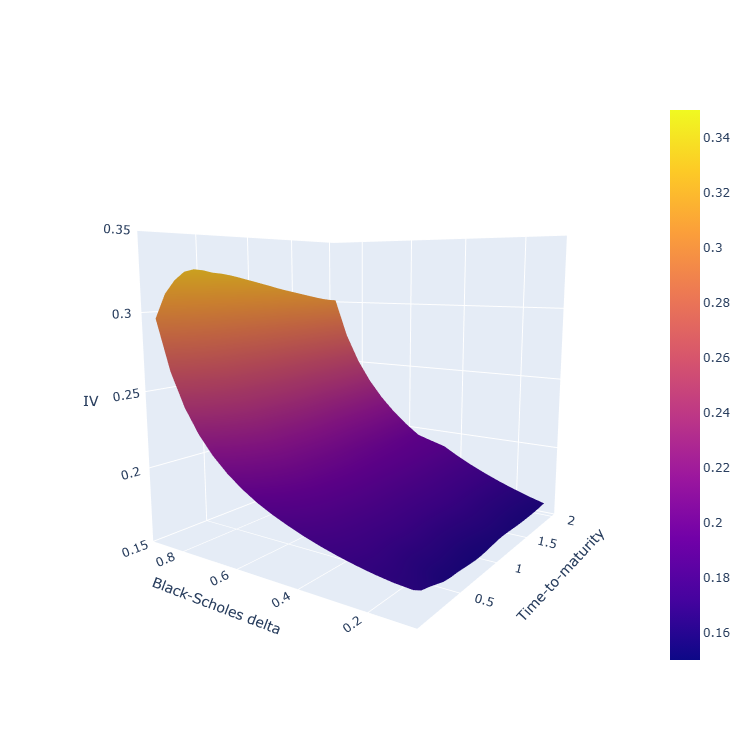}}
    \subfigure[Simulated average IVS of the Euro Stoxx 50]{\includegraphics[width=0.45\linewidth]{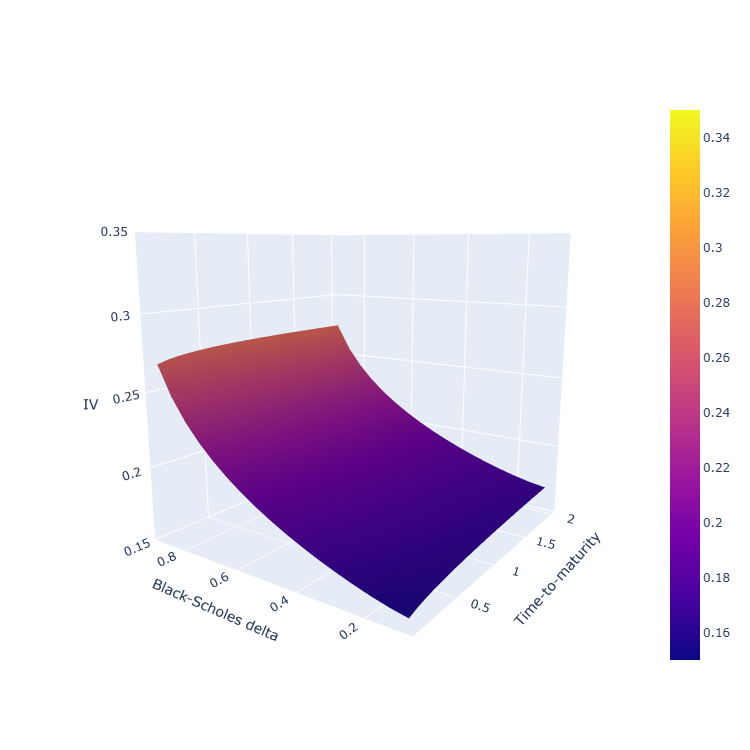}}
    \caption{Comparison of the historical average IVS on the test set and the simulated average IVS. Note that the S\&P 500 IVSs are not rectangular because, for each time-to-maturity, we select the values of the moneyness for which the Black-Scholes delta of a call option is historically always between 0.1 and 0.9 (consistently with the calibration of the parsimonious SSVI parameters, see Appendix \ref{sec:ssvi_calibration_results}). On the other hand, the Euro Stoxx 50 IVS is rectangular because the options in our data set are available on a regular grid of Black-Scholes delta instead of being available on a grid of forward moneynesses (see Section \ref{sec:data}). To simulate IVSs on this Black-Scholes delta grid, we first generate the values of the parsimonious SSVI parameters $a$, $p$, $\rho$, and $\eta$. Then, using the Newton-Raphson method, we find the moneynesses such that the Black-Scholes deltas, calculated with the simulated SSVI volatility, match those in our Black-Scholes delta grid.}
\label{fig:average_ivs}
\end{figure}

We now simulate the complete path-dependent SSVI model, i.e. the underlying asset price is also simulated according to Equations (\ref{eq:underlying_dynamics}). We start by performing simulations initialized using the evolution of the underlying asset price prior to the first date of the train set, namely March 8, 2012, to perform an in-sample comparison of the simulated implied volatilities with the historical ones. More precisely, we simulate 1000 paths of the asset price and the IVS over 2185 days which corresponds to the length over the train data sets. First, we show the evolution of the ATM term structure of two IVS sample paths in Figure \ref{fig:ivs_sample_paths}. As in the previous section, we obtain a very convincing evolution (in particular, we see spikes of the order of magnitude historically observed) which shows that the dynamics of the underlying asset price is also realistic. Then, in Figure \ref{fig:quantiles_envelopes}, we provide the quantile envelopes of the ATM implied volatility for the 1 month, 12 months and 24 months time-to-maturities and we compare them to the historical path of the ATM implied volatility on the train set. These graphs demonstrate that the range of simulated values is reasonable in view of the historical path since the latter exits the 0.5\%-99.5\% quantile envelopes with a frequency that is smaller than 1\%. Note however that the simulated values implied volatilities are in average a bit larger than the historical ones which is purely due to the simulation of the asset price in view of the results presented in Section \ref{sec:cond_simus} where the average implied volatilities are at the right level. We now move to the out-of-sample evaluation of the path-dependent SSVI model. To this end, we perform simulations starting from the evolution of the underlying asset price prior to the start date of the test set, namely January 1, 2021. We consider again 1000 sample paths but we now make the simulations over 504 days corresponding to the length of the test sets. In Figure \ref{fig:quantiles_envelopes_oos}, we provide the quantile envelopes of the ATM implied volatility for the 1 month, 12 months and 24 months time-to-maturities and we compare them to the historical path of the ATM implied volatility on the test set. In this case, the range of the simulated values is again satisfying and the average simulated value is very close the historical one. In Figure \ref{fig:average_ivs}, we compare the historical average IVS on the test set to the simulated average IVS allowing a comparison beyond the ATM implied volatilities. The overall shapes are very close and the average absolute difference between the historical average surface and the simulated one is 0.6\% for the S\&P 500 and 1\% for the Euro Stoxx 50 which shows the simulated away-from-the-money implied volatilities are also consistent with historical data. \\

\begin{figure}
    \centering
    \subfigure[1-month ATM implied volatility (S\&P 500)]{ \includegraphics[width=0.3\linewidth]{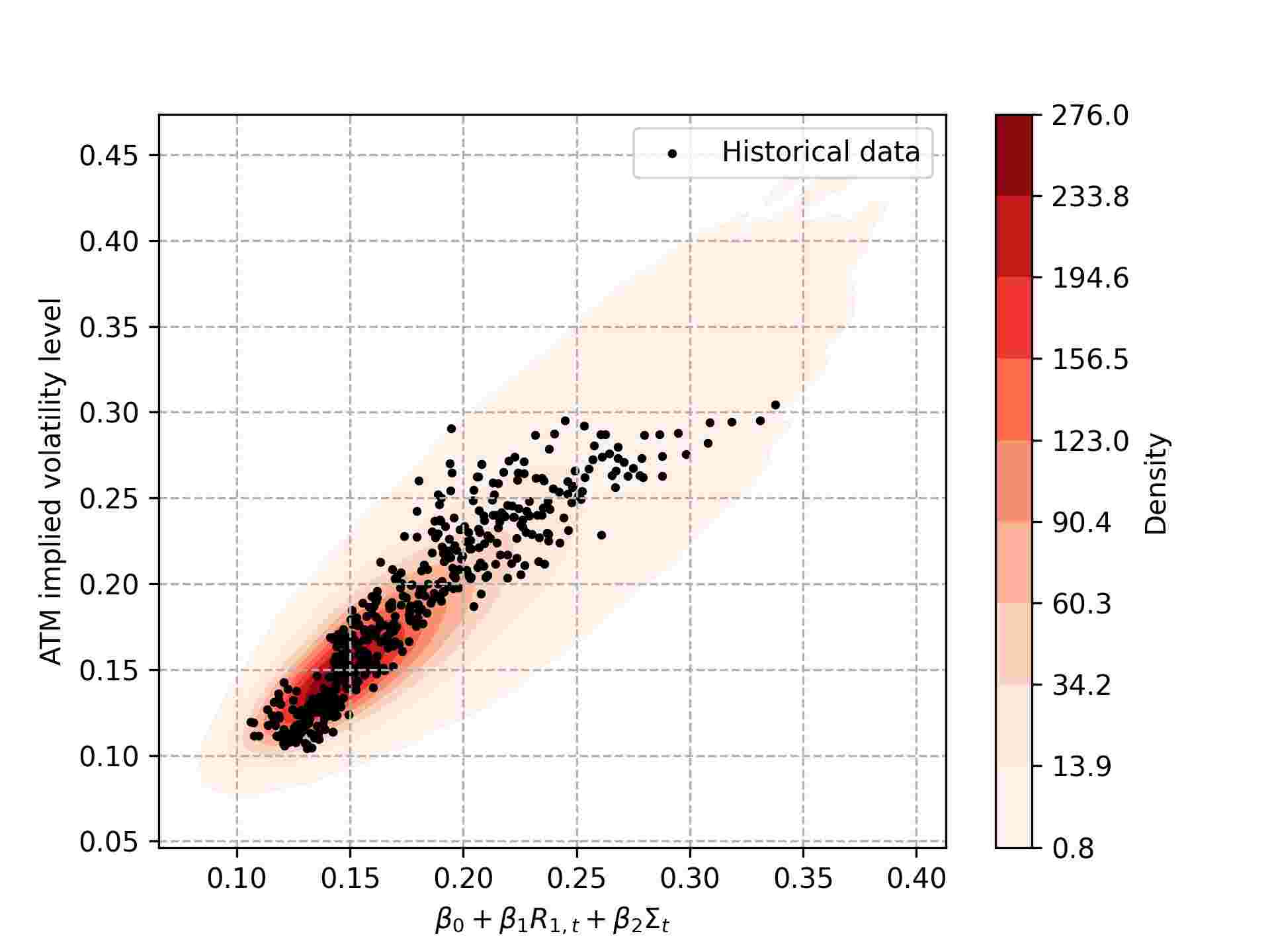}}
    \subfigure[12-months ATM implied volatility (S\&P 500)]{ \includegraphics[width=0.3\linewidth]{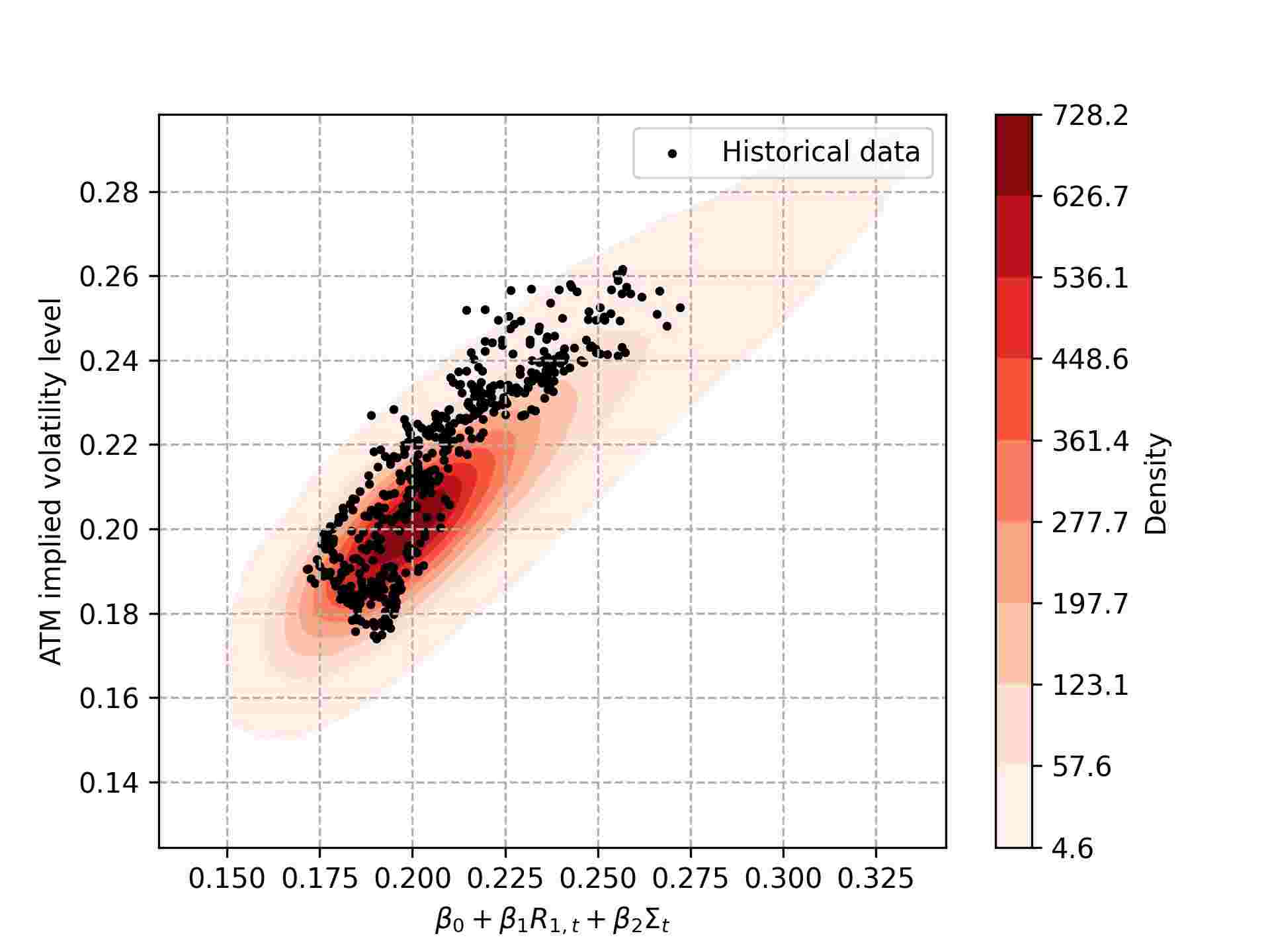}}
    \subfigure[24-months ATM implied volatility (S\&P 500)]{ \includegraphics[width=0.3\linewidth]{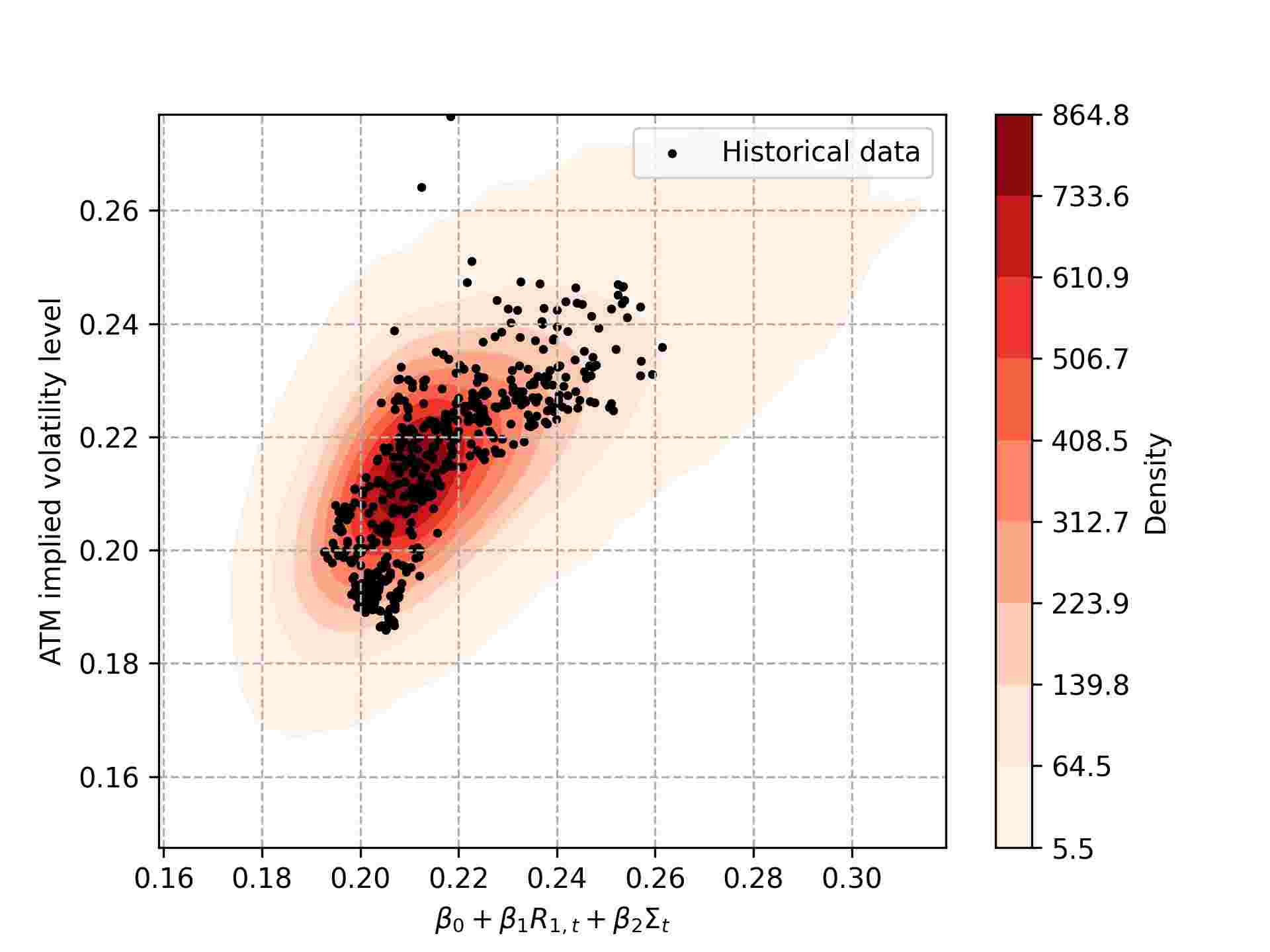}}
    \subfigure[1-month ATM implied volatility (Euro Stoxx 50)]{ \includegraphics[width=0.3\linewidth]{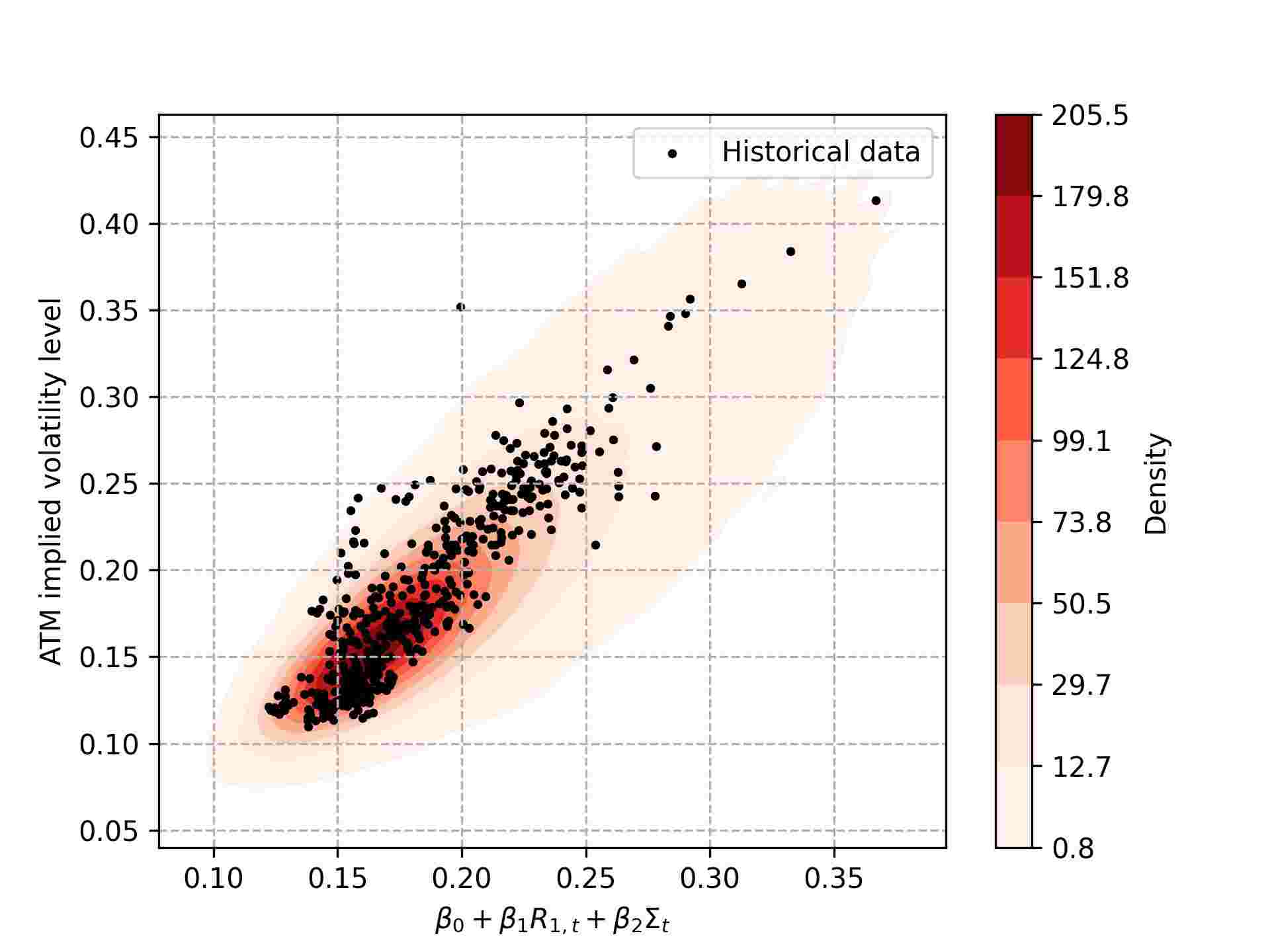}}
    \subfigure[12-months ATM implied volatility (Euro Stoxx 50)]{\includegraphics[width=0.3\linewidth]{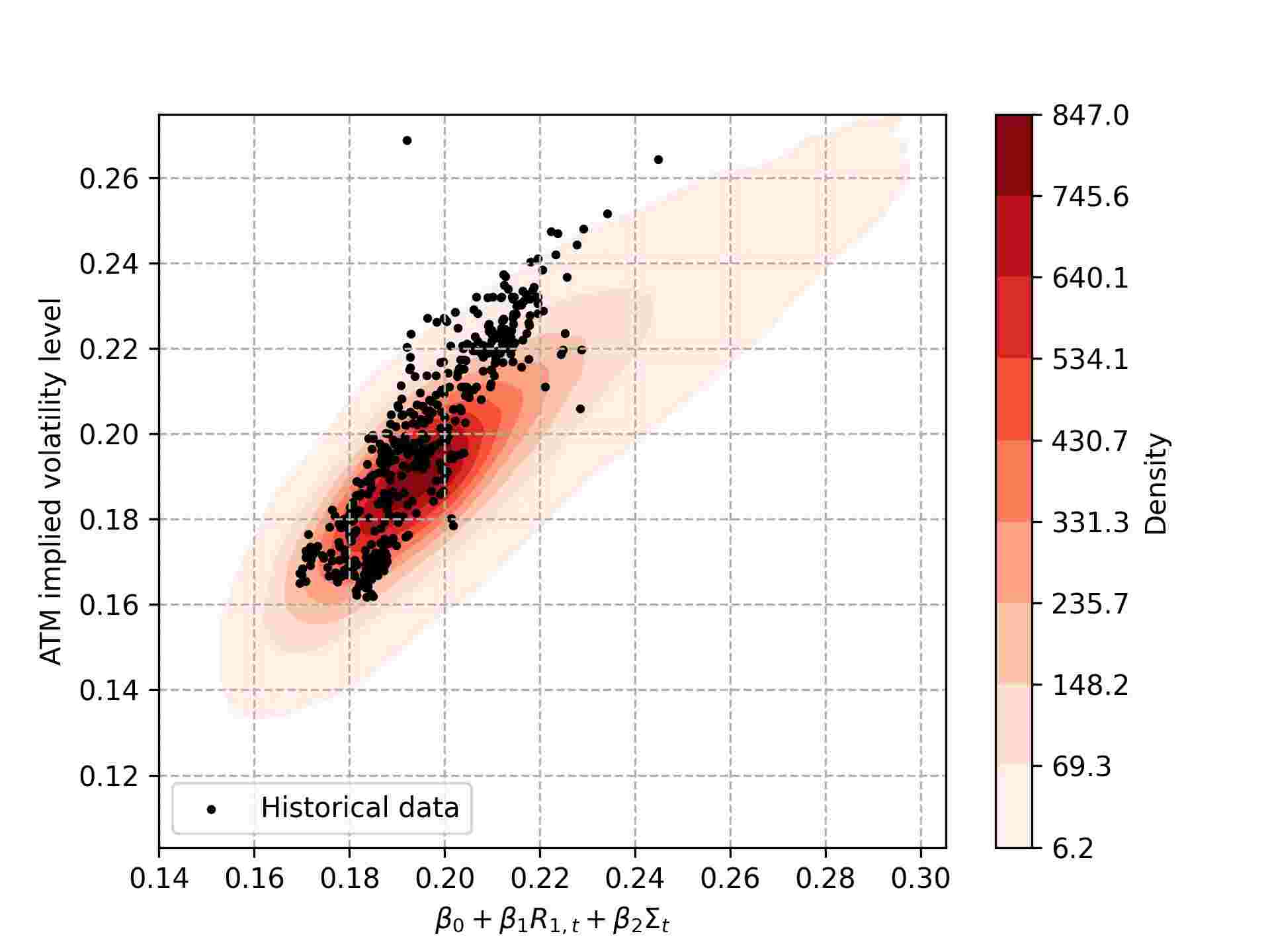}}
    \subfigure[24-months ATM implied volatility (Euro Stoxx 50)]{ \includegraphics[width=0.3\linewidth]{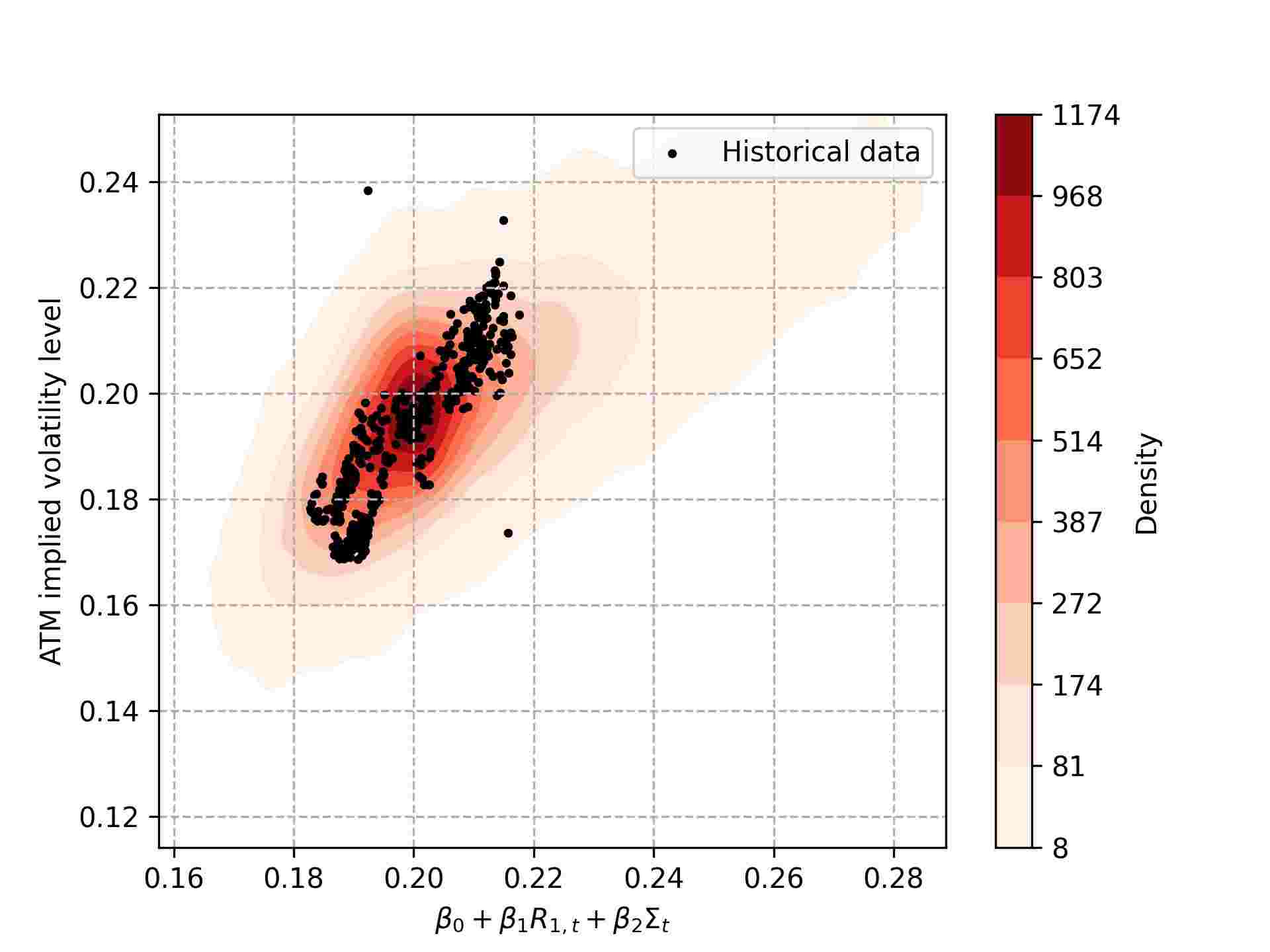}}
    \caption{Joint distribution of $\beta_0+\beta_1R_{1,t}+\beta_2\Sigma_{t}$ and the ATM implied volatility level for several time-to-maturities. The black dots represents the historical realizations of these quantities.   }
\label{fig:kde_plots}
\end{figure}

So far, we have only focused on the distribution of the implied volatility. Since the proposed model allows jointly simulating the implied volatility and the underlying asset price, we propose to study their joint distribution. For this purpose, we use the calibrated PDV parameters of Section \ref{sec:overall_perf} for the 1 month, 12 months and 24 months time-to-maturities, and we plot the density of the pair $(\beta_0+\beta_1R_{1}+\beta_2\Sigma,IV)$ where the features $R_1$ and $\Sigma$ are computed using the simulated returns of the asset price $S$ only and $IV$ is the simulated ATM implied volatility for a given time-to-maturity. This density is compared to the points with coordinates $(\beta_0+\beta_1R_{1}+\beta_2\Sigma,IV)$ where the features $R_1$ and $\Sigma$ are computed using the historical returns of the asset price only and $IV$ is the historical ATM implied volatility for a given time-to-maturity on the test set. The results are displayed in Figure \ref{fig:kde_plots}. Despite the fact that we do not model each ATM implied volatility directly using a PDV model (which is a natural idea in view of the results presented in the empirical study of Section \ref{sec:empirical_study}), we observe that linear relationship between the ATM implied volatility and $\beta_0+\beta_1R_1+\beta_2\Sigma$ is preserved by our model (which only assumes that the parameters $a$ and $p$ are specified using a PDV model). This figure also shows that the simulated dynamics of the asset price (captured through the trend feature $R_1$ and the volatility feature $\Sigma$) is quite consistent with the historical ones. \\

\begin{figure}
    \centering
    \subfigure[Historical (S\&P 500)]{ \includegraphics[width=0.45\linewidth]{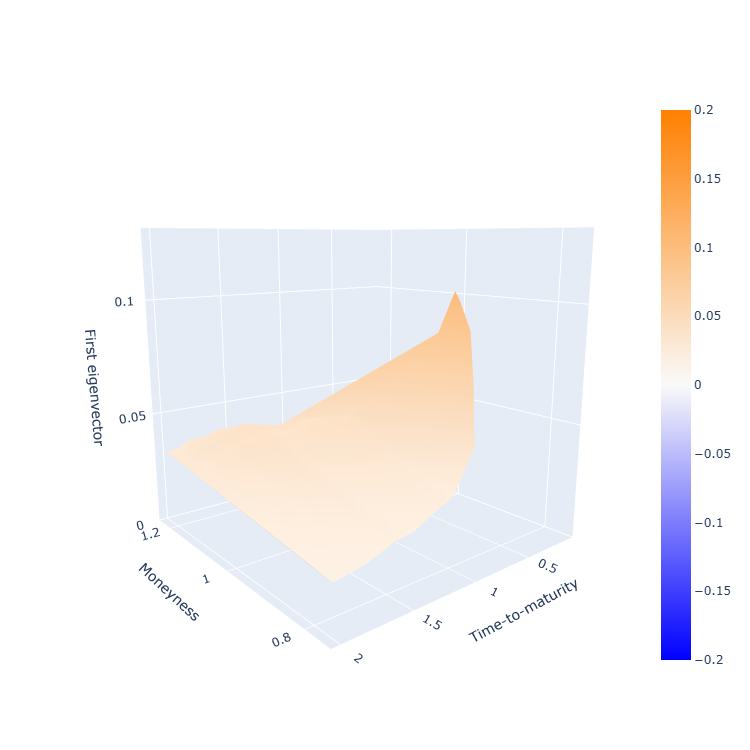}}
    \subfigure[Simulated (S\&P 500)]{ \includegraphics[width=0.45\linewidth]{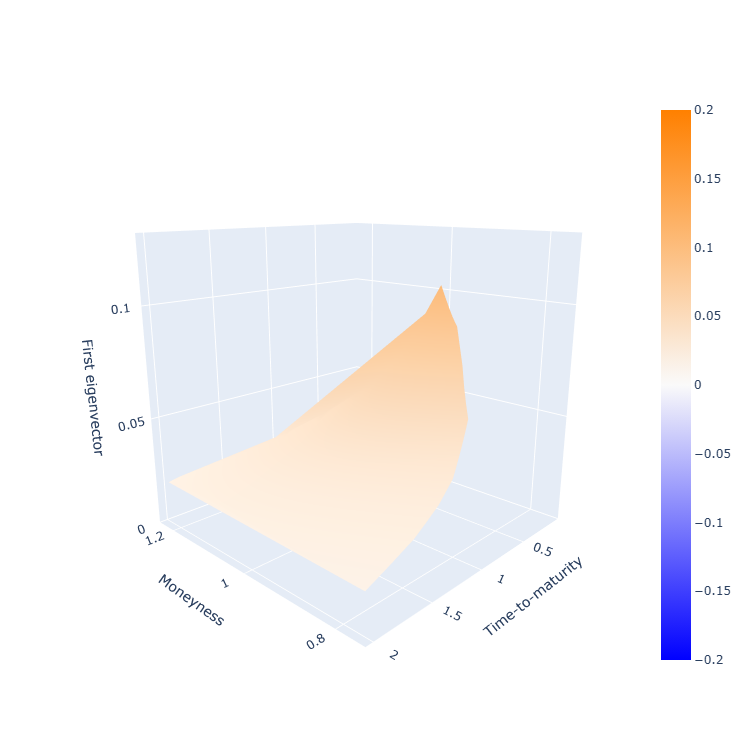}}

    \subfigure[Historical (Euro Stoxx 50)]{ \includegraphics[width=0.45\linewidth]{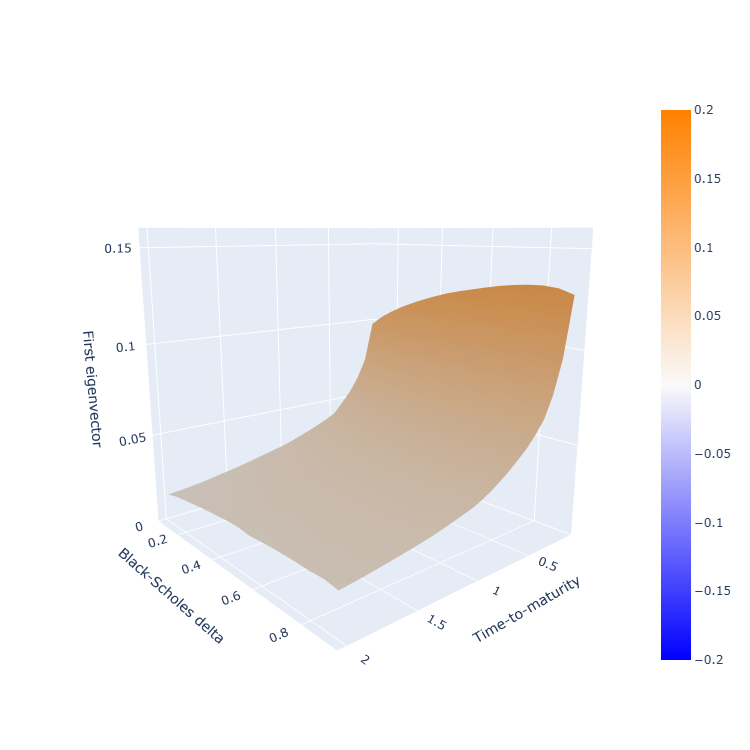}}
    \subfigure[Simulated (Euro Stoxx 50)]{ \includegraphics[width=0.45\linewidth]{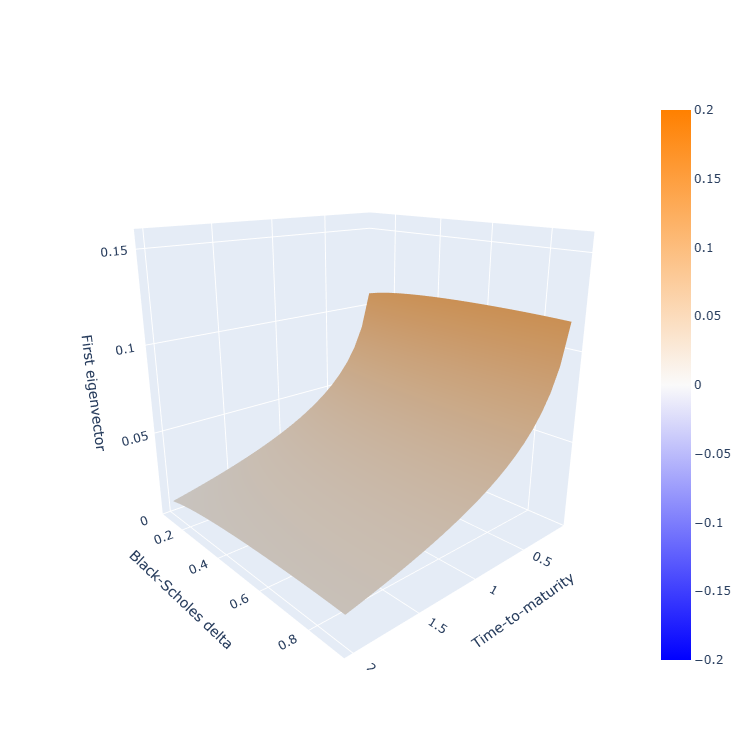}}
    \caption{Comparison between the first eigenvector of the PCA on the historical train set and the average first eigenvector of the PCA on the simulations.  }
\label{fig:first_eigenvector}
\end{figure}

\begin{figure}
    \centering
    \subfigure[Historical (S\&P 500)]{ \includegraphics[width=0.45\linewidth]{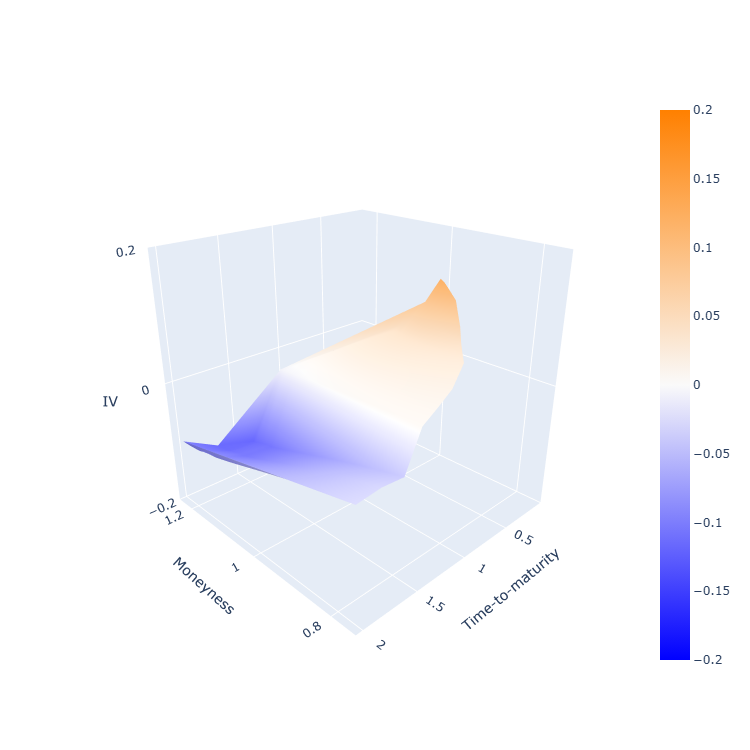}}
    \subfigure[Simulated (S\&P 500)]{ \includegraphics[width=0.45\linewidth]{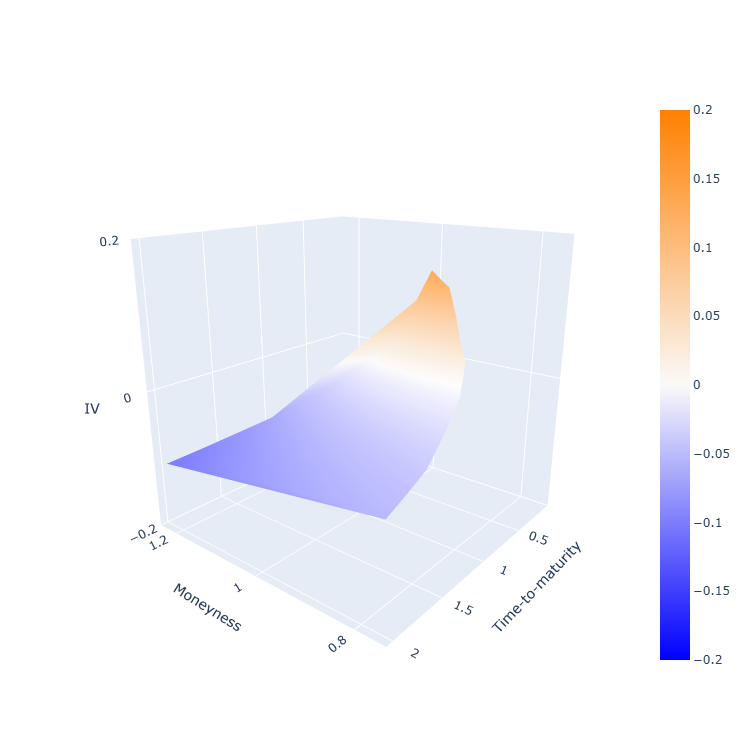}}

    \subfigure[Historical (Euro Stoxx 50)]{ \includegraphics[width=0.45\linewidth]{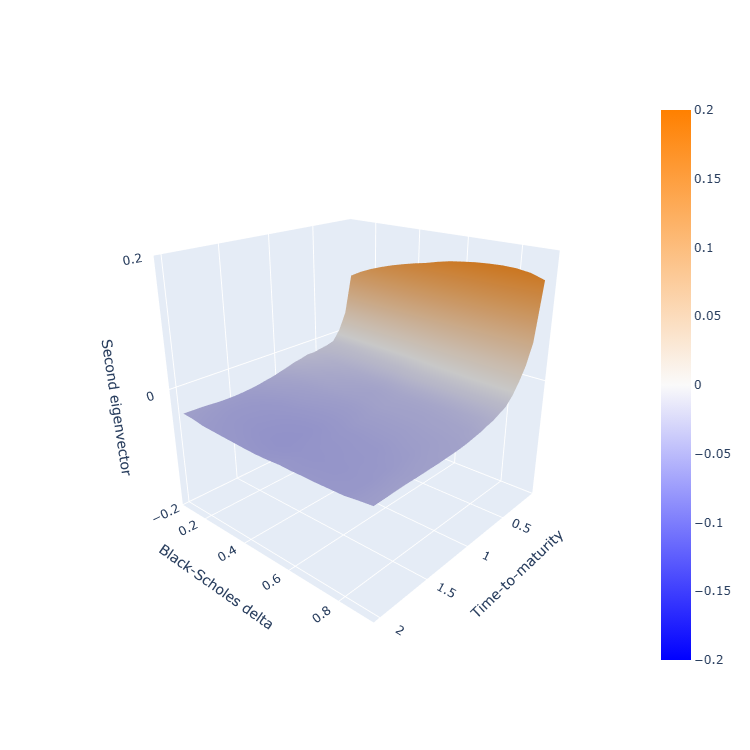}}
    \subfigure[Simulated (Euro Stoxx 50)]{ \includegraphics[width=0.45\linewidth]{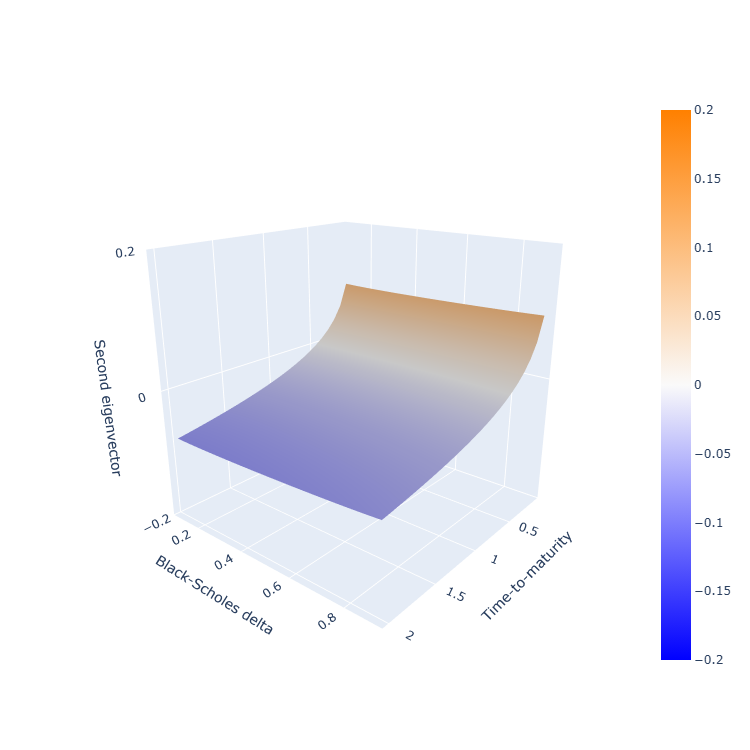}}
    \caption{Comparison between the second eigenvector of the PCA on the historical train set and the average second eigenvector of the PCA on the simulations.} 
\label{fig:second_eigenvector}
\end{figure}

To conclude this section, we compare the results from a principal component analysis (PCA) performed on the historical train set to the ones obtained from a PCA performed on the simulated IVSs. More specifically, in the same vein as \cite{cont2002dynamics}  and \cite{cont2022simulation}, we apply a PCA to the log-variations of the implied volatility on a time-to-maturity $\times$ moneyness grid. The time-to-maturities range from 1 month to 24 months with a monthly timestep while the values of the moneyness depend on each data set. For the S\&P 500, we select the values of the moneyness for which the Black-Scholes delta of a call option is historically always between 0.1 and 0.9 (consistently with the calibration of the parsimonious SSVI parameters, see Appendix \ref{sec:ssvi_calibration_results}). For the Euro Stoxx 50, the moneyness grid consists in the Black-Scholes delta grid on which the historical data is available (see Section \ref{sec:data}). In Table \ref{tab:explained_variance}, we first present the variance explained by the first four eigenvectors of the historical and simulated covariance matrices of the daily log-variations in implied volatilities on the aforementioned grid. To get these results on the simulations of the path-dependent SSVI model, 
\begin{enumerate}
    \item we apply the PCA to each simulated path,
    \item we compute the variance explained by the first four eigenvectors for each path,
    \item we average the variance explained over the 1000 paths. 
\end{enumerate}
We observe that the variance explained by the first eigenvector is similar between the PCA on historical data and the PCA on simulations, whereas the variance explained by the subsequent eigenvectors shows greater differences between the two PCAs. To complete this analysis, we also compare the shape of the two first historical eigenvectors to the two first average simulated eigenvectors in Figures \ref{fig:first_eigenvector} and \ref{fig:second_eigenvector}. We observe very similar shapes in the two Figures. In particular, we remark that:
\begin{enumerate}
    \item the first eigenvector captures the fact that the implied volatility at small time-to-maturities experiences larger variations than the one at larger time-to-maturitie,
    \item the second eigenvector captures the fact that the IVS term structure tends to flatten out when the short-term implied volatility is increasing. 
\end{enumerate}
We also compared the shapes of the third and fourth eigenvectors but they were quite different. However, as we have seen, they represent only a small part of the explained variance of the log-variations covariance matrix. Since, the two first eigenvectors explain more than 86\% (resp. 91\%) of the variance of the S\&P 500 (resp. Euro Stoxx 50) log-variations, we deduce that our model allows to capture a large part of the movements observed historically in the IVS.

\begin{table}[]
    \centering
    \caption{Variance explained by the first four eigenvectors of the covariance matrix of the daily log-variations of the implied volatilities.}
    \label{tab:explained_variance}
    \begin{tabular}{@{}ccccccccc@{}}
    \toprule
     & \multicolumn{4}{c}{S\&P 500} & \multicolumn{4}{c}{Euro Stoxx 50} \\ \midrule
    Eigenvector & 1     & 2        & 3        & 4&  1        &   2       &   3  &4      \\\midrule
    Variance explained (historical) & 76.17\%
     & 9.88\%        & 3.69\%        &  2.14\%       &  87.25\%         &  4.22\%         &   1.54\%& 1.11\%        \\
    Variance explained (simulations) &  71.39\%       &   19.5\%      &   8.96\%      &    0.1\%       &    89.00\%        & 9.50\% & 1.23\% & 0.27\%          \\ \bottomrule
    \end{tabular}
\end{table}

\subsubsection{Computational cost}
 Running 1000 simulations over an horizon of 2 years with a daily time step takes approximately 2.5 minutes for 24 time-to-maturities and 81 values of moneyness on a computer equipped with an Intel Core i7-11850H, 16 cores, 2.5GHz. The model is implemented in the Python programming language.

%% file: conclusion.tex
\section{Concluding remarks}
Using historical time series of implied volatility surfaces for the S\&P 500 and the Euro Stoxx 50, we have shown empirically that a large part of the variability of the at-the-money-forward (ATM) implied volatility for time-to-maturities ranging from 1 month to 24 months can be explained by two features, namely the weighted average of the underlying asset past returns and the weighted average of the past squared returns. As the time-to-maturity increases, the part of variability explained by these two features decreases but remains important. Thus, our empirical study allows to extend the one of \cite{guyon2022volatility}  who focused on implied volatility indices and realized volatility instead of vanilla options implied volatilities. In Section \ref{sec:ssvi}, we have then introduced a parsimonious version (see Definition \ref{def:parsimonious_ssvi}) of the SSVI parameterization of \cite{gatheral2014arbitrage} that depends on four parameters only ($a$, $p$, $\rho$ and $\eta$) and that still achieves a reasonable fit to the implied volatility surfaces in our two data sets. This parsimonious version is essentially obtained by considering a parametric form of the ATM total variance thus avoiding to have one parameter per time-to-maturity. In the last section, we demonstrate that the variations of the two parameters $a$ and $p$ ruling the ATM implied volatility in the parsimonious SSVI parameterization can also be widely explained by the two features that we mentioned earlier. Based on this observation, we introduce a new model for the joint dynamics of the underlying asset price and the full implied volatility surface (there is no restriction in the range of time-to-maturities and strikes that one wants to project) embedding the path-dependency of the implied volatility with respect to the underlying price. On the one hand, the underlying asset price is modelled using the path-dependent volatility model of \cite{guyon2022volatility}. On the other hand, the residuals of the parameters $a$ and $p$ are modelled through Ornstein-Uhlenbeck processes and the parameters $\rho$ and $\eta$ are modelled as Jacobi processes which are guaranteeing the absence of static arbitrage in the simulated surfaces.  Extensive details on how to calibrate this new model are provided. Finally, we show the high consistency of this model with historical data. The study of impact of this new model for applications in asset management, risk management and hedging is left for future research.

%% file: acknowledgements.tex
\section*{Acknowledgements}

The authors are grateful to Julien Guyon for fruitful discussions and to the anonymous referees for their valuable comments.

%% file: disclosure_and_funding.tex
\section*{Disclosure statement}
The authors report there are no competing interests to declare. 

\section*{Funding}
No funding was received.

%% file: appendix.tex
\section{Influence of the cut-off lag on the performance of the PDV model}\label{sec:influence_cutoff}
We mentioned at the beginning of Section \ref{sec:overall_perf} that we truncated the sums in the expressions of $R_1$ and $\Sigma$ after the previous $C=1,000$ business days. In the following, we study the impact of the hyperparameter $C$, which we call the cut-off lag. Remark that if the cut-off lag is too small, there is a risk to lose some information from the past, while if the cut-off lag is too big, there is a risk to capture some information that is actually not relevant to predict the implied volatility. First, let us point out that there is a priori no reason to use the same cut-off lag for $R_1$ and $\Sigma$. Therefore, we consider two different cut-off lag hyperparameters $C_{R_1}$ and $C_{\Sigma}$. In order to measure their influence on the performance of the model, we run a 10-fold cross-validation. Before describing this procedure in more details, we introduce a third hyperparameter $\lambda$ that allows to penalize large values of the kernels parameters $\alpha_1$, $\delta_1$, $\alpha_2$ and $\delta_2$ during the calibration. More specifically, we add a $L^2$ penalization term in the objective function so that Equation (\ref{eq:target_fun}) becomes:
\begin{equation}
    \begin{array}{rrcll}
    \displaystyle\min_{(\alpha_1,\delta_1,\alpha_2,\delta_2,\beta_0,\beta_1,\beta_2) \in \mathbb{R}^7} & \multicolumn{4}{c}{\displaystyle\sum_{t\in \mathcal{T}_{train}} (IV^{mkt}_t - \beta_0-\beta_1R_{1,t}-\beta_2\Sigma_t)^2+\lambda \left(\displaystyle\sum_{j=1}^2 \alpha_j^2 + \displaystyle\sum_{j=1}^2 \delta_j^2 \right) } \\
   \text{s.t.} & \alpha_j, \delta_j &\ge 0 & \text{for } j \in \{1,2\} \\
    \end{array}
\end{equation}  
where $ R_{1,t} = \displaystyle\sum_{t-C_{R_1}\le t_i \le t} \frac{Z_{\alpha_1,\delta_1}}{(t-t_i+\delta_1)^{\alpha_1}} r_{t_i}$ and $\Sigma_{t} = \sqrt{\displaystyle\sum_{t-C_{\Sigma}\le t_i \le t} \frac{Z_{\alpha_2,\delta_2}}{(t-t_i+\delta_2)^{\alpha_2}} r_{t_i}^2}$. The introduction of this penalization is motivated by the fact that we want to avoid overfitting as mentioned in Section \ref{sec:comment_gap}. Note that this modified objective function is minimized using the \texttt{minimize} function with the L-BFGS-B algorithm from the \texttt{scipy} Python package. Let us now describe the 10-fold cross-validation. For each time-to-maturity, the train set is split into 10 adjacent folds of same size (222 days each) and for each triplet $(C_{R_1},C_{\Sigma},\lambda)\in \{5,10,25,50,100,250,500,1000,1500,2000,2500,3000\}^2\times \{10^{-6},10^{-5},\dots,10^{-1}\}$ and for all $i\in \{1,\dots,10\}$, we calibrate on all folds except fold $i$ and we compute the $R^2$ score on the fold $i$. This procedure corresponds to the so-called blocked cross-validation (see e.g. \citeauthor{cerqueira2020evaluating}, \citeyear{cerqueira2020evaluating}). Then, we average the $R^2$ scores over the 10 folds so that we obtain one score per triplet $(C_{R_1},C_{\Sigma},\lambda)$. In Table \ref{tab:best_triplets}, we present the triplet $(C_{R_1},C_{\Sigma},\lambda)$ leading to the best average $R^2$ score for each time-to-maturity. \\

\begin{figure}[h]
    \centering
    \subfigure[S\&P 500]{\includegraphics[width=0.45\linewidth]{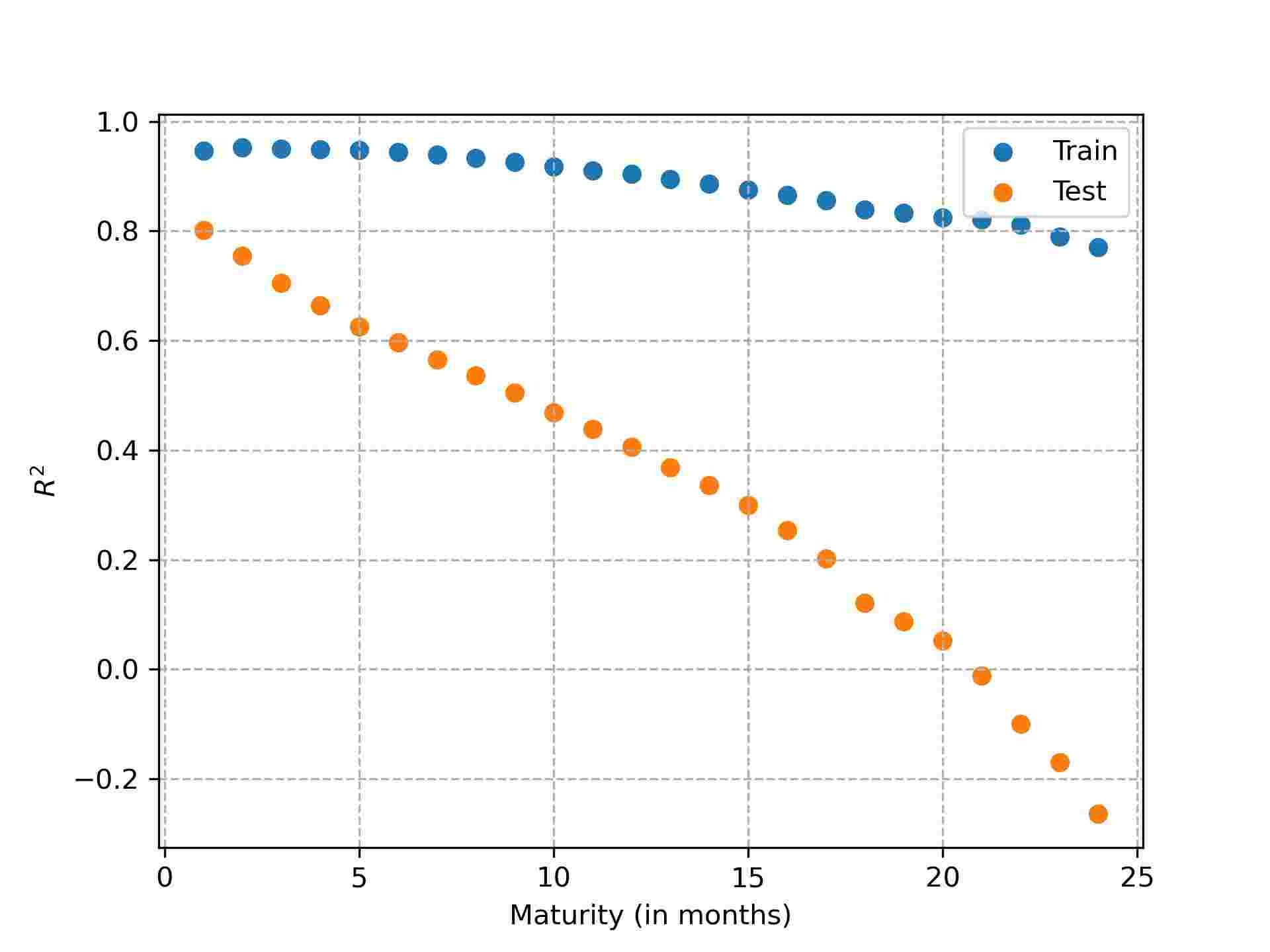}}
    \subfigure[Euro Stoxx 50]{\includegraphics[width=0.45\linewidth]{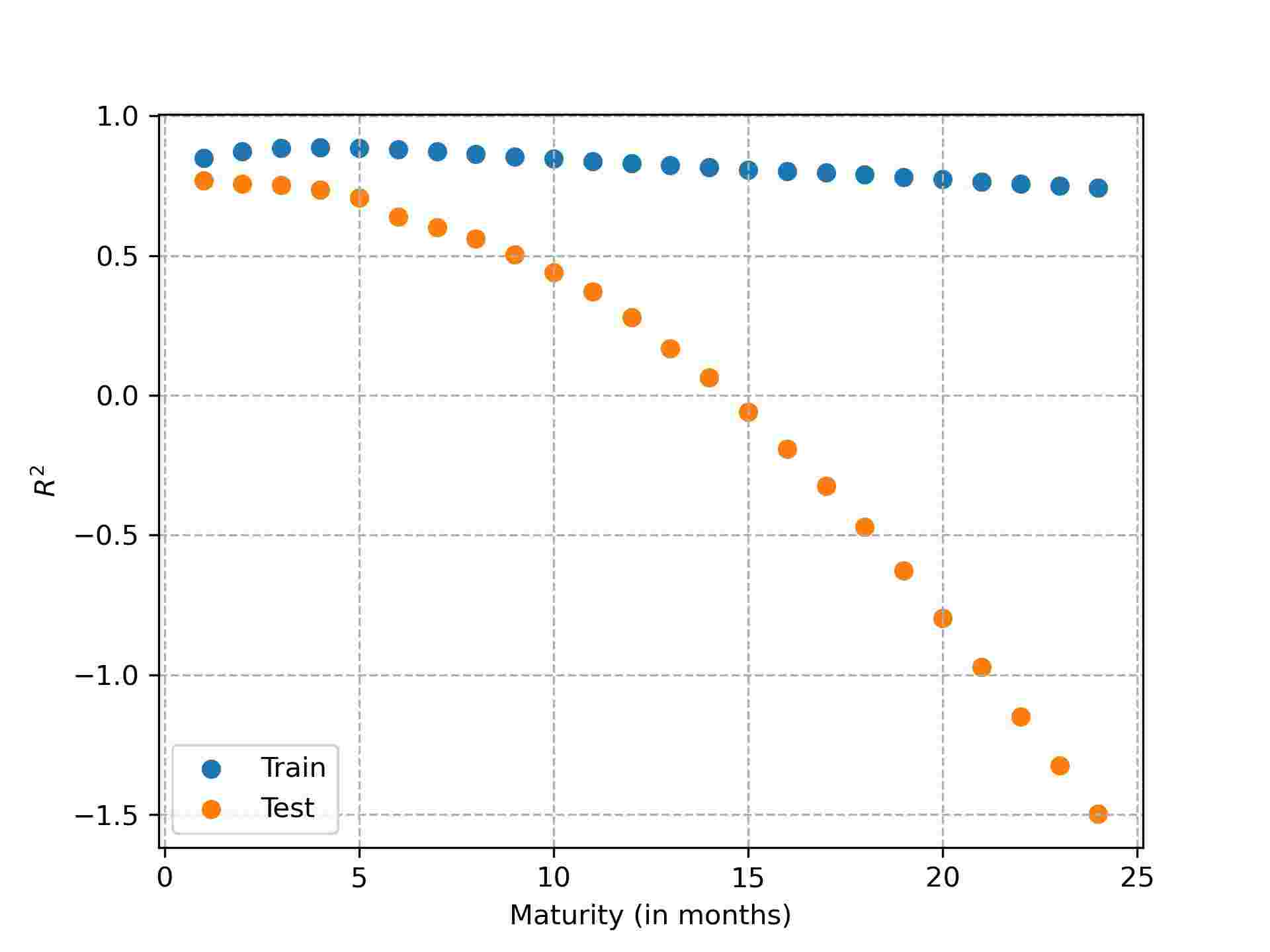}}
    \caption{$R^2$ scores on the train and the test sets as a function of the ATM implied volatility time-to-maturity for $C_{R_1}=C_{\Sigma}=500$ and $\lambda=0$.}
\label{fig:r2_cutoff_500}
\end{figure}

\begin{table}[h]
    \centering
    \caption{Hyperparameters allowing to achieve the highest average $R^2$ scores on the test fold within the 10-fold cross-validation procedure. }
    \label{tab:best_triplets}
    \resizebox{\textwidth}{!}{%
    \begin{tabular}{@{}cccccccccccccccllll@{}}
    \cmidrule(r){1-7} \cmidrule(lr){9-15}
                             & \multicolumn{3}{c}{\textbf{S\&P 500}}                      & \multicolumn{3}{c}{\textbf{Euro Stoxx 50}} & \multicolumn{1}{l}{} &                          & \multicolumn{3}{c}{\textbf{S\&P 500}}                      & \multicolumn{3}{c}{\textbf{Euro Stoxx 50}} & \multicolumn{1}{c}{\textbf{}} & \multicolumn{1}{c}{\textbf{}} & \multicolumn{1}{c}{} & \multicolumn{1}{c}{} \\ \cmidrule(r){1-7} \cmidrule(lr){9-15}
    \textbf{Maturity}        & $C_{R_1}$ & $C_{\Sigma}$ & $\lambda$                        & $C_{R_1}$   & $C_{\Sigma}$   & $\lambda$     &                      & \textbf{Maturity}        & $C_{R_1}$ & $C_{\Sigma}$ & $\lambda$                        & $C_{R_1}$   & $C_{\Sigma}$   & $\lambda$     &                               &                               &                      &                      \\ \cmidrule(r){1-7} \cmidrule(lr){9-15}
    \multicolumn{1}{c|}{1M}  & 50         & 250         & \multicolumn{1}{c|}{$10^{-2}$} & 50           & 250           & $10^{-3}$   &                      & \multicolumn{1}{c|}{13M} & 1000        & 2000        & \multicolumn{1}{c|}{$10^{-1}$} & 10           & 1000           & $10^{-2}$   &                               &                               &                      &                      \\
    \multicolumn{1}{c|}{2M}  &  1500         & 1000         & \multicolumn{1}{c|}{$10^{-2}$} & 50           & 250            & $10^{-2}$   &                      & \multicolumn{1}{c|}{14M} & 1000        & 2000         & \multicolumn{1}{c|}{$10^{-1}$} & 10           & 1000           & $10^{-2}$   &                               &                               &                      &                      \\
    \multicolumn{1}{c|}{3M}  & 50        & 250        & \multicolumn{1}{c|}{$10^{-3}$} & 25           & 1000            & $10^{-5}$   &                      & \multicolumn{1}{c|}{15M} & 1000       & 2000         & \multicolumn{1}{c|}{$10^{-1}$} & 10          & 1000           & $10^{-2}$   &                               &                               &                      &                      \\
    \multicolumn{1}{c|}{4M}  & 1500         & 500         & \multicolumn{1}{c|}{$10^{-2}$} & 25            & 1000               & $10^{-6}$   &                      & \multicolumn{1}{c|}{16M} & 1000        & 2000         & \multicolumn{1}{c|}{$10^{-1}$} & 500           & 2000           & $10^{-1}$   &                               &                               &                      &                      \\
    \multicolumn{1}{c|}{5M}  & 2000         & 500          & \multicolumn{1}{c|}{$10^{-6}$} & 25            & 1000                & $10^{-6}$   &                      & \multicolumn{1}{c|}{17M} & 1000        & 2000         & \multicolumn{1}{c|}{$10^{-1}$} & 500           & 2000            & $10^{-1}$   &                               &                               &                      &                      \\
    \multicolumn{1}{c|}{6M}  & 250         & 2000        & \multicolumn{1}{c|}{$10^{-2}$} & 25           & 1000           & $10^{-6}$   &                      & \multicolumn{1}{c|}{18M} & 1000        & 2000         & \multicolumn{1}{c|}{$10^{-1}$} & 500           & 2000            & $10^{-1}$   &                               &                               &                      &                      \\
    \multicolumn{1}{c|}{7M}  & 250         & 2000           & \multicolumn{1}{c|}{$10^{-2}$} & 10           & 1000           & $10^{-6}$   &                      & \multicolumn{1}{c|}{19M} & 1000       & 2000         & \multicolumn{1}{c|}{$10^{-1}$} & 500           & 2000            & $10^{-1}$   &                               &                               &                      &                      \\
    \multicolumn{1}{c|}{8M}  & 250         & 2000         & \multicolumn{1}{c|}{$10^{-2}$} & 10         & 1000           & $10^{-4}$   &                      & \multicolumn{1}{c|}{20M} & 1000        & 2000         & \multicolumn{1}{c|}{$10^{-1}$} & 500          & 2000          & $10^{-1}$   &                               &                               &                      &                      \\
    \multicolumn{1}{c|}{9M}  & 250         & 2000         & \multicolumn{1}{c|}{$10^{-2}$} & 10           & 1000           & $10^{-4}$   &                      & \multicolumn{1}{c|}{21M} & 1000       & 2000         & \multicolumn{1}{c|}{$10^{-1}$} & 500          & 2000            & $10^{-1}$   &                               &                               &                      &                      \\
    \multicolumn{1}{c|}{10M} & 250         & 2000         & \multicolumn{1}{c|}{$10^{-2}$} & 10           & 1000           & $10^{-3}$   &                      & \multicolumn{1}{c|}{22M} & 1000        & 2000         & \multicolumn{1}{c|}{$10^{-1}$} & 500          & 2000           & $10^{-1}$   &                               &                               &                      &                      \\
    \multicolumn{1}{c|}{11M} & 2000         & 2000         & \multicolumn{1}{c|}{$10^{-2}$} & 10           & 1000           & $10^{-2}$   &                      & \multicolumn{1}{c|}{23M} & 100        & 1000         & \multicolumn{1}{c|}{$10^{-1}$} & 500           &2000          & $10^{-1}$   &                               &                               &                      &                      \\
    \multicolumn{1}{c|}{12M} & 1000         & 2000         & \multicolumn{1}{c|}{$10^{-1}$} & 10           & 1000           & $10^{-4}$   &                      & \multicolumn{1}{c|}{24M} & 100        & 1000         & \multicolumn{1}{c|}{$10^{-1}$} & 500          &2000           & $10^{-1}$   &                               &                               &                      &                      \\ \cmidrule(r){1-7} \cmidrule(lr){9-15}
    \end{tabular}%
    }
\end{table}
First, we observe that the cut-off lags $C_{\Sigma}$ are mostly above 1,000 days and when they are not, they are at least above 250 days and this occurs only for time-to-maturities smaller than 5 months. Second, looking at the average $R^2$ scores for all tested triplets, we observed that the fitting quality is very sensitive to the cut-off lag $C_{\Sigma}$ of the volatility feature while the two other hyperparameters $C_{R_1}$ and $\lambda$ have a smaller influence, which explains the wider range of values obtained for these two hyperparameters in Table \ref{tab:best_triplets}. Choosing a too small cut-off lag $C_{\Sigma}$ can lead to very low $R^2$ scores, especially for the largest time-to-maturities. For example, for $C_{\Sigma}=500$, we obtain $R^2$ scores that are even below 0, as illustrated in Figure \ref{fig:r2_cutoff_500}. Actually any value $C_{\Sigma}$ strictly below 1000 in the grid $\{5,10,25,50,100,250,500,1000,1500,2000,2500,3000\}$ yields overall poor results such as those exhibited in Figure \ref{fig:r2_cutoff_500} and this regardless of the value of $C_{R_1}$. This indicates that considering the squared returns up to 1,000 business days in the past is paramount to predict the implied volatility, particularly for the largest time-to-maturities.\\

\begin{figure}
    \centering
    \subfigure[1-month ATM implied volatility (S\&P 500)]{\includegraphics[width=0.3\linewidth]{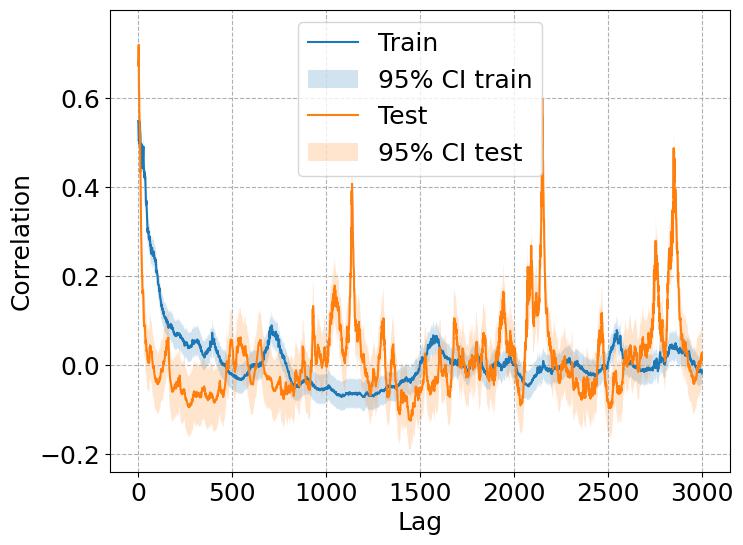}\label{fig:1_month_SP500}}
    \subfigure[12-months ATM implied volatility (S\&P 500)]{\includegraphics[width=0.3\linewidth]{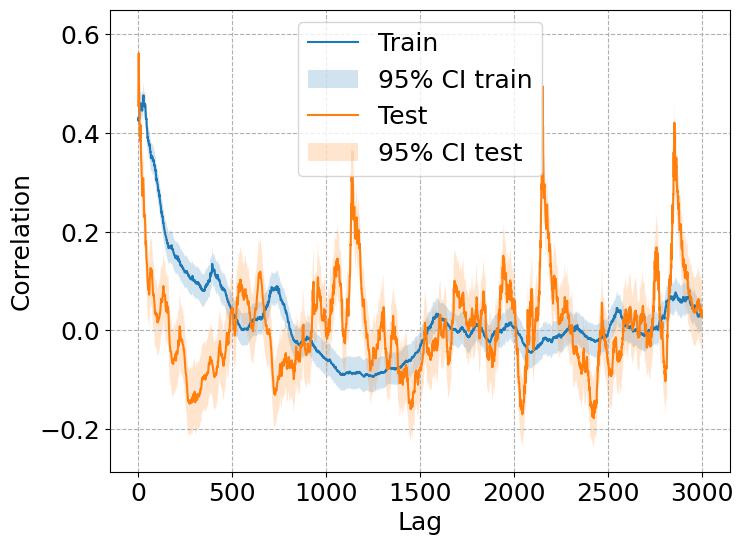}}
    \subfigure[24-months ATM implied volatility (S\&P 500)]{\includegraphics[width=0.3\linewidth]{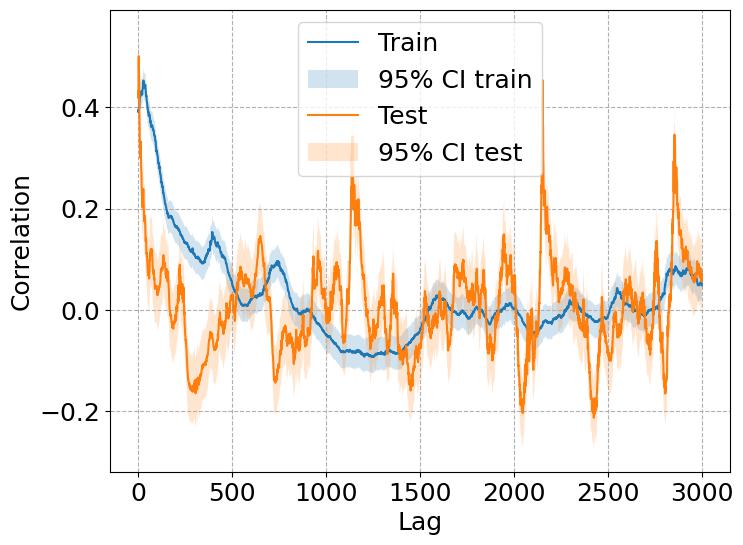}}
    \subfigure[1-month ATM implied volatility (Euro Stoxx 50)]{\includegraphics[width=0.3\linewidth]{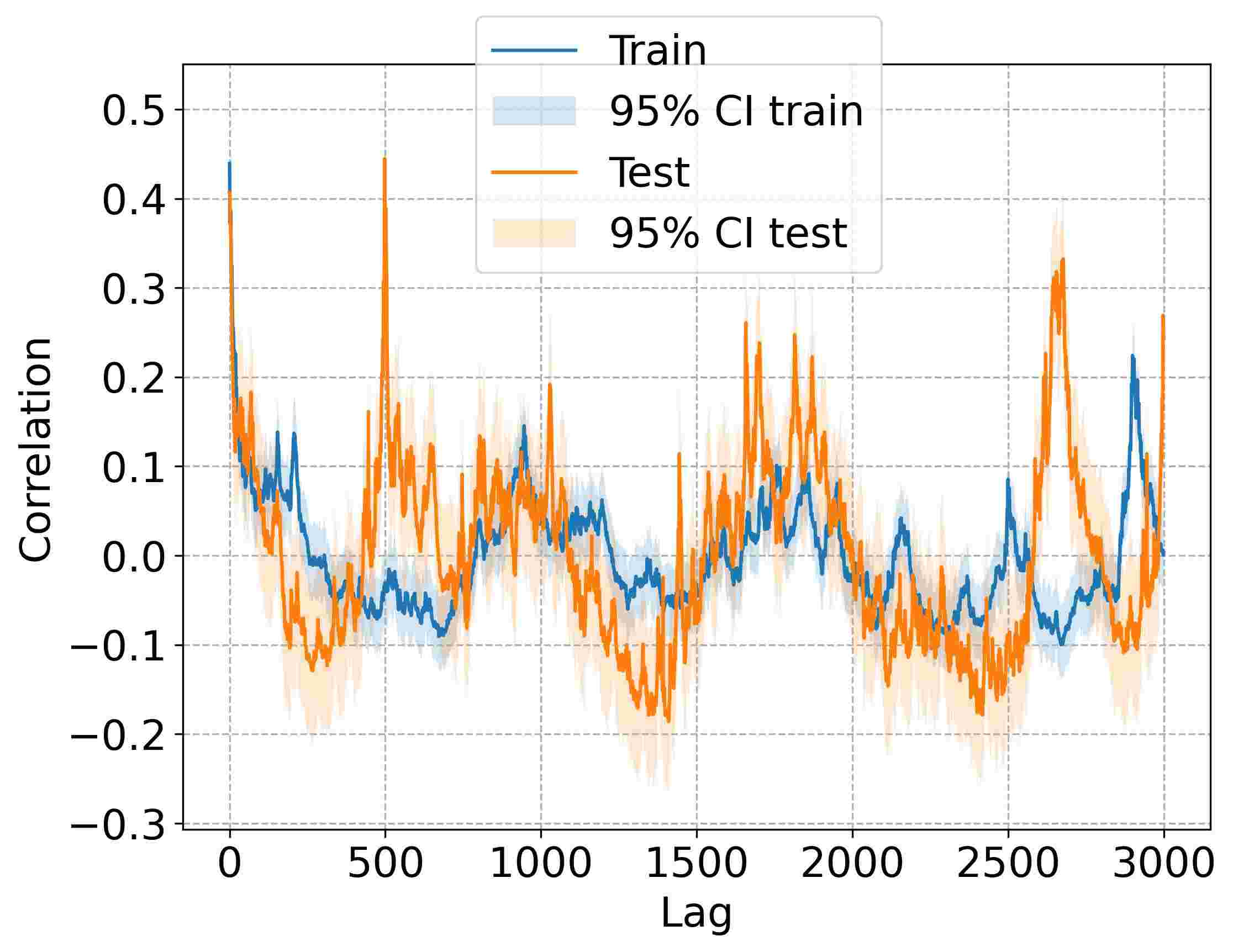}\label{fig:1_month_SX5E}}
    \subfigure[12-months ATM implied volatility (Euro Stoxx 50)]{\includegraphics[width=0.3\linewidth]{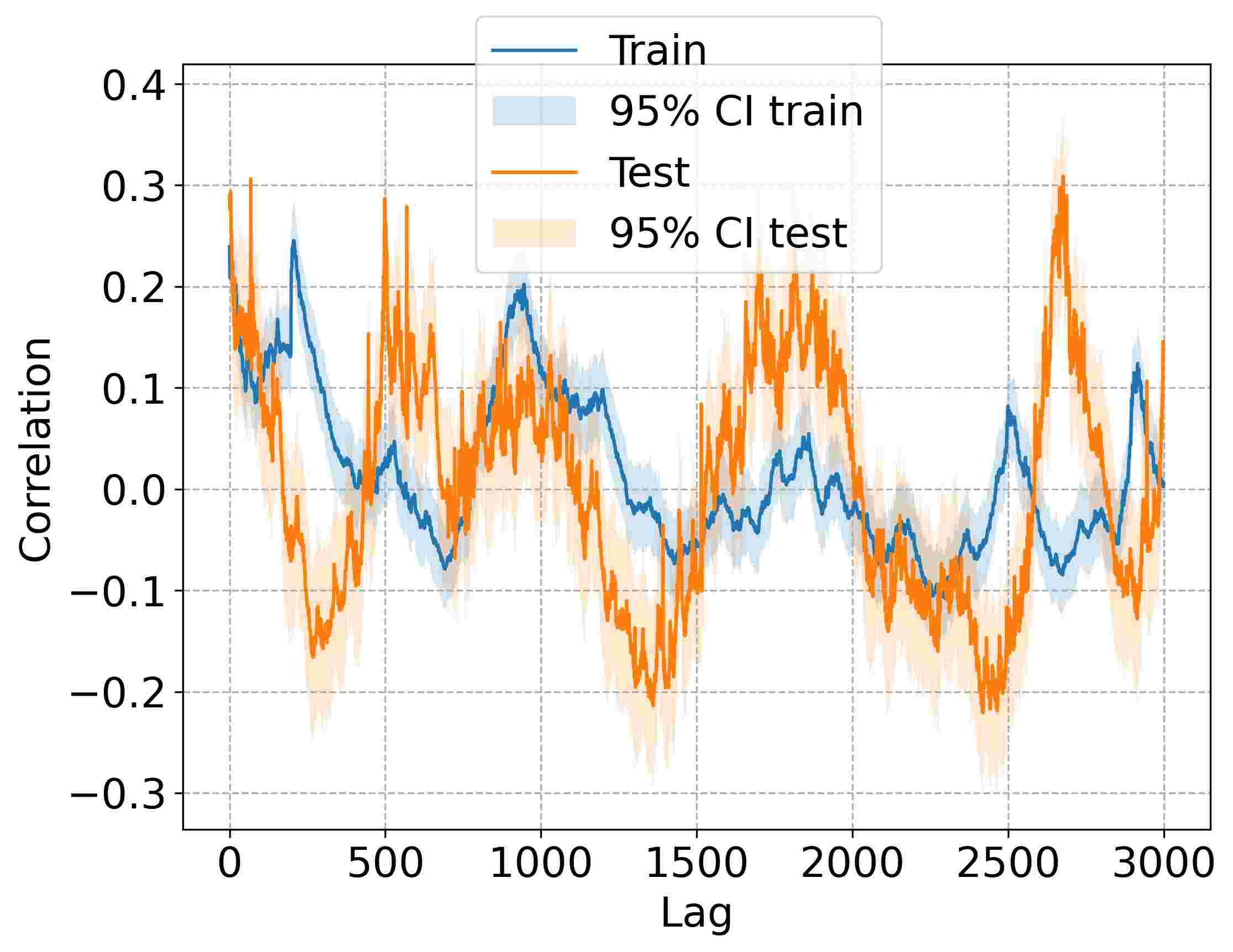}}
    \subfigure[24-months ATM implied volatility (Euro Stoxx 50)]{        \includegraphics[width=0.3\linewidth]{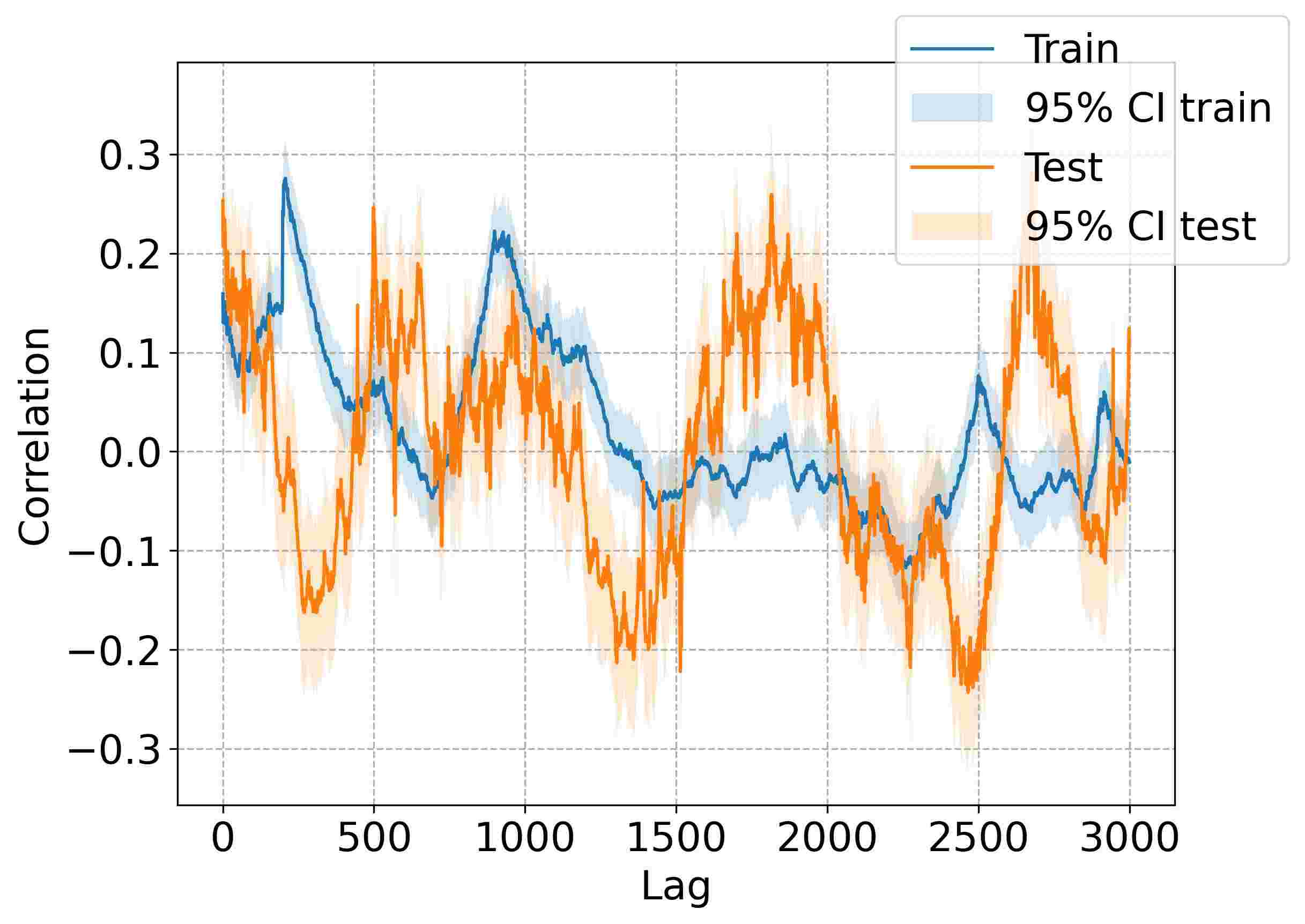}}
    \caption{Correlation between the squared ATM implied volatility and the squared daily returns as a function of the lag both on the train and the test sets. The 95\% confidence intervals are derived from the Fisher transformation. }
\label{fig:corr_structure}
\end{figure}

\begin{figure}
    \centering
    \subfigure[1-month ATM implied volatility (S\&P 500)]{\includegraphics[width=0.3\linewidth]{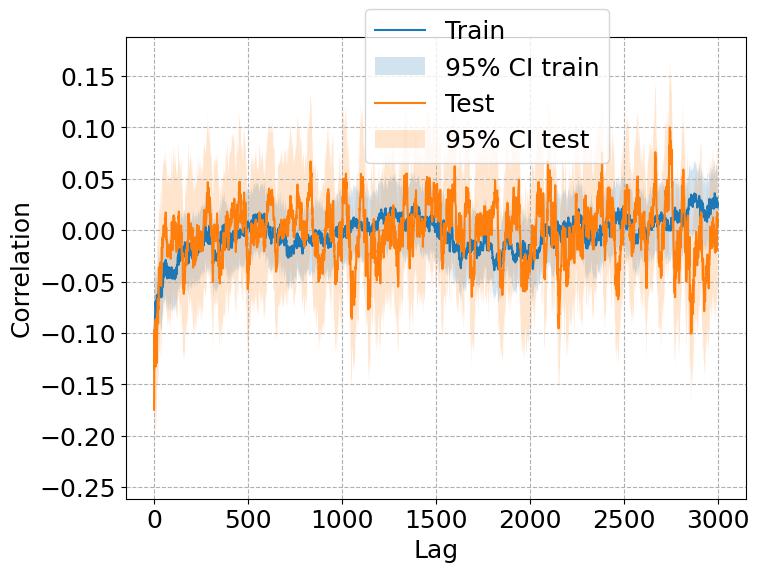}}
    \subfigure[12-months ATM implied volatility (S\&P 500)]{\includegraphics[width=0.3\linewidth]{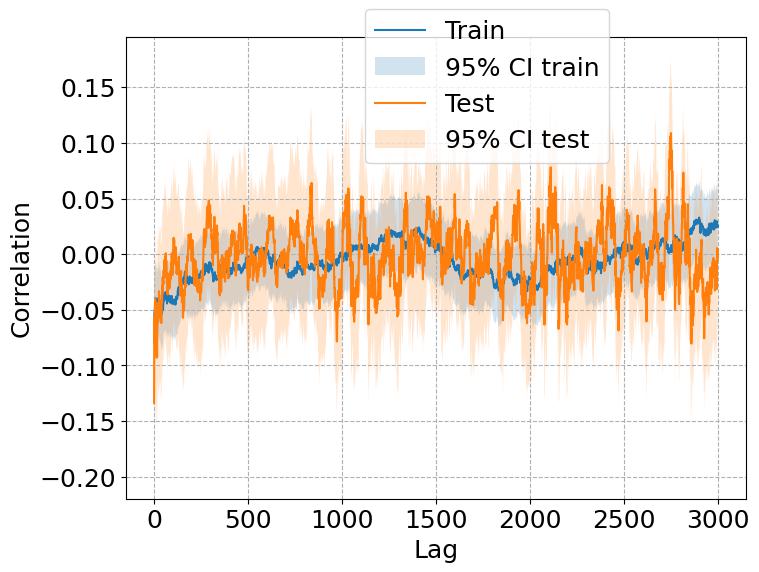}}
    \subfigure[24-months ATM implied volatility (S\&P 500)]{\includegraphics[width=0.3\linewidth]{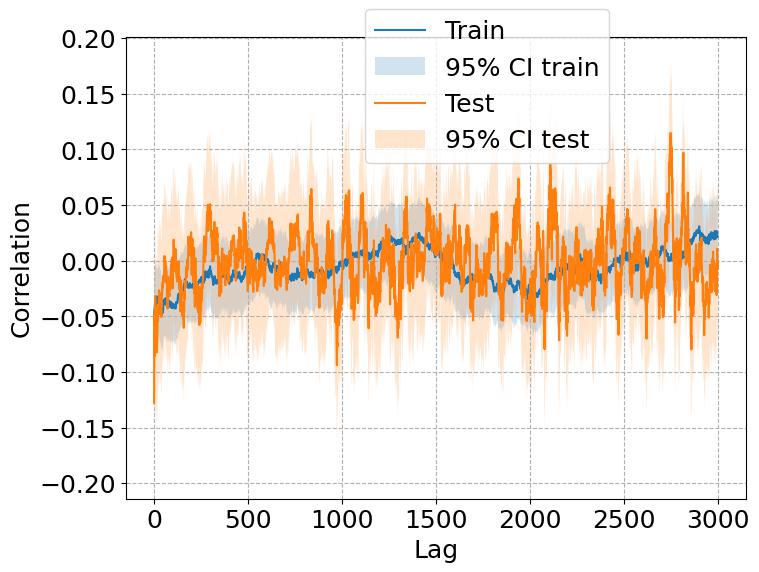}}
    \subfigure[1-month ATM implied volatility (Euro Stoxx 50)]{ \includegraphics[width=0.3\linewidth]{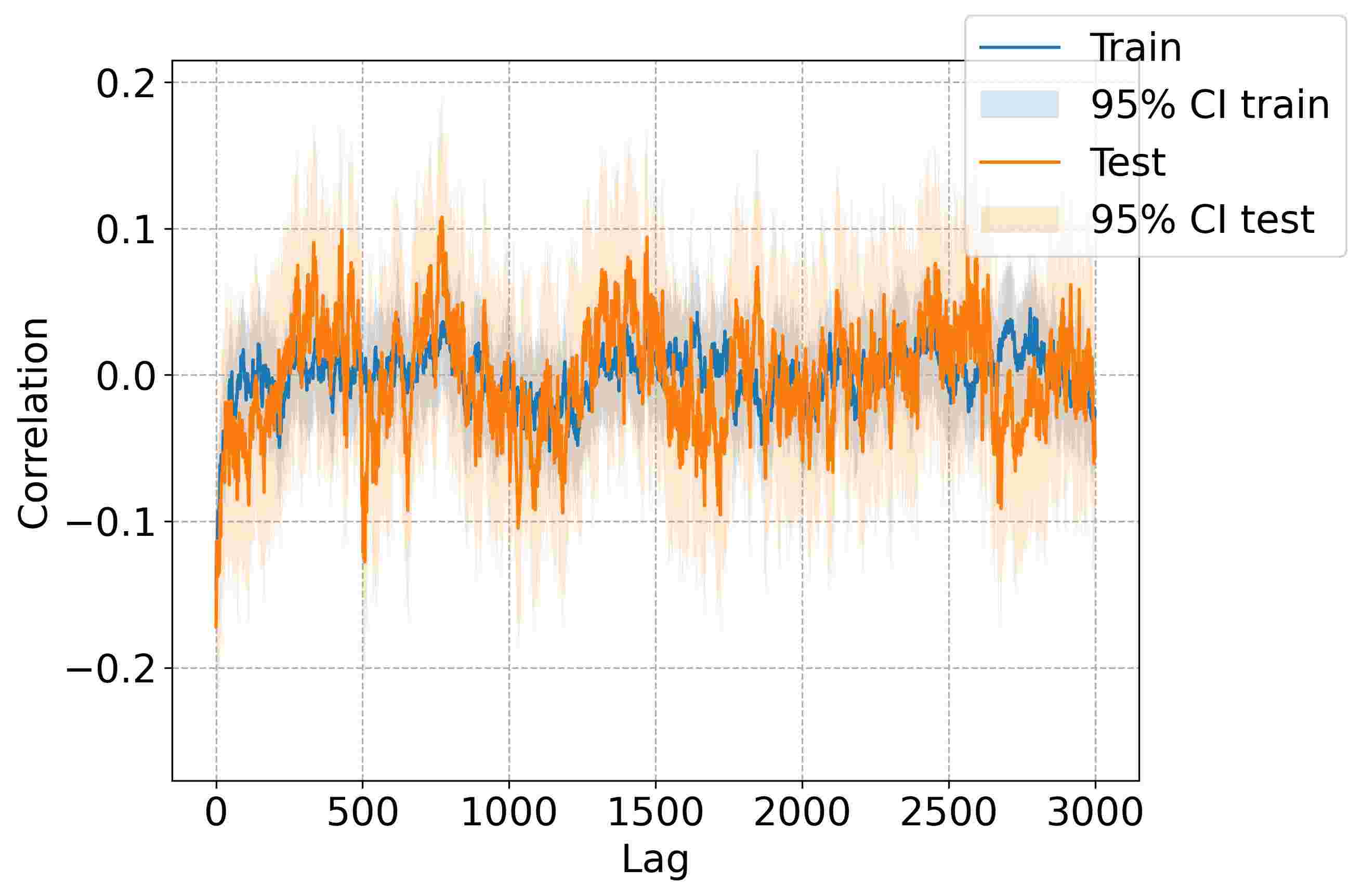}}
    \subfigure[12-months ATM implied volatility (Euro Stoxx 50)]{\includegraphics[width=0.3\linewidth]{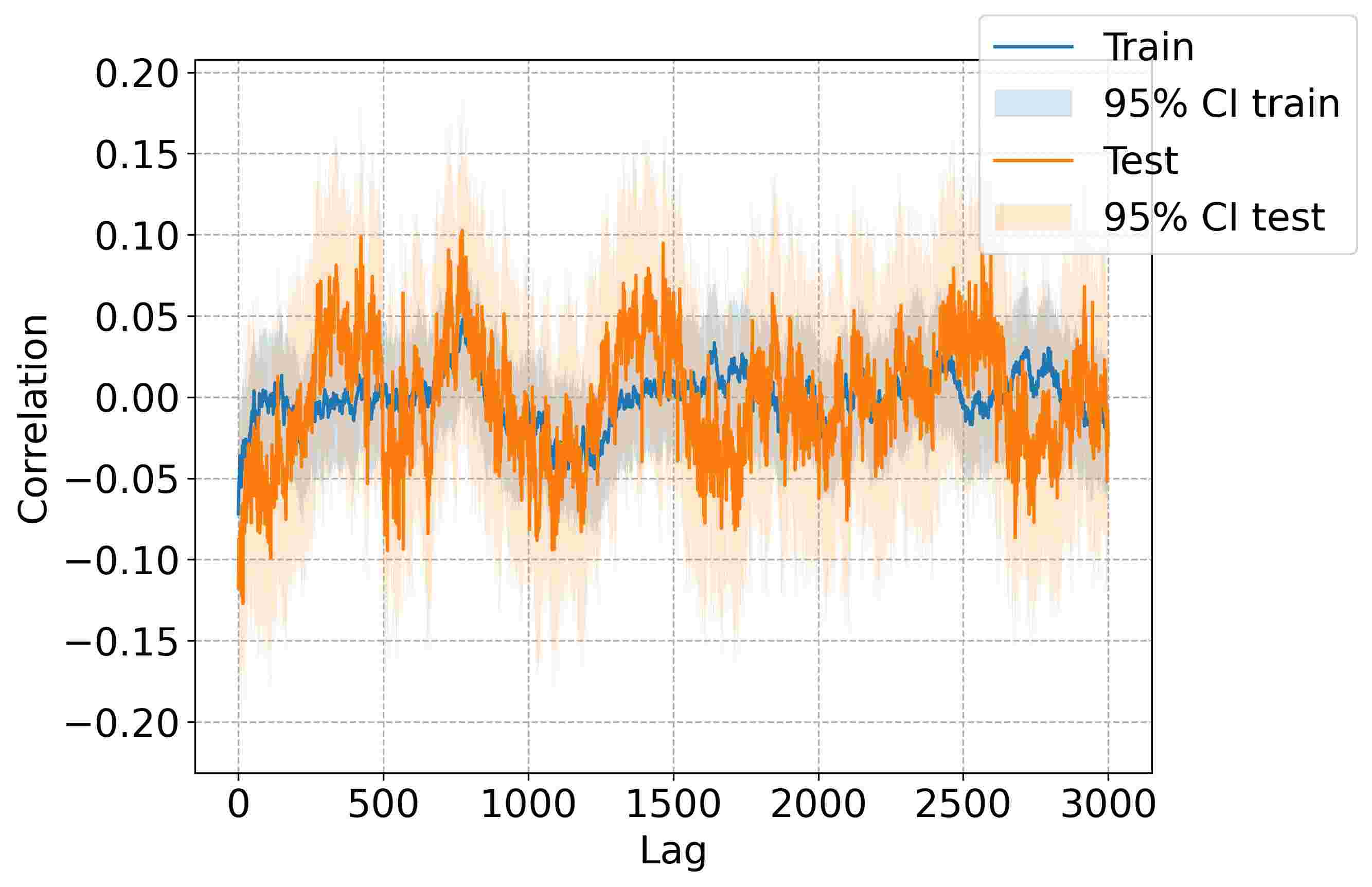}}
    \subfigure[24-months ATM implied volatility (Euro Stoxx 50) ]{        \includegraphics[width=0.3\linewidth]{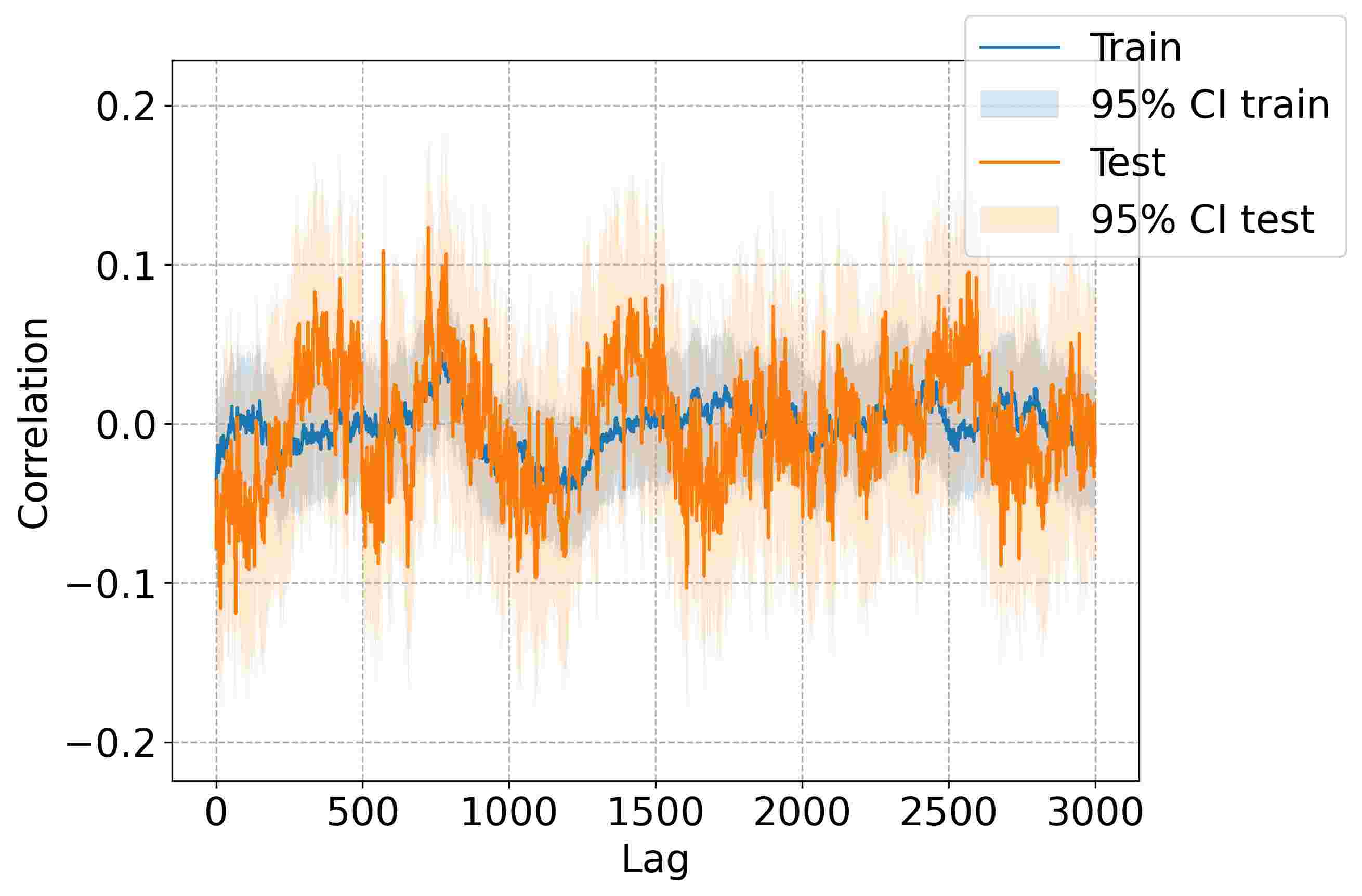}}
    \caption{Correlation between the ATM implied volatility and the daily returns as a function of the lag both on the train and the test sets. The 95\% confidence intervals are derived from the Fisher transformation. }
\label{fig:corr_structure_r1}
\end{figure}

In order to verify that this conclusion is not an artefact of a bad model calibration, we present in Figure \ref{fig:corr_structure} the correlation between the implied variance and the squared daily returns for all lags between 0 and 3,000 days both on the train and the test sets. Note that the estimated correlation $\rho$ is presented along with a 95\% confidence interval derived from the transformation $z=\text{artanh}(\rho)$ introduced by \cite{fisher1915frequency}. Indeed, this transformation is approximately normally distributed when the samples come from a bivariate normal distribution. Although this is not the case here, we consider it as a reasonable proxy of the uncertainty around the correlation estimator. On the train set (blue curve), the graphs show a slow decrease of the correlation with the lag and the decrease becomes slower when the time-to-maturity increases. For the Euro Stoxx 50, we can even observe some spikes at some specific lags that become larger when the time-to-maturity increases. These observations are consistent with the sensitivity of the model to the cut-off lag $C_{\Sigma}$ and the values obtained with the cross-validation that are presented in Table \ref{tab:best_triplets}. Note that these observations have the advantage of not depending on any assumption. Thus, we can consider the long-range dependence of the implied volatility to the past squared returns as a stylized fact of our implied volatility data. An empirical study of a larger set of underlying assets could reveal whether this is a universal property of implied volatility data. To our knowledge, this stylized fact has never been reported in the literature. Let us however mention the work of \cite{hardle2007long} who calibrate a 3-factor model on implied volatility data and show a long-range dependence in the level and absolute returns of the factors loading series. A possible explanation of this phenomenon is that options on widespread equity indices and with relatively large time-to-maturities are presumably traded by long-term investors such as asset managers, pension funds, sovereign funds, etc. who have a low rebalancing frequency of their portfolios and consequently, who base their investment decisions on the previous years returns of the underlying asset rather than the previous days returns. On the other hand, options with shorter time-to-maturities are presumably less traded by long-term investors so that the movements of the implied volatility are more influenced by short-term investors such as hedge funds who base their investment decisions on recent data. This is in line with Figures \ref{fig:1_month_SP500} and \ref{fig:1_month_SX5E} as well as with the smaller values of $C_{\Sigma}$ for the smallest time-to-maturities in Table \ref{tab:best_triplets}. In Figure \ref{fig:corr_structure_r1}, we also present the correlations between the implied volatility and the daily returns for all lags between 0 and 3,000 days both on the train and the test sets. In this case, the correlations fades very quickly with the lag and we do not observe material spikes. \\

So far, we have only described the correlations on the train set but as already discussed extensively, the test set is quite different and the correlations on this set (in orange in Figures \ref{fig:corr_structure} and \ref{fig:corr_structure_r1}) are therefore also distinct from those calculated on the train set. In particular in Figure \ref{fig:corr_structure}, we observe negative correlations with the 250 days lag which can be understood as a consequence of the fact that the implied volatility was decreasing in 2021 due to the leverage effect while one year earlier the squared returns were increasing with the Covid-19 crisis. Conversely, the implied volatility increased in 2022 again due to the leverage effect while one year earlier the squared returns were decreasing with the post-Covid-19 bull market. 

\section{Study of the calibrated parameters}

\begin{figure}[h]
    \centering
    \subfigure[S\&P 500]{\includegraphics[width=0.45\linewidth]{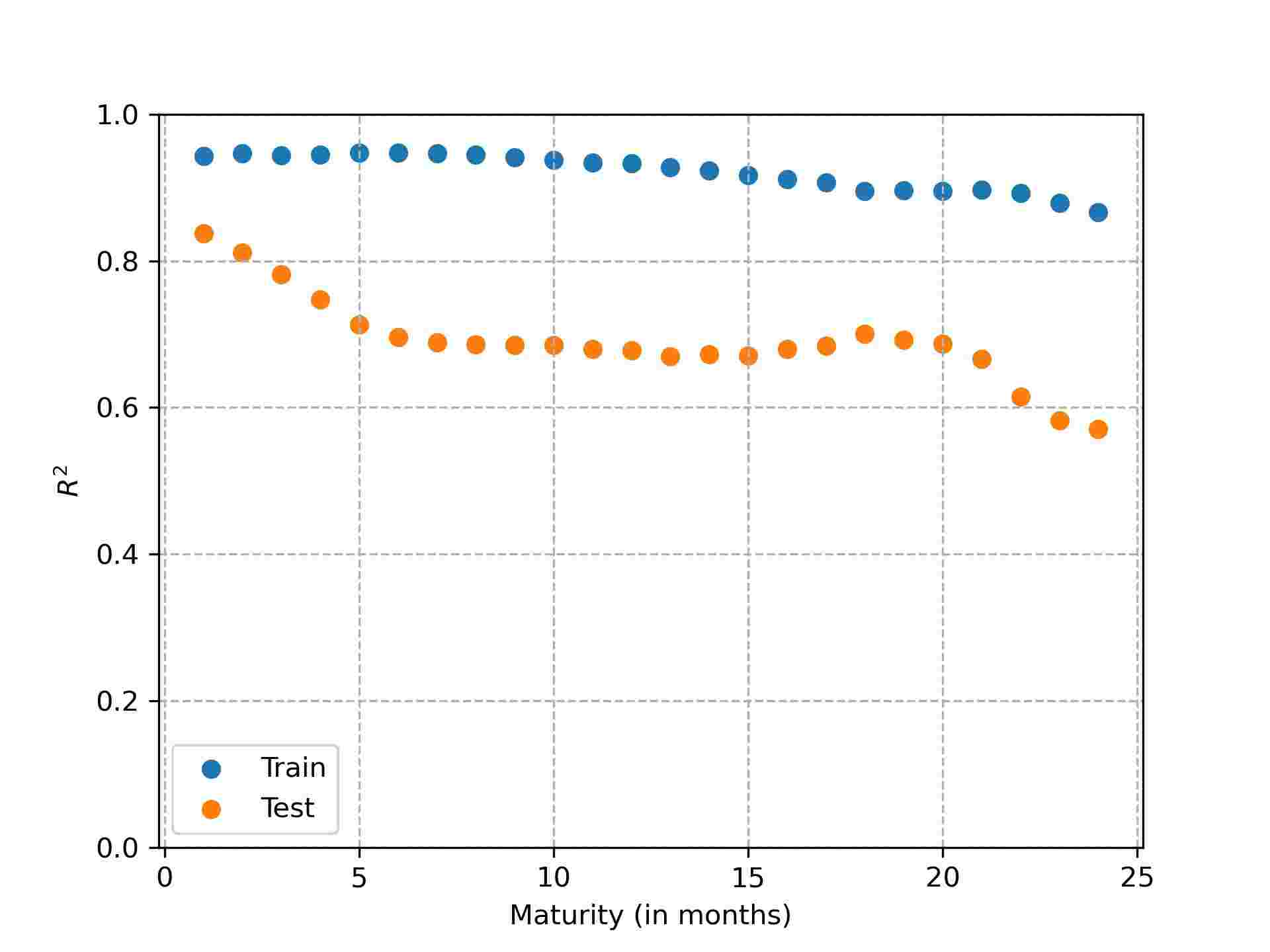}}
    \subfigure[Euro Stoxx 50]{ \includegraphics[width=0.45\linewidth]{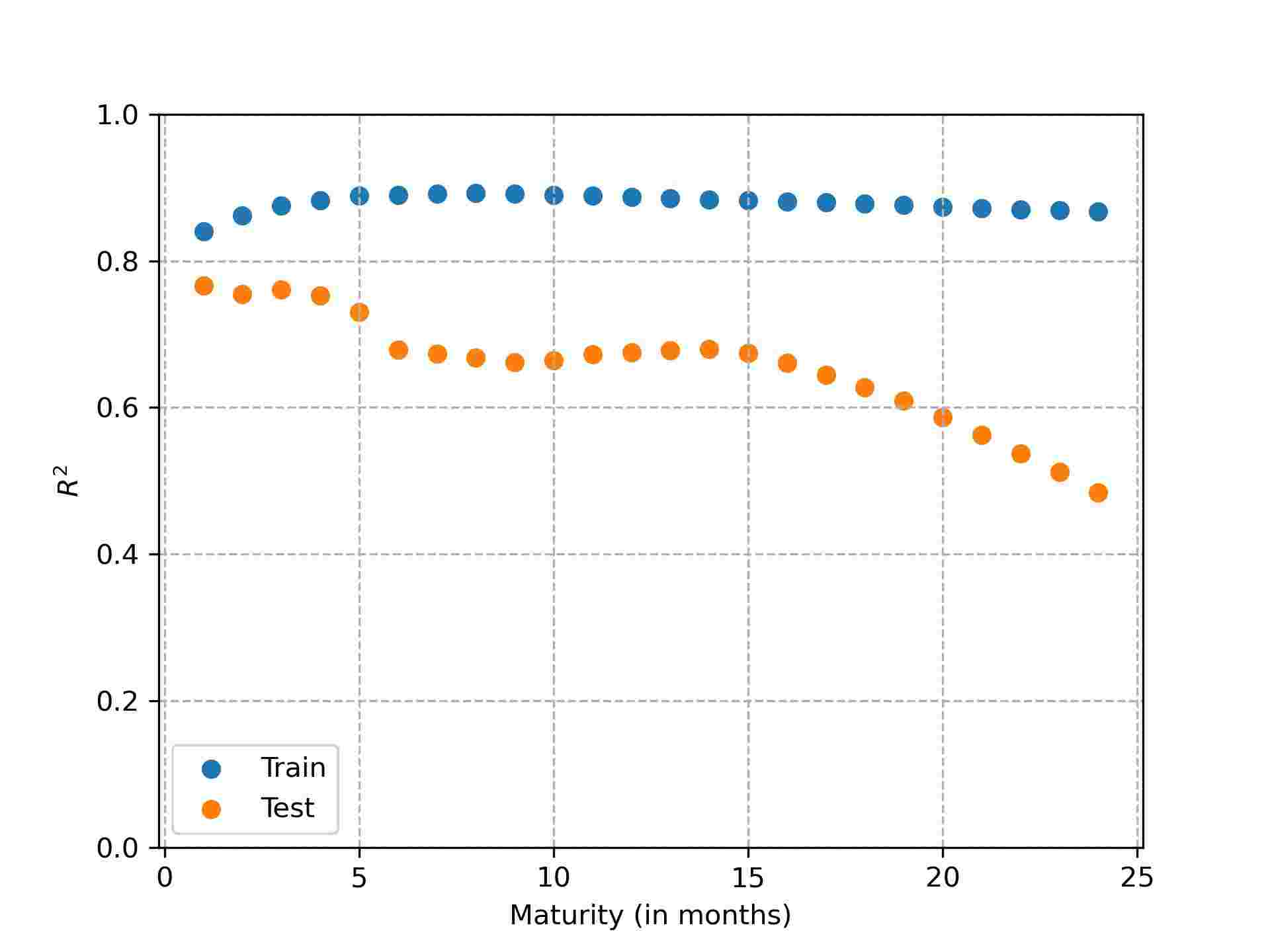}}
    \caption{$R^2$ scores on the train and the test sets as a function of the ATM implied volatility time-to-maturity. The hyperparameters for the S\&P 500 are $(C_{R_1,C_{\Sigma},\lambda)=(250,1000,10^{-3})}$ and those of the Euro Stoxx 50 are $(C_{R_1},C_{\Sigma},\lambda)=(10,1000,10^{-3})$. }
\label{fig:r2_cross_valid}
\end{figure}

We conclude this empirical study by analyzing the calibrated parameters of the PDV model. In order to obtain comparable model parameters between time-to-maturities, we retain a single triplet $(C_{R_1},C_{\Sigma},\lambda)$ for all time-to-maturities. This triplet is selected as follows. For each time-to-maturity, we compute the average $R^2$ score over the 10 folds of each triplet and then, we average these scores over all time-to-maturities. Finally, for each triplet $(C_{R_1},C_{\Sigma},\lambda)$, we average the obtained score with the one of the triplets $(C_{R_1}',C_{\Sigma}',\lambda')$ such that $C'_{\Sigma}=C_{\Sigma}$ and either $C_{R_1}'$ is the closest value above or below $C_{R_1}$ in the grid $\{5,10,25,50,100,250,500,1000,\dots,2500,3000\}$ or $\lambda'$ is the closest value above or below $\lambda$ in the grid $\{10^{-6},10^{-5},\dots,10^{-1}\}$. For example, the score of the triplet $(50,500,10^{-3})$ is averaged with the one of the triplets $(25,500,10^{-3})$, $(100,500,10^{-3})$, $(50,500,10^{-4})$ and $(50,500,10^{-2})$. The triplet that is chosen for all time-to-maturities is the one achieving the higher score through this procedure. This procedure aims at selecting a triplet whose performance is not too sensitive to a modification of $C_{R_1}$ or $\lambda$ and is introduced because we observed that if we consider only the average score over all time-to-maturities, the performance of the obtained triplet was very sensitive to these two hyperparameters (unlike most triplets as underlined earlier) and was quite bad on the test set. This instability is probably due to the small size of the train set which is divided in 10 folds in the blocked cross-validation. With this procedure, we obtain $(C_{R_1},C_{\Sigma},\lambda)=(250,1000,10^{-3})$ for the S\&P 500 and $(C_{R_1},C_{\Sigma},\lambda)=(10,1000,10^{-3})$ for the Euro Stoxx 50. In Figure \ref{fig:r2_cross_valid}, we show the $R^2$ scores obtained on the train and the test sets with these hyperparameters. We notice overall that the obtained $R^2$ scores are very close to those presented in Figure \ref{fig:r2_cutoff_1000} where $(C_{R_1},C_{\Sigma},\lambda)=(1000,1000,0)$. This is due to the fact that the cut-off lag $C_{\Sigma}$ of the volatility did not change between the two figures and, as described in Appendix \ref{sec:influence_cutoff}, the two other hyperparameters $C_{R_1}$ and $\lambda$ have a small influence. \\

\begin{figure}[ht]
    \centering
    \subfigure[$\alpha_1$ (S\&P 500)]{\includegraphics[width=0.23\linewidth]{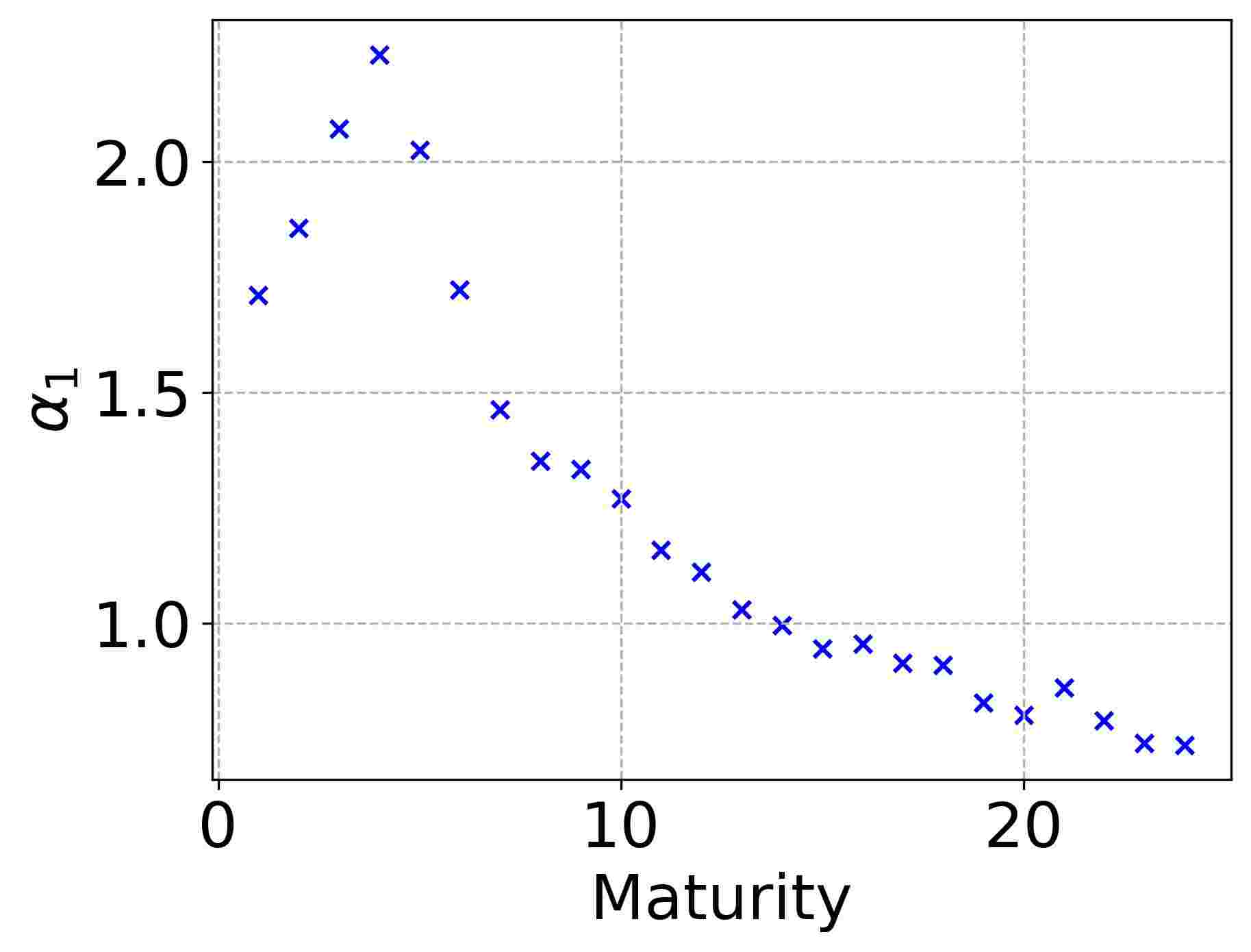}    }
    \subfigure[$\delta_1$ (S\&P 500)]{ \includegraphics[width=0.23\linewidth]{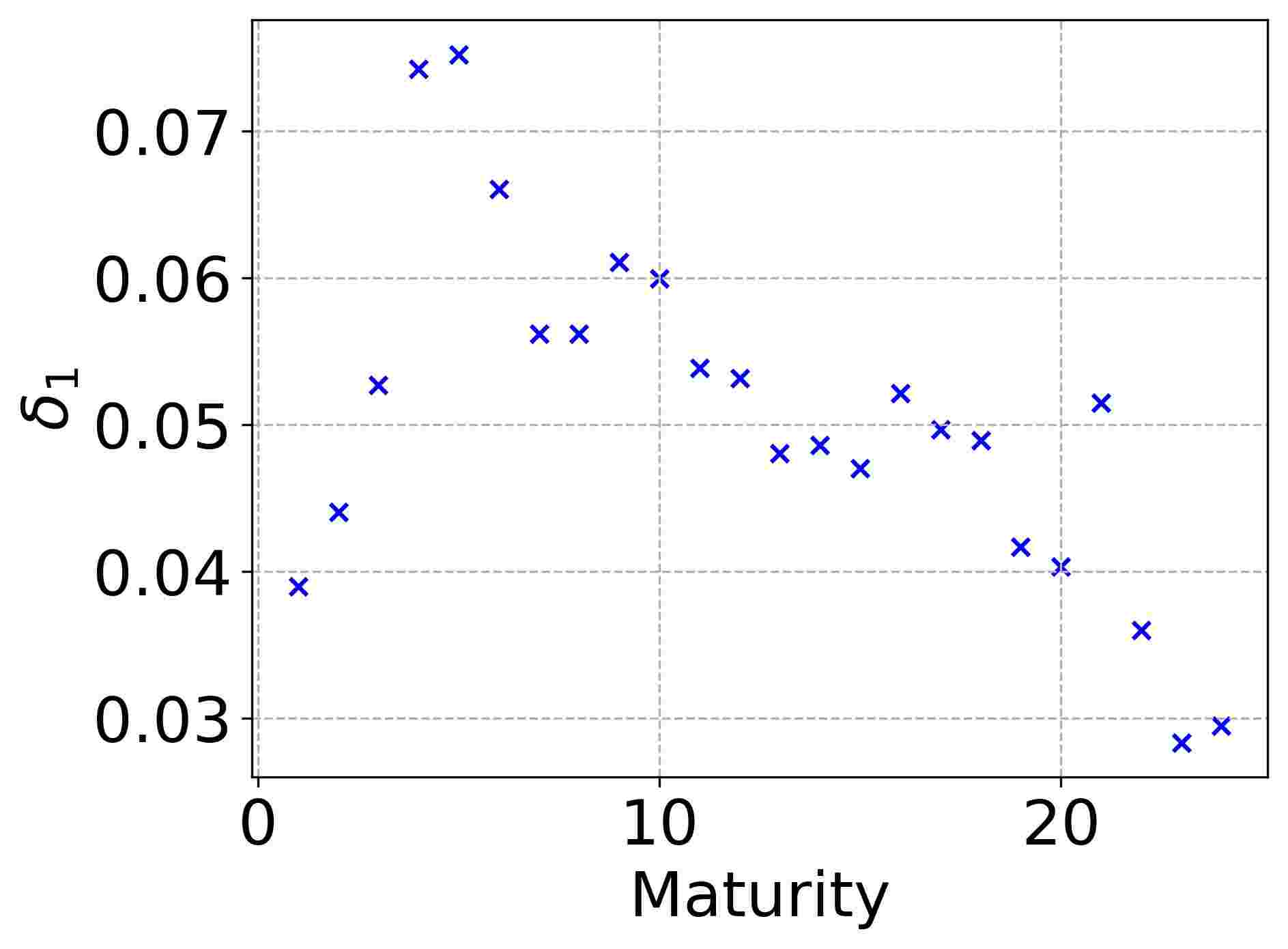}}
    \subfigure[$\alpha_2$ (S\&P 500)]{\includegraphics[width=0.23\linewidth]{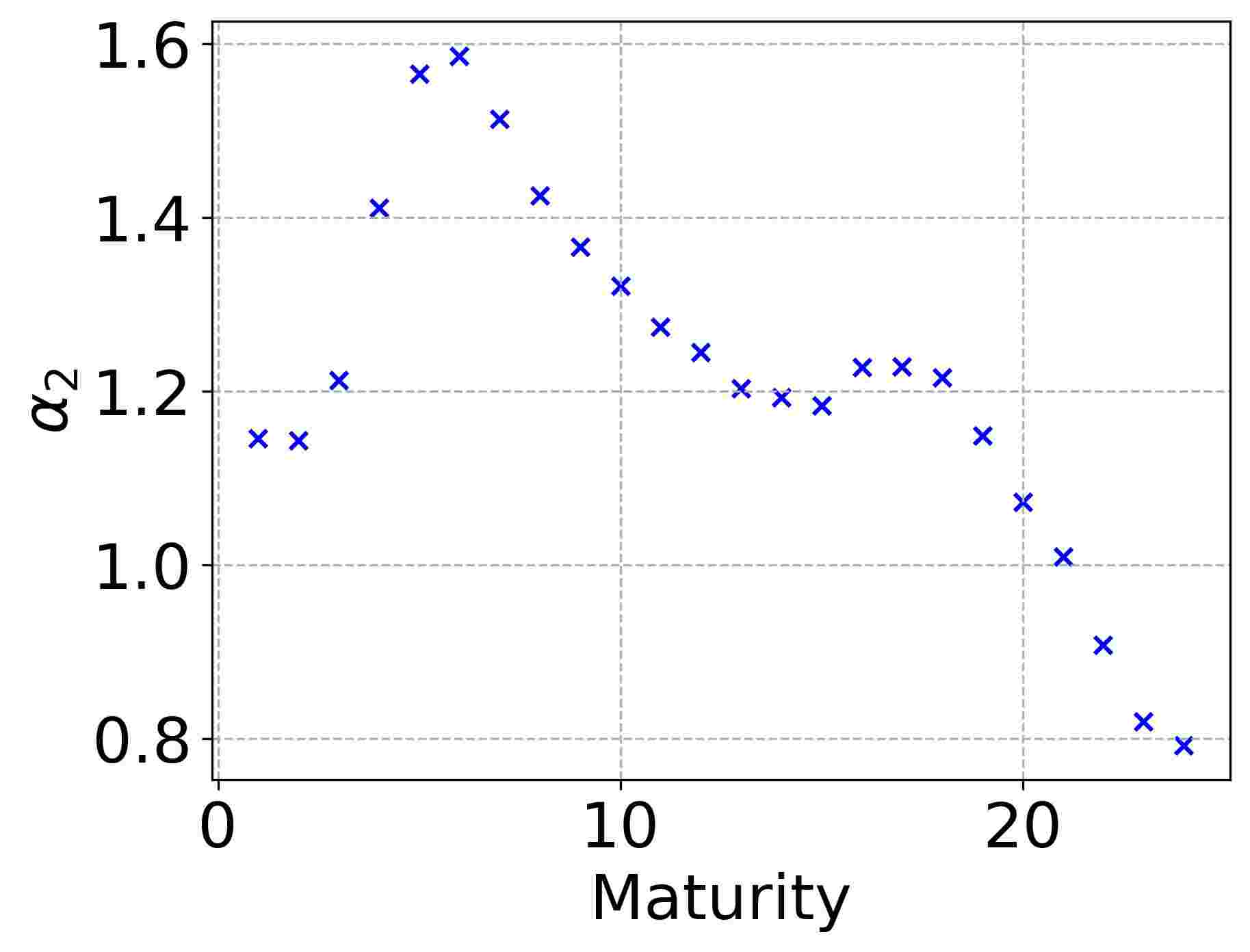}}
    \subfigure[$\delta_2$ (S\&P 500)]{\includegraphics[width=0.23\linewidth]{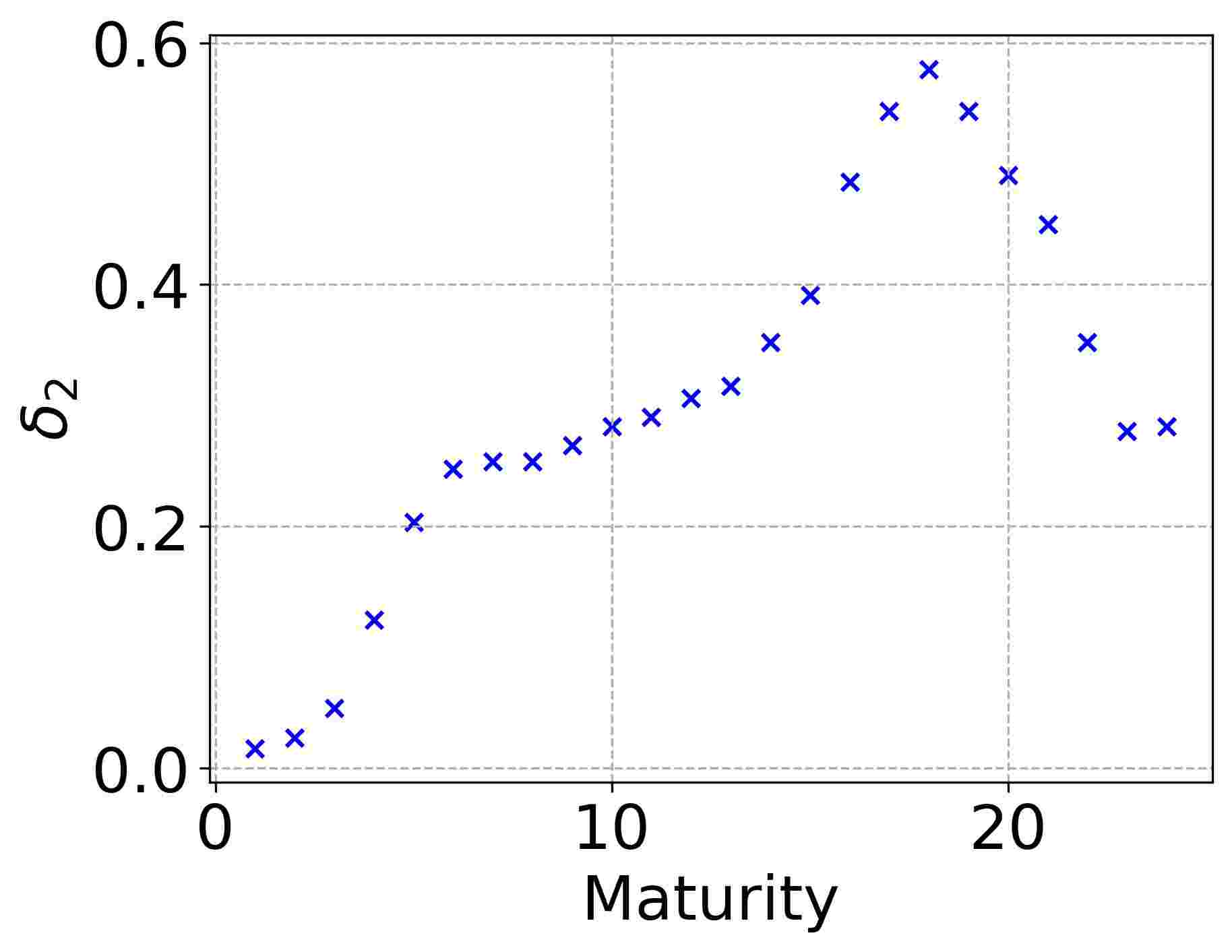}}
    \subfigure[$\alpha_1$ (Euro Stoxx 50)]{ \includegraphics[width=0.23\linewidth]{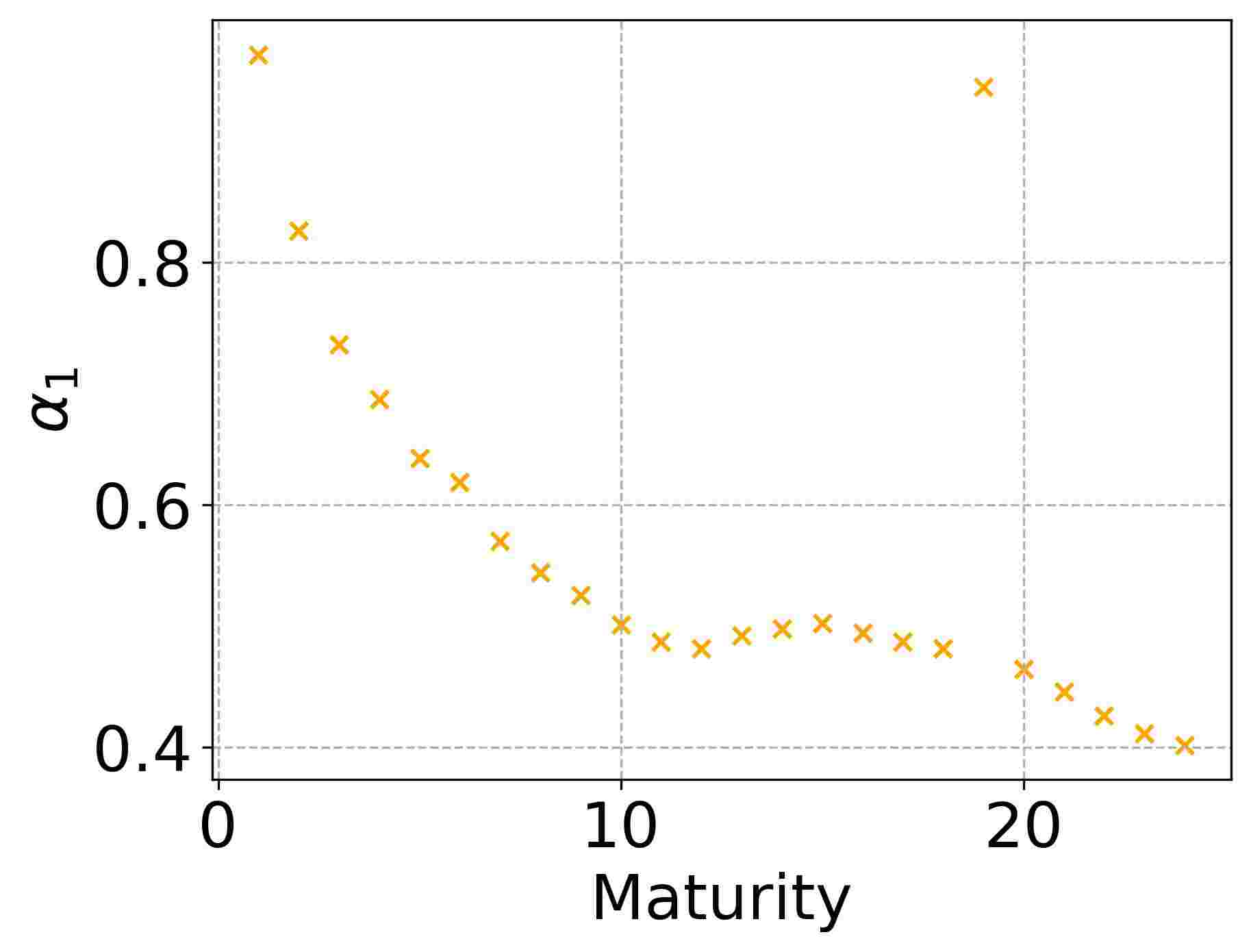}}
    \subfigure[$\delta_1$ (Euro Stoxx 50)]{\includegraphics[width=0.23\linewidth]{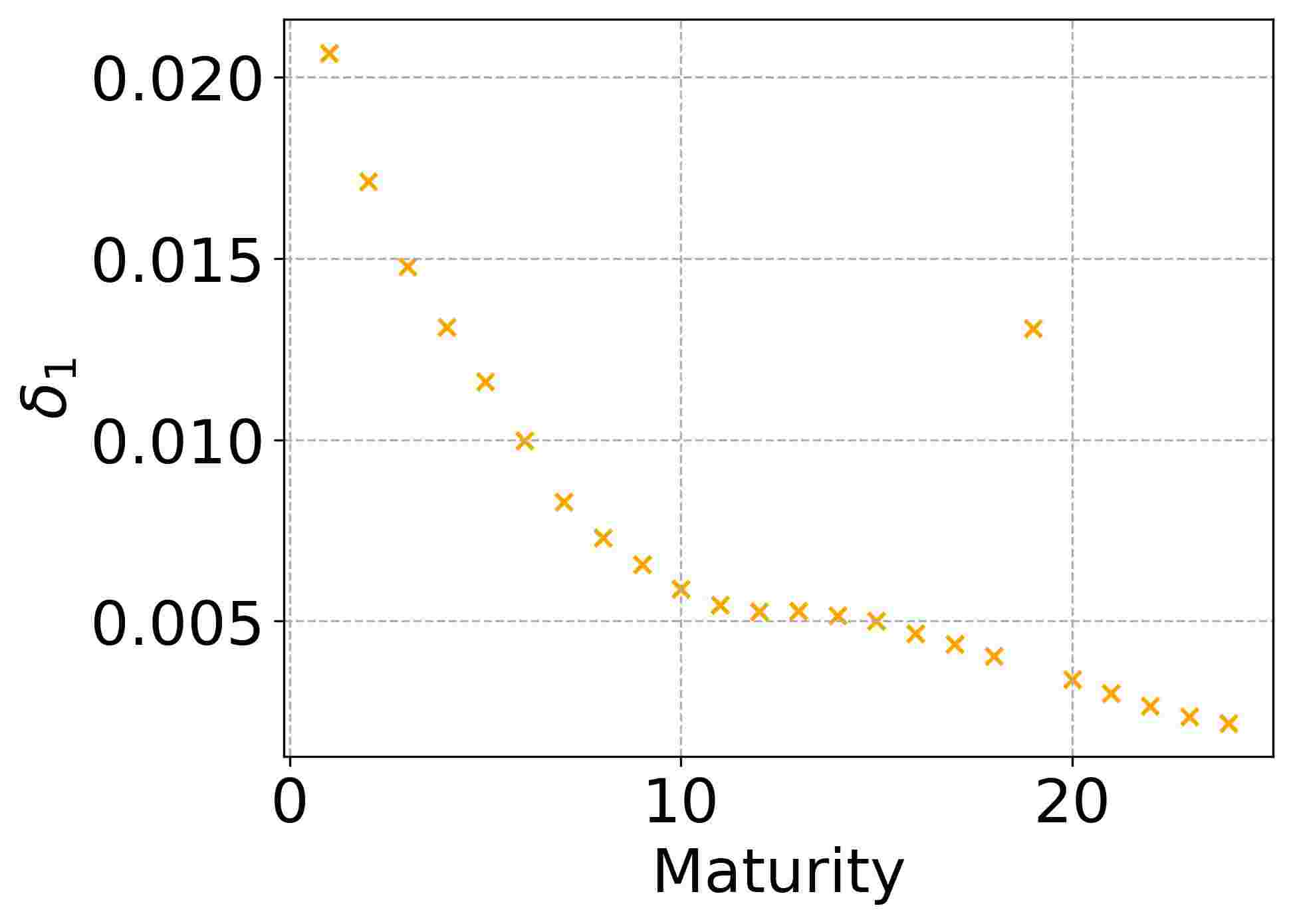}}
    \subfigure[$\alpha_2$ (Euro Stoxx 50)]{\includegraphics[width=0.23\linewidth]{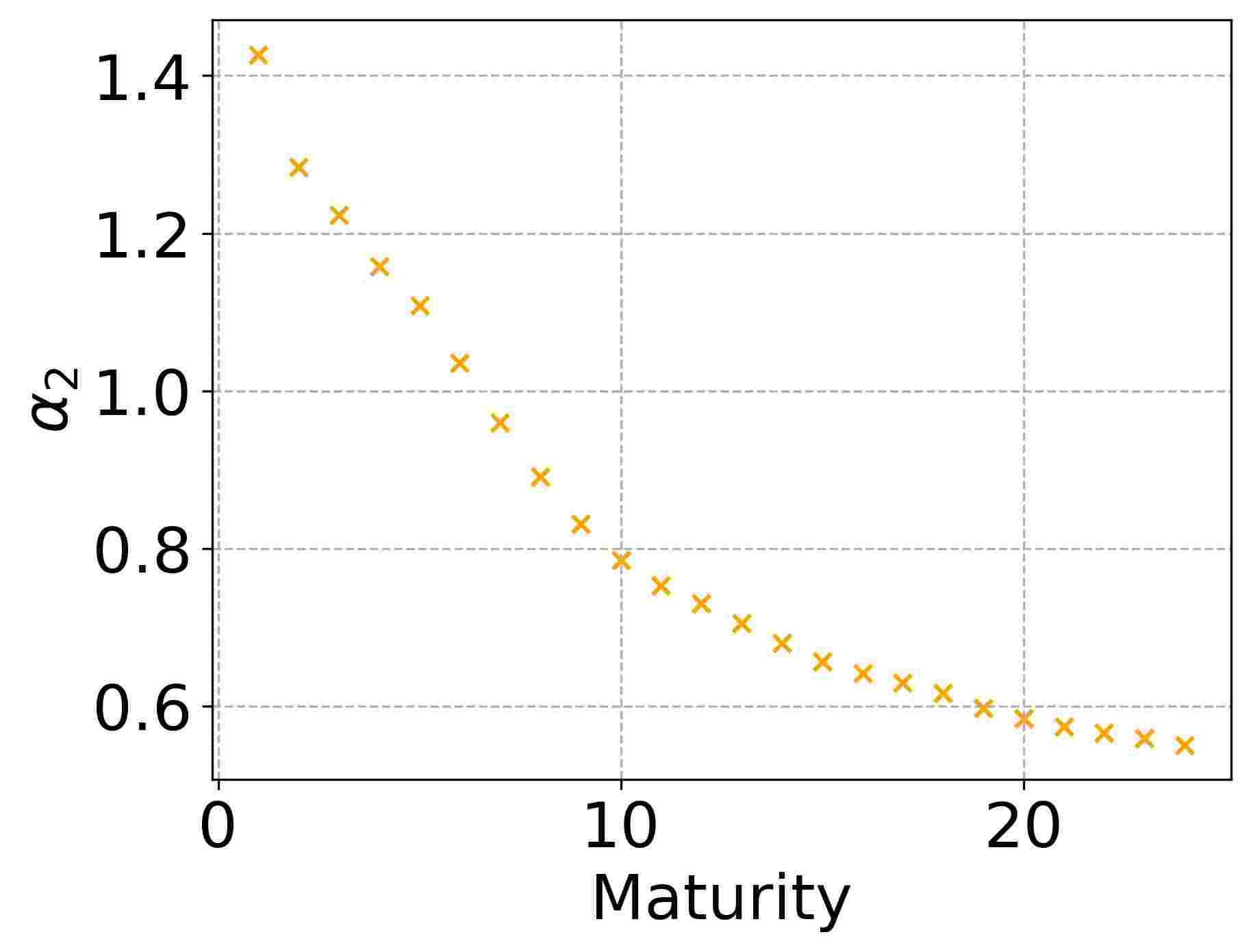}}
    \subfigure[$\delta_2$ (Euro Stoxx 50)]{ \includegraphics[width=0.23\linewidth]{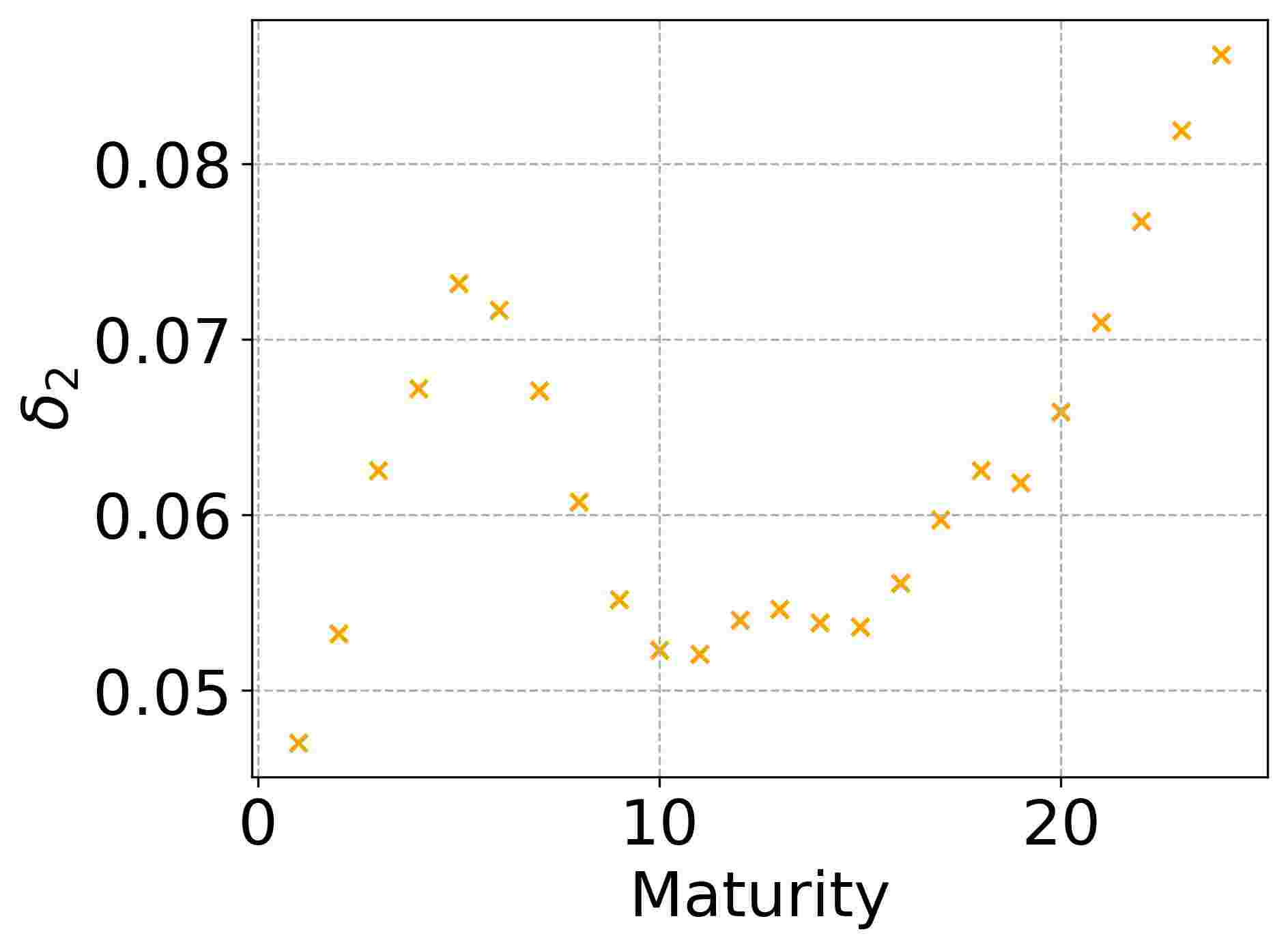}}

    \caption{Calibrated TSPL kernels parameters as a function of the time-to-maturity. The hyperparameters for the S\&P 500 are $(C_{R_1},C_{\Sigma},\lambda)=(250,1000,10^{-3})$ and those of the Euro Stoxx 50 are $(C_{R_1},C_{\Sigma},\lambda)=(10,1000,10^{-3})$.   }
\label{fig:pdv_params_tspl}
\end{figure}

\begin{figure}
    \centering
    \subfigure[$\beta_0$ (S\&P 500)]{\includegraphics[width=0.3\linewidth]{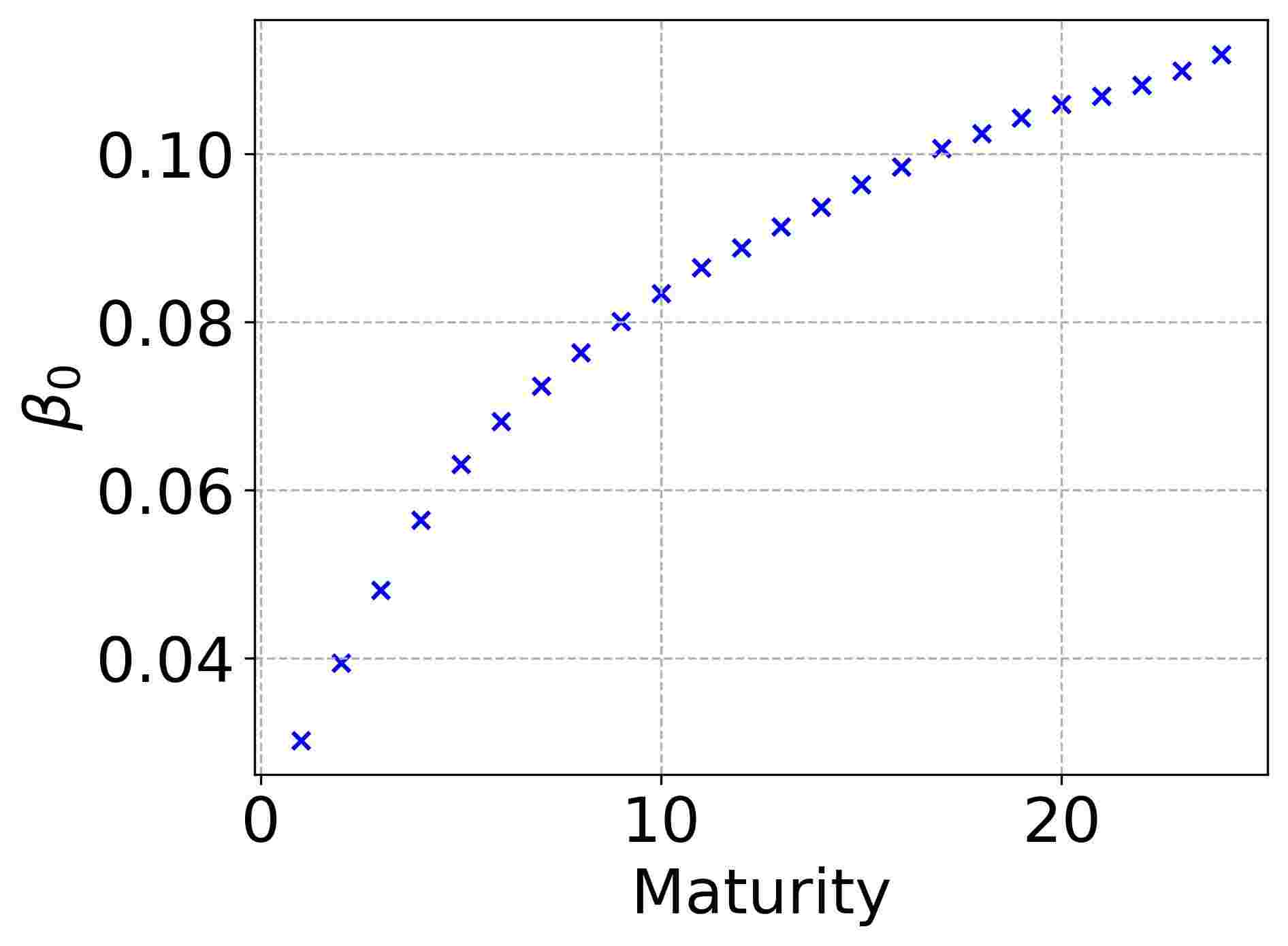}}
    \subfigure[$\beta_1$ (S\&P 500)]{\includegraphics[width=0.3\linewidth]{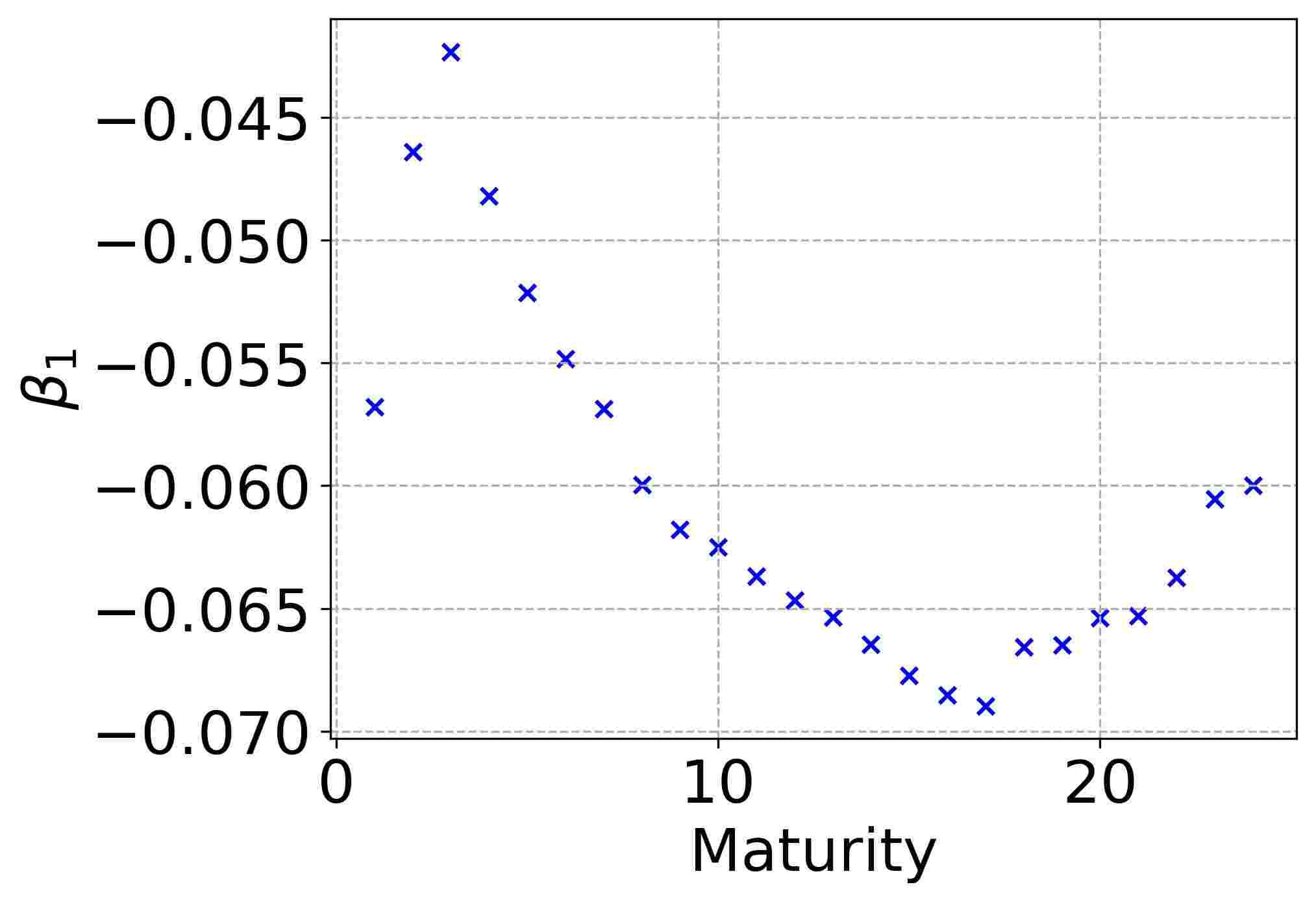}}
    \subfigure[$\beta_2$ (S\&P 500)]{\includegraphics[width=0.3\linewidth]{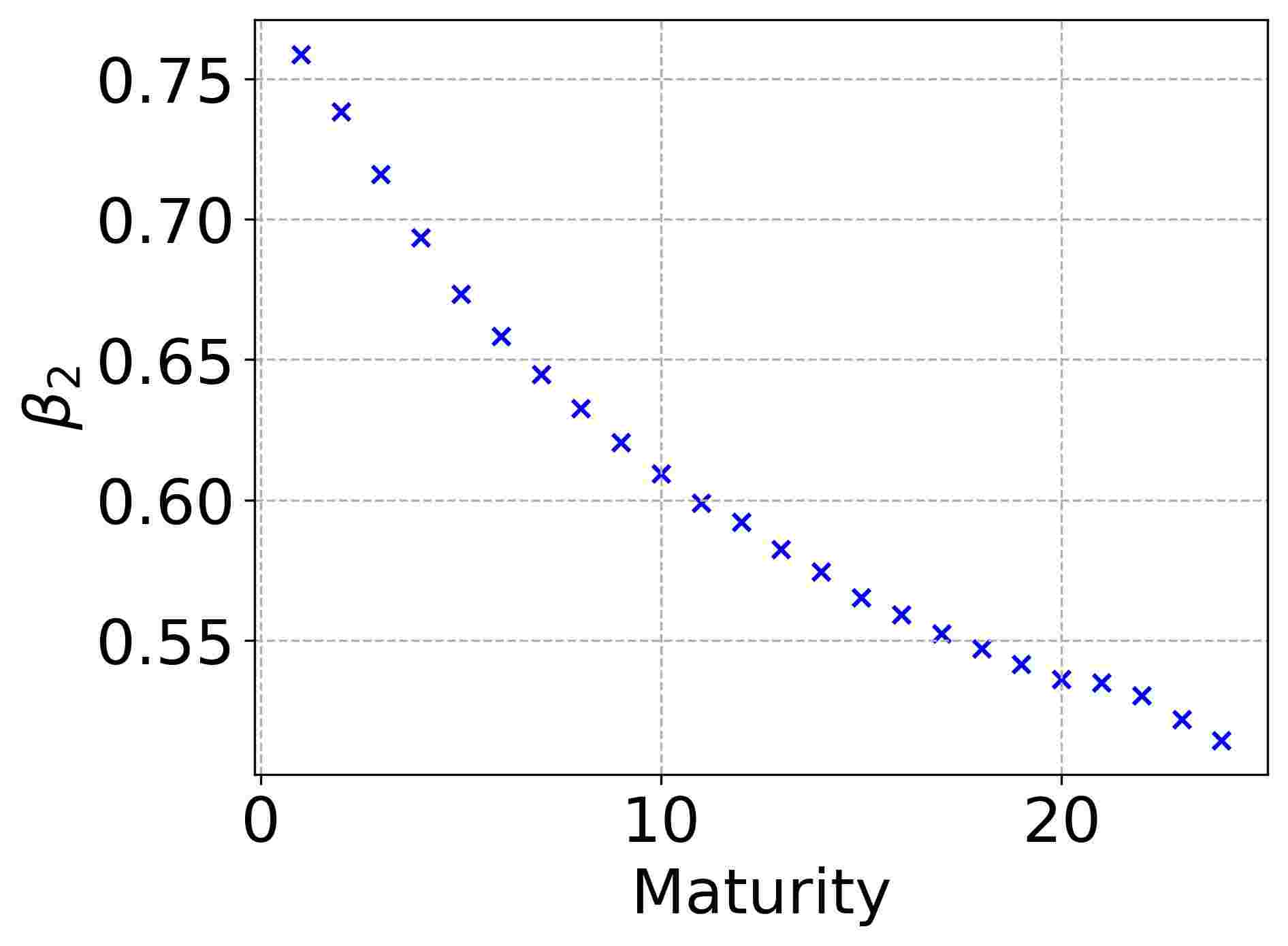}}
    \subfigure[$\beta_0$ (Euro Stoxx 50)]{ \includegraphics[width=0.3\linewidth]{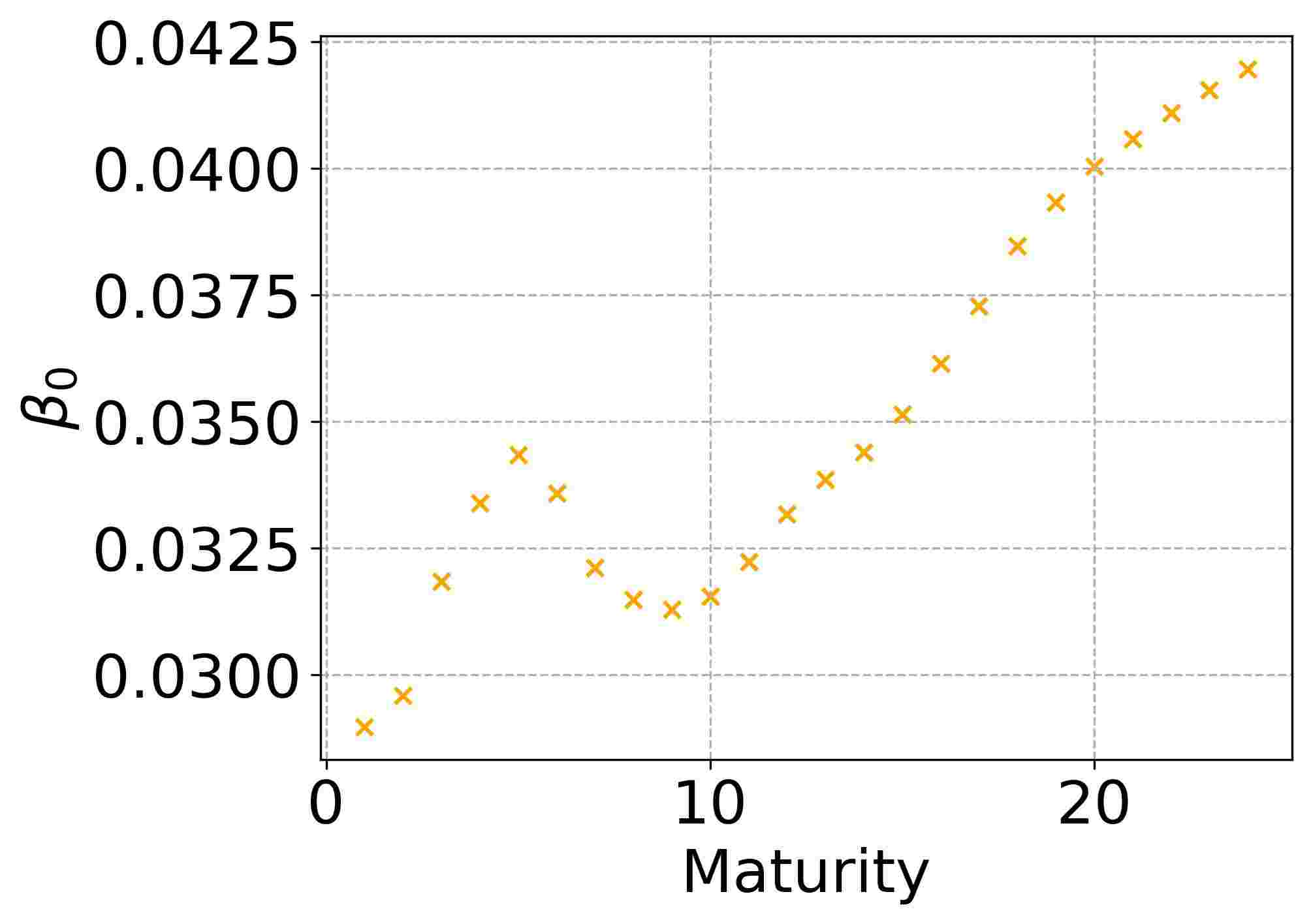}}
    \subfigure[$\beta_1$ (Euro Stoxx 50)]{\includegraphics[width=0.3\linewidth]{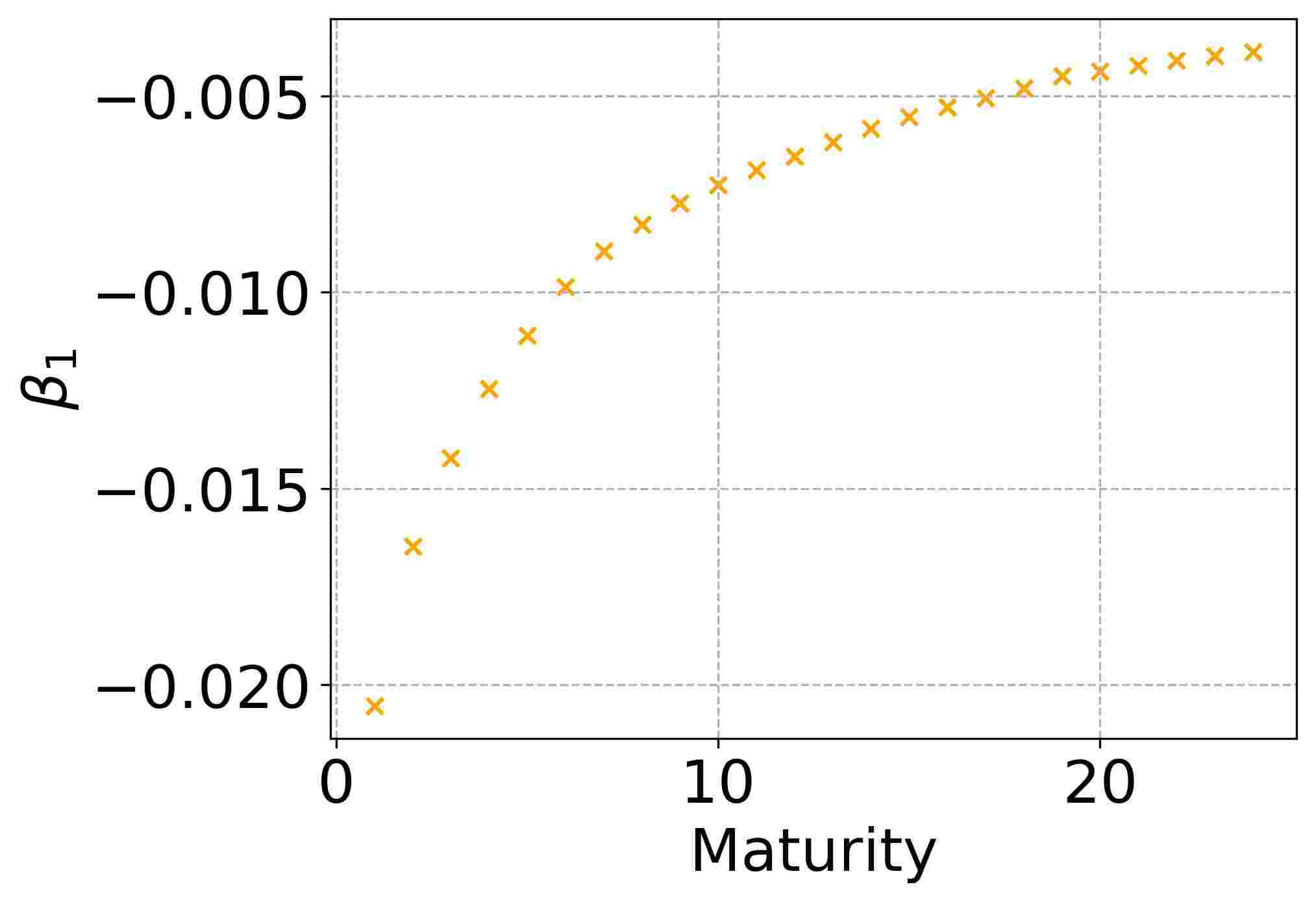}}
    \subfigure[$\beta_2$ (Euro Stoxx 50)]{ \includegraphics[width=0.3\linewidth]{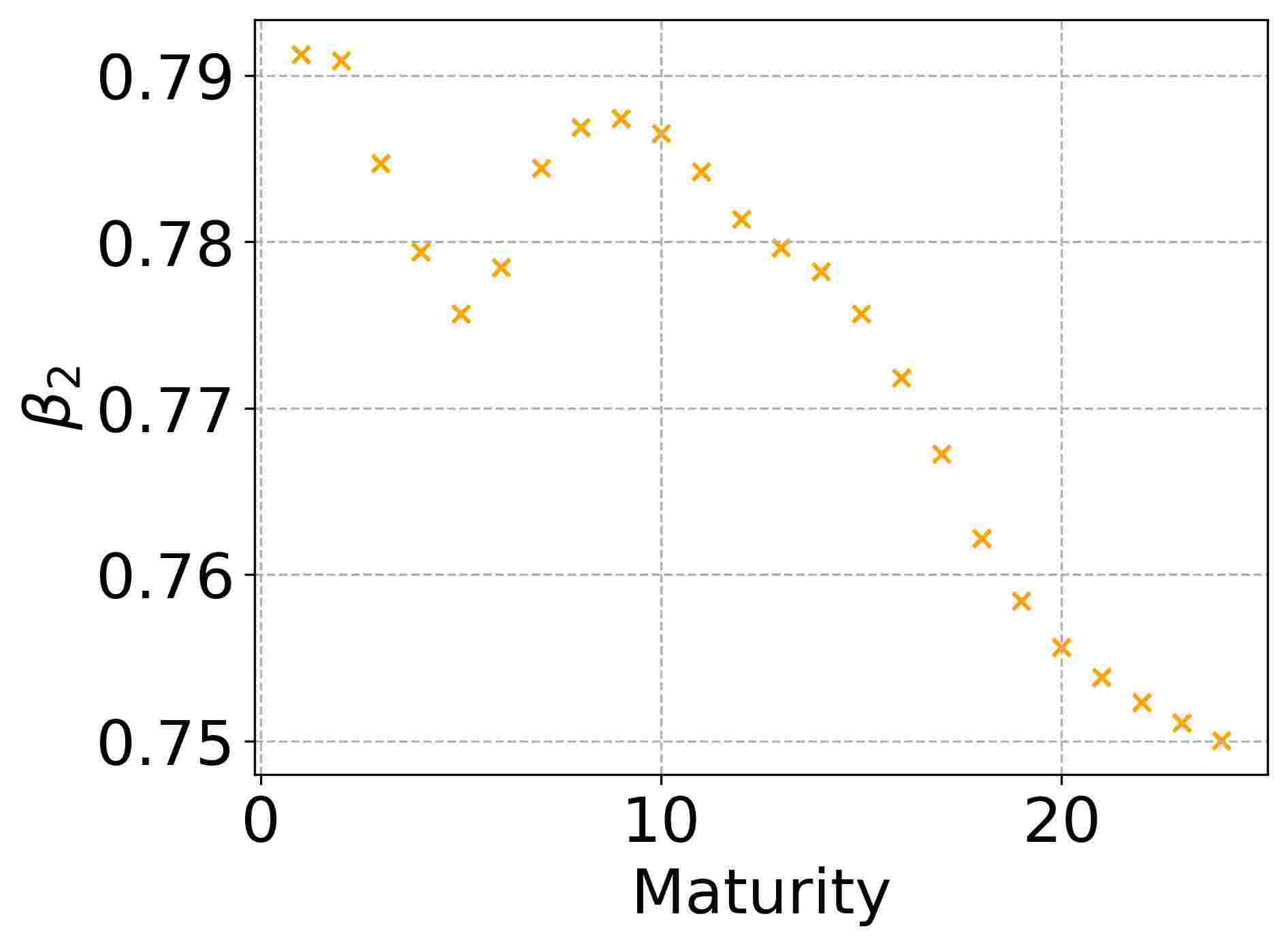}}
    \caption{Calibrated $\beta_0$, $\beta_1$ and $\beta_2$ as a function of the time-to-maturity. The hyperparameters for the S\&P 500 are $(C_{R_1},C_{\Sigma},\lambda)=(250,1000,10^{-3})$ and those of the Euro Stoxx 50 are $(C_{R_1},C_{\Sigma},\lambda)=(10,1000,10^{-3})$. }
\label{fig:pdv_params_lin_reg}
\end{figure}

In Figures \ref{fig:pdv_params_tspl} and \ref{fig:pdv_params_lin_reg}, we plot the evolution of the calibrated parameters associated to the $R^2$ scores presented in Figure \ref{fig:r2_cross_valid} as a function of the time-to-maturity. Regarding the TSPL kernels parameters ($\alpha_1$, $\delta_1$, $\alpha_2$, $\delta_2$), we observe overall a decreasing trend except for $\delta_2$ for which there is no clear trend. This decreasing trend for $\alpha_1$ and $\alpha_2$ indicates that far away past returns explain more and more the ATM implied volatility movements as the time-to-maturity increases. Note that, for the largest time-to-maturities, we obtain values of $\alpha$ that become even lower than 1 (except for $\alpha_1$ for the S\&P 500) which is the critical value below which the integral of the TSPL kernel diverges in continuous time. The decreasing behavior of $\delta_1$ indicates that, as $\alpha_1$ decreases, it still matters to keep a large weight for the more recent returns. Let us now end the study with the analysis of the parameters $\beta_0$, $\beta_1$ and $\beta_2$. First, we notice that we have $\beta_1<0$ and $\beta_2>0$ without imposing any constraint on these parameters. Therefore, a positive (resp. negative) trend in the underlying asset price tends to be followed by a decrease (resp. increase) of the implied volatility (which is consistent with the negative correlation observed by \cite{cont2002dynamics}) while the increase (resp. decrease) of the underlying asset volatility (measured by the squared returns) tends to be followed by an increase (resp. decrease) of the implied volatility. Moreover, the three parameters for the small time-to-maturities are of the same order of magnitude as to those calibrated by \cite{guyon2022volatility}. Regarding the evolution of $\beta_0$, we obtain an overall increase with the time-to-maturity which reflects the fact that, in average, the level of ATM implied volatility increases with the time-to-maturity. The parameter $\beta_1$, which can be interpreted as the influence of the trend feature on the implied volatility, has opposite evolutions for the S\&P 500 and the Euro Stoxx 50 (it is essentially decreasing for the former and increasing for the latter). Thus, it is difficult to draw any conclusion. Finally, the parameter $\beta_2$, which can be interpreted as the influence of the volatility feature on the implied volatility, is mainly decreasing with the time-to-maturity, so it seems that the implied volatility for long time-to-maturities becomes less reactive to the volatility of the underlying index. \\

\section{Absence of static arbitrage in the SSVI parameterization}\label{sec:ssvi_arbitrage}

Gatheral and Jacquier provide sufficient conditions for the SSVI to be free of arbitrage. These conditions are presented in the following theorem. 
\begin{theorem}[Corollary 4.1 from \cite{gatheral2014arbitrage}]
The SSVI is free of static arbitrage if the following conditions are satisfied:
\begin{enumerate}[(i)]
    \item $\partial_T \theta_T \ge 0$ for all $T>0$;
    \item $0\le \partial_{\theta}(\theta \varphi(\theta)) \le \frac{1}{\rho^2}\left(1+\sqrt{1-\rho^2} \right)\varphi(\theta)$ for all $\theta >0$;
    \item $\theta \varphi(\theta)(1+|\rho|)< 4$ for all $\theta >0$;
    \item $\theta\varphi(\theta)^2(1+|\rho|) \le 4$ for all $\theta >0$.
\end{enumerate}
\label{thm:ssvi_arbitrage}
\end{theorem}
\begin{remark}
    The conditions (i) and (ii) actually are necessary and sufficient conditions for the absence of calendar spread arbitrage for the SSVI. The condition (iii) with a non-strict inequality is a necessary condition for the absence of butterfly arbitrage but condition (iv) is only a necessary condition if $\theta \varphi(\theta)(1+|\rho|) = 4$. 
\end{remark}
\begin{remark}\label{rk:ssvi_weak_conditions}
    The conditions (ii), (iii) and (iv) can be weakened as follows:
    \begin{enumerate}[(i)]
        \setcounter{enumi}{1}
        \item $0\le \partial_{\theta}(\theta \varphi(\theta))\rvert_{\theta=\theta_T} \le \frac{1}{\rho^2}\left(1+\sqrt{1-\rho^2} \right)\varphi(\theta_T)$ for all $T >0$;
        \item $\theta_T \varphi(\theta_T)(1+|\rho|)< 4$ for all $T >0$;
        \item $\theta_T\varphi(\theta_T)^2(1+|\rho|) \le 4$ for all $T >0$.
    \end{enumerate}
    Note that these conditions are not necessarily equivalent to the ones in Theorem \ref{thm:ssvi_arbitrage} since $T\mapsto \theta_T$ is not necessarily a bijection from $\mathbb{R}_+^*$ to $\mathbb{R}_+^*$. 
\end{remark}
 The following propositions translate the sufficient conditions of Theorem \ref{thm:ssvi_arbitrage} for the three parameterizations of $\varphi$ introduced in Section \ref{sec:ssvi}. The proofs of Propositions \ref{prop:heston} and \ref{prop:power_law} can be found in Section 4 of \cite{gatheral2014arbitrage} while the one of Proposition \ref{prop:modified_power_law} is provided below. 
\begin{proposition}\label{prop:heston}
The SSVI-HL is free of static arbitrage if $\partial_T \theta_T \ge 0$ for all $T>0$ and $\lambda \ge (1+|\rho|)/4$. 
\end{proposition}
\begin{proposition}\label{prop:power_law}
    Assuming that $\partial_T \theta_T \ge 0$, we have the following cases:
    \begin{enumerate}[(i)]
        \item If $\gamma \in (0,1/2)$, there exists $\theta^* >0$ such that the SSVI-PL is free of static arbitrage if $\theta_T < \theta^*$ for all $T>0$.
        \item If $\gamma \in (1/2,1)$, there exists $\theta_1^*,\theta_2^*>0$ such that the SSVI-PL is free of static arbitrage if $\theta_2^* < \theta_T < \theta_1^*$ for all $T>0$.
        \item If $\gamma = 1/2$ and $\eta^2(1+|\rho|)\le 4$, there exists $\theta_1^*$ such that the SSVI-PL is free of static arbitrage if $\theta_T < \theta_1^*$ for all $T>0$. 
    \end{enumerate}
\end{proposition}
\begin{proposition}\label{prop:modified_power_law}
    Assuming that $\partial_T \theta_T \ge 0$, we have the following cases:
\begin{enumerate}[(i)]
    \item If $\gamma \in (0,1/2)$ and $\eta (1+|\rho|) \le 4$, there exists $\theta^*>0$ such that the SSVI-MPL is free of static arbitrage if, for all $T>0$, $\theta_T\ge \theta^*$.
    \item If $\gamma \in (1/2,1)$, the SSVI-MPL is free of static arbitrage for $\eta(1+|\rho|) \le 4$ and $(1-2\gamma)\varphi(1-2\gamma)^2(1+|\rho|)\le 4$. 
    \item If $\gamma=1/2$, the SSVI-MPL is free of static arbitrage for $\eta^2 (1+|\rho|) \le 4$.\label{prop:pssvi_no_arbitrage}
\end{enumerate}
\begin{proof}
    We have $\partial_{\theta}(\theta\varphi(\theta)) = \frac{1-\gamma}{1+\theta}\varphi(\theta)$, thus $0< \partial_{\theta}(\theta\varphi(\theta)) < \varphi(\theta)$ and condition (ii) of Theorem \ref{thm:ssvi_arbitrage} is satisfied since $\frac{1}{\rho^2}\left(1+\sqrt{1-\rho^2} \right) \ge 1$ for $\rho \in (-1,1)$. This shows also that $\theta \mapsto \theta\varphi(\theta)$ is strictly increasing. Since $\lim_{\theta \to +\infty} \theta\varphi(\theta) = \eta$, we deduce that condition (iii) is equivalent to $\eta(1+|\rho|)\le 4$. Finally, for condition (iv), we have:
\begin{equation}
    \partial_{\theta}(\theta\varphi(\theta)^2) = \frac{\eta^2}{\theta^{2\gamma}(1+\theta)^{3-2\gamma}}(1-\theta-2\gamma).
\end{equation}
Therefore, we have the following cases:
\begin{itemize}
    \item If $\gamma \in (1/2,1)$, then $\theta\mapsto \theta\varphi(\theta)^2$ is stricly decreasing on $\mathbb{R}_+$ with $\lim_{\theta \to 0} \theta\varphi(\theta)^2 = +\infty$ and $\lim_{\theta \to +\infty}\theta\varphi(\theta)^2 = 0$. Thus according to Remark \ref{rk:ssvi_weak_conditions}, if $\eta(1+|\rho|)\le 4$ then the SSVI is free of static arbitrage if $\theta_T \ge \theta^*$ for all $T>0$ where $\theta^*$ satisfies $\theta^*\varphi(\theta^*)^2 = 4/(1+|\rho|)$. 
    \item If $\gamma \in (0,1/2)$, then $\theta\mapsto \theta\varphi(\theta)^2$ is stricly increasing on $(0,1-2\gamma)$ and then stricly decreasing on $(1-2\gamma,+\infty)$, thus it is bounded from above by $(1-2\gamma)\varphi(1-2\gamma)^2$. We deduce that the SSVI is free of static arbitrage for $\eta(1+|\rho|)\le 4$ and $(1-2\gamma)\varphi(1-2\gamma)^2(1+|\rho|)\le 4$.
    \item If $\gamma = 1/2$, then $\theta\mapsto \theta\varphi(\theta)^2$ is stricly decreasing on $\mathbb{R}_+$ and bounded from above by $\eta^2$. We deduce that the SSVI is free of static arbitrage for $\eta(1+|\rho|)\le 4$ and $\eta^2(1+|\rho|) \le 4$ which is equivalent to $\eta^2(1+|\rho|) \le 4$ since for $\eta < 1$, we have $\eta(1+|\rho|)\le 4$ for all $\rho \in [-1,1]$. 
\end{itemize}
\end{proof}
\end{proposition}
\begin{remark}
Note that different sufficient conditions could be found for guaranteeing the absence of static arbitrage by restricting the time-to-maturity $T$ to some subset of $\mathbb{R}_+^*$. Since it is not very satisfying to have an IVS parameterization that could be arbitrable for some time-to-maturities, we looked as much as possible for conditions that do not restrict the values of $T$. 
\end{remark}

\section{SSVI calibration results}\label{sec:ssvi_calibration_results}
In this section, we study the fitting accuracy of the SSVI parameterization for the three choices of the function $\varphi$ introduced in Section \ref{sec:ssvi} on the historical IVSs that we presented in Section \ref{sec:data}. Let us recall that our historical IVSs are already the result of an interpolation of raw data as pointed out in Section \ref{sec:data}. Therefore, the calibration results that we present in this section may differ from those obtained using raw data or data interpolated with a method different from that of our data providers. Moreover, as already noted by \cite{cont2022simulation}, our historical IVSs are not necessarily arbitrage-free because of these interpolations. Procedures such as the ones of \cite{davis2007range} or \cite{cohen2020repairing} allow to detect arbitrages in a finite set of prices of European call options given the forward prices. Our database does not contain short rates data or forward prices data so we cannot use these procedures. Since calendar spread arbitrages are equivalent to the total implied variance being non-decreasing in time-to-maturity, we remove the IVSs (we recall that our data sets contain one IVS per business day) such that there is at least one crossing between the linearly interpolated total implied variances of two adjacent time-to-maturities. IVSs with butterfly arbitrages (if any) are not removed as the condition in terms of total implied variance is much more complicated to verify. In total, 0.8\% (resp. 5.6\%) of the IVSs are removed from the S\&P 500 (resp. Euro Stoxx 50) data set.  \\

To measure the fitting accuracy, we calibrate the SSVI for each day of our data sets without calendar spread arbitrages and for each parametric form of $\varphi$ by solving the following minimization problem using the \texttt{minimize} function with the SLSQP algorithm from the \texttt{scipy} Python package:
\begin{equation}\label{eq:ssvi_calibration}
    \begin{array}{rrrcllr}
    \displaystyle\min_{\Theta=((\theta_{T_i})_{i=1,\dots,M},\rho,\Pi_{\varphi})} & \multicolumn{5}{c}{\displaystyle\sum_{i=1}^M\sum_{k\in\mathcal{K}_{T_i}} \phi(k) \left(\sigma_{Mkt}(k,T_i) - \sigma_{SSVI}(k,T_i;\Theta)\right)^2 } \\
        \text{s.t.} & \theta_{T_1} & \ge & 0 & & \\
    &\theta_{T_{i+1}} &\ge &\theta_{T_i} & \text{for } 1\le i \le M-1 &  \\
    & (\rho,\Pi_{\varphi}) & \in & C_{\varphi}. &&
        \end{array}
\end{equation}

\begin{figure}[h]
    \centering
    \subfigure[S\&P 500]{\includegraphics[width=0.45\linewidth]{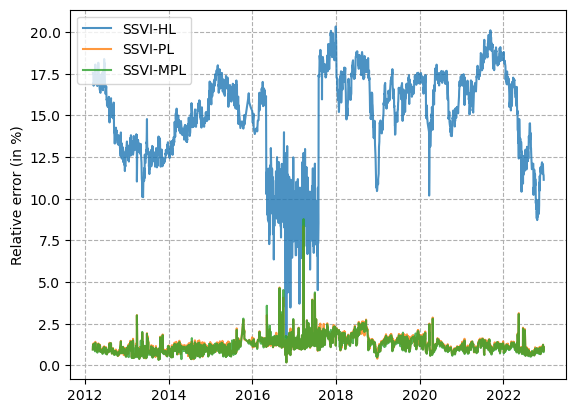}}
    \subfigure[Euro Stoxx 50]{\includegraphics[width=0.45\linewidth]{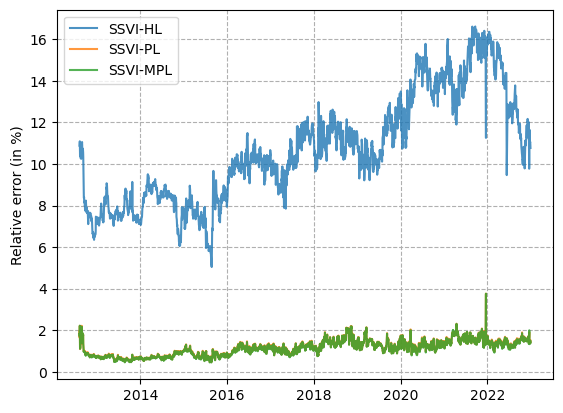}}
    \caption{Average relative errors between the market implied volatilities and the SSVI implied volatilities after the calibration.}
\label{fig:ssvi_relative_error}
\end{figure}

We used the following notations:
\begin{itemize}
    \item $\Pi_{\varphi}$ is the vector of the parameters in $\varphi$: only $\lambda$ for the SSVI-HL and $(\eta,\gamma)$ for the SSVI-PL and the SSVI-MPL. 
    \item $T_1<T_2<\dots < T_M$ is the set of time-to-maturities and $\mathcal{K}_T$ is the set of log-strikes for the time-to-maturity $T$. Note that for the S\&P 500 data set, we have the implied volatilities for forward moneynesses between 0.6 and 1.4. In order to keep only the most liquid options and for consistency with the Euro Stoxx 50 data set, we only consider the forward moneynesses for which the (call) Black-Scholes delta is between 0.1 and 0.9.
    \item $\phi(k)$ is the standard normal density function evaluated in the log-strike $k$. This weighting function allows to give more weight to the replication of the implied volatilities that are close to the money. Another choice based on the Black-Scholes vega has been considered but we did not notice any improvement.
    \item $\sigma_{Mkt}$ is the market implied volatility. 
    \item  $\sigma_{SSVI}(\cdot,\cdot;\Theta)$ is the SSVI implied volatility associated to the parameter vector $\Theta=((\theta_{T_i})_{i=1,\dots,M},\rho,\Pi_{\varphi})$.
    \item $C_{\varphi}$ is given by: 
    \begin{itemize}
        \item[--]  $C_{\varphi} = \{ (\rho,\lambda) \in (-1,1)\times \mathbb{R}_+^* : \lambda \ge (1+|\rho|)/4 \}$ for the SSVI-HL,
        \item[--]  $C_{\varphi} = \{(\rho,\eta,\gamma) \in (-1,1)\times \mathbb{R}_+^*\times (0,1) : \eta^2(1+|\rho|) \le 4 \text{ if } \gamma = 1/2 \}$ for the SSVI-PL and,
        \item[--] \begin{equation*}
            C_{\varphi} = \left\{
            \begin{array}{ll}
                \multicolumn{2}{l}{(\rho,\eta,\gamma)\in(-1,1)\times \mathbb{R}_+^*\times (0,1) :} \\
                \eta (1+|\rho|) \le 4 & \text{ if } \gamma < 1/2 \\
                \eta (1+|\rho|) \le 4 \text{ and } (1-2\gamma)\varphi(1-2\gamma)^2(1+|\rho|)\le 4 & \text{ if } \gamma >1/2 \\
                \eta^2(1+|\rho|) \le 4 & \text{ if } \gamma = 1/2 
            \end{array}
            \right\}
        \end{equation*}
        for the SSVI-MPL. 
    \end{itemize}
\end{itemize}
\begin{remark}
The constraints in the above minimization problem do not guarantee the absence of static arbitrage for the power-law and the modified power-law parameterizations since only the constraints involving $\rho$, $\eta$ and $\gamma$ are included as well as the non-decreasing property of $T\mapsto \theta_T$. The reason for this choice is the fact that there is no closed-form expression for $\theta^*$, $\theta_1^*$ and $\theta_2^*$ so the addition of the constraints involving these terms would strongly complexify the numerical optimization. By the end of this section, the study will be restricted to the SSVI-MPL parameterization with $\gamma=1/2$ (for which there is no constraint involving $\theta^*$, $\theta_1^*$ or $\theta_2^*$) so this choice is impact-free. 
\end{remark}
\begin{remark}
We use in practice the change of variable $(\tilde{\theta}_T)_{i=1,\dots,M}$ where $\tilde{\theta}_{T_1}=\theta_{T_1}$ and $\tilde{\theta}_{T_{i}} = \theta_{T_i}-\theta_{T_{i-1}}$ for $i \in \{2,\dots,M\}$ since it allows to transform the second inequality constraints in the minimization problem (\ref{eq:ssvi_calibration}) in bounds constraints: $\tilde{\theta}_{T_i} \ge 0$ for $2\le i \le M$. 
\end{remark}
The initial guess for each parameter is provided in Table \ref{tab:initial_guess}. The average relative errors (without weighting) between the market implied volatilities and the SSVI implied volatilities for each day of our data sets are presented in Figure \ref{fig:ssvi_relative_error}. It appears clearly that the Heston-like parameterization performs very poorly in comparison to the two others parameterizations. This results from the fact that, in the Heston-like parameterization, the function $\varphi$ is bounded from above by $1/2$ which constraints strongly the shape of the IVS. The power-law and the modified power-law parameterizations achieve essentially the same accuracy which is very stable over time. In particular, we do not observe a decrease of the fitting quality during the Covid-19 crisis.
\begin{table}[h]
    \centering
    \caption{Initial guesses provided to the optimization routine. }
    \label{tab:initial_guess}
    \begin{tabular}{@{}ccccc@{}}
    \toprule
    $(\theta_{T_i})_{i=1,\dots,M}$ & $\rho$ & $\lambda$ & $\eta$ & $\gamma$ \\ \midrule
    $(\sigma_{Mkt}(0,T_i)^2T_i)_{i=1,\dots,M}$ & -0.5 & 1      & 1   & 0.25   \\ \bottomrule
    \end{tabular}
\end{table}

\section{Hyperparameters of the PDV model calibrated on the parsimonious SSVI parameters}\label{sec:hyperparams_ssvi}
Similarly to the study in Appendix \ref{sec:influence_cutoff}, for each parameter of the parsimonious SSVI, we run a 10-fold blocked cross-validation on the train set to determine the optimal hyperparameters $C_{R_1}$, $C_{\Sigma}$ and $\lambda$ in the grid $\{5,10,25,50,100,250,500,1000,1500,2000,2500\}^2\times \{10^{-6},10^{-5},\dots,10^{-1}\}$. The obtained hyperparameters are given in Table \ref{tab:pssvi_hyperparameters}
\begin{table}[h]
    \centering
    \caption{Cut-off lags and penalizations resulting from the 10-fold cross-validation which are used to obtain the $R^2$ scores presented in Table \ref{tab:ssvi_r2_scores}}
    \label{tab:pssvi_hyperparameters}
    \begin{tabular}{@{}ccccccc@{}}
    \toprule
    & \multicolumn{3}{c}{S\&P 500} & \multicolumn{3}{c}{Euro Stoxx 50} \\ \toprule
        & $C_{R_1}$ & $C_{\Sigma}$ & $\lambda$ & $C_{R_1}$ & $C_{\Sigma}$ & $\lambda$   \\ \midrule
    $a$   &     1500 &  1000 & $10^{-4}$            & 10 & 1500 & $10^{-3}$        \\
    $p$   &     50 & 100 & $10^{-1}$           & 250 & 50 & $10^{-1}$        \\
    $\rho$ &  500 & 100 & $10^{-5}$        &   10 & 25 & $10^{-3}$             \\
    $\eta$ &    1500 & 1500 & $10^{-2}$ &    1000 & 1500 & $10^{-3}$             \\ \bottomrule
\end{tabular}
\end{table}

%% file: main.bbl
\begin{thebibliography}{56}
\providecommand{\natexlab}[1]{#1}
\providecommand{\url}[1]{\texttt{#1}}
\expandafter\ifx\csname urlstyle\endcsname\relax
  \providecommand{\doi}[1]{doi: #1}\else
  \providecommand{\doi}{doi: \begingroup \urlstyle{rm}\Url}\fi

\bibitem[Amrani et~al.(2021)Amrani, Jacquier, and Martini]{amrani2021dynamics}
M.~E. Amrani, A.~Jacquier, and C.~Martini.
\newblock Dynamics of symmetric {SSVI} smiles and implied volatility bubbles.
\newblock \emph{SIAM Journal on Financial Mathematics}, 12\penalty0
  (2):\penalty0 SC1--SC15, 2021.

\bibitem[Andr{\`e}s and Jourdain(2024)]{andres2024existence}
H.~Andr{\`e}s and B.~Jourdain.
\newblock Existence, uniqueness and positivity of solutions to the
  guyon-lekeufack path-dependent volatility model with general kernels.
\newblock \emph{arXiv preprint arXiv:2408.02477}, 2024.

\bibitem[Bakshi et~al.(2000)Bakshi, Cao, and Chen]{bakshi2000call}
G.~Bakshi, C.~Cao, and Z.~Chen.
\newblock Do call prices and the underlying stock always move in the same
  direction?
\newblock \emph{The Review of Financial Studies}, 13\penalty0 (3):\penalty0
  549--584, 2000.

\bibitem[Black and Scholes(1973)]{black1973pricing}
F.~Black and M.~Scholes.
\newblock The pricing of options and corporate liabilities.
\newblock \emph{J. Polit. Econ.}, 81\penalty0 (3):\penalty0 637--654, 1973.

\bibitem[Bloch and B{\"o}{\"o}k(2021)]{bloch2021deep}
D.~A. Bloch and A.~B{\"o}{\"o}k.
\newblock Deep learning based dynamic implied volatility surface.
\newblock \emph{Available at SSRN 3952842}, 2021.

\bibitem[Brace et~al.(1997)Brace, Gatarek, and Musiela]{brace1997market}
A.~Brace, D.~Gatarek, and M.~Musiela.
\newblock The market model of interest rate dynamics.
\newblock \emph{Math. Finance}, 7\penalty0 (2):\penalty0 127--155, 1997.

\bibitem[Brace et~al.(2001)Brace, Goldys, Klebaner, and
  Womersley]{brace2001market}
A.~Brace, B.~Goldys, F.~Klebaner, and R.~Womersley.
\newblock Market model of stochastic implied volatility with application to the
  bgm model.
\newblock \emph{Preprint, available at the page http://www. maths. unsw. edu.
  au/statistics/files/preprint-2001-01. pdf}, 2001.

\bibitem[Cao et~al.(2020)Cao, Chen, and Hull]{cao2020}
J.~Cao, J.~Chen, and J.~Hull.
\newblock A neural network approach to understanding implied volatility
  movements.
\newblock \emph{Quant. Finance}, 20\penalty0 (9):\penalty0 1405--1413, 2020.
\newblock ISSN 1469-7688,1469-7696.

\bibitem[Carmona and Nadtochiy(2011)]{carmon2011tangent}
R.~Carmona and S.~Nadtochiy.
\newblock Tangent models as a mathematical framework for dynamic calibration.
\newblock \emph{Int. J. Theor. Appl. Finance}, 14\penalty0 (1):\penalty0
  107--135, 2011.

\bibitem[Carmona et~al.(2017)Carmona, Ma, and Nadtochiy]{carmona2017levy}
R.~Carmona, Y.~Ma, and S.~Nadtochiy.
\newblock Simulation of implied volatility surfaces via tangent {L}\'{e}vy
  models.
\newblock \emph{SIAM J. Financial Math.}, 8\penalty0 (1):\penalty0 171--213,
  2017.

\bibitem[Carr and Madan(2001)]{carr2001formula}
P.~Carr and D.~Madan.
\newblock Towards a theory of volatility trading.
\newblock In \emph{Option pricing, interest rates and risk management}, Handb.
  Math. Finance, pages 458--476. Cambridge Univ. Press, Cambridge, 2001.

\bibitem[{CBOE}(2023)]{vixwp}
{CBOE}.
\newblock Volatility index methodology: Cboe volatility index.
\newblock
  \url{https://cdn.cboe.com/api/global/us_indices/governance/Volatility_Index_Methodology_Cboe_Volatility_Index.pdf},
  2023.
\newblock Accessed on: 2023-07-21.

\bibitem[Cerqueira et~al.(2020)Cerqueira, Torgo, and
  Mozeti{\v{c}}]{cerqueira2020evaluating}
V.~Cerqueira, L.~Torgo, and I.~Mozeti{\v{c}}.
\newblock Evaluating time series forecasting models: An empirical study on
  performance estimation methods.
\newblock \emph{Machine Learning}, 109:\penalty0 1997--2028, 2020.

\bibitem[Choudhary et~al.(2024)Choudhary, Jaimungal, and
  Bergeron]{choudhary2024}
V.~Choudhary, S.~Jaimungal, and M.~Bergeron.
\newblock Funvol: multi-asset implied volatility market simulator using
  functional principal components and neural sdes.
\newblock \emph{Quantitative Finance}, 24\penalty0 (8):\penalty0 1077--1103,
  2024.

\bibitem[Cohen et~al.(2020)Cohen, Reisinger, and Wang]{cohen2020repairing}
S.~N. Cohen, C.~Reisinger, and S.~Wang.
\newblock Detecting and repairing arbitrage in traded option prices.
\newblock \emph{Appl. Math. Finance}, 27\penalty0 (5):\penalty0 345--373, 2020.
\newblock ISSN 1350-486X,1466-4313.

\bibitem[Cohen et~al.(2021)Cohen, Reisinger, and Wang]{cohen2021arbitrage}
S.~N. Cohen, C.~Reisinger, and S.~Wang.
\newblock Arbitrage-free neural-sde market models.
\newblock \emph{arXiv preprint arXiv:2105.11053}, 2021.

\bibitem[Cont and da~Fonseca(2002)]{cont2002dynamics}
R.~Cont and J.~da~Fonseca.
\newblock Dynamics of implied volatility surfaces.
\newblock volume~2, pages 45--60. 2002.
\newblock Special issue on volatility modelling.

\bibitem[Cont and Vuleti{\'c}(2023)]{cont2022simulation}
R.~Cont and M.~Vuleti{\'c}.
\newblock Simulation of arbitrage-free implied volatility surfaces.
\newblock \emph{Applied Mathematical Finance}, pages 1--28, 2023.

\bibitem[Cont et~al.(2002)Cont, da~Fonseca, and Durrleman]{cont2002stochastic}
R.~Cont, J.~da~Fonseca, and V.~Durrleman.
\newblock Stochastic models of implied volatility surfaces.
\newblock \emph{Economic Notes}, 31\penalty0 (2):\penalty0 361--377, 2002.

\bibitem[Corbetta et~al.(2019)Corbetta, Cohort, Laachir, and
  Martini]{corbetta2019robust}
J.~Corbetta, P.~Cohort, I.~Laachir, and C.~Martini.
\newblock Robust calibration and arbitrage-free interpolation of {SSVI} slices.
\newblock \emph{Decisions in Economics and Finance}, 42\penalty0 (2):\penalty0
  665--677, 2019.

\bibitem[Davis and Hobson(2007)]{davis2007range}
M.~H.~A. Davis and D.~G. Hobson.
\newblock The range of traded option prices.
\newblock \emph{Math. Finance}, 17\penalty0 (1):\penalty0 1--14, 2007.

\bibitem[Delemotte et~al.(2023)Delemotte, Marco, and Segonne]{delemotte2023yet}
J.~Delemotte, S.~D. Marco, and F.~Segonne.
\newblock Yet another analysis of the {SP}500 at-the-money skew: Crossover of
  different power-law behaviours.
\newblock \emph{Available at SSRN 4428407}, 2023.

\bibitem[Fengler et~al.(2003)Fengler, H{\"a}rdle, and
  Villa]{fengler2003dynamics}
M.~R. Fengler, W.~K. H{\"a}rdle, and C.~Villa.
\newblock The dynamics of implied volatilities: A common principal components
  approach.
\newblock \emph{Review of Derivatives Research}, 6:\penalty0 179--202, 2003.

\bibitem[Fengler et~al.(2007)Fengler, H{\"a}rdle, and
  Mammen]{fengler2007semiparametric}
M.~R. Fengler, W.~K. H{\"a}rdle, and E.~Mammen.
\newblock A semiparametric factor model for implied volatility surface
  dynamics.
\newblock \emph{Journal of Financial Econometrics}, 5\penalty0 (2):\penalty0
  189--218, 2007.

\bibitem[Fisher(1915)]{fisher1915frequency}
R.~A. Fisher.
\newblock Frequency distribution of the values of the correlation coefficient
  in samples from an indefinitely large population.
\newblock \emph{Biometrika}, 10\penalty0 (4):\penalty0 507--521, 1915.

\bibitem[François et~al.(2023)François, Galarneau-Vincent, Gauthier, and
  Godin]{francois2023joint}
P.~François, R.~Galarneau-Vincent, G.~Gauthier, and F.~Godin.
\newblock Joint dynamics for the underlying asset and its implied volatility
  surface: a new methodology for option risk management.
\newblock \emph{Available at SSRN 4319972}, 2023.

\bibitem[Gatheral(2004)]{gatheral2004parsimonious}
J.~Gatheral.
\newblock A parsimonious arbitrage-free implied volatility parameterization
  with application to the valuation of volatility derivatives.
\newblock \emph{Presentation at Global Derivatives \& Risk Management, Madrid},
  page~0, 2004.

\bibitem[Gatheral and Jacquier(2011)]{gatheral2011convergence}
J.~Gatheral and A.~Jacquier.
\newblock Convergence of {H}eston to {SVI}.
\newblock \emph{Quant. Finance}, 11\penalty0 (8):\penalty0 1129--1132, 2011.

\bibitem[Gatheral and Jacquier(2014)]{gatheral2014arbitrage}
J.~Gatheral and A.~Jacquier.
\newblock Arbitrage-free svi volatility surfaces.
\newblock \emph{Quantitative Finance}, 14\penalty0 (1):\penalty0 59--71, 2014.

\bibitem[Gatheral et~al.(2018)Gatheral, Jaisson, and
  Rosenbaum]{gatheral2018volatility}
J.~Gatheral, T.~Jaisson, and M.~Rosenbaum.
\newblock Volatility is rough.
\newblock \emph{Quantitative finance}, 18\penalty0 (6):\penalty0 933--949,
  2018.

\bibitem[Gazzani and Guyon(2024)]{gazzani2024pricing}
G.~Gazzani and J.~Guyon.
\newblock Pricing and calibration in the 4-factor path-dependent volatility
  model.
\newblock \emph{arXiv preprint arXiv:2406.02319}, 2024.

\bibitem[Guyon and El~Amrani(2023)]{guyon2023does}
J.~Guyon and M.~El~Amrani.
\newblock Does the term-structure of equity at-the-money skew really follow a
  power law?
\newblock \emph{Risk Magazine}, 2023.

\bibitem[Guyon and Lekeufack(2023)]{guyon2022volatility}
J.~Guyon and J.~Lekeufack.
\newblock Volatility is (mostly) path-dependent.
\newblock \emph{Quant. Finance}, 23\penalty0 (9):\penalty0 1221--1258, 2023.
\newblock ISSN 1469-7688,1469-7696.

\bibitem[Hafner and Schmid(2005)]{hafner2005factor}
R.~Hafner and B.~Schmid.
\newblock A factor-based stochastic implied volatility model.
\newblock \emph{Technical report}, 2005.

\bibitem[H{\"a}rdle and Mungo(2007)]{hardle2007long}
W.~K. H{\"a}rdle and J.~Mungo.
\newblock Long memory persistence in the factor of implied volatility dynamics.
\newblock \emph{Available at SSRN 2894358}, 2007.

\bibitem[Heath et~al.(1992)Heath, Jarrow, and Morton]{heath1992bond}
D.~Heath, R.~Jarrow, and A.~Morton.
\newblock Bond pricing and the term structure of interest rates: A new
  methodology for contingent claims valuation.
\newblock \emph{Econometrica: Journal of the Econometric Society}, pages
  77--105, 1992.

\bibitem[Heber et~al.(2009)Heber, Lunde, Shephard, and
  Sheppard]{heber2009oxford}
G.~Heber, A.~Lunde, N.~Shephard, and K.~Sheppard.
\newblock Oxford-man institute’s realized library.
\newblock \emph{Version 0.1, Oxford\&Man Institute, University of Oxford},
  2009.

\bibitem[Hendriks and Martini(2017)]{hendriks2017extended}
S.~Hendriks and C.~Martini.
\newblock The extended {SSVI} volatility surface.
\newblock \emph{Available at SSRN 2971502}, 2017.

\bibitem[Jacod and Protter(2010)]{jacod2010risk}
J.~Jacod and P.~Protter.
\newblock Risk-neutral compatibility with option prices.
\newblock \emph{Finance and Stochastics}, 14\penalty0 (2):\penalty0 285--315,
  2010.

\bibitem[Ledoit and Santa-Clara(1998)]{ledoit1998relative}
O.~Ledoit and P.~Santa-Clara.
\newblock Relative pricing of options with stochastic volatility.
\newblock \emph{University of California-Los Angeles finance working paper},
  pages 9--98, 1998.

\bibitem[Lee(2004)]{lee2004moment}
R.~W. Lee.
\newblock The moment formula for implied volatility at extreme strikes.
\newblock \emph{Math. Finance}, 14\penalty0 (3):\penalty0 469--480, 2004.

\bibitem[Lord et~al.(2010)Lord, Koekkoek, and Van~Dijk]{lord2010}
R.~Lord, R.~Koekkoek, and D.~Van~Dijk.
\newblock A comparison of biased simulation schemes for stochastic volatility
  models.
\newblock \emph{Quant. Finance}, 10\penalty0 (2):\penalty0 177--194, 2010.
\newblock ISSN 1469-7688,1469-7696.

\bibitem[Martini and Mingone(2022)]{martini2022svi}
C.~Martini and A.~Mingone.
\newblock No arbitrage {SVI}.
\newblock \emph{SIAM J. Financial Math.}, 13\penalty0 (1):\penalty0 227--261,
  2022.

\bibitem[Martini and Mingone(2023)]{martini2023refined}
C.~Martini and A.~Mingone.
\newblock Refined analysis of the no-butterfly-arbitrage domain for {SSVI}
  slices.
\newblock \emph{Journal of Computational Finance}, 27\penalty0 (2), 2023.

\bibitem[Mingone(2022)]{mingone2022no}
A.~Mingone.
\newblock No arbitrage global parametrization for the e{SSVI} volatility
  surface.
\newblock \emph{Quantitative Finance}, 22\penalty0 (12):\penalty0 2205--2217,
  2022.

\bibitem[Morel et~al.(2024)Morel, Mallat, and Bouchaud]{morel2024path}
R.~Morel, S.~Mallat, and J.-P. Bouchaud.
\newblock Path shadowing monte carlo.
\newblock \emph{Quantitative Finance}, 24\penalty0 (9):\penalty0 1199--1225,
  2024.

\bibitem[Nutz and Riveros~Valdevenito(2024)]{nutz2024}
M.~Nutz and A.~Riveros~Valdevenito.
\newblock On the {G}uyon--{L}ekeufack volatility model.
\newblock \emph{Finance Stoch.}, 28\penalty0 (4):\penalty0 1203--1223, 2024.
\newblock ISSN 0949-2984,1432-1122.
\newblock \doi{10.1007/s00780-024-00544-2}.
\newblock URL \url{https://doi.org/10.1007/s00780-024-00544-2}.

\bibitem[Sch\"{o}nbucher(1999)]{schonbucher1999marketmodel}
P.~J. Sch\"{o}nbucher.
\newblock A market model for stochastic implied volatility.
\newblock \emph{R. Soc. Lond. Philos. Trans. Ser. A Math. Phys. Eng. Sci.},
  357\penalty0 (1758):\penalty0 2071--2092, 1999.

\bibitem[Schweizer and Wissel(2008)]{schweizer2008term}
M.~Schweizer and J.~Wissel.
\newblock Term structures of implied volatilities: Absence of arbitrage and
  existence results.
\newblock \emph{Mathematical Finance: An International Journal of Mathematics,
  Statistics and Financial Economics}, 18\penalty0 (1):\penalty0 77--114, 2008.

\bibitem[Skiadopoulos et~al.(2000)Skiadopoulos, Hodges, and
  Clewlow]{skiadopoulos2000dynamics}
G.~Skiadopoulos, S.~Hodges, and L.~Clewlow.
\newblock The dynamics of the {S}\&{P} 500 implied volatility surface.
\newblock \emph{Review of derivatives research}, 3:\penalty0 263--282, 2000.

\bibitem[{STOXX}(2023)]{vstoxxwp}
{STOXX}.
\newblock Stoxx strategy index guide.
\newblock
  \url{https://www.stoxx.com/document/Indices/Common/Indexguide/stoxx_strategy_guide.pdf},
  2023.
\newblock Accessed on: 2023-07-21.

\bibitem[Tang and Chen(2009)]{tang2009parameter}
C.~Y. Tang and S.~X. Chen.
\newblock Parameter estimation and bias correction for diffusion processes.
\newblock \emph{Journal of Econometrics}, 149\penalty0 (1):\penalty0 65--81,
  2009.

\bibitem[Wei et~al.(2016)Wei, Shu, and Liu]{wei2016}
C.~Wei, H.~Shu, and Y.~Liu.
\newblock Gaussian estimation for discretely observed {C}ox-{I}ngersoll-{R}oss
  model.
\newblock \emph{Int. J. Gen. Syst.}, 45\penalty0 (5):\penalty0 561--574, 2016.
\newblock ISSN 0308-1079,1563-5104.

\bibitem[Wen et~al.(2024)Wen, Zhai, Wang, and Cao]{wen2024implied}
C.~Wen, J.~Zhai, Y.~Wang, and Y.~Cao.
\newblock Implied volatility is (almost) past-dependent: Linear vs non-linear
  models.
\newblock \emph{International Review of Financial Analysis}, 95:\penalty0
  103406, 2024.

\bibitem[Wiese et~al.(2019)Wiese, Bai, Wood, and Buehler]{wiese2019deep}
M.~Wiese, L.~Bai, B.~Wood, and H.~Buehler.
\newblock Deep hedging: learning to simulate equity option markets.
\newblock \emph{arXiv preprint arXiv:1911.01700}, 2019.

\bibitem[Zhang et~al.(2023)Zhang, Li, and Zhang]{zhang2023}
W.~Zhang, L.~Li, and G.~Zhang.
\newblock A two-step framework for arbitrage-free prediction of the implied
  volatility surface.
\newblock \emph{Quant. Finance}, 23\penalty0 (1):\penalty0 21--34, 2023.
\newblock ISSN 1469-7688,1469-7696.

\end{thebibliography}
